\documentclass[conference]{IEEEtran}
\usepackage{cite}
\usepackage{amsmath,amssymb,amsfonts}
\usepackage{algorithmic}
\usepackage{graphicx}
\usepackage{textcomp}
\usepackage{xcolor}
\def\BibTeX{{\rm B\kern-.05em{\sc i\kern-.025em b}\kern-.08em
    T\kern-.1667em\lower.7ex\hbox{E}\kern-.125emX}}

\usepackage{booktabs}
\usepackage[ruled, vlined, linesnumbered]{algorithm2e}
\usepackage{caption}
\usepackage{subcaption}
\usepackage{algorithmic}
\usepackage{diagbox}
\usepackage{balance}
\usepackage{multirow}
\usepackage{epstopdf}
\usepackage{float}
\usepackage{bm}
\usepackage{url}
\usepackage{enumerate}
\usepackage{afterpage}
\usepackage{mathrsfs}
\usepackage{diagbox}
\usepackage{makecell}
\usepackage{microtype}
\usepackage{mdframed}
\usepackage{enumitem}
\usepackage{threeparttable}

\newenvironment{tipbox}{
  \begin{mdframed}[
    backgroundcolor=blue!10,
    linecolor=blue!50!black,
    innerleftmargin=5pt,
    innerrightmargin=5pt,
    innertopmargin=5pt,
    innerbottommargin=5pt,
    linewidth=0pt
  ]
}{\end{mdframed}}

\newtheorem{definition}{\bfseries Definition}

\newtheorem{example}{\bfseries Example}
	
\newlength{\oldtextfloatsep}
\newlength{\oldfloatsep}

\newcommand{\ie}{\emph{i.e.},\xspace}
\newcommand{\eg}{\emph{e.g.},\xspace}

\newcommand{\etal}{\emph{et al.}\xspace}
\newcommand\figref[1]{Fig.~\ref{#1}}

\newcommand\tabref[1]{Table~\ref{#1}}

\newcommand\secref[1]{Sec.~\ref{#1}}
\newcommand\equref[1]{Eq.~(\ref{#1})}

\newcommand\algref[1]{Alg.~\ref{#1}}

\newcommand\defref[1]{Def.~\ref{#1}}
\newcommand{\fakeparagraph}[1]{\vspace{1mm}\noindent\textbf{#1.}}

\newcommand{\Dis}{{dis}}
\newcommand{\FDScanning}{\textsf{FDScanning}\xspace}
\newcommand{\PDScanning}{\textsf{PDScanning}\xspace}
\newcommand{\RPDScanning}{\textsf{PDScanning+}\xspace}
\newcommand{\ADSampling}{\textsf{ADSampling}\xspace}
\newcommand{\DADE}{\textsf{DADE}\xspace}
\newcommand{\DDCres}{\textsf{DDCres}\xspace}
\newcommand{\DDCpca}{\textsf{DDCpca}\xspace}
\newcommand{\DDCopq}{\textsf{DDCopq}\xspace}

\newcommand{\LoopTitle}{{Simple Scanning Based}\xspace}
\newcommand{\Loop}{\text{simple}\xspace\text{scanning}\xspace\text{based}\xspace}
\newcommand{\LOOP}{\text{Simple}\xspace\text{scanning}\xspace\text{based}\xspace}
\newcommand{\Class}{\text{classification}\xspace\text{based}\xspace}
\newcommand{\CLASS}{\text{Classification}\xspace\text{based}\xspace}
\newcommand{\ClassTitle}{{Classification Based}\xspace}
\newcommand{\HypothesisTitle}{Hypothesis Testing Based\xspace}
\newcommand{\Hypothesis}{\text{hypothesis}\xspace\text{testing}\xspace\text{based}\xspace}
\newcommand{\HYPOTHESIS}{\text{Hypothesis}\xspace\text{testing}\xspace\text{based}\xspace}

\newcommand{\Sift}{\text{SIFT}\xspace}
\newcommand{\Gist}{\text{GIST}\xspace}
\newcommand{\Glove}{\text{GloVe}\xspace}
\newcommand{\Deep}{\text{Deep}\xspace}

\newcommand{\Trevi}{\text{Trevi}\xspace}
\newcommand{\Wikipedia}{\text{Wikipedia}\xspace}
\newcommand{\Openai}{\text{OpenAI}\xspace}
\newcommand{\TextImage}{\text{Text2Image}\xspace}
\newcommand{\Laion}{\text{Laion}\xspace}
\newcommand{\Msmacro}{\text{XUltra}\xspace}

\ifodd 0

\newcommand{\zheng}[1]{{\color{blue}{#1}}}

\else

\newcommand{\zheng}[1]{#1}
\fi

\makeatother

\begin{document}
\bstctlcite{IEEEexample:BSTcontrol}

\title{Distance Comparison Operations Are Not Silver Bullets in Vector Similarity Search: A Benchmark Study on Their Merits and Limits}

\author{\IEEEauthorblockN{Zhuanglin Zheng, Yuxiang Zeng, Chenchen Liu, Yunzhen Chi, Binhan Yang, Yongxin Tong}
\IEEEauthorblockA{
SKLCCSE Lab, BDBC and IRI, Beihang University, Beijing, China \\
\{zzlin, yxzeng, 23371020, chiyz, yangbh, yxtong\}@buaa.edu.cn}}

% \author{\IEEEauthorblockN{1\textsuperscript{st} Given Name Surname}
% \IEEEauthorblockA{\textit{dept. name of organization (of Aff.)} \\
% \textit{name of organization (of Aff.)}\\
% City, Country \\
% email address or ORCID}
% \and
% \IEEEauthorblockN{2\textsuperscript{nd} Given Name Surname}
% \IEEEauthorblockA{\textit{dept. name of organization (of Aff.)} \\
% \textit{name of organization (of Aff.)}\\
% City, Country \\
% email address or ORCID}
% \and
% \IEEEauthorblockN{3\textsuperscript{rd} Given Name Surname}
% \IEEEauthorblockA{\textit{dept. name of organization (of Aff.)} \\
% \textit{name of organization (of Aff.)}\\
% City, Country \\
% email address or ORCID}
% \and
% \IEEEauthorblockN{4\textsuperscript{th} Given Name Surname}
% \IEEEauthorblockA{\textit{dept. name of organization (of Aff.)} \\
% \textit{name of organization (of Aff.)}\\
% City, Country \\
% email address or ORCID}
% }

\maketitle

\microtypesetup{protrusion=false, expansion=false}

\pagestyle{plain}
\thispagestyle{plain}
\begin{abstract}
Distance Comparison Operations (DCOs), which decide whether the distance between a data vector and a query is within a threshold, are a critical performance bottleneck in vector similarity search. 
Recent DCO methods that avoid full-dimensional distance computations promise significant speedups, but their readiness for production vector database systems remains an open question.
To address this, we conduct a comprehensive benchmark of 8 DCO algorithms across 10 datasets (with up to 100M vectors and 12,288 dimensions) and diverse hardware configurations (CPUs with/without SIMD, and GPUs).
Our study reveals that \textit{these methods are not silver bullets}: their efficiency is highly sensitive to data dimensionality, degrades under out-of-distribution queries, and is unstable across hardware.
Yet, our evaluation also demonstrates often-overlooked merits: they can accelerate index construction and data updates.
Despite these benefits, their unstable performance, which can be slower than a full-dimensional scan, leads us to conclude that \textit{\zheng{recent algorithmic advancements in DCO are not yet ready for production deployment.}}
\end{abstract}
\vspace{-1ex}

\section{Introduction}

\setlength{\tabcolsep}{3.5pt}
\begin{table*}[t]
\centering
\begin{threeparttable}
\captionsetup{skip=2.2pt}
\caption{Comparisons of evaluation setup in prior work on Distance Comparison Operation (DCO)}\label{tab:exp-compare}
\begin{tabular}{c|cccccccccc}
\hline
\textbf{\makecell[c]{Related\\ Work}} & 
\textbf{\#(Dataset)} & 
\textbf{Dimensionality} & 
\textbf{Scalability} &
\textbf{\makecell[c]{\#(Algorithm)\\ Tested}} & 
\textbf{\makecell[c]{Compared\\ Work}} &
\textbf{\makecell[c]{Hardware\\ Environment}} & 
\textbf{\makecell[c]{Application\\ Scenario}} & 
\textbf{\makecell[c]{Distance\\ Metric}} & 
\textbf{\makecell[c]{OOD\\ Query}} \\
\hline 
\cite{DBLP:journals/pacmmod/GaoL23} & $6$ & $256 \sim 960$ & $5$ million & $3$ & N/A (first work) & CPU w/o SIMD & Query  & Euclidean & $\times$ \\
\cite{DBLP:journals/pvldb/DengCZWZZ24} & $6$ & $256 \sim 960$ & $5$ million & $3$ & \cite{DBLP:journals/pacmmod/GaoL23} & CPU w/o SIMD & Query & Euclidean & $\times$ \\
\cite{yang2025effective} & $8$ & $128 \sim 960$ & $100$ million & $5$ & \cite{DBLP:journals/pacmmod/GaoL23} & CPU w/o SIMD & Query  & Euclidean, IP & $\checkmark$ \\
\cite{DBLP:journals/debu/0007X0H0P024} & $6$ & $96 \sim 4096$ & $1$ million & $3$ & \cite{DBLP:journals/pacmmod/GaoL23} & CPU $\pm$ SIMD & Query & Euclidean & $\times$ \\
\hline
\textbf{Ours} & $\mathbf{10}$ & $\mathbf{96 \sim 12288}$ & $\mathbf{100}$ \textbf{million} & $\mathbf{8}$ & \cite{DBLP:journals/pacmmod/GaoL23,DBLP:journals/pvldb/DengCZWZZ24,yang2025effective,DBLP:journals/debu/0007X0H0P024} & \textbf{\makecell[c]{CPU $\pm$ SIMD\\ \& GPU}} & \textbf{\makecell[c]{Query, Update,\\ Index Construction}} &  \textbf{Euclidean, IP} & \textbf{$\checkmark$} \\
\hline
\end{tabular}
\vspace{-2pt}
\begin{tablenotes}
    \item
    SIMD: Single Instruction Multiple Data; IP: Inner Product; OOD: Out-of-Distribution, Ref. \cite{yang2025effective} uses only synthetic data for OOD evaluation.
\end{tablenotes}
\end{threeparttable}
\vspace{-1ex}
\end{table*}

Vector similarity search is a fundamental component of modern data-intensive applications, including recommendation systems, search engines, and Retrieval-Augmented Generation (RAG). A central challenge in this domain is improving search efficiency. While the majority of research has focused on designing index structures, a recent and promising line of work optimizes efficiency at a more granular operator level.

% 添加更多文章，检查引用格式
These studies \cite{DBLP:journals/pacmmod/GaoL23, DBLP:journals/pvldb/DengCZWZZ24, yang2025effective, DBLP:journals/debu/0007X0H0P024, DBLP:conf/isca/LiJTZ025, DBLP:journals/tkde/LiSW24} identify the \textit{Distance Comparison Operation (DCO)} (which determines whether the distance between a data vector and a query vector is within a given threshold) as a critical efficiency bottleneck. 
Traditional vector database systems \cite{DBLP:conf/sigmod/WangYGJXLWGLXYY21, weaviate, qdrant} often perform DCOs by exhaustively computing the full-dimensional distance, a naive method denoted as \underline{F}ull \underline{D}imension \underline{Scanning} (``\FDScanning'' as short). 
In contrast, state-of-the-art methods \cite{DBLP:journals/pacmmod/GaoL23, DBLP:journals/pvldb/DengCZWZZ24, yang2025effective, DBLP:journals/debu/0007X0H0P024} leverage techniques like hypothesis testing or prediction to scan only partial dimensions and infer the comparison result. These approaches have been reported to improve query throughput by 2--4$\times$ over \FDScanning with almost no loss in recall.

The impressive results raise a natural yet critical question: \textit{\zheng{Are these new DCO algorithms ready for deployment in production vector databases?}}
To answer this, we analyzed the evaluation protocols of prior research (see \tabref{tab:exp-compare}) and identified several key limitations in their evaluations \cite{DBLP:journals/pacmmod/GaoL23, DBLP:journals/pvldb/DengCZWZZ24, yang2025effective, DBLP:journals/debu/0007X0H0P024}:
\begin{itemize}[noitemsep, topsep=1pt, parsep=0pt]
    \item \textbf{Narrow Dimensionality Coverage}: They are largely confined to the dimensionality range of 128--960, neglecting lower-dimensional and, particularly, ultra-high-dimensional data common in embeddings from LLMs.
    
    \item \textbf{Insufficient Direct Comparison}: They lack empirical comparisons between the state-of-the-art DCO methods.
    
    \item \textbf{Restricted Hardware Configuration}: They typically use CPUs with SIMD disabled, which does not reflect the computing capabilities of a modern server.

    \item \textbf{Limited Application Scenario}: They mainly focus on in-distribution queries, overlooking practical scenarios like Out-of-Distribution (OOD) queries and index building.
\end{itemize}
\vspace{-0.1ex}

These limitations prevent a definitive answer to the previous question and motivate our work. To this end, we conduct a comprehensive benchmark of 8 DCO methods across 10 datasets and \zheng{diverse hardware}. Our evaluation reveals that \textit{these methods are not silver bullets}. Their effectiveness is highly contingent, and their adoption requires careful consideration. Specifically, we have the following findings:

(1) \textbf{Dimensionality Sensitivity}: Their performance is highly sensitive to data dimensionality. While they excel within a moderate range,  their advantage degrades significantly outside of it, at times falling behind \FDScanning by up to 20\%--42\%.
    
(2) \textbf{Robustness Dilemma}: Existing methods face a dual challenge. On \zheng{the} one hand, they are vulnerable to OOD queries that are common in applications like multimodal retrieval. On the other hand, their efficiency gains are unstable across hardware. For instance, enabling SIMD can reduce or even reverse their advantage over \FDScanning.

(3) \textbf{Benefit Beyond Query}: An often-overlooked merit of DCO methods lies in accelerating index construction and data insertions by up to 64\% and 63\%, respectively.

(4) \textbf{No Universal Winner}: There is no single dominant method. Their performance ranking varies drastically with data characteristics, query distribution, and hardware.

\textbf{Contribution.}
We make the following major contributions:
\begin{itemize}[noitemsep, topsep=1pt, parsep=0pt]
    \item We present the first comprehensive benchmark for evaluating DCOs in vector similarity search.
    
    \item Through extensive evaluation, we provide a detailed analysis of the merits and limitations of existing DCO methods. 
    The in-depth analysis leads to the final answer: \textit{the \zheng{recent algorithmic advancements in DCO}, while promising, are not yet ready for production deployment.}

    \item We derive practical guidelines for selecting the optimal DCO method under different scenarios and highlight future research directions.
    To foster further work, we have open-sourced the benchmark on GitHub \cite{code}.
\end{itemize}
\vspace{-0.2ex}

\fakeparagraph{Road Map} 
The rest of this paper is structured as follows.
\secref{sec:definition-problem} defines the DCO problem, and \secref{sec:exist-method} introduces existing algorithms. 
Next, \secref{sec:setup} and \secref{sec:result} detail the benchmark setup and analyze evaluation results. 
Finally, \secref{sec:related} reviews related work and \secref{sec:conclusion} concludes the paper.

\vspace{-0.5ex}
\section{Preliminary}\label{sec:definition-problem}
This section defines the \textit{\underline{D}istance \underline{C}omparison \underline{O}peration (DCO)} problem. 
\tabref{tab:notations} summarizes the major notations.

\vspace{-1ex}
\subsection{Basic Concepts}
Before defining the DCO, we first introduce two basic concepts: vector data and vector similarity search.

\begin{definition}[Vector Data] 
\label{def:data-object}
A vector data (``vector'' as short) $o$ is defined as an ordered sequence $(x_1, x_2, \cdots, x_{D})$. Here, $D$ is the vector's dimension, and $x_i$ represents the $i$-th coordinate.
The function $\Dis(o_1, o_2)$ denotes the distance between two vectors $o_1$ and $o_2$, which is a measure of their similarity.
\end{definition}

We use $\mathcal{O}$ to denote a dataset containing $N$ vectors, each with $D$ dimensions.
In practice, vector datasets can be classified into three kinds based on their dimensionality \cite{sun2025gaussdb, han2023comprehensive}:
\begin{itemize}[noitemsep, topsep=1pt, parsep=0pt]
    \item \textbf{Low-Dimensional Vector Dataset}: These are defined by a moderate dimensionality ($D \in [10, 100]$). They are prevalent in classical statistics and traditional machine learning (ML) applications \cite{casella2024statistical, james2013introduction}.
    
    \item \textbf{High-Dimensional Vector Dataset}: These datasets exhibit a large number of dimensions ($D \in (100, 1000]$). They usually consist of embedding vectors generated by deep learning models in domains like computer vision and natural language processing \cite{szeliski2022computer, manning1999foundations}.

    \item \textbf{Ultra-High-Dimensional Vector Dataset}: This kind encompasses datasets with dimensionality on the order of $10^3$ or higher.
    They often arise from raw sensor data and embedding layers of LLMs \cite{min2023recent, pecher2024survey}.
\end{itemize}

Distance functions are diverse, and commonly used examples include Euclidean distance, inner product, and cosine similarity.
For simplicity, we use Euclidean distance as the default distance function in the rest of the paper.

\begin{definition}[Vector Similarity Search] 
\label{def:vector-similarity-search}
Given a vector dataset $\mathcal{O}$, a query vector $q \in \mathbb{R}^D$, and a positive integer $k$, vector similarity search aims to find the $k$ nearest neighbor (KNN) vectors $\mathcal{S} \subseteq \mathcal{O}$ to the query vector $q$ satisfying
\vspace{-0.3\baselineskip}
\begin{equation*}
|\mathcal{S}| = k \text{ and } \forall v \in \mathcal{S}, \forall u \in (\mathcal{O} \setminus \mathcal{S}),\, \Dis(v, q) \leq \Dis(u,q)
\end{equation*}
\vspace{-0.3\baselineskip}
\end{definition}

\begin{table}[t]
\centering
\captionsetup{skip=2.4pt}
\caption{Summary of frequently used notations}\label{tab:notations} 
\begin{tabular}{ll}
\hline
\textbf{Notation} & \textbf{Description} \\
\hline
$\mathcal{O}$ & A dataset $\mathcal{O}$ of $N$ vectors, each of dimensionality $D$ \\
$\mathcal{S}$ & Query answer \\
$q, k$ & Query vector $q$ and the number $k$ of nearest neighbors \\
$\Dis(o_1, o_2)$ & Distance between two vectors $o_1$ and $o_2$ \\
$\tau$, $\widetilde{\Dis}$ & A given distance threshold and estimated distance \\
$d$ & Number of scanned dimensions during performing DCO \\
\hline
\end{tabular}
\vspace{-3ex}
\end{table}

\vspace{-2ex}
\fakeparagraph{Remark}
The scalability of exact solutions to vector similarity search is severely limited by the curse of dimensionality \cite{toth2017handbook}.
This has led to a rich line of research focused on \textit{approximate algorithms} that trade exactness for efficiency, with the primary goal of \textit{maximizing the recall}.
The recall of the query answer $\mathcal{S}$ against the ground truth $\mathcal{S}^*$ is defined as:
\begin{equation}\label{equ:recall}
    \textbf{recall} = \frac{|\mathcal{S} \cap \mathcal{S}^*|}{k}
\end{equation}

\subsection{Definition of Distance Comparison Operation}

\fakeparagraph{Overview of Vector Distance Operations}
Prior research \cite{DBLP:journals/debu/0007X0H0P024,DBLP:journals/pvldb/WangXY021,pan2024survey} in vector similarity search relies heavily on two types of vector distance operations: \textit{distance computation operations} and \textit{distance comparison operations}.
\begin{itemize}
    \item \textbf{Distance Computation Operation}: This operation calculates the \textit{exact distance} between two vectors. 
    It is computationally intensive but can often be accelerated by hardware like SIMD instructions and GPUs.
    
    \item \textbf{Distance Comparison Operation (DCO)}: 
    This operation determines whether the distance between two vectors exceeds a given threshold.
    Its efficiency enhancement stems from a key insight: \textit{computing exact distance is often unnecessary}, since the comparison result can be derived by scanning only a subset of the dimensions.
    We formally define this operation below.
\end{itemize}

\begin{definition}[Distance Comparison Operation (DCO) \cite{DBLP:journals/pacmmod/GaoL23}] 
\label{def:DCO}
Given two vectors $o, q$ and a distance threshold $\tau$, the distance comparison operation (DCO) determines the true value of the statement $\Dis(o,q) \leq \tau$.
If true, it also returns the exact distance. 
Otherwise, it returns \zheng{false alone, without the distance}.
\end{definition}

\zheng{The distance threshold $\tau$ is usually set to the distance of candidate vectors to the query vector.
As \cite{DBLP:journals/pacmmod/GaoL23,DBLP:journals/pvldb/DengCZWZZ24,yang2025effective} identifies, the vast majority of DCOs within a vector search return false. 
This makes the overall cost of exact distance calculation relatively low, as it is incurred only for the small subset of DCOs that return true.}
The following example illustrates this operation. 

\begin{figure}[t]
	\centering
    \includegraphics[width=0.43\textwidth]{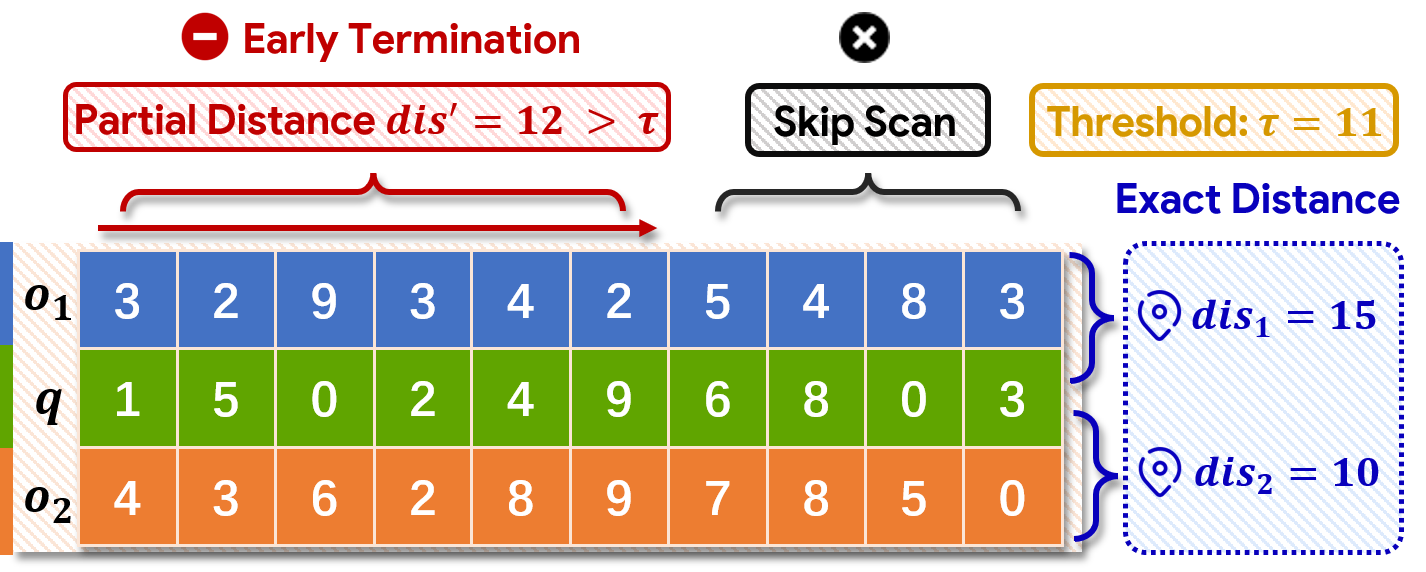}
    \vspace{-1ex}
	\caption{Instances of Distance Comparison Operation (DCO)}\label{fig:Example1}
    \vspace{-1ex}
\end{figure}

\begin{example}
\figref{fig:Example1} shows two instances of DCO with data vectors $o_1$ and $o_2$, a query vector $q$, and a distance threshold $\tau = 11$.
A naive solution is to scan all dimensions, compute the Euclidean distances $\Dis(o_1,q) = 15$ and $\Dis(o_2,q) = 10$, and then compare them with $\tau$.
\textit{However, a full-dimensional scan is not always necessary.}
For $o_1$, after scanning just the first 6 coordinates, the partial distance reaches 12, which already exceeds $\tau$.
This eliminates the need to scan the remaining dimensions of $o_1$ and thereby reduces computational cost.
\end{example}

\vspace{-0.5ex}
\section{Existing DCO Methods}\label{sec:exist-method}
In this section, we first introduce a taxonomy of existing DCO methods, and categorize them as \textit{\Loop}, \textit{\Hypothesis}, and \textit{\Class} methods.
We then review each category and finally present a comparative analysis of these DCO methods.

\vspace{-0.5ex}
\subsection{Method Taxonomy}
The design of high-performance DCOs follows two paths: \textit{optimized algorithms} and \textit{dedicated hardware}. Hardware-based approaches usually leverage GPUs \cite{johnson2019billion} for massive thread-level parallelism or SIMD instructions \cite{douze2024faiss} for dimension-level parallelism.
However, as recent research focuses on algorithmic optimization strategies \cite{DBLP:journals/pacmmod/GaoL23, DBLP:journals/pvldb/DengCZWZZ24, yang2025effective, DBLP:conf/www/ChenCJYDH23}, these hardware-centric methods are not our primary concern.

\begin{figure}[t]
	\centering
    \includegraphics[width=0.42\textwidth]{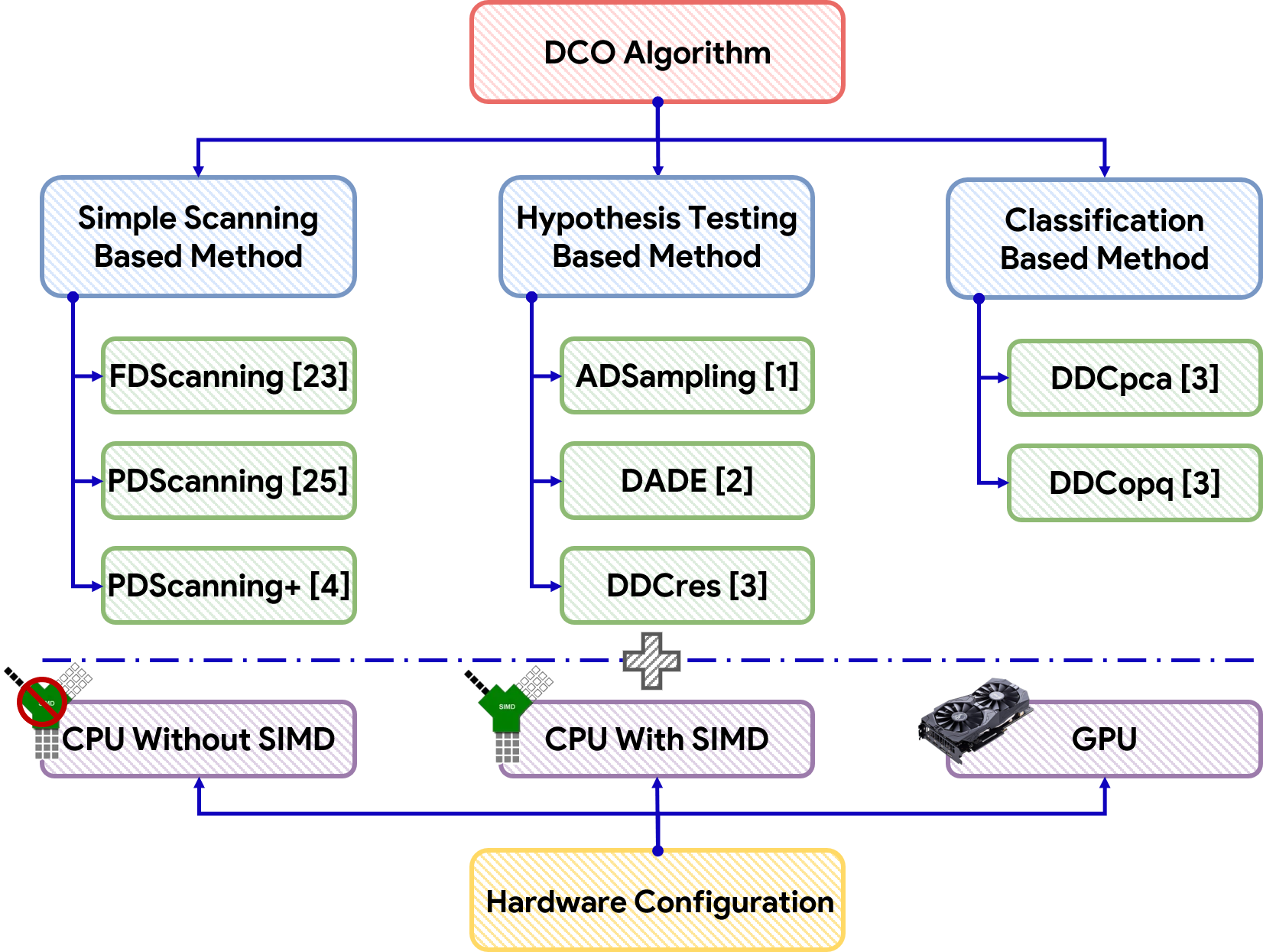}
    \vspace{-1ex}
	\caption{Taxonomy of existing DCO methods}\label{fig:taxonomy}
    \vspace{-0.5ex}
\end{figure}

\begin{figure*}[t]
    \centering
    \begin{subfigure}{0.27\textwidth}
        \centering
        \includegraphics[width=\textwidth]{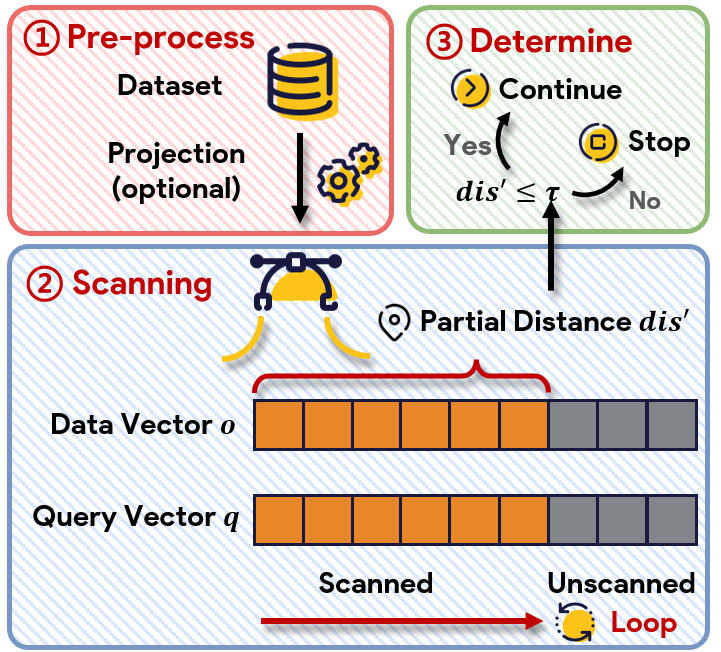}
        \vspace{-3.0ex}
        \caption{\LOOP method}
    \end{subfigure}
    ~
    \begin{subfigure}{0.285\textwidth}
        \centering
        \includegraphics[width=\textwidth]{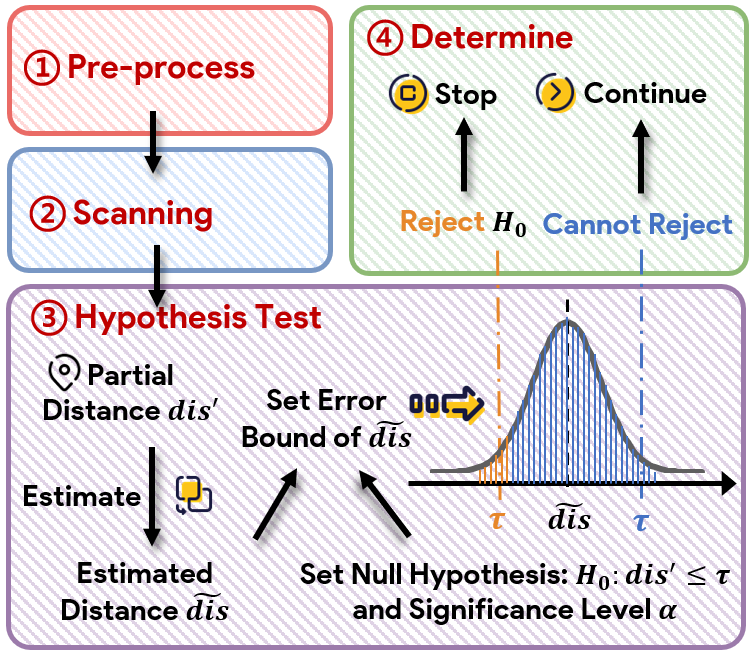}
        \vspace{-3.0ex}
        \caption{\HYPOTHESIS method}
    \end{subfigure}
    ~
    \begin{subfigure}{0.29\textwidth}
        \centering
        \includegraphics[width=\textwidth]{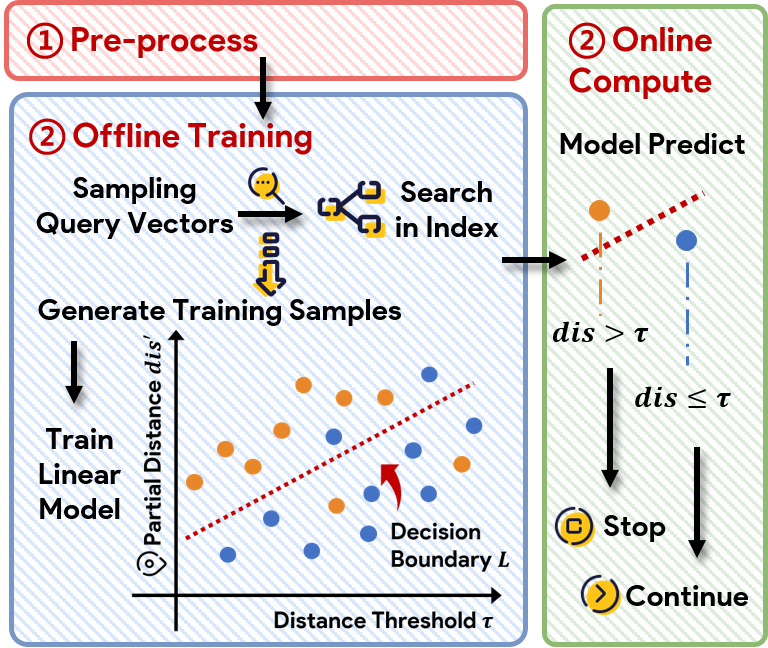}
        \vspace{-3.0ex}
        \caption{\CLASS method}
    \end{subfigure}
    \vspace{-1ex}
    \caption{The three categories of DCO methods: core ideas and main workflows}\label{fig:dco-idea}
    \vspace{-3ex}
\end{figure*}

As shown in \figref{fig:taxonomy}, we categorize these optimization algorithms into three kinds based on their core ideas: \textit{\Loop}, \textit{\Hypothesis}, and \textit{\Class}.
These methods are complementary to hardware accelerators like GPUs and SIMD instructions, which will be evaluated later.
The subsequent subsections will delve into each of these categories in detail, and \figref{fig:dco-idea} provides a high-level overview of their respective ideas and workflows.

\vspace{-0.5ex}
\subsection{\LoopTitle Method}\label{subsec:simple-loop}

\LOOP methods solve the DCO problem by scanning  dimensions of vectors $o$ and $q$, through two strategies: \textit{\underline{F}ull \underline{D}imension \underline{Scanning} (\FDScanning)} and \textit{\underline{P}artial \underline{D}imension \underline{Scanning} (\PDScanning)}. 
\begin{itemize}
    \item \FDScanning \cite{douze2024faiss} scans all dimensions to compute the exact distance and then compares it to the threshold $\tau$.
    
    \item \PDScanning \cite{muja2009fast} scans dimensions incrementally while maintaining a running partial distance.
    It terminates the scan as soon as the intermediate result proves that $\Dis(o,q) > \tau$, avoiding unnecessary computation.
\end{itemize}

\setlength{\textfloatsep}{1ex}
\setlength{\floatsep}{1ex}
\begin{algorithm}[t]
	\caption{\LoopTitle Framework}\label{alg:Simple-loop}
	\KwIn{Two vectors $o, q$ and a distance threshold $\tau$}
	\KwOut{Whether $\Dis(o,q) \leq \tau$}
    Initialize the number of currently scanned dimensions $d \gets 0$, partial distance $\Dis' \gets 0$\;
    \While{the number of scanned dimensions $d < D$}{
        Update partial distance $\Dis'$ and increase  $d$\;
        \If(\tcp*[f]{\textcolor{blue}{\textbf{only for} \PDScanning}}){$\Dis' > \tau$}{
            \KwRet{False};
        }
    }
    \KwRet{True \zheng{and $\Dis \gets \Dis'$}}\;
\end{algorithm}
\afterpage{\global\setlength{\textfloatsep}{\oldtextfloatsep}}
\afterpage{\global\setlength{\floatsep}{\oldfloatsep}}

\fakeparagraph{General Framework}
\algref{alg:Simple-loop} presents the general framework of these DCOs. 
% In lines 2-5, the framework scans both vectors' dimensions from scratch and maintains the partial distance $\Dis'$.
% A key distinction emerges in the termination condition:
% \PDScanning immediately exits the loop when $\Dis'$ exceeds the threshold $\tau$, whereas \FDScanning continues through all dimensions regardless of the intermediate results.
\zheng{It starts scanning dimensions from scratch, keeping track of the partial distance $\Dis'$ so far. The main difference between \FDScanning and \PDScanning lies in when they stop: \PDScanning quits early as soon as $\Dis'$ exceeds the threshold $\tau$, while \FDScanning scans all dimensions.}

% For \RPDScanning, vectors $o$ and $q$ need to be transformed using the matrix obtained by applying PCA to the dataset $\mathcal{O}$.

\vspace{-0.5ex}
\fakeparagraph{Optimization}
The efficiency of this framework can be further enhanced by two techniques: Single Instruction Multiple Data (SIMD) and \zheng{Principal Component Analysis (PCA)~\cite{abdi2010principal}}.

\textbf{(1) Optimization via SIMD}: The partial distance computation can be parallelized at the dimension-level using SIMD.
For example, both \FDScanning and \PDScanning can partition the dimensions of a vector into contiguous blocks that align with the SIMD register width (\eg 128-bit or wider).
This enables the CPU to simultaneously perform arithmetic operations across multiple dimensions within a single instruction cycle.
In this way, more than 4 dimensions can be processed together in a batch, which saves computational cost.

\textbf{(2) Optimization via PCA}: 
\zheng{PCA projects the original vector data into a new coordinate system, where dimensions are ordered by their variance contributions.
By rotating the original vector space (rather than performing dimensionality reduction), this transformation preserves the overall Euclidean distances between vectors while prioritizing the dominant influence of the leading dimensions}.
\PDScanning can also leverage this property to enable early termination in lines 4-5 of \algref{alg:Simple-loop}.
\zheng{We denote this method as \RPDScanning~\cite{DBLP:journals/debu/0007X0H0P024} in our work. Despite its simplicity, it has not been considered as a candidate DCO method in prior studies~\cite{DBLP:journals/pacmmod/GaoL23, DBLP:journals/pvldb/DengCZWZZ24, yang2025effective}, primarily because \FDScanning and \PDScanning are prevalent in vector databases. However, we treat it as a distinct baseline, since it offers unique advantages according to our evaluations.}
% and conduct comprehensive experimental comparisons with other baselines in  \secref{sec:result}.

%  In the context of \PDScanning, it is natural to consider applying an orthogonal transformation to reproject the original space and then perform \PDScanning in the projected space, which allows us to determine termination condition with fewer dimensions. A widely used method for such dimensionality reduction is Principal Component Analysis (PCA). 

% \textbf{Why PCA}. Because it identifies the orthogonal directions of maximum variance in the data, enabling effective dimensionality reduction while preserving the most informative features for subsequent analysis. We refer to this variant as \textit{rotated based partial dimension scanning (\RPDScanning)}. \RPDScanning aims to operate in a lower-dimensional space with minimal loss of structural information, thereby enabling earlier termination.

\subsection{\HypothesisTitle Method}\label{subsec:projection-based}

% \PDScanning is anticipated to be more efficient than \FDScanning due to scanning less dimensions.
% To traverse even less dimensions, one natural idea is to estimate the distance (denoted by $\tilde{\Dis}$) based on the partial distance $\Dis'$.
% The algorithms belong to this category are named as \Correct methods.
% On this basis, we categorize the existing studies into two families: (1) \textit{\Hypothesis} and (2) \textit{\Error method}.

The anticipated efficiency gain of \PDScanning over \FDScanning stems from scanning fewer dimensions.
This intuition motivates a class of approximate algorithms that estimate the full distance (denoted by $\widetilde{\Dis}$) from a partial distance ($\Dis'$) to scan even fewer dimensions.
Using the estimated distance within a probability inequality, these algorithms then perform a hypothesis test to accept or reject $\Dis(o,q) \leq \tau$.
They are classified into two kinds based on the estimation strategies: (1) \textit{residual distance estimator} and (2) \textit{cross-term estimator}.

% Compared to \Loop methods, projection-based methods improve the comparison step by correcting the partial distance $dis'$ onto an approximate distance $\hat{dis}$ that is closer to the exact distance. The decision to terminate is then based on the relationship between $\hat{dis}$ and threshold $\tau$. This allows for correct judgments to be made using fewer dimensions, with high confidence. 

\fakeparagraph{Strategy I. Residual Distance Estimator}
This class of algorithms estimates the residual distance (\ie the distance contributed by unscanned dimensions) based on the partial distance $\Dis'$ from the currently scanned dimensions.
There are two general methods for such estimators.

% \textbf{(1) Assuming Equal Contribution per Dimension.}
% Under this assumption, the residual distance is estimated as $\sqrt{\frac{D-d}{d}}\Dis'$, leading to a full distance estimate $\sqrt{\frac{D}{d}}\Dis'$.
% To enforce this assumption in practice, \ADSampling~\cite{DBLP:journals/pacmmod/GaoL23} employs a random projection matrix $P \in \mathbb{R}^{d \times D}$ to map original vectors into a new coordinate system. 
\textbf{(1) Assuming Equal Contribution per Dimension.}
%Under this assumption, the residual distance is estimated as $\sqrt{\frac{D-d}{d}}\Dis'$, leading to a full distance estimate $\sqrt{\frac{D}{d}}\Dis'$.
\zheng{Under this assumption, the full distance can be estimated by simply scaling up the partial distance proportionally to the ratio of total dimensions to scanned dimensions.}
To enforce this assumption in practice, \ADSampling~\cite{DBLP:journals/pacmmod/GaoL23} employs a random projection matrix $P \in \mathbb{R}^{d \times D}$ to map original vectors into a new coordinate system. 
\zheng{This work also proves an error bound on the estimated distance as follows:}

\vspace{-2.5ex}
\begin{small}
\begin{equation*}
    \mathbb{P} \left\{ \left|\sqrt{\frac{D}{d}} \Dis(Po,Pq) - \Dis(o,q) \right| \leq \epsilon \Dis(o,q) \right\} \geq 1 - 2 e^{-c \cdot d \cdot \epsilon^2}
\end{equation*}
\end{small}
\vspace{-0.5ex}

In this inequality, $\sqrt{\frac{D}{d}}\Dis(Po,Pq)$ is the estimated distance $\widetilde{\Dis}$, $c$ is a constant factor, and $\epsilon$ is the error tolerance.
Thus, the inequality implies that 
$\widetilde{\Dis} \le (1+\epsilon) \cdot \Dis$ holds with high probability $1 - 2 e^{-c d \epsilon^2}$.
Moreover, this estimator ensures that $\widetilde{\Dis} = \Dis$ when all original dimensions are preserved in the projection (\ie $d = D$).

\textbf{(2) Assuming Differing Contribution per Dimension.}
In contrast, \DADE \cite{DBLP:journals/pvldb/DengCZWZZ24} is designed for scenarios where dimensions contribute differently to the distance, and treats dimensions with different weights.
Accordingly, it performs a PCA over the vector dataset to obtain the loading matrix $W \in \mathbb{R}^{D \times D}$ as the projection matrix. 
% Here, $W_d$ consists of the eigenvectors of the data covariance matrix corresponding to the largest $d$ eigenvalues.
This ensures that the initial dimensions capture the largest possible contribution to the distance.
Consequently, the result of DCOs can be quickly decided after scanning only a few leading dimensions.

\zheng{Similarly, Deng \etal \cite{DBLP:journals/pvldb/DengCZWZZ24} also propose an error bound on the estimated distance as follows:}

\vspace{-2ex}
\begin{small}
\begin{equation}\label{equ:DADE}  
    \mathbb{P} \left\{ \sqrt{ \frac{\sum_{k=1}^D\lambda_k}{\sum_{k=1}^d\lambda_k}} \Dis(W_d^To,W_d^Tq) > (1+\epsilon_d) \Dis(o,q) \right\} < \alpha
\end{equation}
\end{small}

\noindent{where $\lambda_k$ is the $k$-th largest eigenvalue of the vector dataset's covariance matrix, $\alpha$ is the significance level, and $\epsilon_d$ is the error tolerance after scanning $d$ dimensions. Both $\alpha$ and $\epsilon_d$ are user-defined parameters whose values are set empirically.}

\fakeparagraph{Strategy II. Cross-Term Estimator}
This strategy employs an algebraic decomposition of the squared Euclidean distance into a sum of square-terms and cross-terms:

\vspace{-3.0ex}
\begin{align}
    \Dis(o, q) &= \Vert o \Vert^2 + \Vert q \Vert^2 - 2 \langle o, q \rangle \label{equ:decomposition} \\
    &= \Vert o \Vert^2 + \Vert q \Vert^2 - 2 \langle od, qd \rangle - 2 \langle or, qr \rangle \label{equ:decompose-2}
\end{align}
\vspace{-3.0ex}

\noindent where $\Vert \cdot \Vert$ is the $L_2$-norm, and $\langle \cdot, \cdot \rangle$ is the inner product.
The sub-vectors $od, qd$ and $or, qr$ contain the scanned and unscanned dimensions, respectively. 
In \equref{equ:decompose-2}, both $\Vert o \Vert$ and $\Vert q \Vert$ can be pre-processed, and $\langle od, qd \rangle$ can be computed directly from the scanned dimensions.
Their collective partial result is denoted by $\Dis'$.
This method thus focuses on estimating the remaining cross-term, $\langle or, qr \rangle$.

\zheng{Yang \etal~\cite{yang2025effective} propose a new method called \DDCres to bound this term.}
\DDCres~\cite{yang2025effective} first centers the data vectors to have zero mean.
It then assumes that both query and data vectors follow a Gaussian distribution,
where $\sigma_i^2$ denotes the variance of the $i$-th dimension.
Accordingly, the expectation and variance of the remaining cross-term are derived as:

\vspace{-2.0ex}
\begin{small}
\begin{align} 
    \mathbb{E}[\langle o_r, q_r \rangle] &= \sum_{i=d+1}^D \mathbb{E}[o_i \cdot q_i] = \sum_{i=d+1}^D \big( \mathbb{E}[o_i] \cdot \mathbb{E}[q_i] \big) = 0 \label{equ:e} \\
    \mathrm{Var}[\langle o_r, q_r \rangle] &= \sum_{i=d+1}^D (q_i \cdot \sigma_i)^2 \label{equ:var}
\end{align}
\end{small}
\vspace{-1.0ex}

Consequently, the exact distance $\Dis$ lies within a confidence interval $\Dis' \pm m \cdot 2\sqrt{\mathrm{Var}[\langle o_r, q_r \rangle]}$ with high probability, where $m$ is a deviation multiplier.
To test if $\Dis > \tau$, the lower bound of this  interval is used as the estimated distance $\widetilde{\Dis}$, \ie
\begin{equation}\label{equ:rule}  
    \widetilde{\Dis} = \Dis' - m \cdot 2\sqrt{\mathrm{Var}[\langle o_r, q_r \rangle]}
\end{equation}
\zheng{To tighten this lower bound, \DDCres leverages PCA to minimize \equref{equ:var}, since PCA reorders data dimensions in descending order of $\sigma_i^2$.}

\begin{algorithm}[t]
	\caption{\HypothesisTitle Framework}\label{alg:Hypothesis-testing-based}
	\KwIn{Two vectors $o, q$ and a distance threshold $\tau$}
	\KwOut{Whether $\Dis(o,q) \leq \tau$}
    Set the null hypothesis $H_0: dis \leq \tau$ and its alternative hypothesis $H_1: dis > \tau$\;
    \While{the number of scanned dimensions $d < D$}{
        Update partial distance $\Dis'$ and increase $d$\;
        Set significance level $\alpha$ by their parameter settings\;
        \tcp*[f]{\textcolor{blue}{$\epsilon=\epsilon_d$ for \DADE, $\epsilon=0$ for \DDCres}}\\
        Compute a relaxed distance threshold $(1+\epsilon)\cdot \tau$\;
        Compute the estimated distance $\widetilde{\Dis}$ based on $dis'$\;
        \If{$\widetilde{\Dis} > (1+\epsilon) \cdot \tau$}{
            Reject $H_0$ with confidence and \KwRet{False}\;
        }
    }
    \KwRet{True \zheng{and $\Dis \gets \Dis'$}}\;
\end{algorithm}

\fakeparagraph{General Framework}
\algref{alg:Hypothesis-testing-based} presents the main workflow of \Hypothesis methods. 
% Compared to \Loop methods, \Hypothesis methods include an additional step in lines 4-6, which utilizes hypothesis testing to determine whether the null hypothesis holds.
\zheng{Unlike \Loop methods, these methods incorporate an extra step (lines 4-6) in which hypothesis testing is used to check if the null hypothesis is valid.}
Correspondingly, the condition in line 7 is replaced by a verification on whether the estimated distance  $\widetilde{\Dis}$ exceeds a statistically derived error bound. If it does, the null hypothesis is rejected and the loop terminates early.

\fakeparagraph{Beyond Euclidean Distance}
The \Hypothesis methods also support inner product and cosine similarity by transforming them into Euclidean distance.
This requires the vector dataset to be normalized first.
Then, the inner product $\langle o, q \rangle$  is derived from the Euclidean distance $\Dis(o,q)$ as:
\vspace{-0.2\baselineskip}
\begin{equation}\label{equ:L2toIP}
    \langle o, q \rangle = 1 - 0.5 \cdot \left( \Dis(o,q) \right)^2
\vspace{-0.2\baselineskip}
\end{equation}
Since cosine similarity is equivalent to the inner product for normalized vectors, this extension applies to both metrics.

\vspace{-0.5ex}
\subsection{\ClassTitle Method}

\fakeparagraph{Main Idea}
Other research \cite{yang2025effective} formulates the DCO problem as a binary classification task.
Given the partial distance $\Dis'$ and threshold $\tau$, the goal of this task is to predict whether the full distance satisfies $\Dis \le \tau$.
These methods typically assume the query workload is known a priori (\eg often modeled using the vector dataset's distribution).
Under this assumption, they utilize a fixed vector index and sampled queries (including both an integer $k$ and a query vector $q$) to generate sufficient training samples.
Subsequently, a distinct linear model $M_{k,d}$ is trained for each combination of the integer $k$ and the number $d$ of scanned dimensions.

\begin{algorithm}[t]
	\caption{\ClassTitle Method \DDCpca}\label{alg:Distance-classification-based}
	\KwIn{Two vectors $o, q$ and a distance threshold $\tau$}
	\KwOut{Whether $\Dis(o,q) \leq \tau$}
    \textcolor{blue}{\tcp{\textbf{Offline Phase}}}
    Sample query vectors from dataset and parameters $k$\;
    Generate training samples through vector similarity search for sampled queries using a specific index\;
    Train linear models $M_{k,d}$, denoting the model for query parameter $k$ after $d$ dimensions are scanned.\;
    \textcolor{blue}{\tcp{\textbf{Online Phase}}}
    \While{the number of scanned dimensions $d < D$}{
        Update partial distance $\Dis'$ and increase $d$\;
        \lIf{$M_{k,d}$ predicts $\Dis(o,q) > \tau$}{
            \KwRet{False}
        }
    }
    \KwRet{True \zheng{and $\Dis \gets \Dis'$}}\;
\end{algorithm}

\fakeparagraph{Representative Algorithm}
\algref{alg:Distance-classification-based} illustrates a representative algorithm \DDCpca to this kind. 
% The offline phase generates training samples by performing similarity searches with varying $k$ on a fixed vector index. 
\zheng{In the offline phase, it runs similarity searches with different $k$ values on a fixed vector index to generate training samples.}
These samples are used to train a set of linear models $M_{k,d}$, each corresponding to a query parameter $k$ after $d$ dimensions are scanned. 
During the online phase, \DDCpca feeds the partial distance $dis'$ and distance threshold $\tau$ into the model $M_{k,d}$.
The process terminates early if the model predicts $\Dis(o,q) > \tau$.

\fakeparagraph{Variant}
Yang \etal~\cite{yang2025effective} propose another \Class method, named \DDCopq.
Unlike \DDCpca, \DDCopq trains a single linear model $M_{k}$ for each query parameter $k$, using approximate distances derived from Product Quantization (PQ).
During a DCO, this model first predicts if $\Dis(o,q) > \tau$. If the result is negative, it performs a distance computation operation by scanning all dimensions to verify the result.

\vspace{-1ex}
\subsection{Summary}

\tabref{tab:methods} compares DCO methods on the following aspects:

\textbf{(1) Exactness}: \LOOP algorithms are exact, while the others trade slight accuracy for efficiency.
Besides, only \DADE provides an unbiased distance estimator.

\textbf{(2) Assumption}: DCO methods vary in their  assumptions.
Most DCO methods are limited to Euclidean distance (or transformable metrics), whereas \PDScanning and \RPDScanning also support monotonic distances.
\DADE, \DDCres, \DDCpca, and \DDCopq assume queries and data follow the same distribution. \DDCres further assumes this distribution is Gaussian.
\CLASS methods require prior knowledge of the query parameter $k$ and are coupled to a specific index.

\textbf{(3) Time Complexity}:
Only \FDScanning, \PDScanning, and \ADSampling provide explicit time complexity analyses. 
However, \ADSampling underestimates the $O(D^2)$ time cost of projecting a query vector into a new coordinate system.
We categorize this per-query operation as \textit{online pre-processing}, a cost also incurred by \RPDScanning, \Hypothesis, and \Class methods.
Our analysis introduces $d$ to denote the number of dimensions scanned and $p_i$ as the probability of a negative prediction. 
We also include the time cost of \textit{offline pre-processing}, which includes PCA computation and model training.
Our results show that \textbf{online pre-processing can become the dominant bottleneck at sufficiently high dimensions}, a finding that will be validated empirically later.

\begin{table*}[ht]
\centering
\begin{threeparttable}
\captionsetup{skip=2.2pt}
\caption{Comparisons of existing methods for Distance Comparison Operation (DCO)}\label{tab:methods}
\begin{tabular}{cccccccc}
\toprule
\multirow{2}{*}[-2ex]{\textbf{Category}} & \multirow{2}{*}[-2ex]{\textbf{Method}} & \multirow{2}{*}[-2ex]{\textbf{Exactness}}  & \multirow{2}{*}[-2ex]{\textbf{Assumption$^1$}} & \multicolumn{3}{c}{\textbf{Time Complexity Breakdown$^2$}} \\
 \cmidrule(lr){5-7}
 & & & &  \textbf{\makecell[c]{Offline \\ Pre-Processing}} & \textbf{\makecell[c]{Online \\ Pre-Processing}} & \textbf{\makecell[c]{Online \\ Computation}} \\
\midrule
\multirow{3}{*}{\makecell[c]{Simple \\ Scanning}} & \FDScanning \cite{douze2024faiss} & $\checkmark$  & None & N/A & N/A & $O(D)$ \\
& \PDScanning \cite{muja2009fast} & $\checkmark$ & Monotonic Distance & N/A & N/A & $O(\min(D,D\frac{\tau^2}{\Dis^2}))$ \\
& \RPDScanning \cite{DBLP:journals/debu/0007X0H0P024} & $\checkmark$ & Monotonic Distance &$ O(ND^2)$ & $O(D^2)$ & $O(d)$\\
\midrule
\multirow{3}{*}{\makecell[c]{Hypothesis \\ Testing}} & \ADSampling \cite{DBLP:journals/pacmmod/GaoL23} & $\times$  & Euclidean & $O(ND^2)$ & $O(D^2)$ & $O(\min (D, \frac{\tau^2}{(\Dis-\tau)^2} \log {\frac{D}{\delta}}))$ \\
& \DADE \cite{DBLP:journals/pvldb/DengCZWZZ24} & $\times$  & Euclidean, Query & $O(ND^2)$ & $O(D^2)$ & $O(d)$ \\
& \DDCres \cite{yang2025effective} & $\times$ & Euclidean, Data, Query & $O(ND^2)$ & $O(D^2)$ & $O(d)$ \\
\midrule
\multirow{2}{*}{Classification} & \DDCpca \cite{yang2025effective} & $\times$ & Euclidean, Index, Query & $O(ND^2)$ & $O(D^2)$ & $O(\sum_{j=1}^D \prod_{i=1}^{j-1}(1-p_i))$\\
& \DDCopq \cite{yang2025effective} & $\times$ & Euclidean, Index, Query &  $O(ND(D+2^{b}))$ & $O(D^2+D \cdot 2^{b})$ & $O(c+(1-p)D)$ \\
\bottomrule
\end{tabular}
\begin{tablenotes}
    \item[1] 
    ``Euclidean'': only applicable to Euclidean distance or distances that can be transformed into it.; 
    ``Data'': data vectors follow a known distribution (\eg Gaussian); 
    ``Query'': query vectors follow the data distribution; ``Index'': requires building a vector index for training.
    
    \item[2] 
    $\delta$: failure probability for $(1+\epsilon)$-approximation; $b$, $c$: product quantization parameters for subvector encoding length and codebook count.
\end{tablenotes}
\end{threeparttable}
\vspace{-3ex}
\end{table*}

\vspace{-1ex}
\section{Benchmark Setup}\label{sec:setup}
This section introduces the detailed benchmark setup. We make our benchmark suite publicly available on GitHub \cite{code}.

\vspace{-1ex}
\subsection{Overview}
Our benchmark covers the following critical scenarios:
\begin{itemize}[noitemsep, topsep=1pt, parsep=0pt]
    \item \textbf{Query Processing}: Evaluating how DCOs impact vector similarity search performance across diverse datasets.

    \item \textbf{Out-of-Distribution (OOD) Queries}: Generalization when query distribution differs from the data distribution.

    \item \textbf{Distance Metrics}: \zheng{Evaluating DCO performance on Euclidean distance, inner product, and cosine similarity.}

    \item \textbf{Index Construction}: The efficacy of DCOs in building  mainstream vector indexes: HNSW \cite{DBLP:journals/pami/MalkovY20} and IVF \cite{DBLP:journals/pami/BabenkoL15}.

    \item \textbf{Dynamic Updates}: The robustness and effectiveness of DCOs when handling dynamic data.

    % \item \textbf{Common Parameters}: The impact of DCOs' parameters.

    \item \textbf{Diversified Hardware Condition}: The performance of DCOs under varied hardware configurations.
\end{itemize}

% Due to page limits, the parameter study on the number of scan dimensions per round is provided in our full paper \cite{fullpaper}.

\vspace{-1ex}

\begin{table}[ht]
\centering
\captionsetup{skip=2.2pt}
\caption{Dataset statistics}\label{tab:dataset}
\begin{tabular}{c|ccccc}
\hline
\textbf{Dataset} & \textbf{Dim.} & \textbf{Category} & \textbf{Raw Data} & \textbf{Cardinality} & \textbf{\zheng{Size (GB)}} \\
\hline
\Deep & 96 & Low-D & Image & 100,000,000 & \zheng{36} \\
\Glove & 100 & Low-D & Text & 400,000 & \zheng{0.15}\\
\Sift & 128 & High-D & Image & 1,000,000 & \zheng{0.5}\\
\TextImage & 200 & High-D & Multimodal & 10,000,000 & \zheng{7.5}\\
\Laion & 512 & High-D & Multimodal & 1,000,448 & \zheng{2.0}\\
\Wikipedia & 768 & High-D & Text & 1,000,000 & \zheng{2.9}\\
\Gist & 960 & High-D & Image & 1,000,000 & \zheng{3.6}\\
\Openai & 1536 & Ultra-High-D & Text & 2,321,096 & \zheng{14}\\
\Trevi & 4096 & Ultra-High-D & Image & 99,900 & \zheng{1.6}\\
\Msmacro & 12288 & Ultra-High-D & Text & 100,000 & \zheng{4.6}\\
\hline
\end{tabular}
\vspace{-2ex}
\end{table}

\vspace{-1ex}
\subsection{Dataset and Query Workload}
Our benchmark employs 10 public datasets \cite{bigann,glove,DBLP:journals/corr/abs-2111-02114,nguyen2016ms} for evaluating vector similarity search with varying dimensions, sizes, and raw data types (see \tabref{tab:dataset}). 
Most datasets contain in-distribution queries, which either share a  distribution with the dataset or are sampled from it.
In contrast, the multimodal datasets (\Laion and \TextImage) exhibit an inherent distribution shift, as their data and query vectors are embedded from images and text, respectively, making their queries OOD. 

To simulate the e\underline{X}tremely \underline{Ultra}-high dimensionality of embeddings in modern LLMs (\eg OpenAI's davinci-001 model with 12,288 dimensions~\cite{openai2022improved}), we also construct a synthetic dataset named \Msmacro. 
It is generated by concatenating token-level embeddings from the real-world dataset MSMARCO~\cite{nguyen2016ms} until the target dimensionality of 12,288 is reached.
In this dataset, each resulting vector represents the aggregation of tokens in a complete text phrase. 
\zheng{The ground truth is obtained by exhaustive linear scan of the full dataset.}

% \subsection{Query Workload}
% We follow established practices from prior work \TODO{ref}. For the datasets that do not provide query vectors, we randomly sample from the dataset to form the query set. For others, we use the provided query vectors \TODO{wiki}. Among these, for multimodal datasets \Laion and \TextImage, we build the index using image embeddings and then use image embeddings as ID (in-distribution) queries and text embeddings as OOD (out-of-distribution) queries. Furthermore, we integrate these DCO methods into widely used vector similarity search algorithms, \eg HNSW \cite{DBLP:journals/pami/MalkovY20} and IVF \cite{DBLP:journals/pami/BabenkoL15}, to test the performance of DCOs.

% We implement eight DCO methods: \FDScanning, \PDScanning, \RPDScanning, \ADSampling, \DADE, \DDCres, \DDCpca, and \DDCopq \cite{douze2024faiss, muja2009fast, DBLP:journals/debu/0007X0H0P024, DBLP:journals/pacmmod/GaoL23, DBLP:journals/pvldb/DengCZWZZ24, yang2025effective}. The first three, categorized as \Loop methods, serve as our baselines. To ensure a fair comparison, all algorithms are implemented within a unified framework with the following considerations:

\vspace{-0.5ex}
\subsection{Compared DCO Algorithms and Their Implementations}
We implement 8 DCO methods from prior research \cite{DBLP:journals/pacmmod/GaoL23, DBLP:journals/pvldb/DengCZWZZ24, yang2025effective, DBLP:journals/debu/0007X0H0P024}: \FDScanning, \PDScanning, \RPDScanning, \ADSampling, \DADE, \DDCres, \DDCpca, and \DDCopq.
The first three, serve as \textit{baselines}, widely adopted in industrial vector databases.
The other five are the \textit{state-of-the-art} algorithms (``SOTA'' as short).
\zheng{For a fair comparison, we implement all algorithms from scratch within a unified framework:}
\begin{itemize}[noitemsep, topsep=1pt, parsep=0pt]
    \item \zheng{\textbf{Index Selection}: Each method is integrated into two indexes using identical data layouts: HNSW \cite{DBLP:journals/pami/MalkovY20} and IVF \cite{DBLP:journals/pami/BabenkoL15}, with HNSW running on CPUs and IVF on GPUs.} 
    
    \item \zheng{\textbf{Target Hardware}: We implement SIMD-optimized and GPU-accelerated versions for all algorithms.
    On CPUs, we evaluate these algorithms with and without SIMD optimizations, which accelerate distance computations via intra-vector parallelism. 
    On GPUs, we pre-loaded IVF into device memory and adopt vector-level parallelism, launching a CUDA kernel for each IVF partition with each thread assigned to process one candidate vector.}

    \item \zheng{\textbf{Parameter Settings}: DCO methods are configured with parameters recommended in their original papers.}
    
    \item \zheng{\textbf{Batch Query}: Batch queries are processed sequentially.}
\end{itemize}

\zheng{We implement all methods in both Python and C++, following their open-source implementations.
Online query processing is written in C++, while offline pre-processing uses both C++ and Python, where Python is mainly employed for model training and PCA computation.
By default, we enable SIMD via the SSE instruction set and keep multi-threading disabled. 
Please refer to \cite{fullpaper} for more implementation details.}

\begin{figure*}[t]
    \centering
    \includegraphics[width=0.9\textwidth]{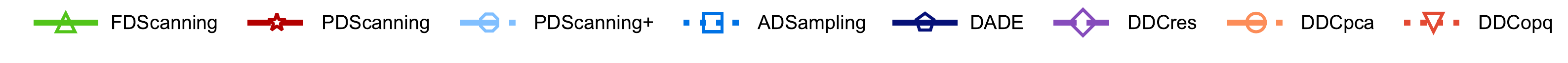}
    \begin{subfigure}{0.24\textwidth}
        \centering
        \includegraphics[width=\textwidth]{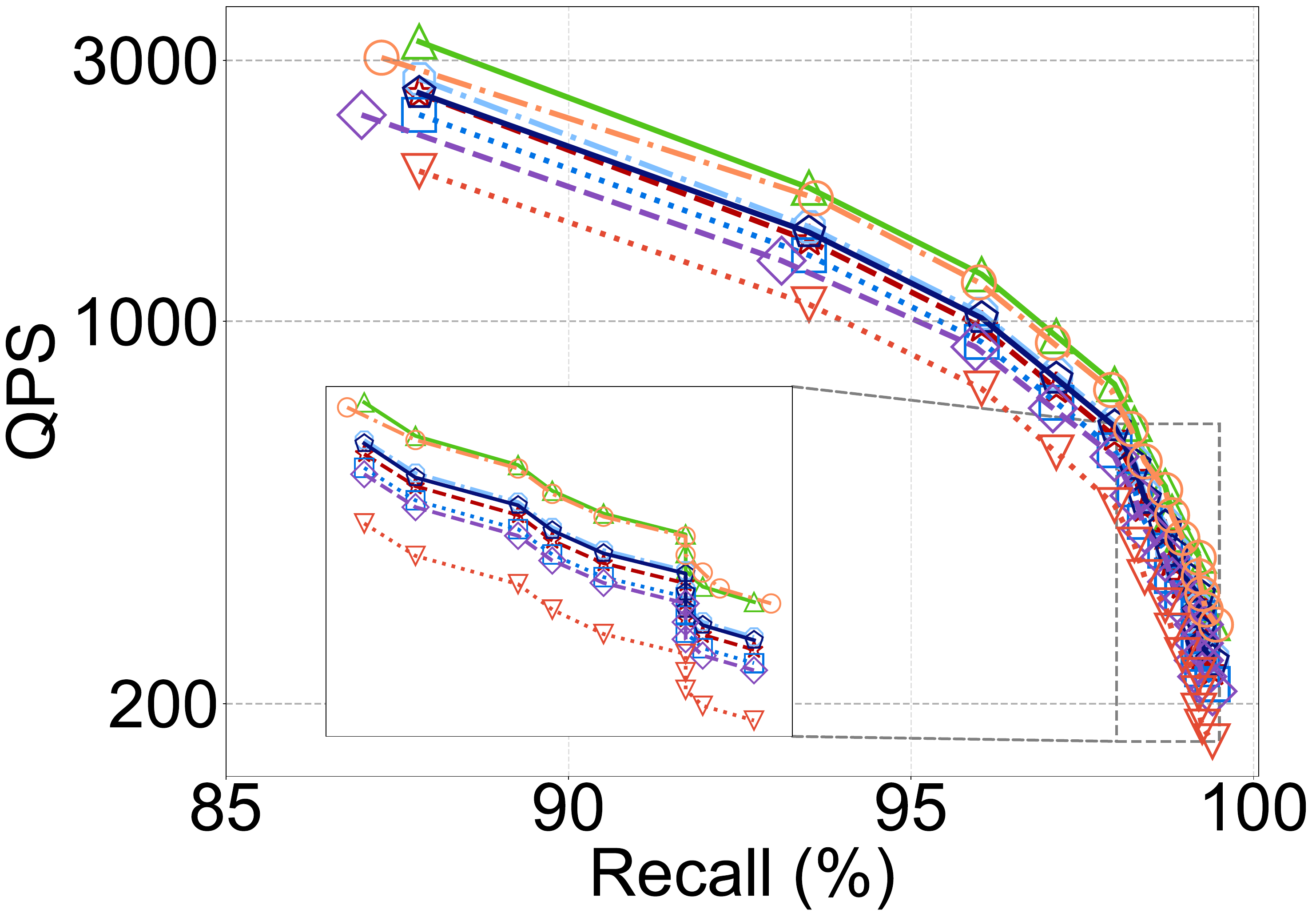}\vspace{-1.0ex}
        \caption{\Deep ($k=20$)}
        \label{fig:deep-qk-20}
    \end{subfigure}
    \begin{subfigure}{0.24\textwidth}
        \centering
        \includegraphics[width=\textwidth]{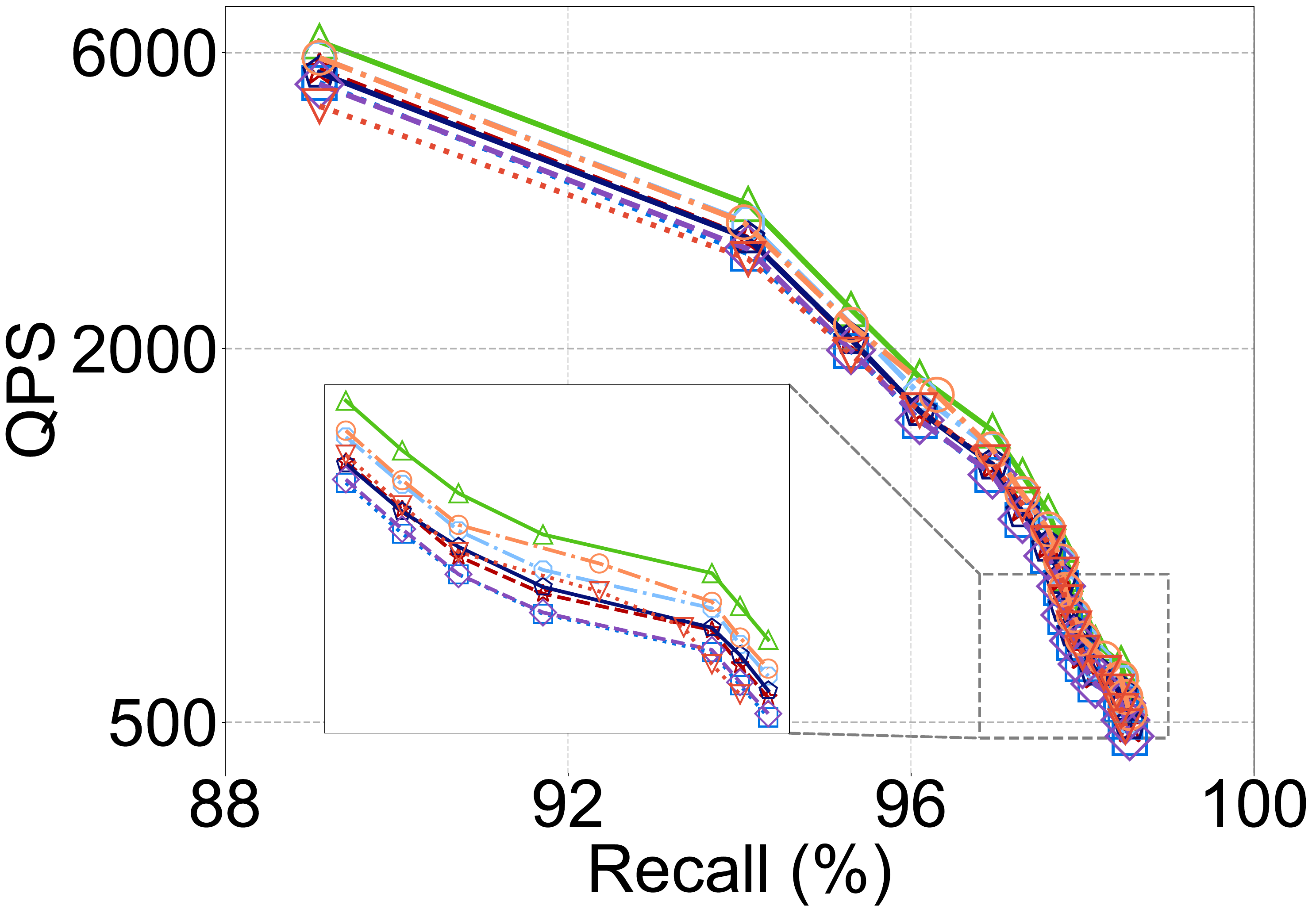}\vspace{-1.0ex}
        \caption{\Glove ($k=20$)}
        \label{fig:glove-qk-20}
    \end{subfigure}
    \begin{subfigure}{0.24\textwidth}
        \centering
        \includegraphics[width=\textwidth]{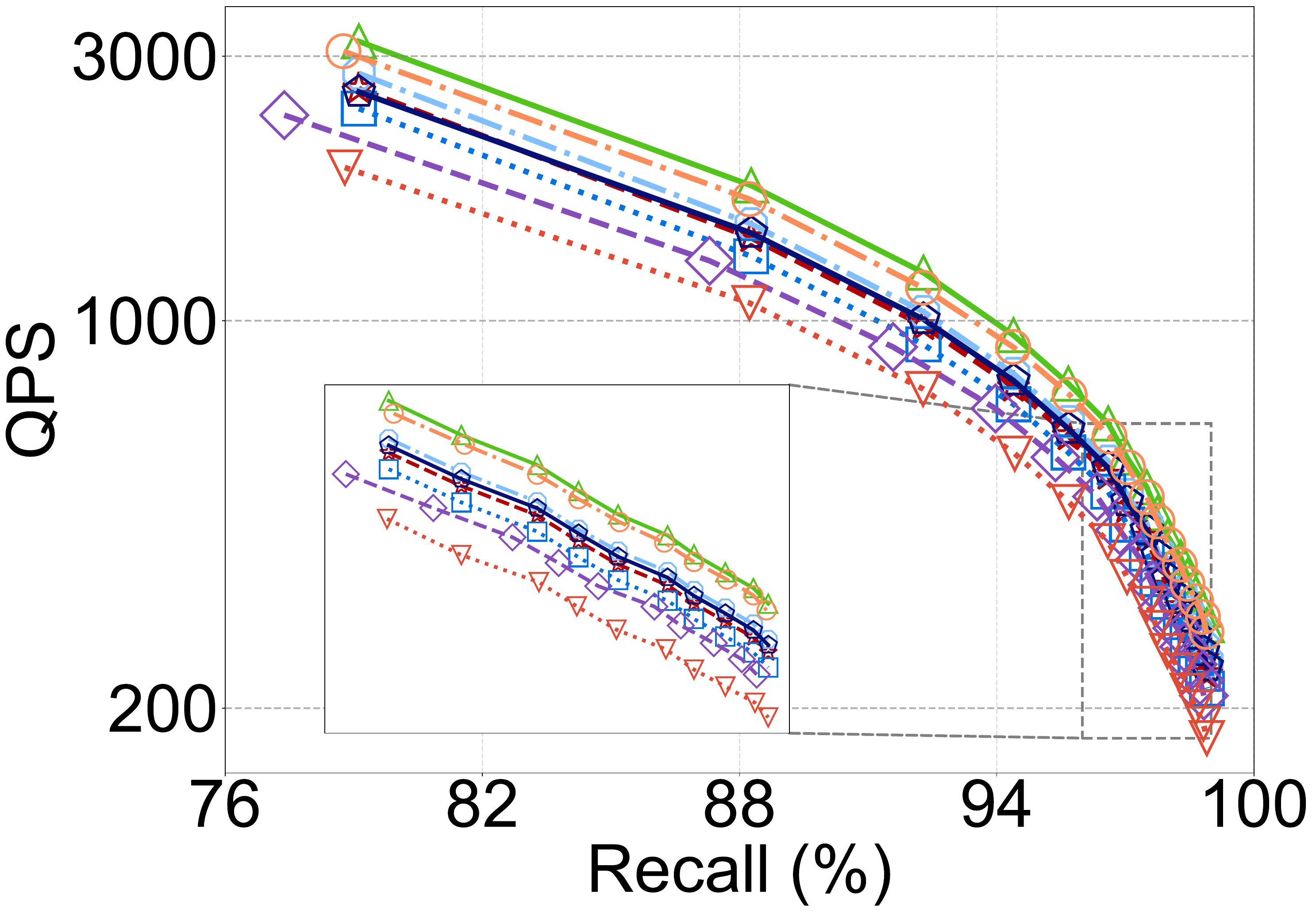}\vspace{-1.0ex}
        \caption{\Deep ($k=100$)}
        \label{fig:deep-qk-100}
    \end{subfigure}
    \begin{subfigure}{0.24\textwidth}
        \centering
        \includegraphics[width=\textwidth]{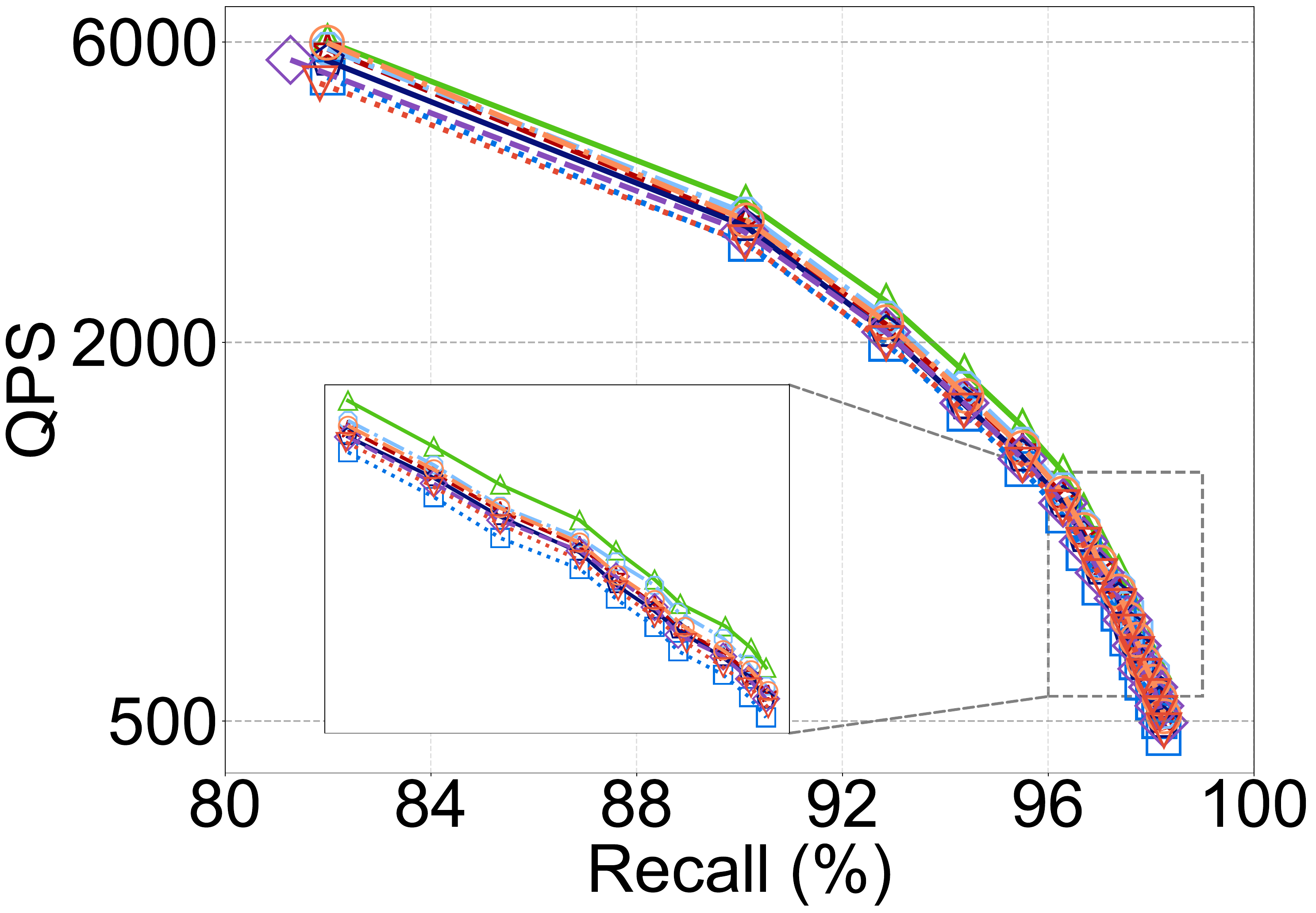}\vspace{-1.0ex}
        \caption{\Glove ($k=100$)}
        \label{fig:glove-qk-100}
    \end{subfigure}
    \begin{subfigure}{0.24\textwidth}
        \centering
        \includegraphics[width=\textwidth]{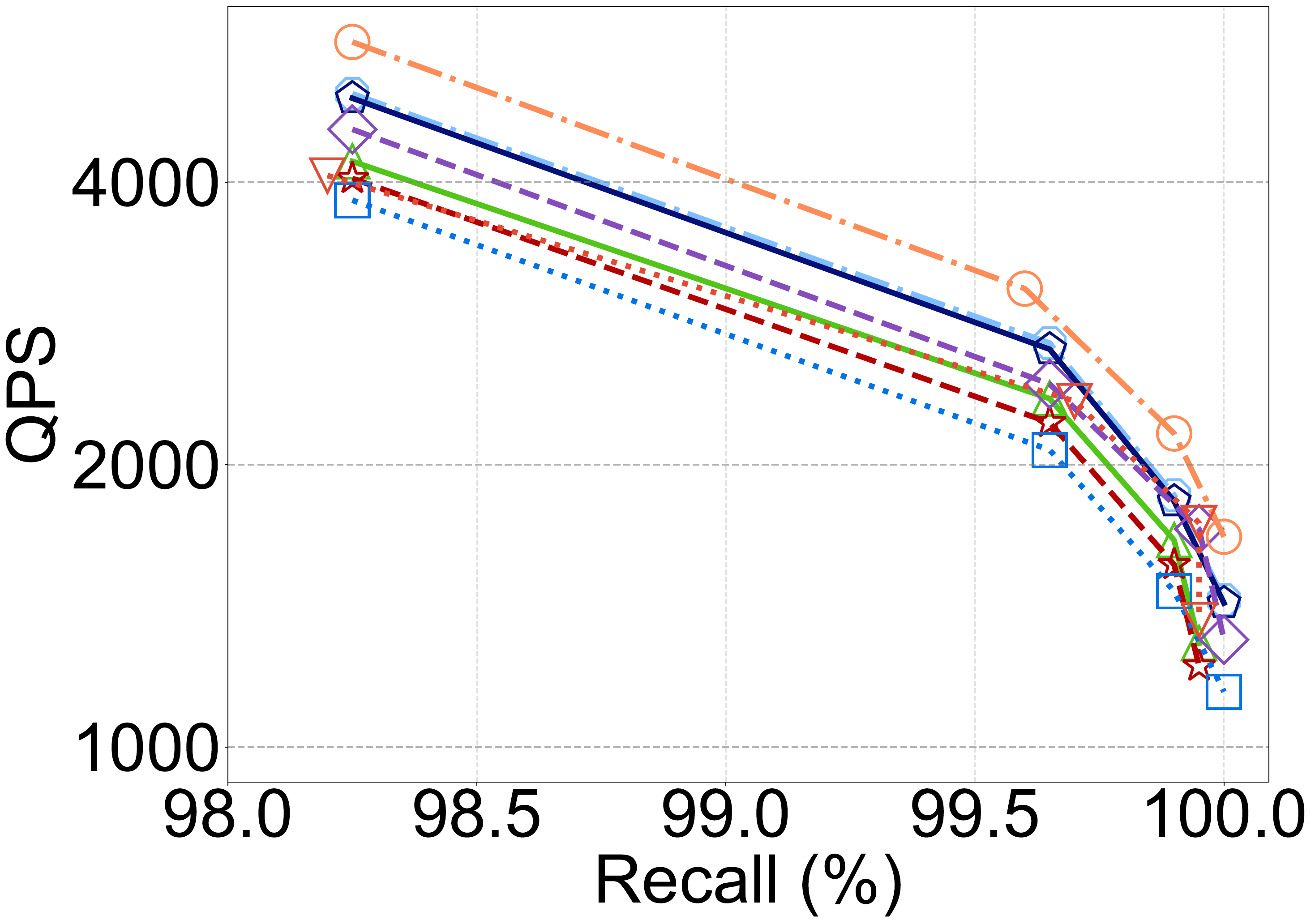}\vspace{-1.0ex}
        \caption{\Sift ($k=20$)}
        \label{fig:sift-qk-20}
    \end{subfigure}
    \begin{subfigure}{0.24\textwidth}
        \centering
        \includegraphics[width=\textwidth]{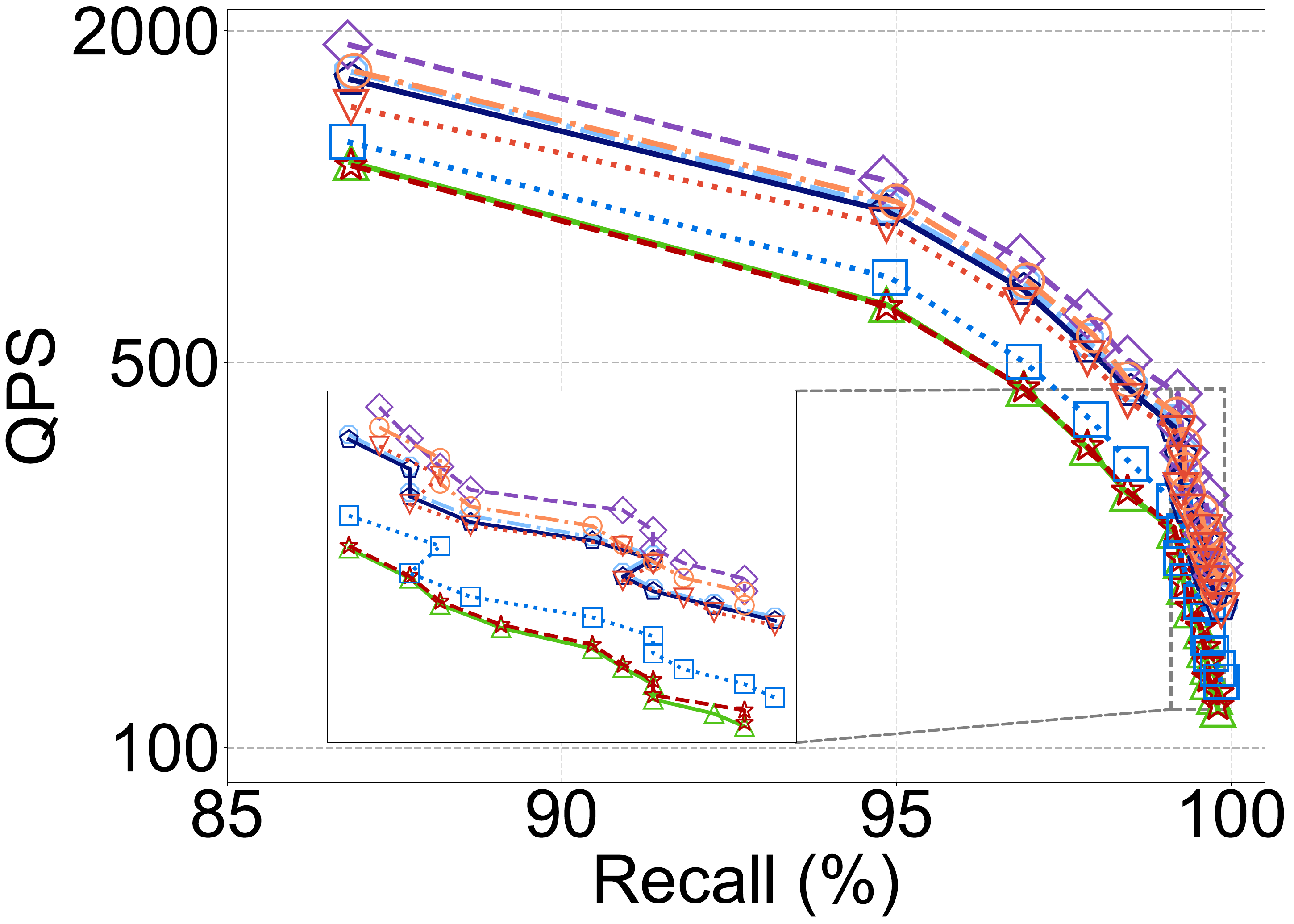}\vspace{-1.0ex}
        \caption{\Gist ($k=20$)}
        \label{fig:GIST-qk-20}
    \end{subfigure}
    \begin{subfigure}{0.24\textwidth}
        \centering
        \includegraphics[width=\textwidth]{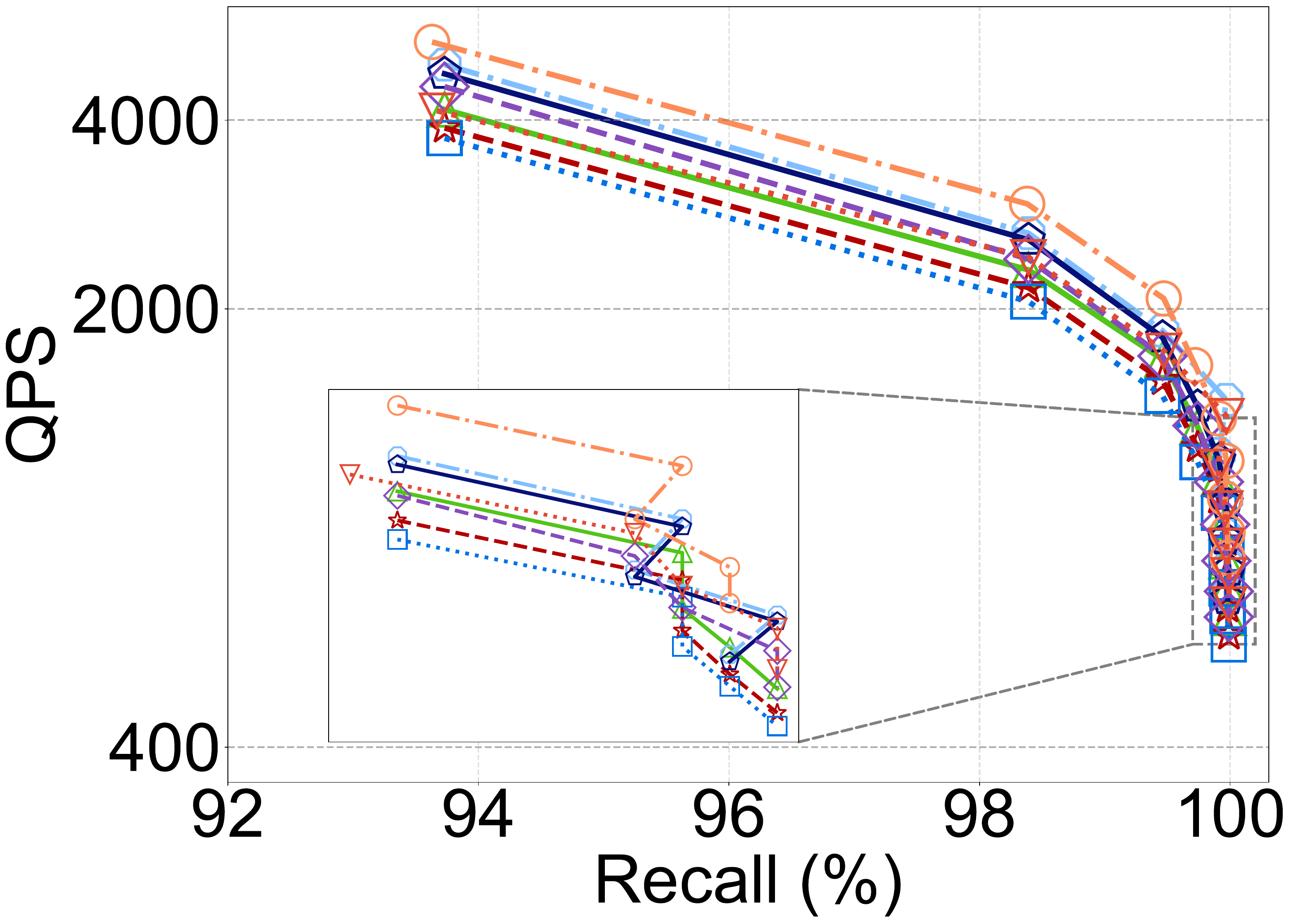}\vspace{-1.0ex}
        \caption{\Sift ($k=100$)}
        \label{fig:sift-qk-100}
    \end{subfigure}
    \begin{subfigure}{0.24\textwidth}
        \centering
        \includegraphics[width=\textwidth]{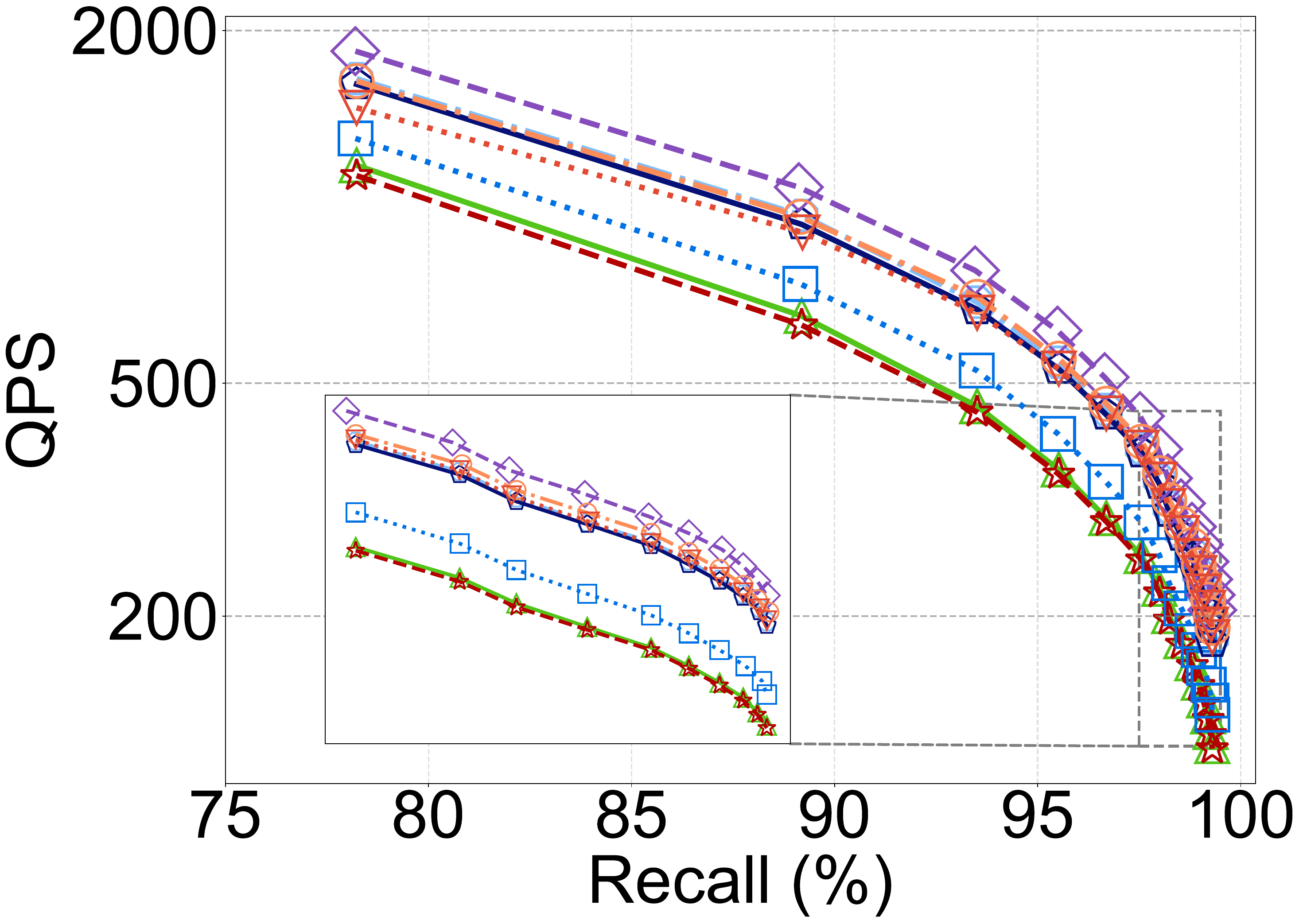}\vspace{-1.0ex}
        \caption{\Gist ($k=100$)}
        \label{subfig:GIST-k100-HNSW}
    \end{subfigure}
    \begin{subfigure}{0.25\textwidth}
        \centering
        \includegraphics[width=\textwidth]{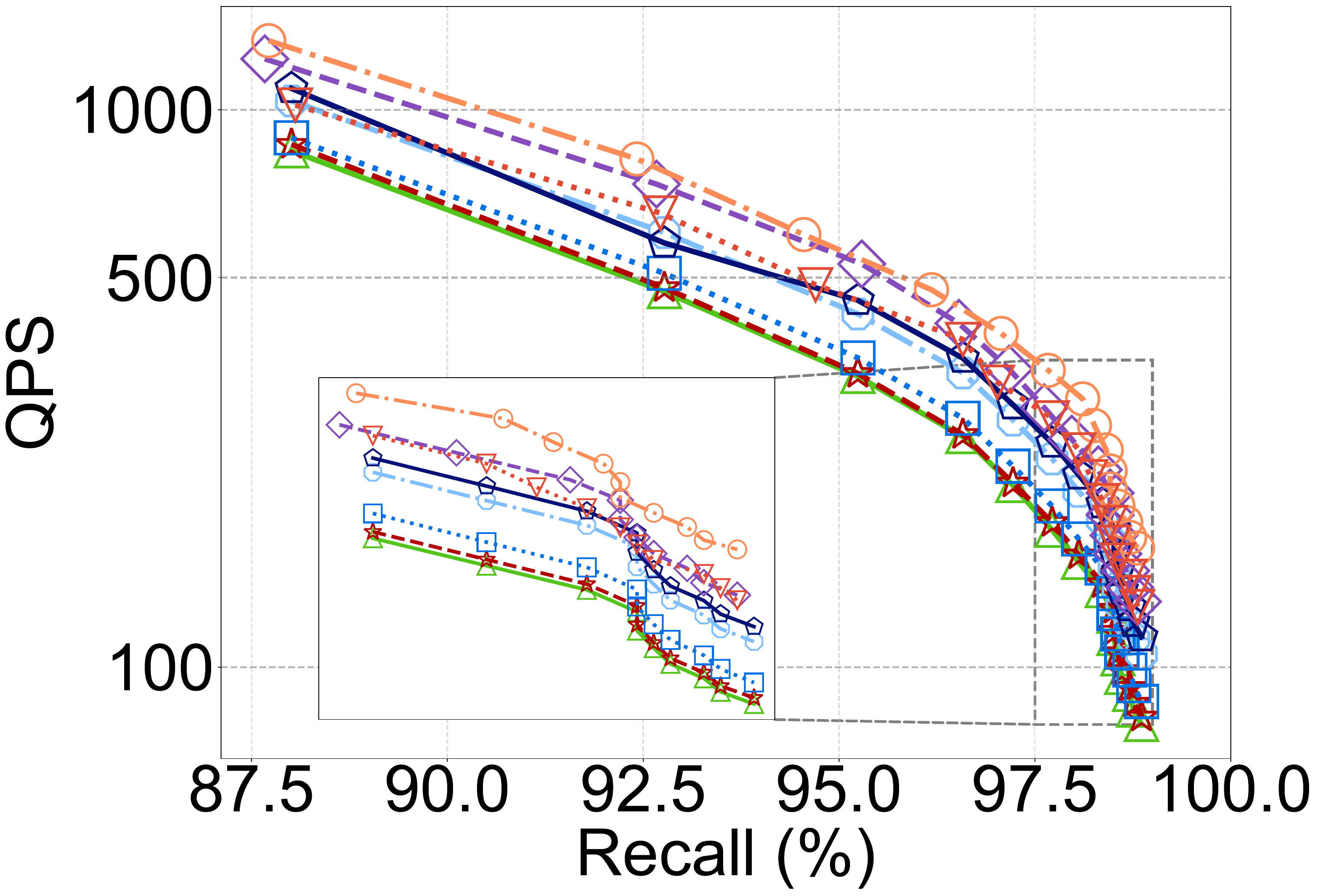}\vspace{-1.0ex}
        \caption{\Openai ($k=20$)}
    \end{subfigure}
    \begin{subfigure}{0.24\textwidth}
        \centering
        \includegraphics[width=\textwidth]{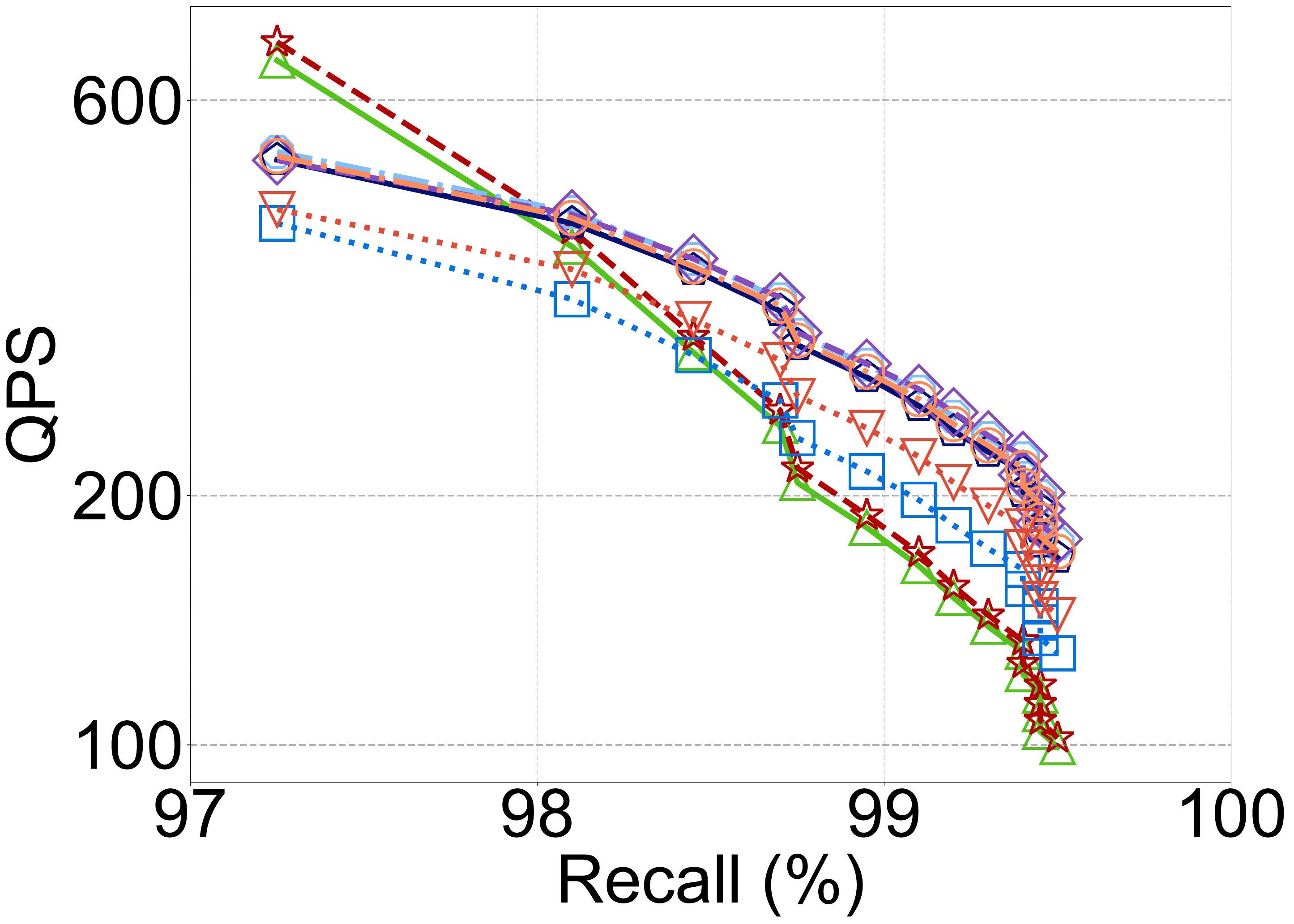}\vspace{-1.0ex}
        \caption{\Trevi ($k=20$)}
        \label{fig:trevi-qk-20}
    \end{subfigure}
    \begin{subfigure}{0.245\textwidth}
        \centering
        \includegraphics[width=\textwidth]{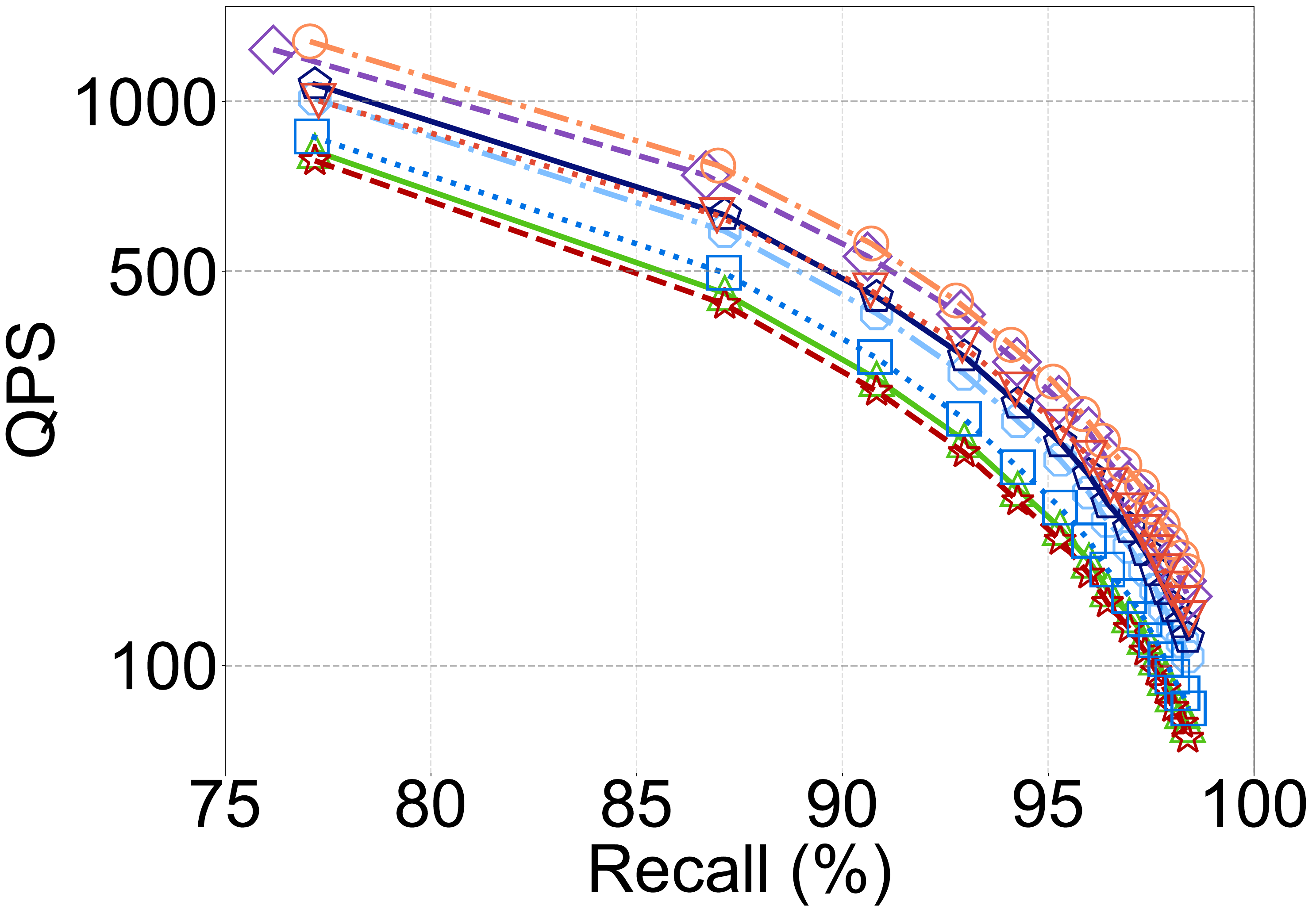}\vspace{-1.0ex}
        \caption{\Openai ($k=100$)}
    \end{subfigure}
    \begin{subfigure}{0.24\textwidth}
        \centering
        \includegraphics[width=\textwidth]{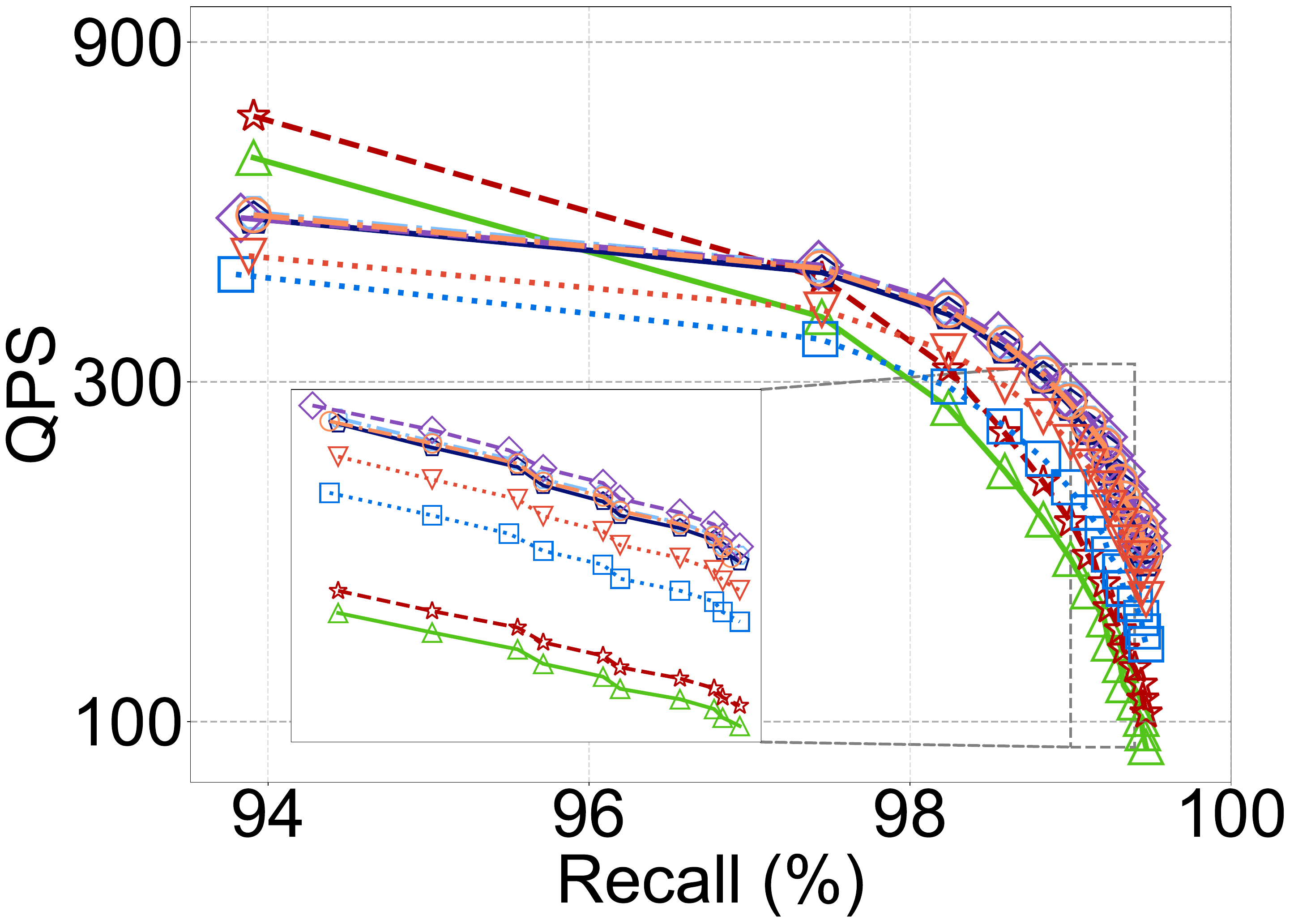}\vspace{-1.0ex}
        \caption{\Trevi ($k=100$)}
        \label{fig:trevi-qk-100}
    \end{subfigure}
    \vspace{-1.0ex}
    \caption{Query performance when DCOs are applied to vector similarity search}\label{fig:time-accuracy}
    \vspace{-2.5ex}
\end{figure*}

\vspace{-0.5ex}
\subsection{Parameter Setting}
The benchmark setup involves three types of parameters:

\textbf{Index Parameter.}
We configure HNSW with \zheng{maximum node connections} $M = 16$ and \zheng{construction candidate list size} $efConstruction = 500$, and IVF with 4096 partitions, following recommendations from Faiss \cite{douze2024faiss}.

\textbf{Search Parameter.}
For HNSW, we vary the \zheng{search candidate list size} $efSearch$ from 100 to 1500 in steps of 100.
For IVF, we vary \zheng{the probed partition count} $nprobe$ from 20 to 400 in steps of 20.
By default, $efSearch = 200$ and $nprobe = 80$.

\textbf{Query Parameter.}
We set the integer $k$ to two common values, 20 and 100, as used in prior work \cite{jayaram2019diskann, DBLP:journals/pvldb/WangXY021, DBLP:conf/nips/ZhangWCCZMHDMWP23}.

\vspace{-0.5ex}
\subsection{Evaluation Metric}
We assess DCOs from the following four perspectives:
% , where \textit{higher values indicate better performance}:
\begin{itemize}
    \item \textbf{Effectiveness}: Evaluated via \textbf{recall} defined in \defref{equ:recall}.
    
    \item \textbf{Query Throughput}: Measured by queries per second (\textbf{QPS}). As each query typically invokes the DCO multiple times, QPS represents the overall efficiency of DCOs.
    
    \item \textbf{Pruning Capability}: Quantified by the \textbf{dimension pruning ratio}, \ie the fraction of dimensions saved.

    \item \textbf{Construction \& Insertion Overhead}: Measured by \textbf{index construction time} and \textbf{data insertion time}.
\end{itemize}
Following prior work \cite{DBLP:journals/pacmmod/GaoL23, DBLP:journals/pvldb/DengCZWZZ24, yang2025effective}, we plot QPS–recall curves by varying the index search parameters $efSearch$ and $nprobe$ for HNSW and IVF, respectively. 
Increasing these parameters raises recall while concurrently decreasing QPS.

\vspace{-0.5ex}
\subsection{Experimental Environment}
Experiments are conducted on a server with an AMD Ryzen 9 5950X CPU, 128GB RAM, and an NVIDIA GeForce RTX 3090 GPU.
The server ran Ubuntu 24.04.2 with CUDA 12.4.

\vspace{-2.5ex}
\section{Experimental Evaluation}\label{sec:result}
\vspace{-0.5ex}

This section presents our experimental results and analysis.
Due to page limits, we show only selected results here.
Please refer to the full paper \cite{fullpaper} for  complete results (\eg the parameter study on the number of scan dimensions per round and the evaluation using the AVX2 SIMD instruction set).

% Due to page limits, the parameter study on the number of scan dimensions per round is provided in our full paper \cite{fullpaper}.

\vspace{-1.0ex}
\subsection{Query Processing}\label{sec:query-processing}
\begin{figure}[t]
    \centering
    \includegraphics[width=0.44\textwidth]{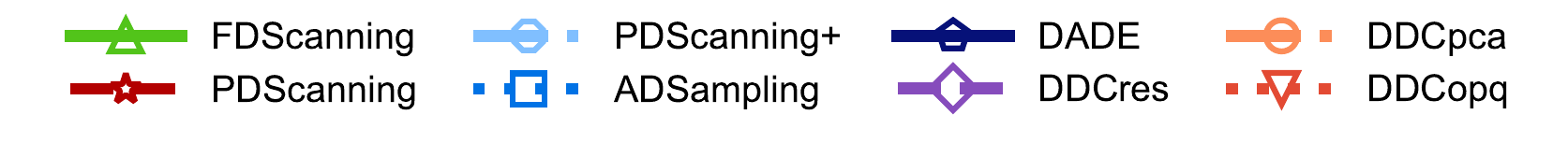}
    \vspace{-3ex}
    \begin{subfigure}{0.22\textwidth}
        \centering
        \includegraphics[width=\textwidth]{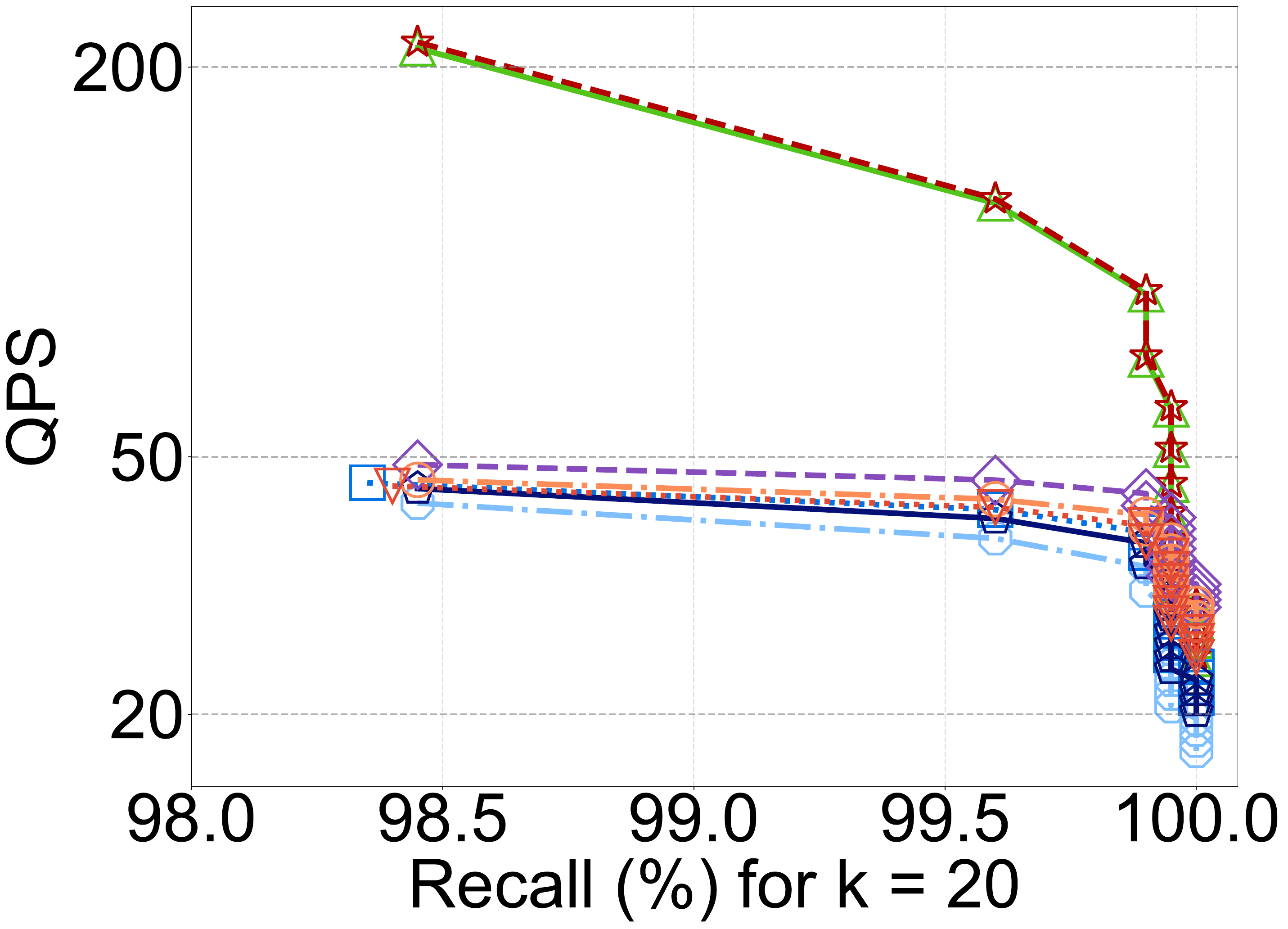}
        % \vspace{-4.5ex}
        % \caption{$k=20$}
    \end{subfigure}
    \begin{subfigure}{0.22\textwidth}
        \centering
        \includegraphics[width=\textwidth]{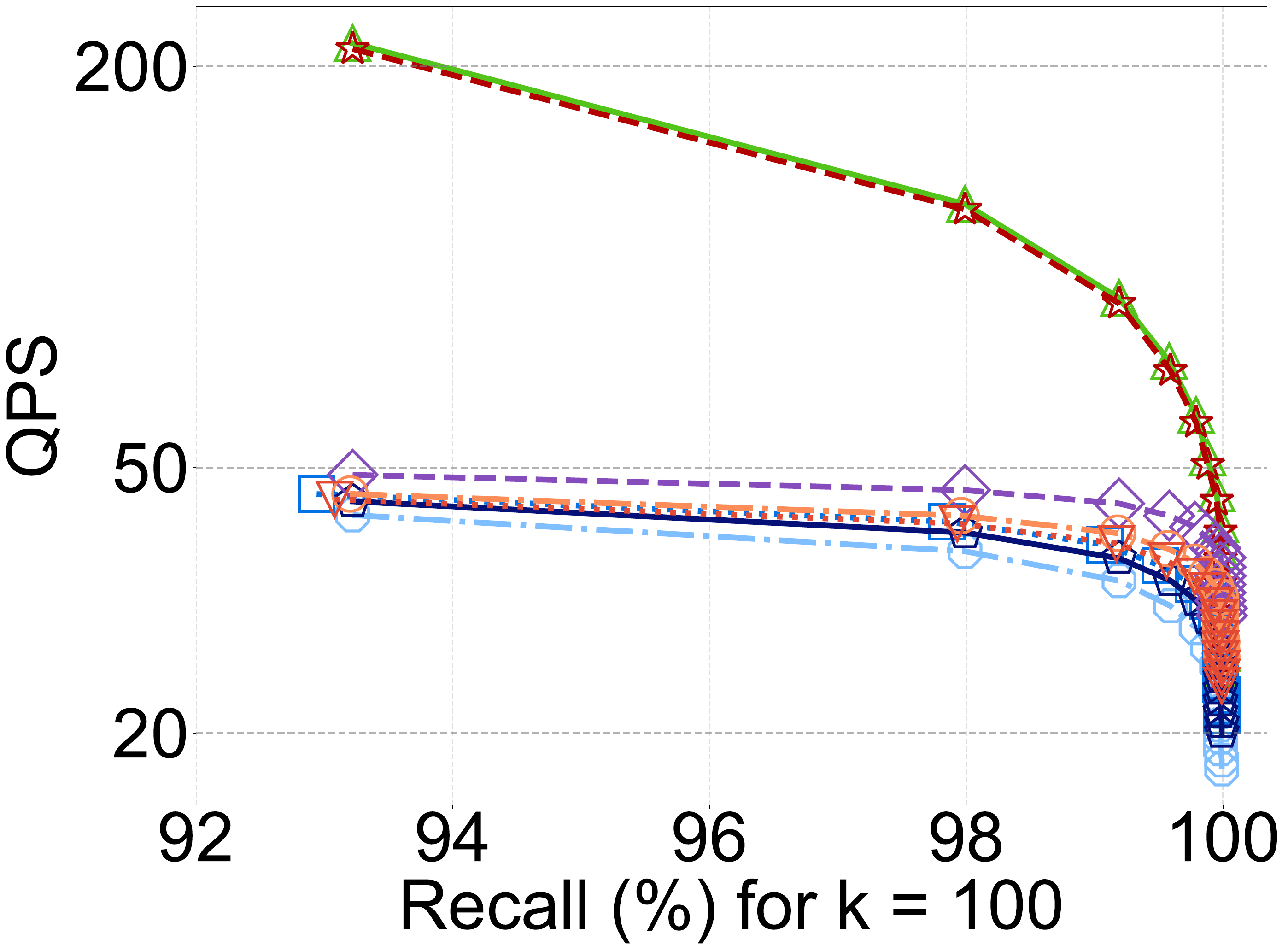}
        % \vspace{-4.5ex}
        % \caption{$k=100$}
    \end{subfigure}
    \vspace{1.5ex}
    \caption{Performance on ultra-high-dimensional dataset \Msmacro}\label{fig:msmacro-qk}
    % \vspace{-0.5ex}
\end{figure}

This experiment evaluates the performance improvement from applying DCOs to vector similarity search for $k = 20$ and $k = 100$.
We use seven datasets with varying dimensions (96 to 12,288) and cardinalities (99,900 to 100 million) to ensure a comprehensive assessment.
\figref{fig:time-accuracy} and \ref{fig:msmacro-qk} present time--recall curves to demonstrate the effectiveness of DCO methods. 

\fakeparagraph{Overall Query Performance}
The results show that in only 4 out of the 14 subfigures, the \Hypothesis and \Class methods \textit{consistently outperform} the baseline \FDScanning.
Specifically, compared to \FDScanning, the \Hypothesis methods improve QPS by up to $1.4$--$1.9\times$, while \DDCpca and \DDCopq achieve improvements of up to $1.6$--$2.1\times$.
These results demonstrate the superior enhancement in time efficiency when applying DCOs to vector similarity search, while their recall remains largely stable with at most $2\%$ change.

\begin{figure}[t]
    \centering
    \begin{subfigure}{0.22\textwidth}
        \centering
        \includegraphics[width=\textwidth]{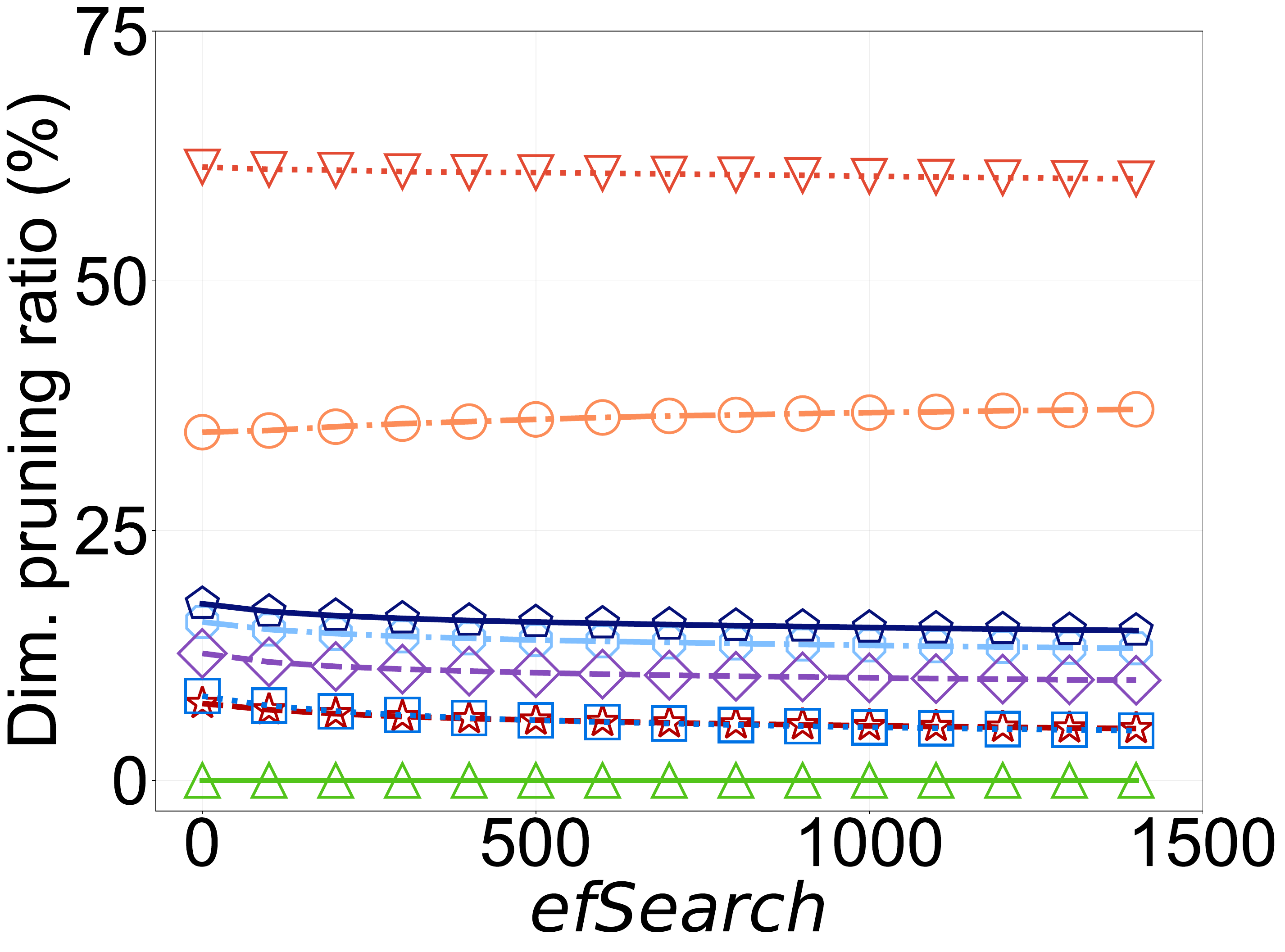}
        \vspace{-3.5ex}
        \caption{\Deep}
    \end{subfigure}
    \begin{subfigure}{0.23\textwidth}
        \centering
        \includegraphics[width=\textwidth]{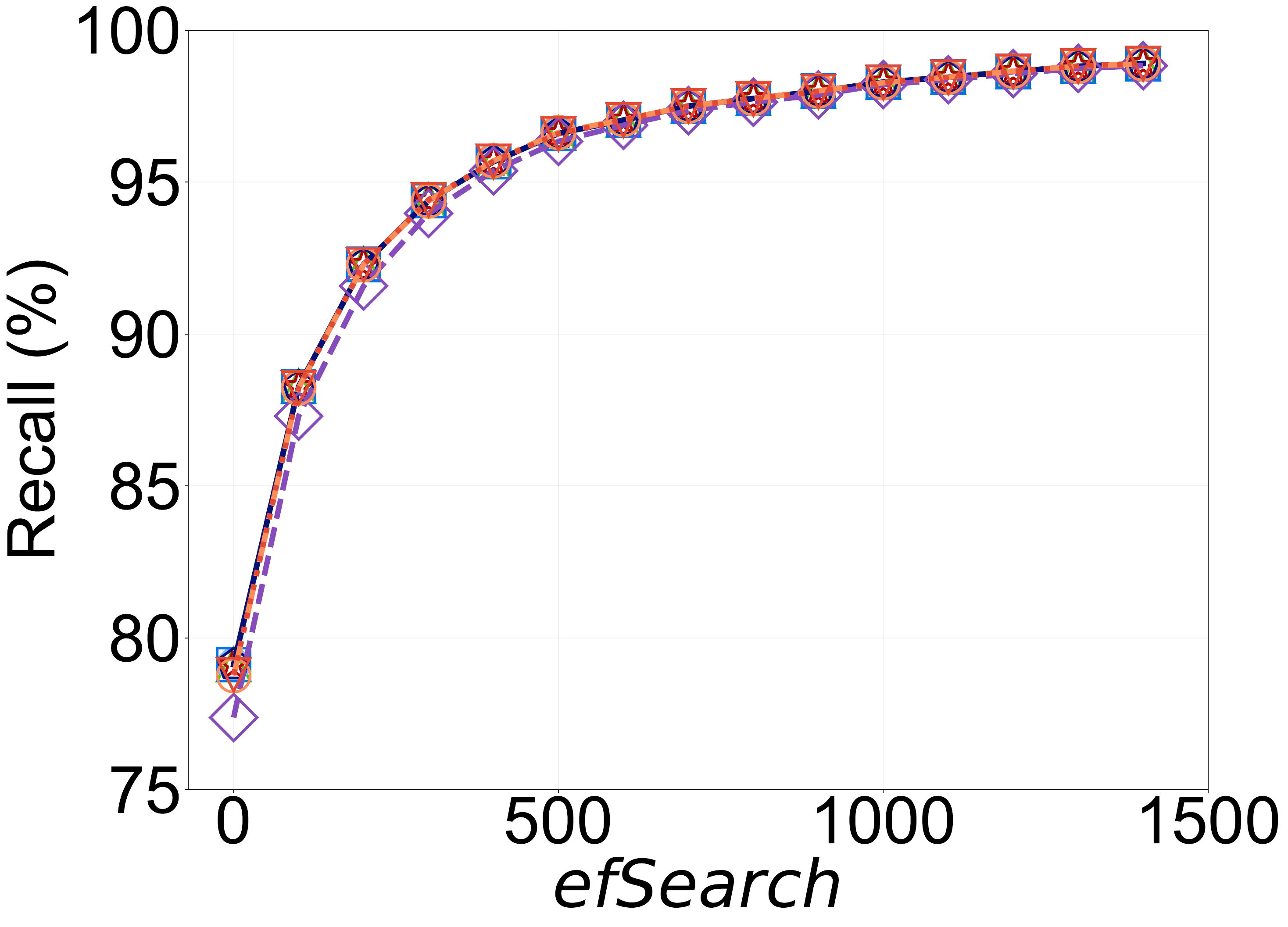}
        \vspace{-3.5ex}
        \caption{\Deep}
    \end{subfigure}
    \begin{subfigure}{0.22\textwidth}
        \centering
        \includegraphics[width=\textwidth]{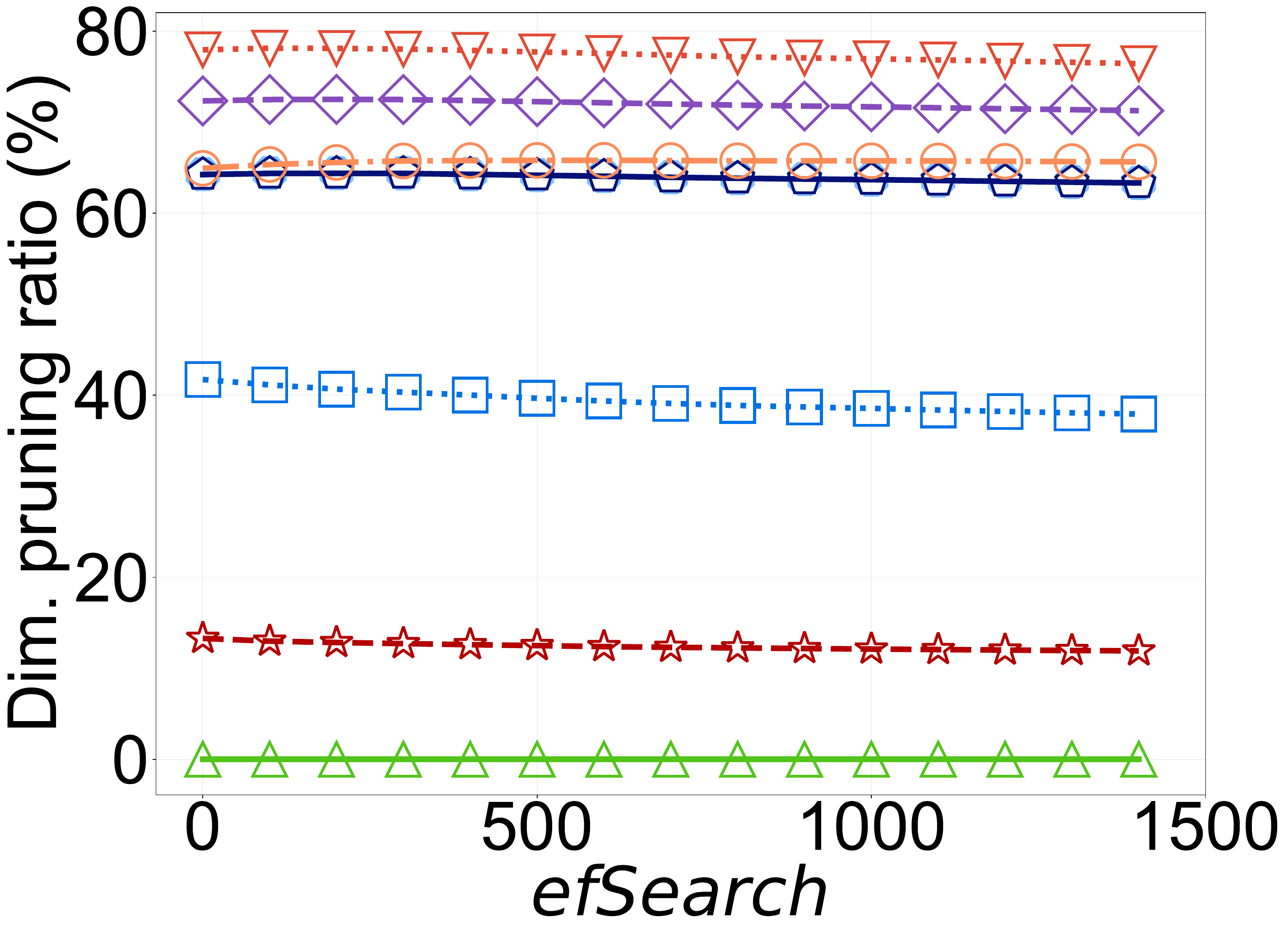}
        \vspace{-3.5ex}
        \caption{\Gist}
    \end{subfigure}
    \begin{subfigure}{0.23\textwidth}
        \centering
        \includegraphics[width=\textwidth]{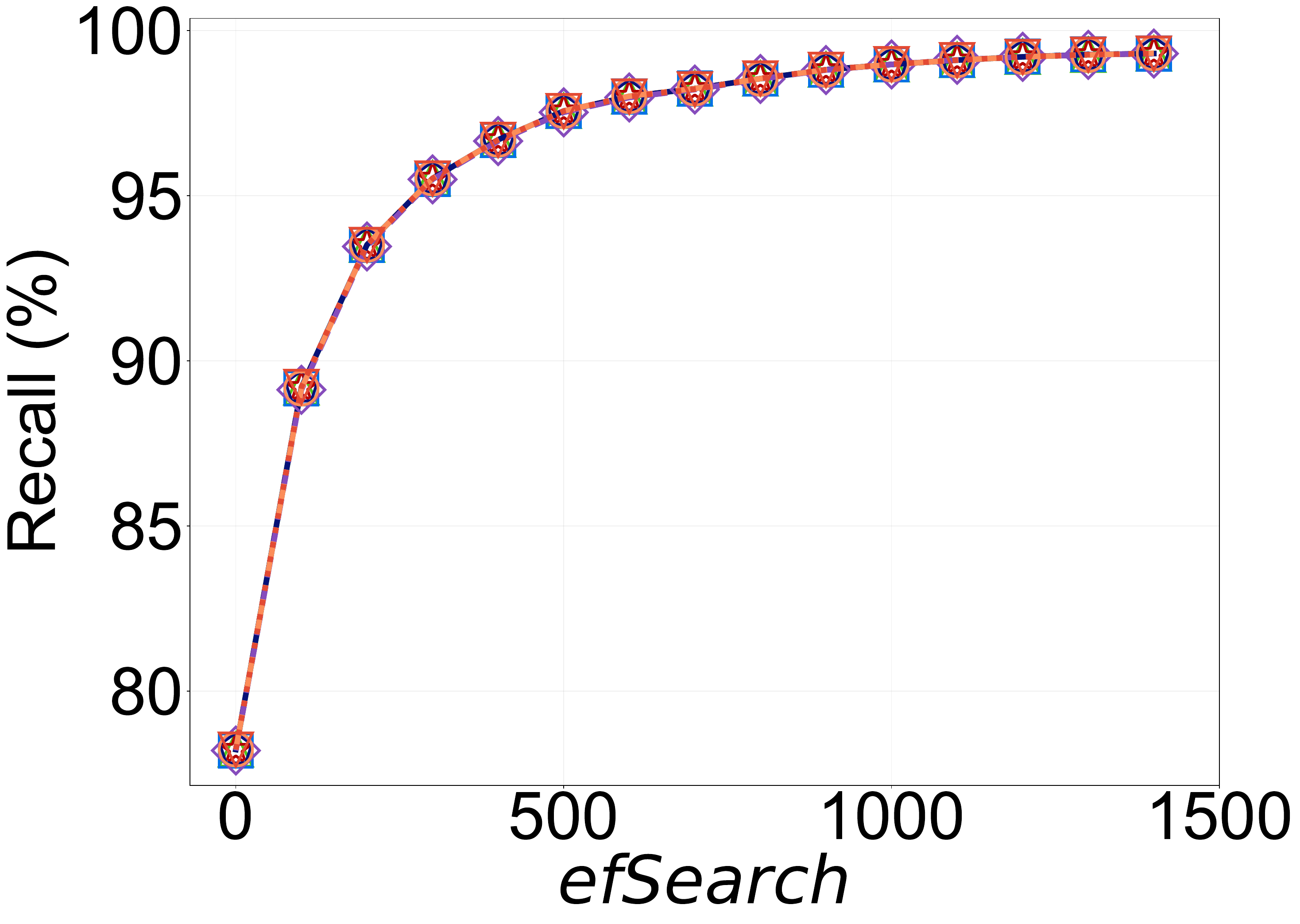}
        \vspace{-3.5ex}
        \caption{\Gist}
    \end{subfigure}
    \begin{subfigure}{0.22\textwidth}
        \centering
        \includegraphics[width=\textwidth]{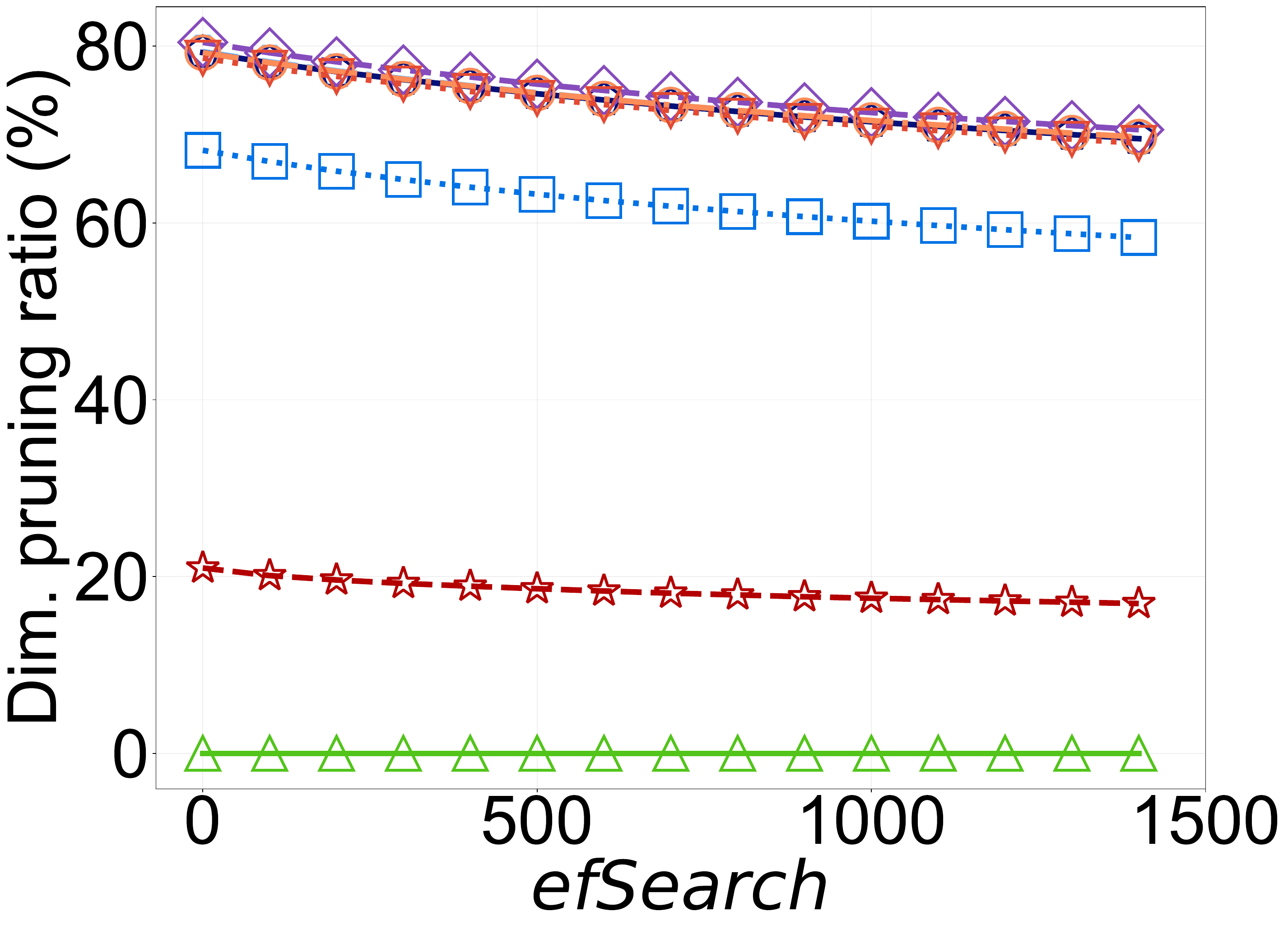}
        \vspace{-3.5ex}
        \caption{\Trevi}
    \end{subfigure}
    \begin{subfigure}{0.23\textwidth}
        \centering
        \includegraphics[width=\textwidth]{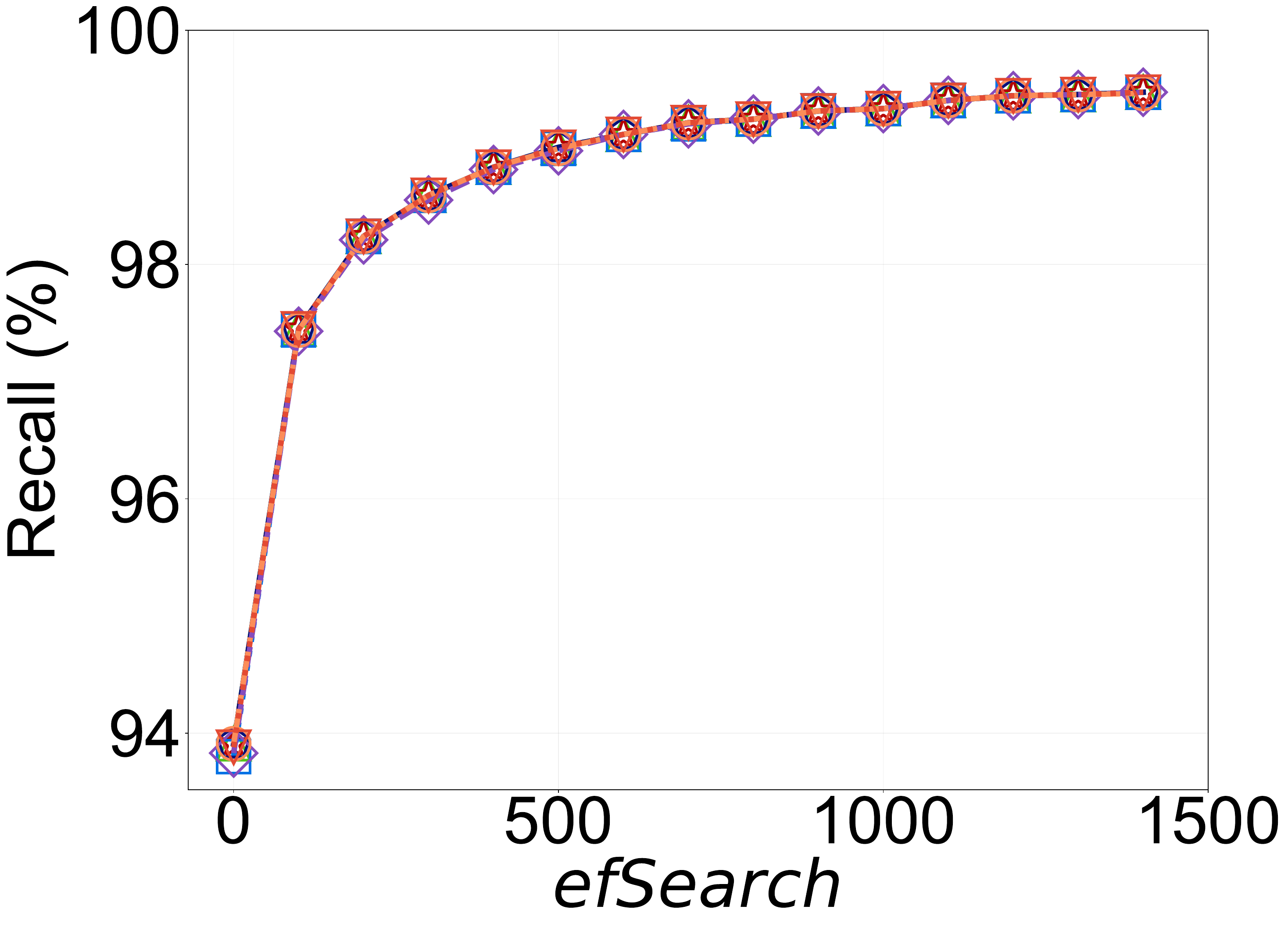}
        \vspace{-3.5ex}
        \caption{\Trevi}
    \end{subfigure}
    \vspace{-1ex}
    \caption{Dimension pruning via DCO and its impact on recall}\label{fig:dco-pruning}
    % \vspace{-1ex}
\end{figure}

We also observe cases where naive baselines (\eg \FDScanning) perform better than the others.
For instance, \figref{fig:deep-qk-20} and \ref{fig:deep-qk-100} show that \FDScanning usually achieves the highest QPS on the \Deep dataset. 
Similarly, \figref{fig:trevi-qk-20} and \ref{fig:trevi-qk-100} show that \PDScanning obtains the optimal efficiency for recall until 97\% and 94\% respectively on the \Trevi dataset.

The SOTA methods become ineffective on the \Deep and \Trevi datasets for different reasons.
On the low-dimensional dataset \Deep, the time saved by skipping $6$\%--$36$\% of dimensions is marginal (except for \DDCopq), 
\zheng{since the accumulated overhead of these methods including distance estimation, threshold checking, and conditional branching outweighs the benefits of dimension pruning.}
On the ultra-high-dimensional dataset \Trevi, the $O(D^2)$ online pre-processing time per query becomes the efficiency bottleneck.
This cost is not amortized effectively before the recall reaches 94\%, where DCOs are only invoked for $O(\log{N})$ candidate vectors on HNSW \cite{DBLP:journals/pami/MalkovY20}.

This phenomenon becomes increasingly pronounced at higher dimensionalities.
\figref{fig:msmacro-qk} shows the result on the \Msmacro dataset which has the highest dimensionality.
Here, \FDScanning and \PDScanning achieve higher QPS than all SOTA methods for both $k$ values. % when recall is blew 99.95\%
The QPS of these SOTA methods decreases by up to 77\%--79\%$\times$ compared to \FDScanning.

% We can also observe that the performance of DCO methods remains stable across different $k$ values. 
% For example, on the \Openai dataset, \DDCres achieves up to a $1.7\times$ QPS improvement over \FDScanning for both $k=20$ and $k=100$.

% @ while \DDCres performs best on the other two datasets with higher dimensions.
\fakeparagraph{Pruning Capability of DCO}
\figref{fig:dco-pruning} illustrates the dimension pruning ratios of DCOs and their recall across three datasets of varying dimensionality. 
The results demonstrate that the pruning capabilities of DCOs are dimension-dependent: \DDCopq achieves the highest pruning ratio on the \Deep and \Gist datasets, while \DDCres performs best on the other dataset with higher dimensions.
On \Deep, the pruning capability of \ADSampling is nearly equivalent to that of \PDScanning, which is consistent with its suboptimal performance in \figref{fig:time-accuracy}.
Moreover, all methods maintain a comparable level of recall. 
This consistent result indicates that the minor approximation errors of the DCO solutions do not compromise recall.
% By a way, \DDCopq incurs additional overhead from computing quantized distances, which results in relatively poor actual performance despite its high dimension pruning ratio. Aside from \DDCopq, 

\newcounter{takeaway}
\setcounter{takeaway}{1}
\begin{tipbox}
\textbf{Takeaway \#\thetakeaway: DCO is no silver bullet.} 
While adept at skipping unnecessary dimensions with minimal recall loss, the SOTA DCO methods do not universally boost efficiency. 
Our evaluation shows that they can even be slower than the native baseline \FDScanning, especially on low- and ultra-high-dimensional datasets.
\end{tipbox}

\begin{figure}[t]
    \centering
    \includegraphics[width=0.44\textwidth]{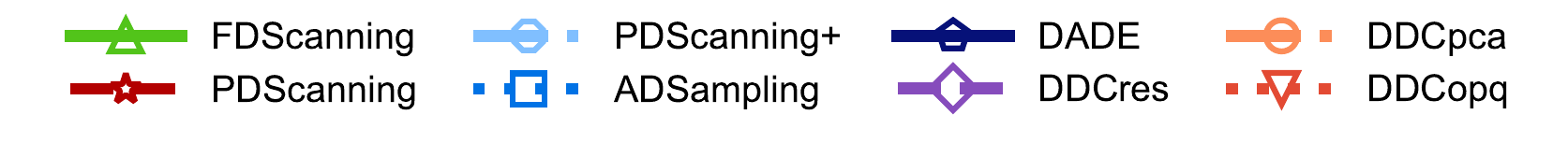}
    % \vspace{-2.5ex}
    \begin{subfigure}{0.24\textwidth}
        \centering
        \includegraphics[width=\textwidth]{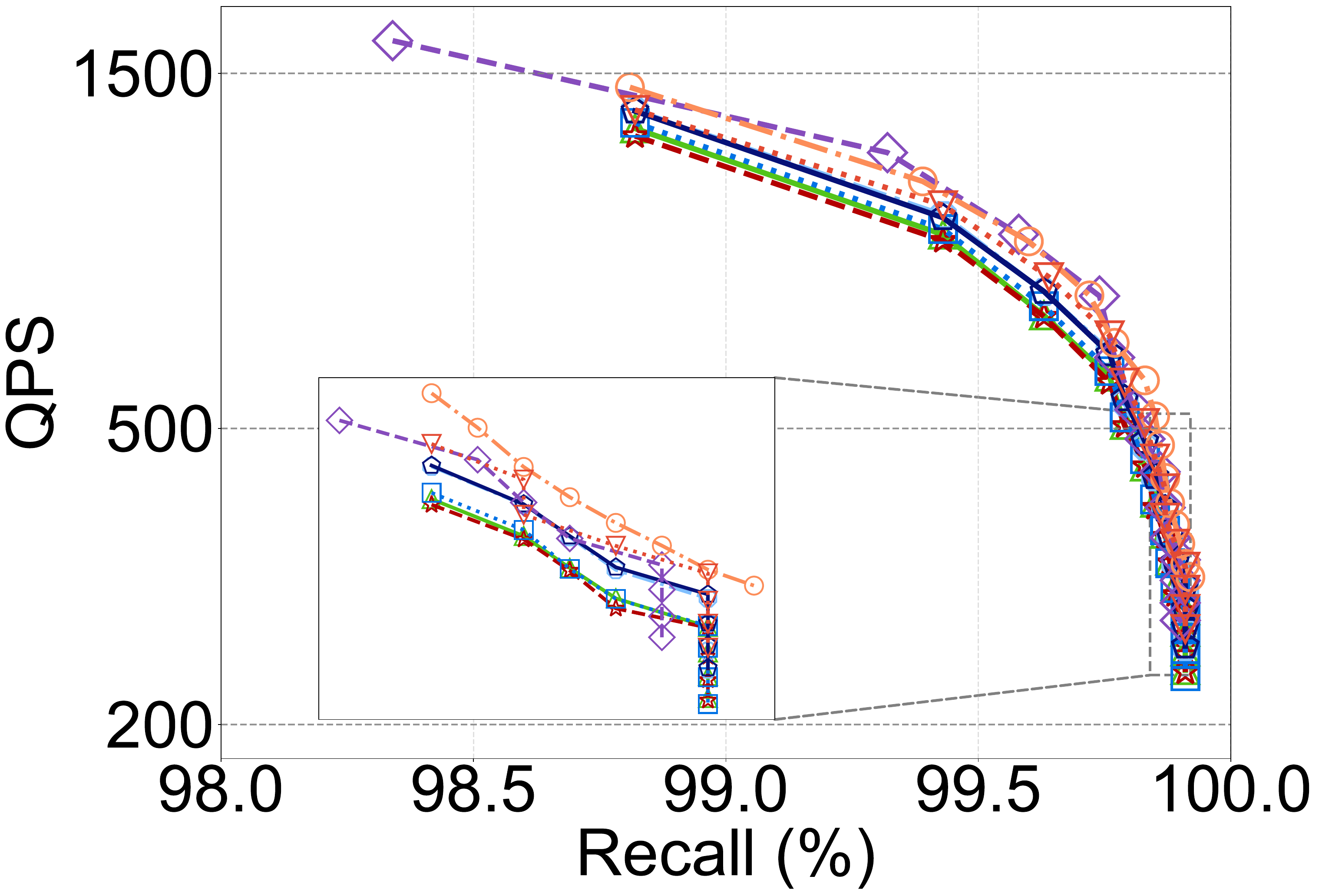}
        \vspace{-3.5ex}
        \caption{\Laion (in-distribution)}\label{fig:laion-id}
    \end{subfigure}
    \begin{subfigure}{0.238\textwidth}
        \centering
        \includegraphics[width=\textwidth]{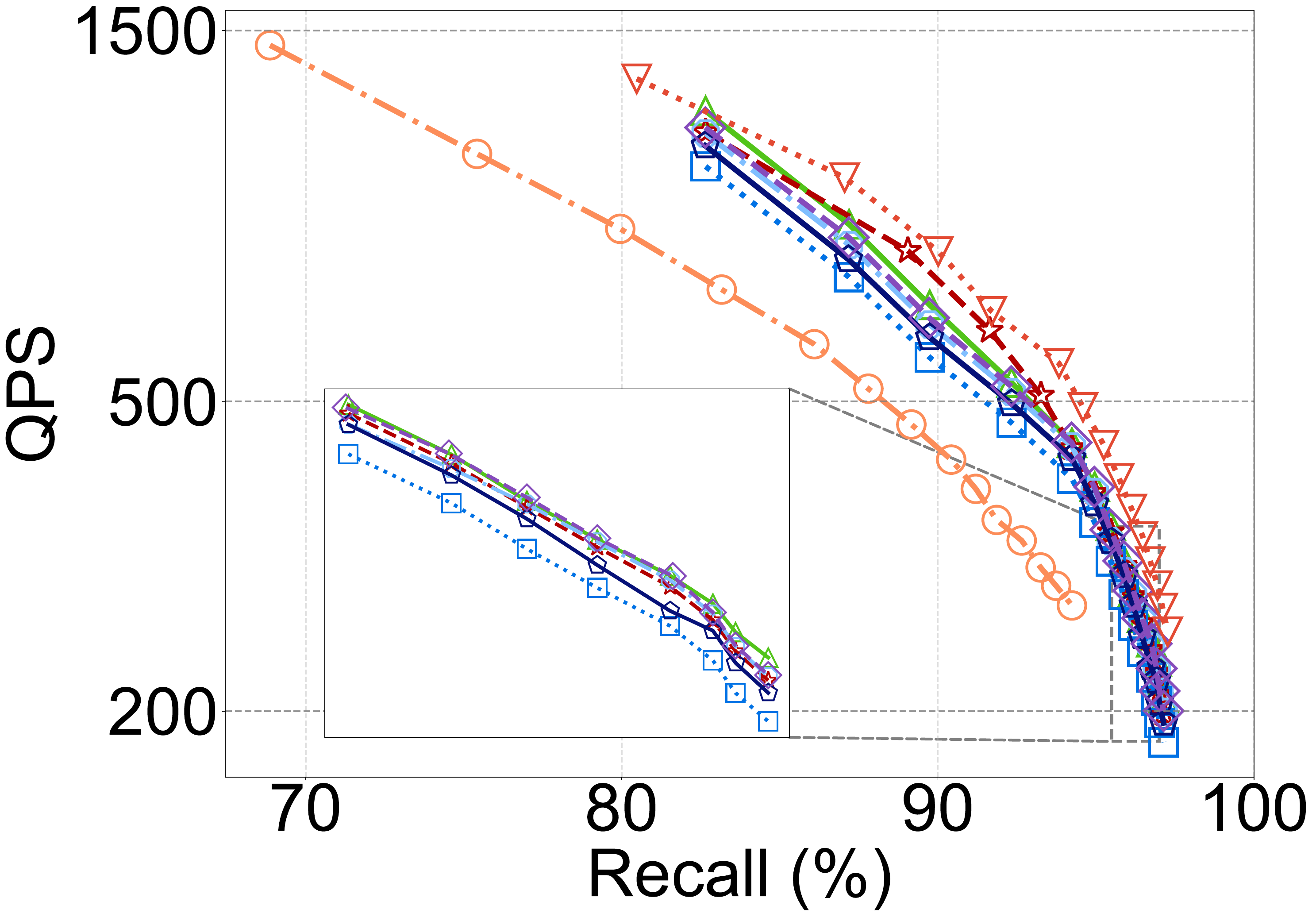}
        \vspace{-3.5ex}
        \caption{\Laion (OOD)}\label{fig:laion-ood}
    \end{subfigure}
    \begin{subfigure}{0.24\textwidth}
        \centering
        \includegraphics[width=\textwidth]{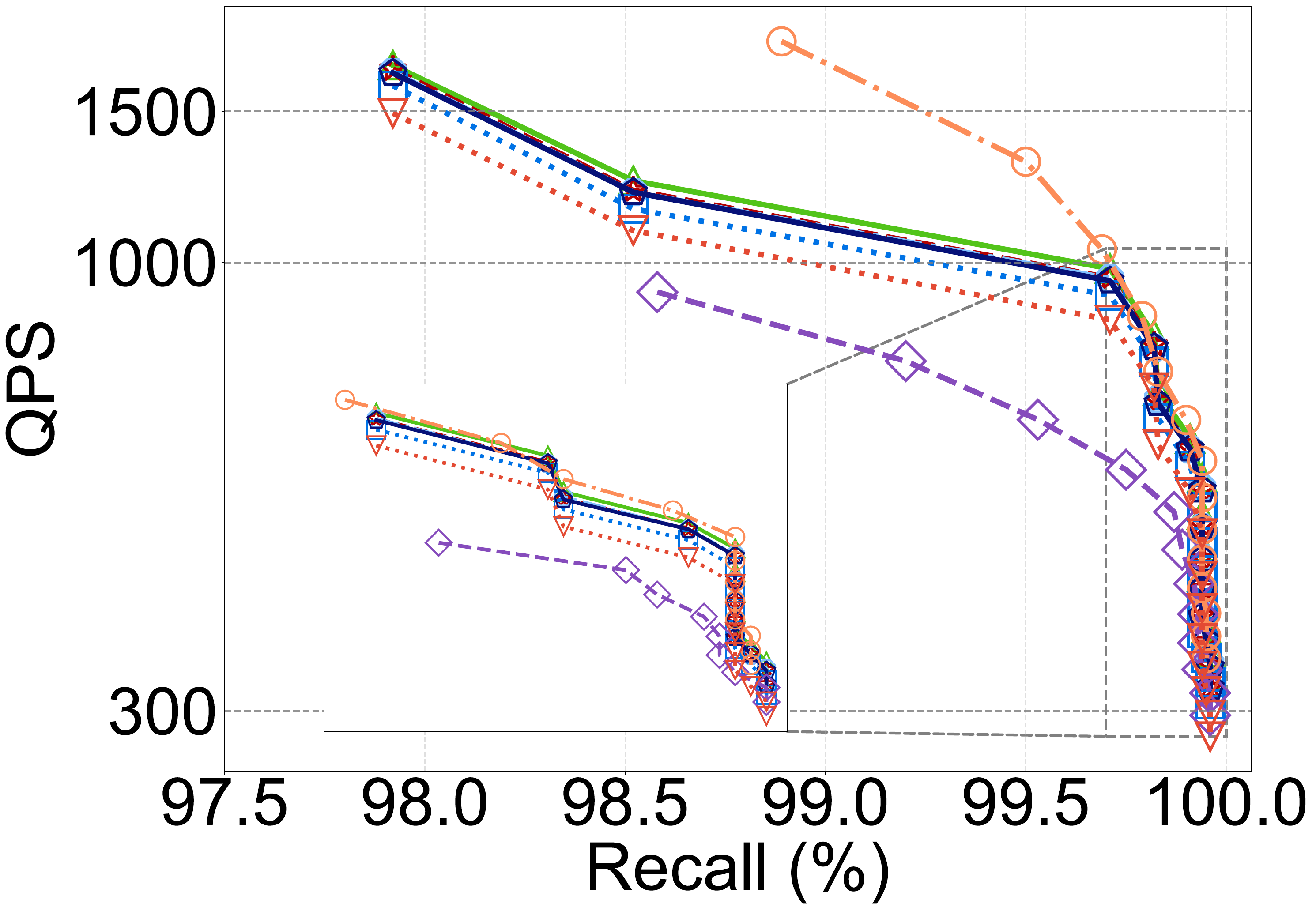}
        \vspace{-3.5ex}
        \caption{\TextImage (in-distribution)}\label{fig:textimage-id}
    \end{subfigure}
    \begin{subfigure}{0.24\textwidth}
        \centering
        \includegraphics[width=\textwidth]{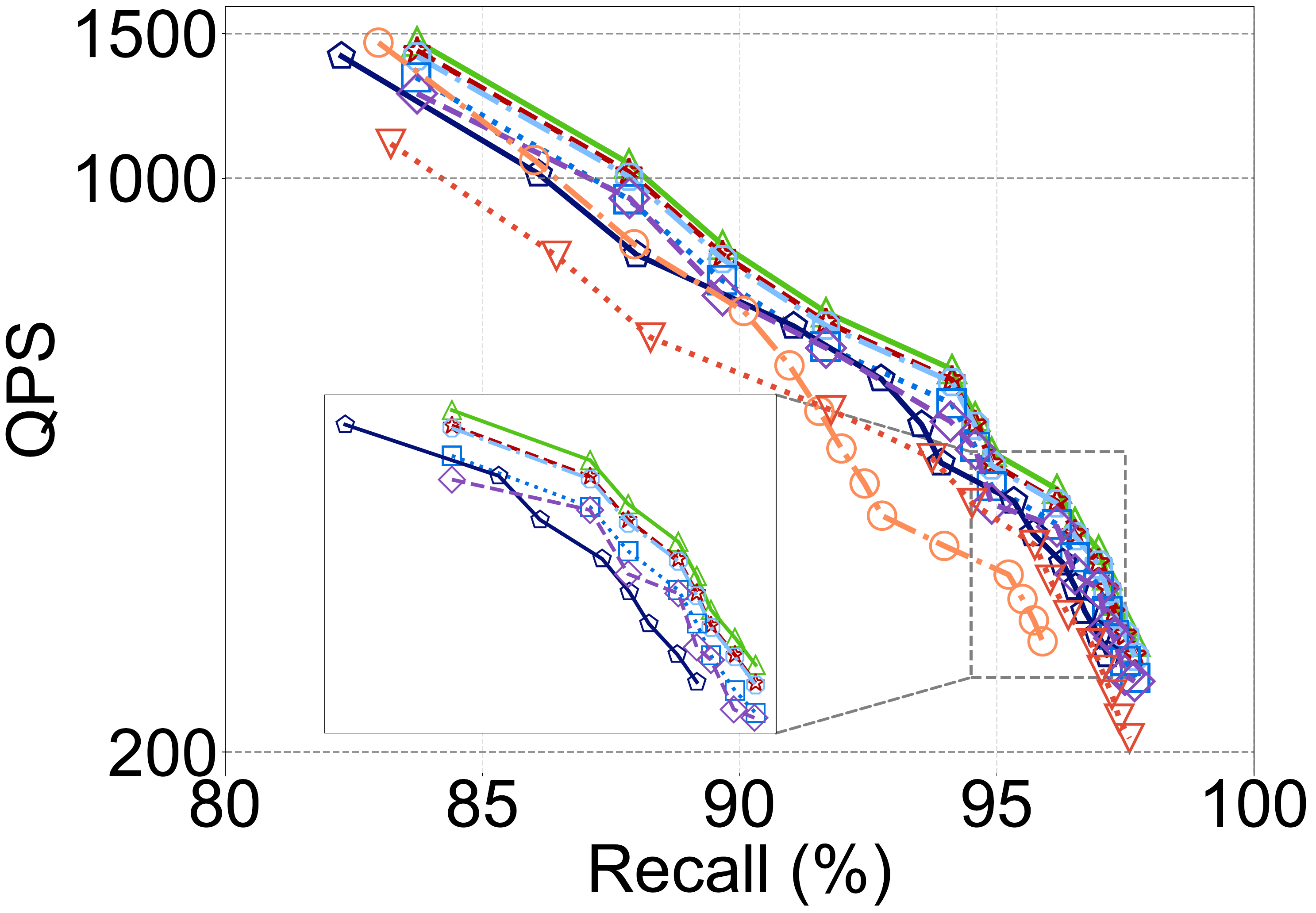}
        \vspace{-3.5ex}
        \caption{\TextImage (OOD)}\label{fig:textimage-ood}
    \end{subfigure}
    \vspace{-3.5ex}
    \caption{Results on in-distribution and OOD queries}\label{fig:out-of-distribution}
    \vspace{-0.5ex}
\end{figure}

\vspace{-1ex}
\subsection{Out-of-Distribution (OOD) Queries}
% Out-of-distribution (OOD) queries are commonly seen in real-world applications like multimodal retrieval.
To evaluate the robustness of DCOs for OOD queries, we conduct experiments on two text-to-image datasets: \Laion and \TextImage.
We also include in-distribution queries by sampling query vectors directly from image embeddings.

In \figref{fig:out-of-distribution}, \FDScanning and \PDScanning, consistently achieve higher QPS than most SOTA methods on OOD queries.
For example, \DDCpca achieves up to $1.5\times$ higher QPS than \FDScanning on in-distribution queries, its performance degrades significantly on OOD queries, becoming up to $1.6\times$ slower than it. 
\zheng{This is primarily due to two factors: (1) the reduced prediction accuracy of \DDCpca necessitates more candidate vectors to maintain recall, and (2) the dimension pruning ratio of most DCO methods drops by over 60\% on OOD queries.}
By contrast, \DDCopq achieves the highest QPS on the \Laion dataset.
This advantage stems from its use of quantized distances, which maintain high prediction accuracy (over 95\%) while also being accelerated by SIMD instructions.

Furthermore, even for in-distribution queries, \FDScanning and \PDScanning outperform \RPDScanning, \ADSampling, \DADE, \DDCres, and \DDCopq in QPS.
This result confirms the sensitivity of most DCO methods to data dimensionality.

\stepcounter{takeaway}
\vspace{-0.5ex}
\begin{tipbox}
\textbf{Takeaway \#\thetakeaway: SOTA methods often fail to maintain efficiency gains on Out-of-Distribution (OOD) queries.} 
In particular, no SOTA algorithms consistently achieve higher QPS than \FDScanning on multimodal datasets. 
\end{tipbox}
\vspace{-3ex}

\begin{figure}[htbp]
    \centering
    \includegraphics[width=0.44\textwidth]{figure/GPU/legend.pdf}
    \begin{subfigure}{0.23\textwidth}
        \centering
        \includegraphics[width=\textwidth]{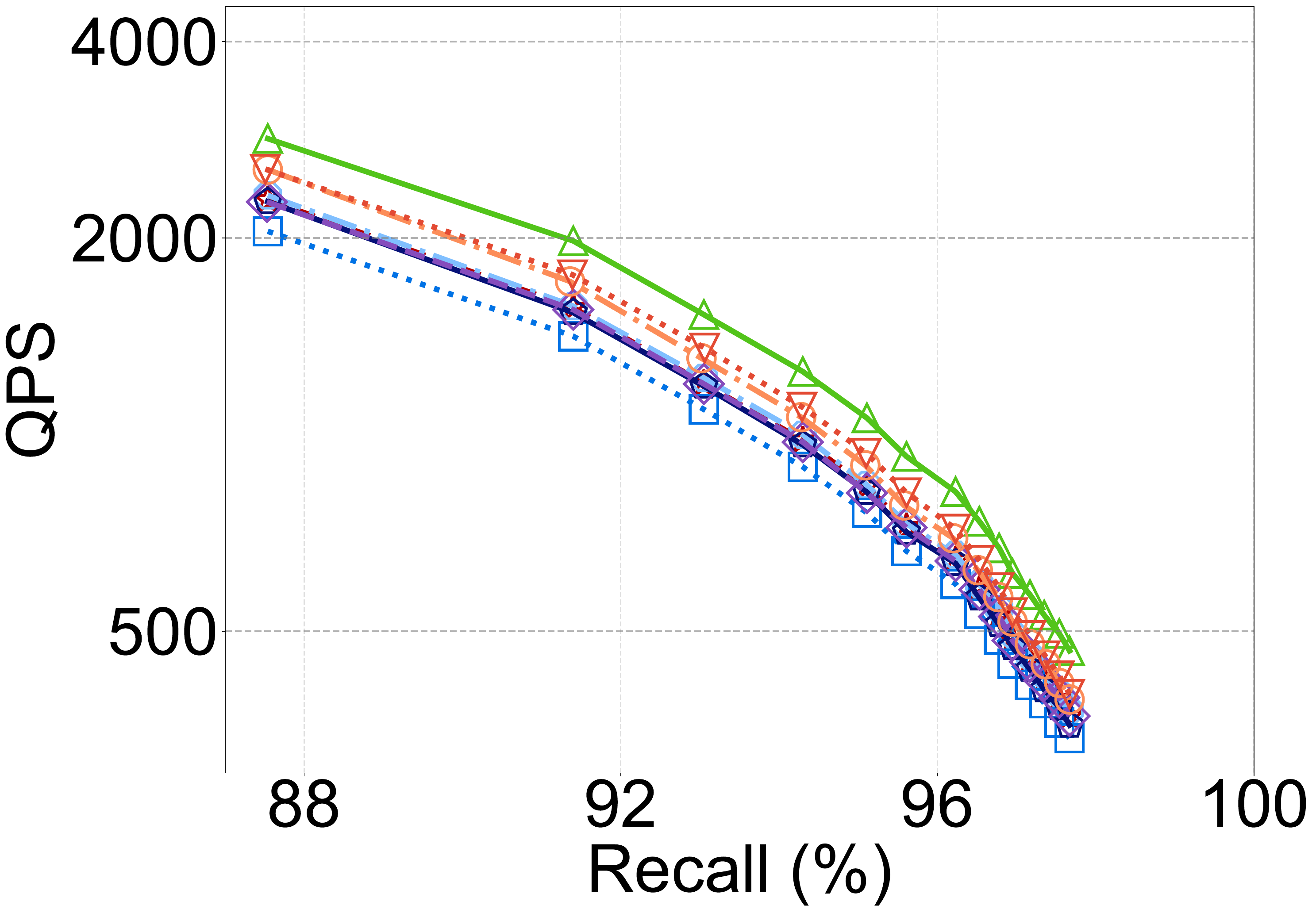}
        \vspace{-3.5ex}
        \caption{\Glove}\label{fig:IP-glove}
    \end{subfigure}
    \begin{subfigure}{0.23\textwidth}
        \centering
        \includegraphics[width=\textwidth]{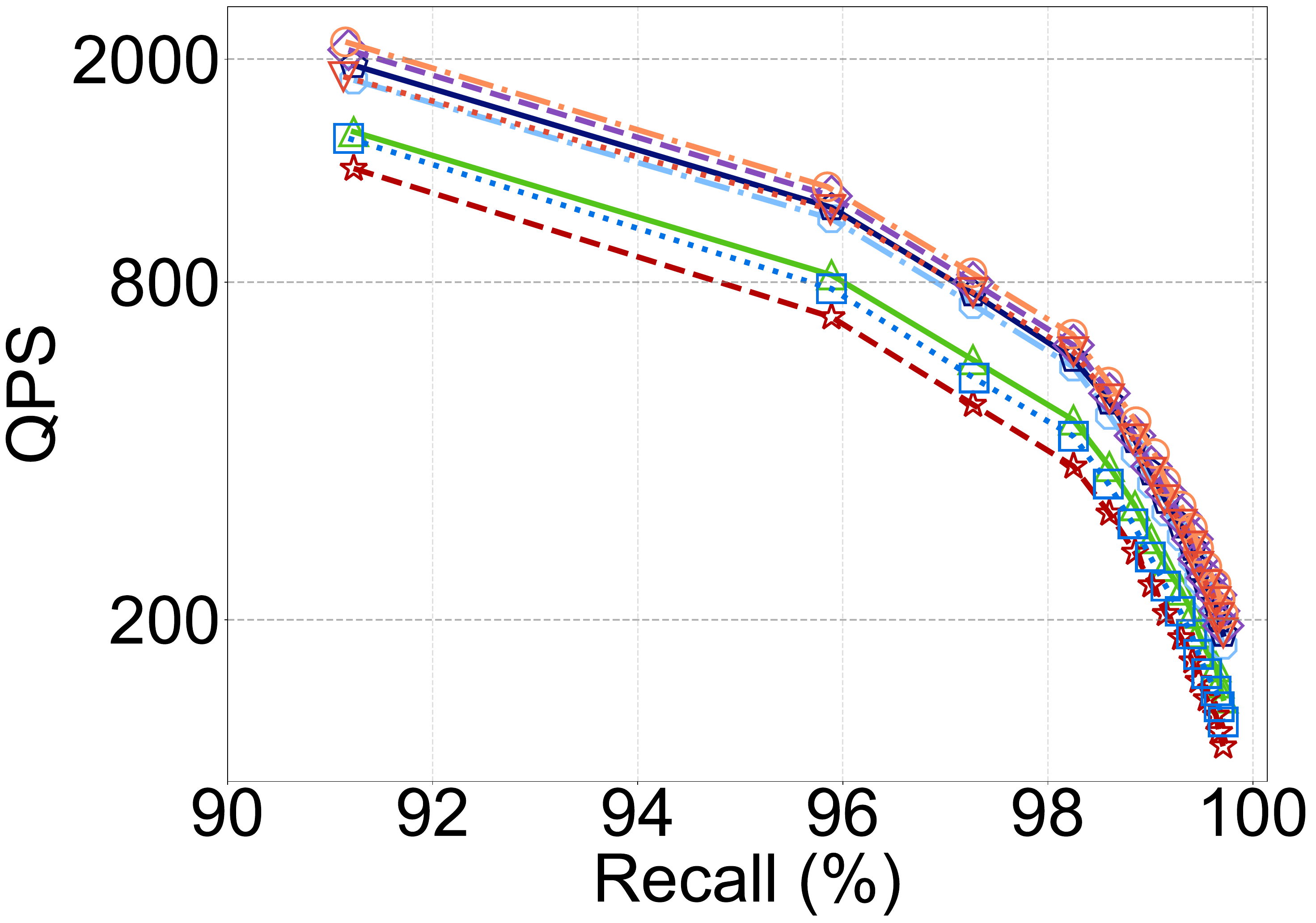}
        \vspace{-3.5ex}
        \caption{\Wikipedia}
    \end{subfigure}
    \begin{subfigure}{0.23\textwidth}
        \centering
        \includegraphics[width=\textwidth]{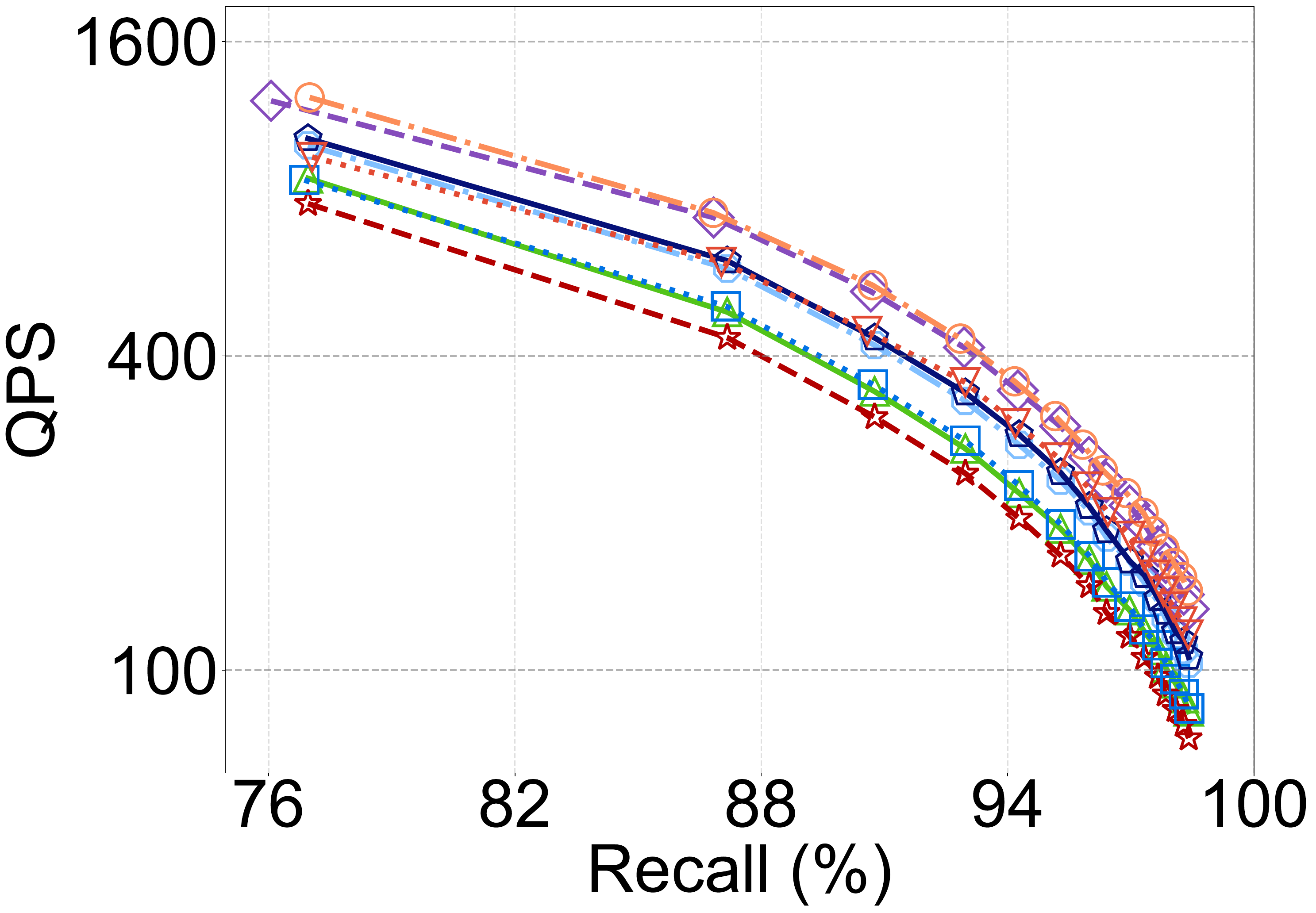}
        \vspace{-3.5ex}
        \caption{\Openai}
    \end{subfigure}
    \begin{subfigure}{0.218\textwidth}
        \centering
        \includegraphics[width=\textwidth]{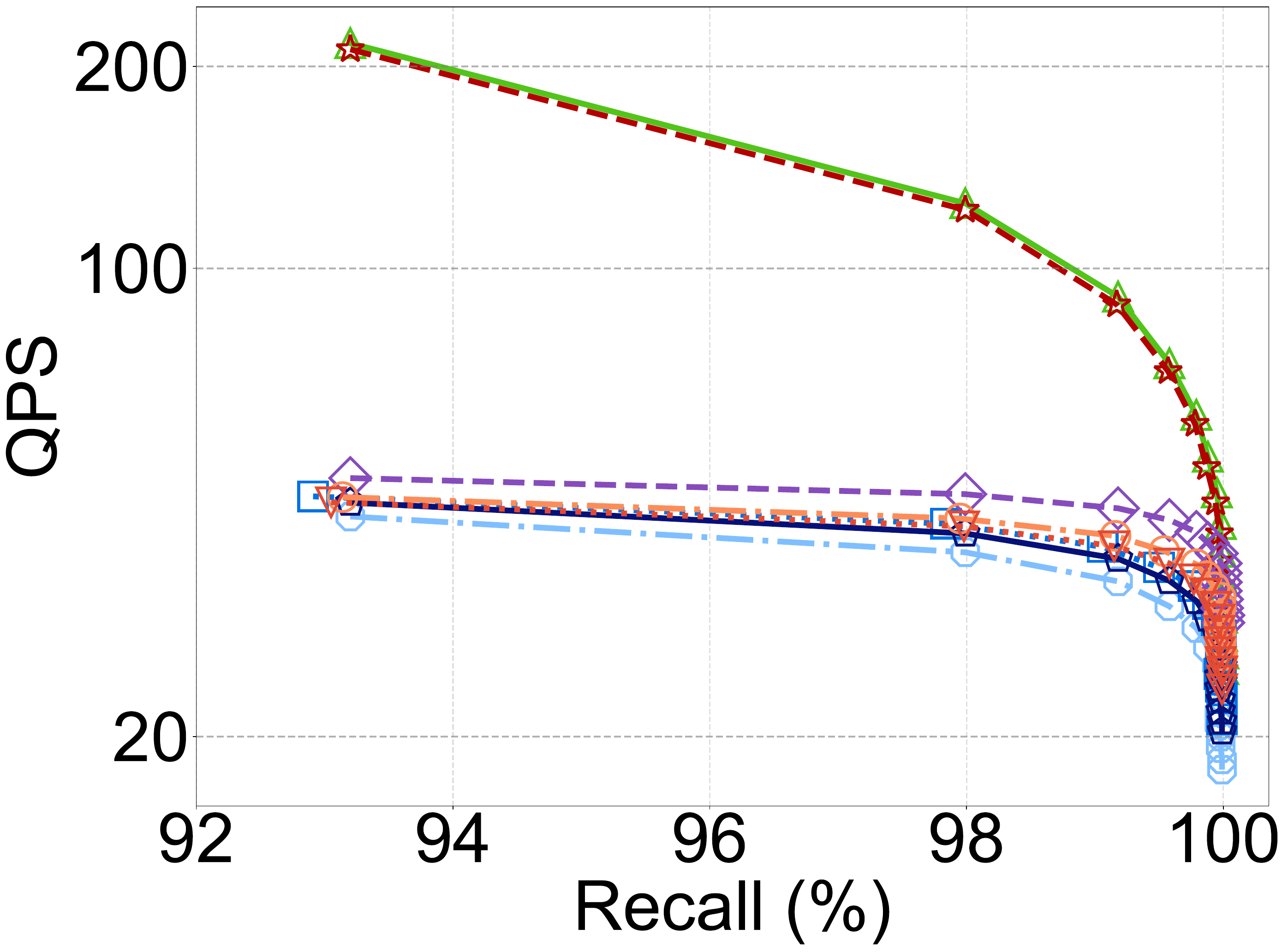}
        \vspace{-3.5ex}
        \caption{\Msmacro}\label{fig:IP-msmacro}
    \end{subfigure}
    \vspace{-0.5ex}
    \caption{Performance on distance metric of Inner Product (IP)}\label{fig:different-metric}
    \vspace{-1.0ex}
\end{figure}

\vspace{-1.5ex}
\subsection{Beyond Euclidean Distance}
This experiment evaluates extensions of DCOs to non-Euclidean distance (inner product), as all DCOs support it.
% We use the \Wikipedia and \Msmacro datasets, which natively employ IP, alongside two other datasets whose vectors were pre-normalized for IP-based vector search.
\zheng{As shown in \figref{fig:different-metric}, we conduct comprehensive evaluations using four datasets with distinct dimensionality.}

The similar trends observed in \figref{fig:time-accuracy} and \ref{fig:different-metric} are expected, as most DCO methods convert IP into a Euclidean distance computation.
On the \Openai dataset, \DADE, \DDCres, \DDCpca, and \DDCopq improve QPS by up to 1.3--1.7$\times$ compared to \FDScanning, which demonstrates their ability to handle IP. 
However, there are also cases on \Glove and \Msmacro datasets, where \FDScanning outperforms all the other methods. 
For instance, it achieves the highest QPS on \Glove, outperforming the SOTA methods by up to 1.2--1.4$\times$.
Meanwhile, the bottleneck caused by online preprocessing time persists: the QPS of these SOTA methods drops by up to 78\%--79\% compared to \FDScanning across these datasets.
\zheng{Notably, a similar trend is observed for cosine similarity. Please refer to our full paper \cite{fullpaper} for detailed results of cosine similarity.}

%These phenomena indicate that DCOs have similar performance in both Euclidean distance and IP. This can be attributed to the fact that most DCO algorithms transform IP into Euclidean distance to generalize across distance metrics.

\vspace{-0.5ex}
\stepcounter{takeaway}
\begin{tipbox}
\textbf{Takeaway \#\thetakeaway: 
The similar performance of DCOs under IP and Euclidean metrics stems from their reliance on converting the former to the latter, but this comes at the cost of generality.}
\end{tipbox}

\begin{figure}[t]
	\centering
    \includegraphics[width=0.30\textwidth]{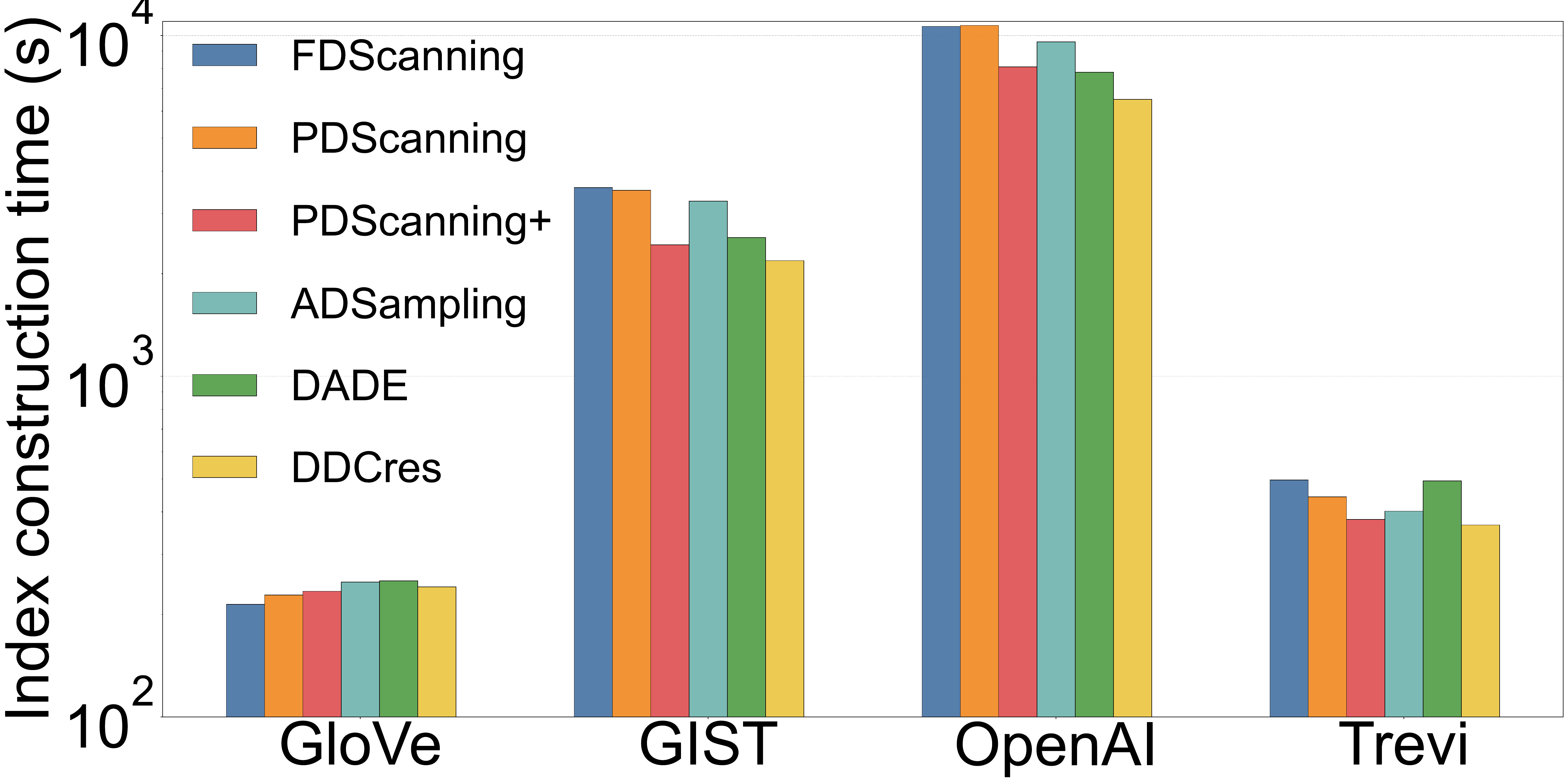}
    \vspace{-1.5ex}
	\caption{Index construction time using DCO}\label{fig:Index-construction-time} 
    \vspace{-1.5ex}
\end{figure}

\vspace{-1.0ex}
\subsection{Index Construction}
While most DCOs have the potential to accelerate index construction, this capability has not been systematically evaluated in prior work.
We fill this gap by evaluating their performance, excluding the \Class methods (\DDCpca and \DDCopq) that themselves require an index for training.

\fakeparagraph{Index Construction Time}
\figref{fig:Index-construction-time} shows the index construction time with DCOs.
On the low-dimensional dataset \Glove, construction time even increases with the SOTA methods. 
This further confirms that most DCO methods do not perform well in datasets with comparably lower dimensions.
By contrast, on high-dimensional datasets (\Gist and \Openai), most DCO methods achieve acceleration, including \RPDScanning, \ADSampling, \DADE, and \DDCres.
For example, \DDCres reduces index construction time by up to 39\% across these datasets.
% Theses patterns are consistent with our earlier findings, confirming a critical dependency between the efficiency gain of DCOs and data dimensionality.
On the \Trevi dataset, most DCO methods also achieve acceleration.
%\ADSampling and \DADE can also reduce the construction time, but are still longer than \RPDScanning.
\zheng{The pre-processing time for PCA is lower than that required for HNSW index construction, the latter of which typically constitutes over 60\% of the total runtime.}

\vspace{-1.0ex}
\begin{figure}[h]
    \centering
    \begin{subfigure}{0.24\textwidth}
        \centering
        \includegraphics[width=\textwidth]{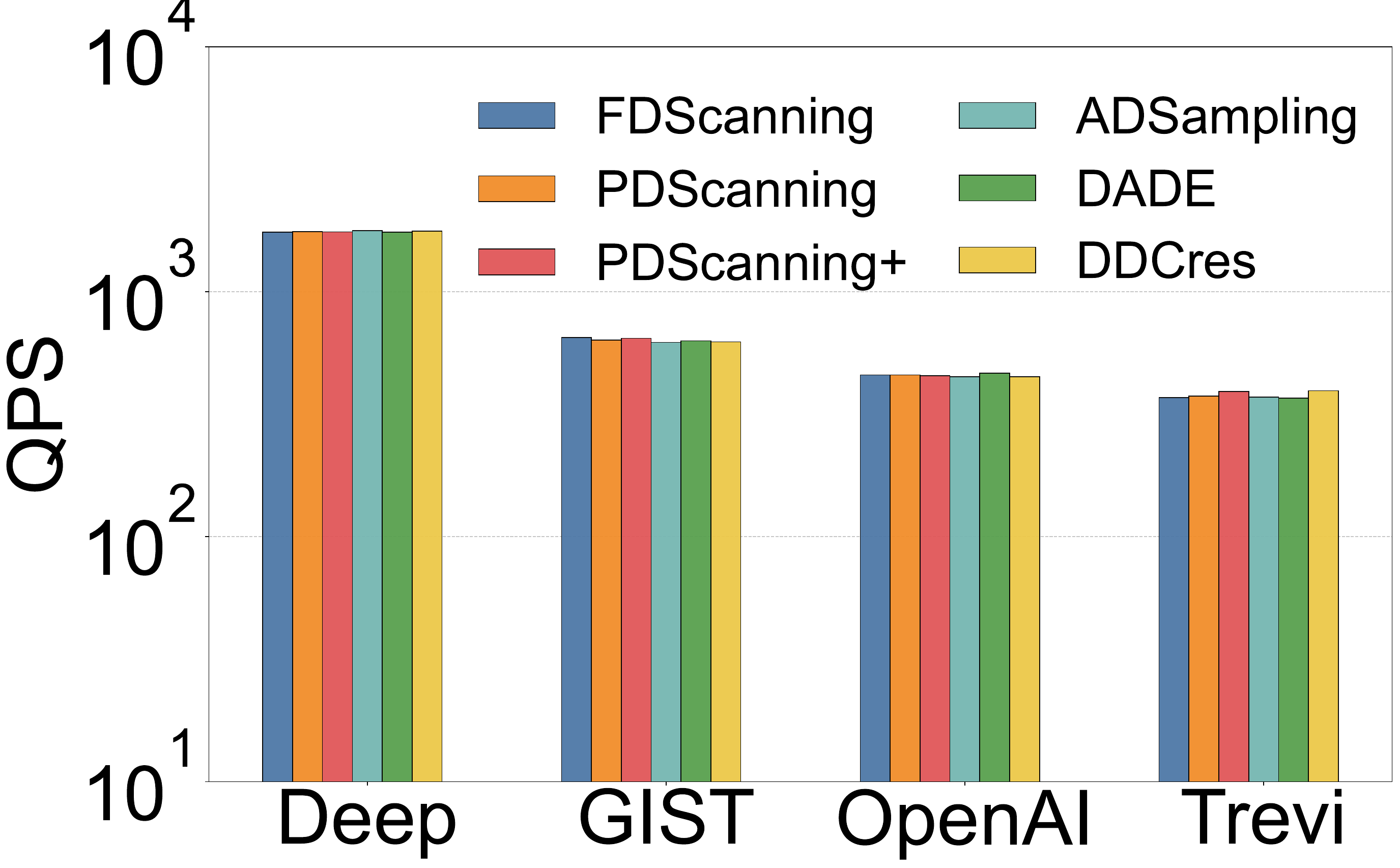}
    \end{subfigure}
    \begin{subfigure}{0.24\textwidth}
        \centering
        \includegraphics[width=\textwidth]{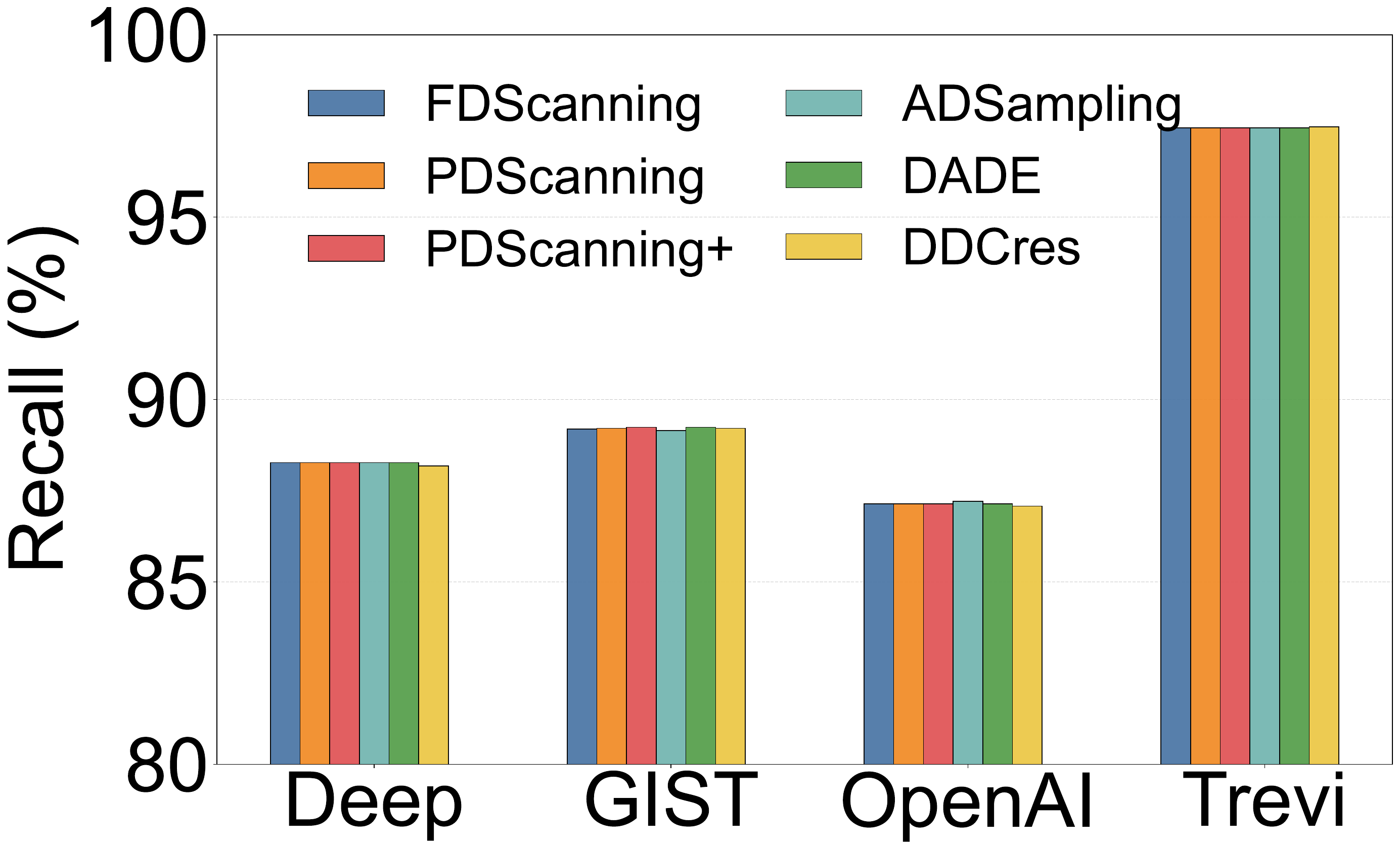}
    \end{subfigure}
    \vspace{-3.0ex}
    \caption{Performance evaluation of different indexes}
    \label{fig:index-quality-evaluation}
    \vspace{-2.0ex}
\end{figure}

\fakeparagraph{Vector Similarity Search Performance}
As most DCOs are approximate, they may result in different indexes. 
We evaluate the constructed indexes on vector similarity search using a fixed query processing method (\FDScanning) for a fair comparison.
\figref{fig:index-quality-evaluation} shows that all indexes achieve nearly identical average recall and QPS across different datasets, where their recall gap to \FDScanning is no more than 0.1\%.
This demonstrates that indexes built with DCOs effectively maintain end-to-end vector search performance.

\stepcounter{takeaway}
\begin{tipbox}
\textbf{Takeaway \#\thetakeaway: DCOs can accelerate  index construction without compromising search performance, but their efficacy mainly depends on data dimensionality.} 
\end{tipbox}

\subsection{Robustness on Dynamic Data}
While DCO methods have demonstrated strong performance on certain static datasets, their behavior on dynamic data remains unknown due to limited evaluations on this scenario.
We fill this gap by conducting experiments on incrementally inserting data.
We focus solely on insertion because deletion mechanisms are implemented independently of the DCO.

\begin{figure}[t]
	\centering
    \begin{subfigure}{0.42\textwidth}
        \centering
        \includegraphics[width=\textwidth]{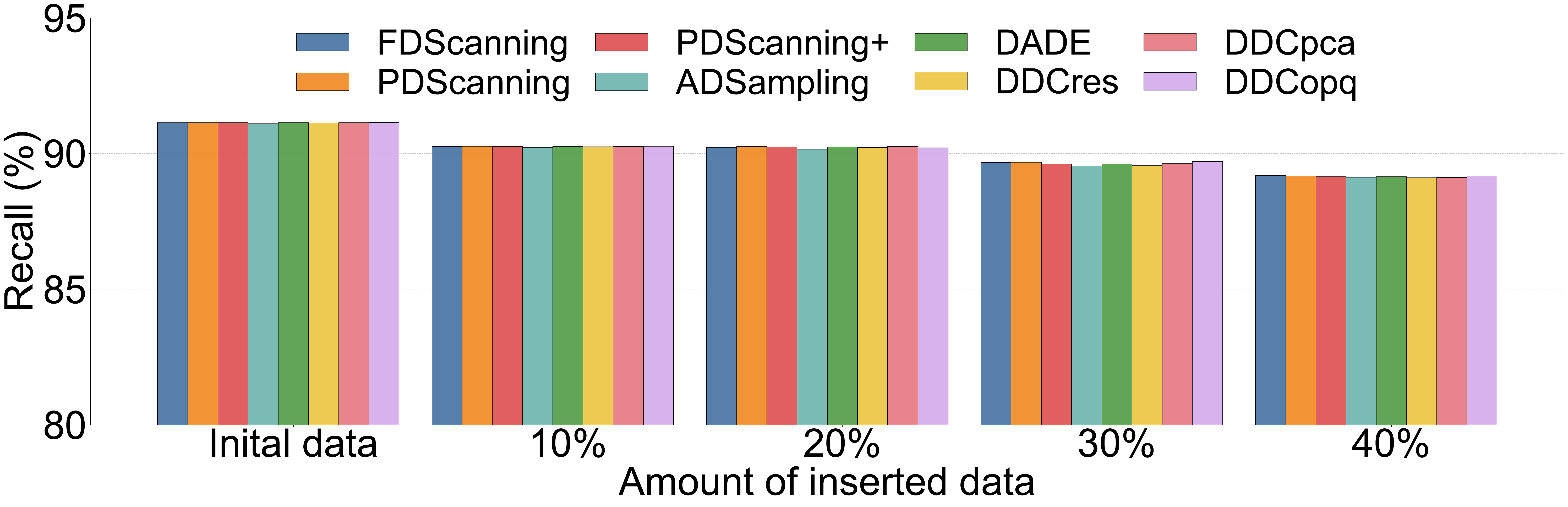}
    \end{subfigure}
    \begin{subfigure}{0.42\textwidth}
        \centering
        \includegraphics[width=\textwidth]{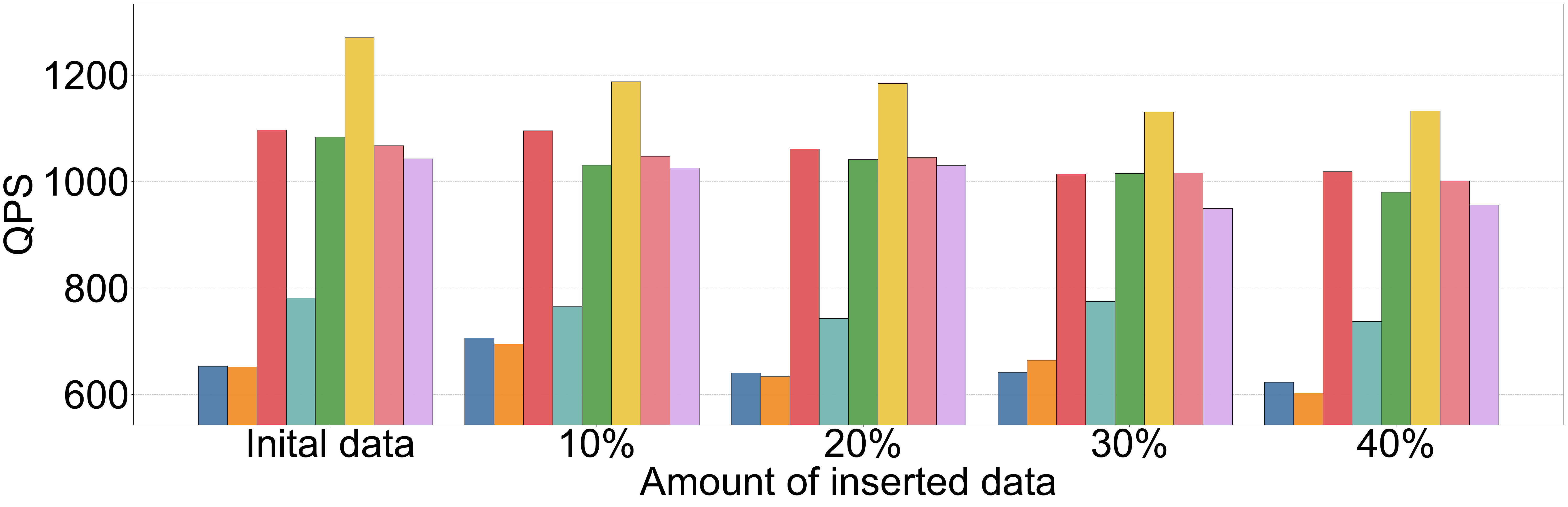}
    \end{subfigure}
    \begin{subfigure}{0.42\textwidth}
        \centering
        \includegraphics[width=\textwidth]{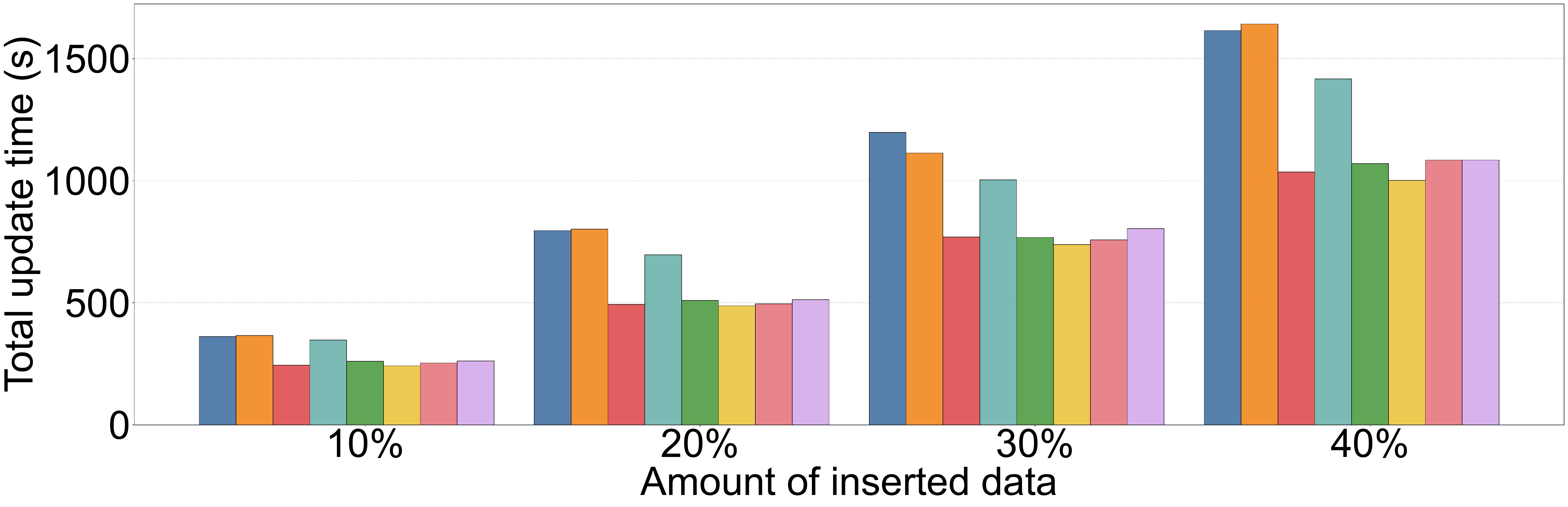}
    \end{subfigure}
    \vspace{-1ex}
	\caption{QPS, recall, and total update time of DCO methods as new vectors are incrementally inserted into HNSW}
    \label{fig:index-maintenance}
    \vspace{-0.5ex}
\end{figure}

\fakeparagraph{Stability Under Sequential Data Insertion}
We evaluate DCO methods during incremental insertions on the \Gist dataset.
\zheng{This dataset is split into a 60\% base set for initial HNSW construction and a 40\% incremental set inserted in four batches.
When inserting a new vector, HNSW must search for neighbors at each layer. 
During the search, we replace \FDScanning with various DCO methods to evaluate their impact on insertion performance.
The resulting QPS, recall, and total update time after each batch are reported in \figref{fig:index-maintenance}.}

As the dataset grows, the recall of all methods declines due to the increased search space. Similarly, the QPS of some methods (\eg \DDCres) decreases, while others (\eg \ADSampling) remain relatively stable.
Among the compared algorithms, \DDCres exhibits the most noticeable drop in QPS, reaching up to 11\%.
Meanwhile, most DCO methods reduce the data insertion time.
For example, \DDCres reduces the total update time by up to 39\%.
This shows that DCOs provide the flexibility to maintain efficiency in dynamic data scenarios.

% During insertion, most methods achieve acceleration compared to \FDScanning. For instance, \DDCres reduces the update time by up to 0.4$\times$.
% Meanwhile, after updating, the SOTA methods maintain comparable recall and achieve higher QPS. This suggests that DCOs can speed up vector insertion without degrading search performance.

%As the dataset grows, the recall of all methods declines due to the increased search space, while DCO methods maintain comparable recall.
%Similarly, the QPS of some methods (\eg \DDCres) decreases, while others (\eg \ADSampling) remain relatively stable.
%Among the compared algorithms, \DDCres exhibits the most noticeable drop in QPS, reaching up to 11\%.
%This indicates that it may be less robust in the scenario of dynamic data insertions.

\vspace{-2ex}
\begin{figure}[h]
	\centering
    \begin{subfigure}{0.24\textwidth}
        \centering
        \includegraphics[width=\textwidth]{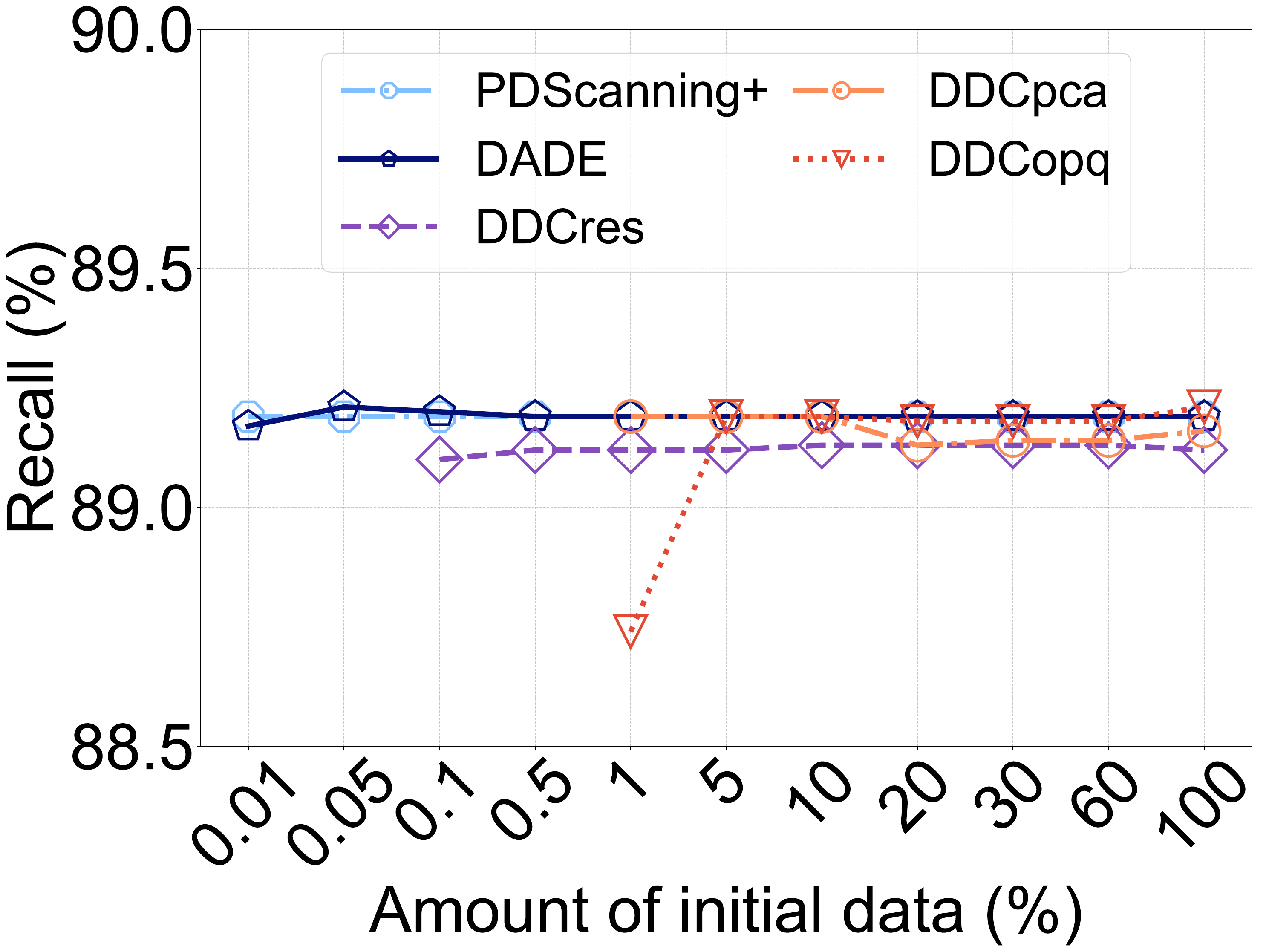}
    \end{subfigure}
    \begin{subfigure}{0.238\textwidth}
        \centering
        \includegraphics[width=\textwidth]{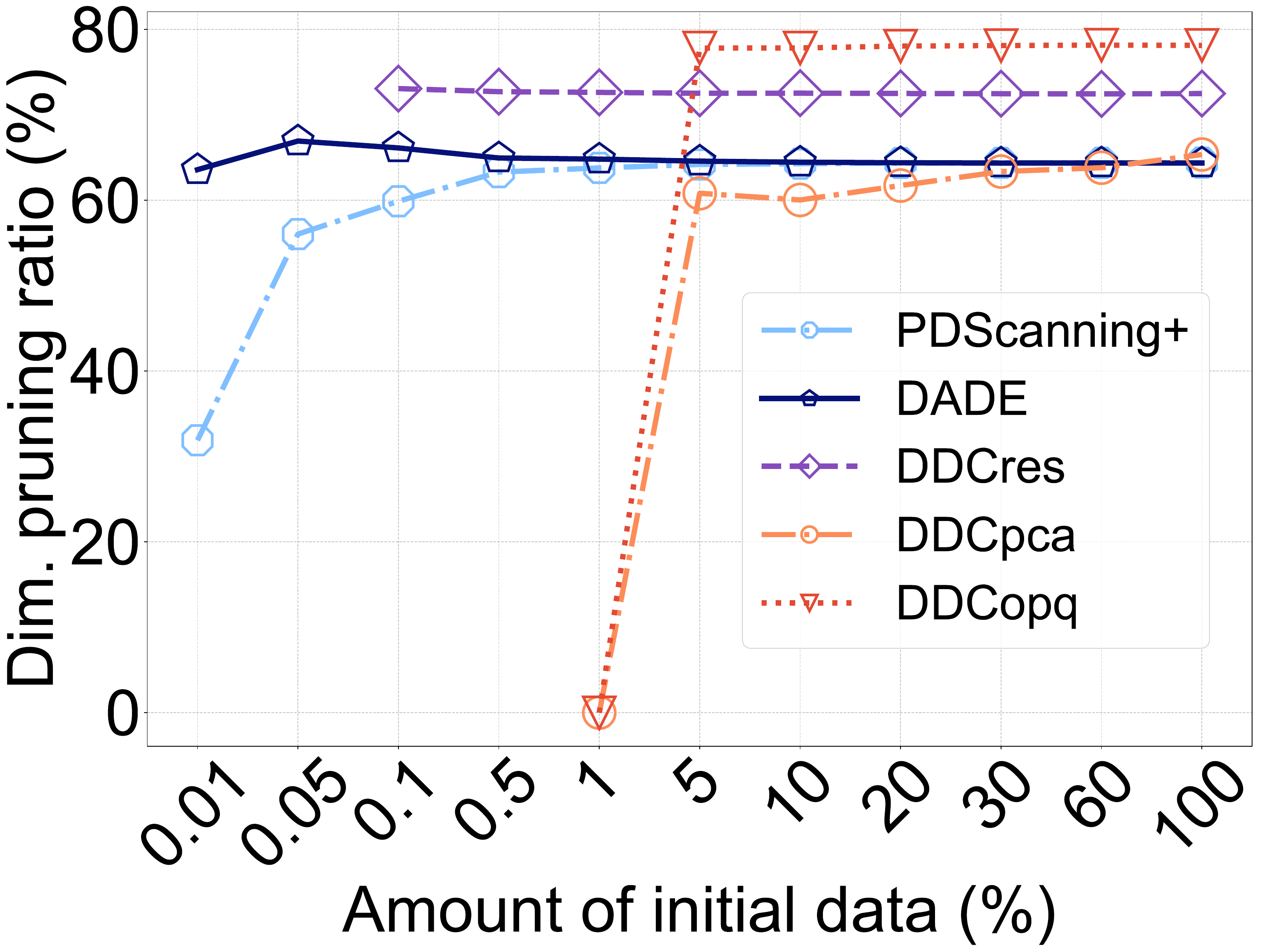}
    \end{subfigure}
    \vspace{-3.5ex}
	\caption{Performance with limited initial data}\label{fig:data-aware}
    \vspace{-2.5ex}
\end{figure}

\fakeparagraph{Performance with Limited Initial Dataset}
Most methods, except for \FDScanning and \PDScanning, rely heavily on the initial data for training prediction models or computing PCA matrices.
To evaluate their sensitivity, we test their performance across a wide range of initial dataset sizes (\ie from 0.01\% to 100\% of the full dataset) and report the resulting recall and dimension pruning ratio in \figref{fig:data-aware}.
% The upper bound of 60\% corresponds to the initial data size used in the previous insertion experiment, enabling direct comparison with those results.

Specifically, with only 1\% of the full data as the training set, \RPDScanning, \DADE, and \DDCres achieve effective pruning performance.
They demonstrate strong adaptability to real-world scenarios, where only a small fraction of data is available at the start and more is appended over time. 
In contrast, \DDCopq and \DDCpca require at least 5\% of the full dataset to reach a relatively stable recall and pruning ratio.
Moreover, both methods require a minimum of 10,000 vectors for training, which explains why their results are absent when the initial dataset is below 1\%. 
%It is worth noting that the pruning ratio of \DDCpca keeps increasing as the initial dataset size grows.

\stepcounter{takeaway}
\begin{tipbox}
\textbf{Takeaway \#\thetakeaway: Most DCO methods maintain their performance under dynamic datasets and reduce insertion time, but  \Class methods are suboptimal with limited initial training data.}
\end{tipbox}

\begin{figure*}[t]
    \centering
    \includegraphics[width=0.9\textwidth]{figure/query_performance/legend.pdf}
    \begin{subfigure}{0.24\textwidth}
        \centering
        \includegraphics[width=\textwidth]{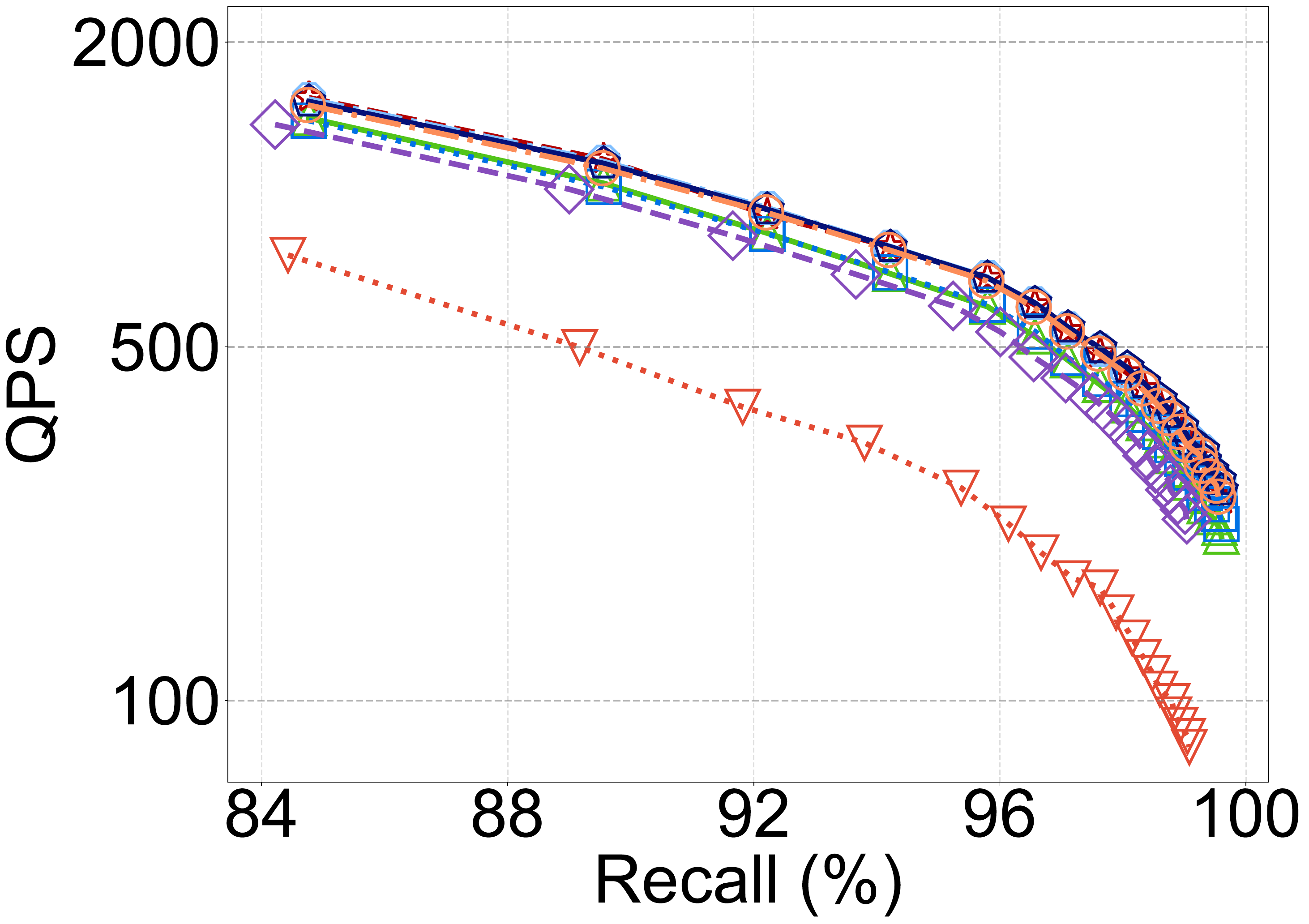}
        \vspace{-3.5ex}
        \caption{\Glove (without SIMD)}
    \end{subfigure}
    \begin{subfigure}{0.24\textwidth}
        \centering
        \includegraphics[width=\textwidth]{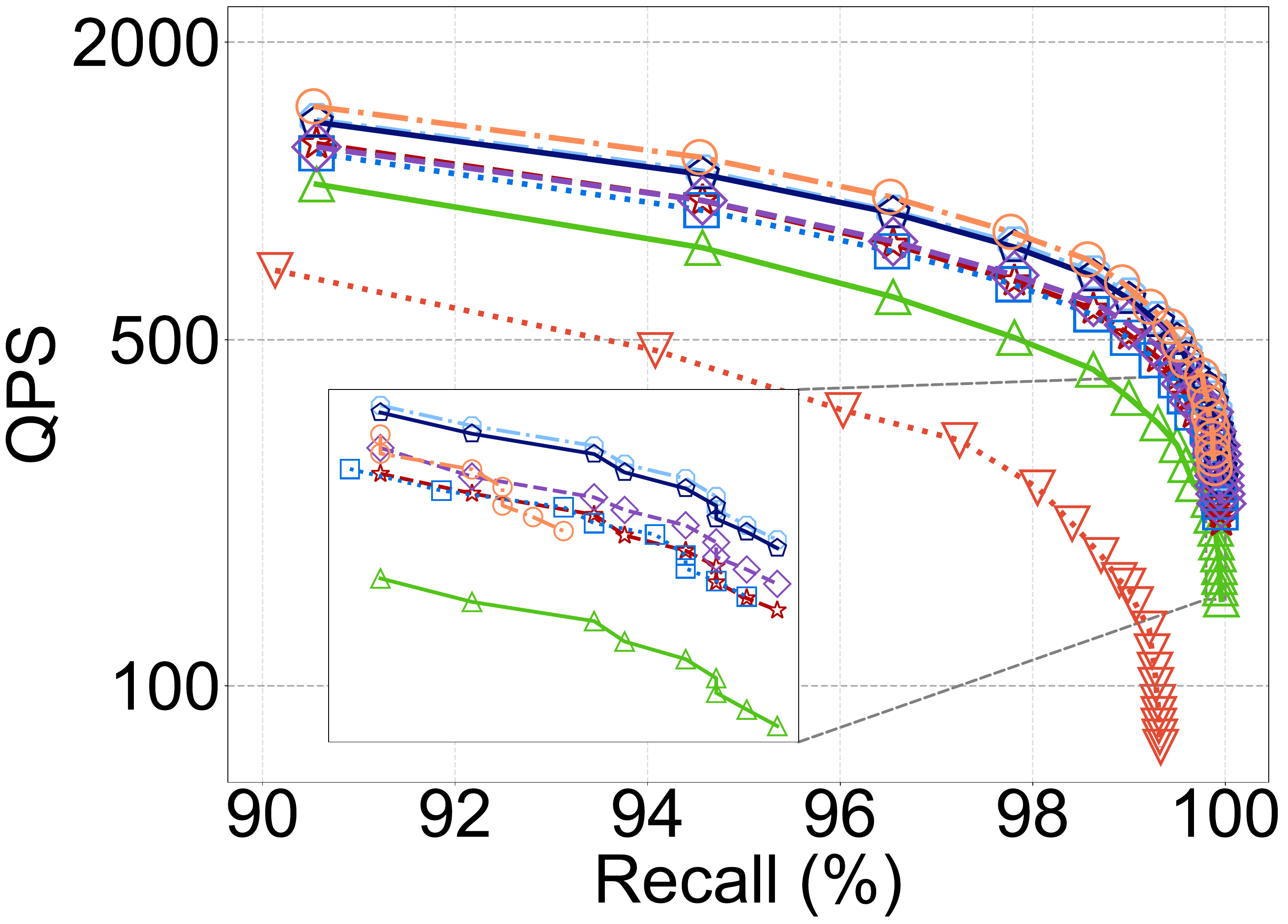}
        \vspace{-3.5ex}
        \caption{\Sift (without SIMD)}\label{fig:without-simd-sift}
    \end{subfigure}
    \begin{subfigure}{0.238\textwidth}
        \centering
        \includegraphics[width=\textwidth]{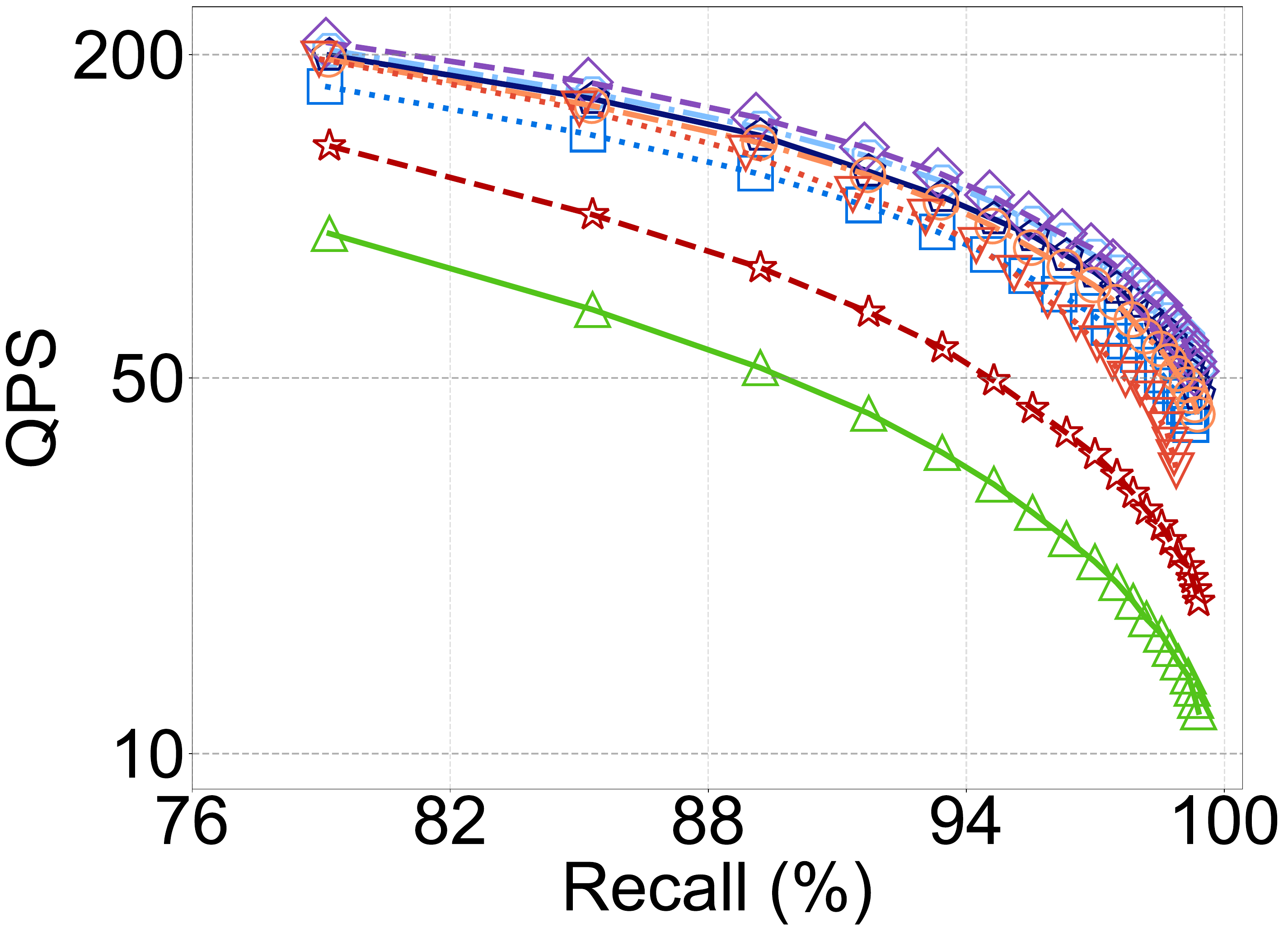}
        \vspace{-3.5ex}
        \caption{\Gist (without SIMD)}
    \end{subfigure}
    \begin{subfigure}{0.235\textwidth}
        \centering
        \includegraphics[width=\textwidth]{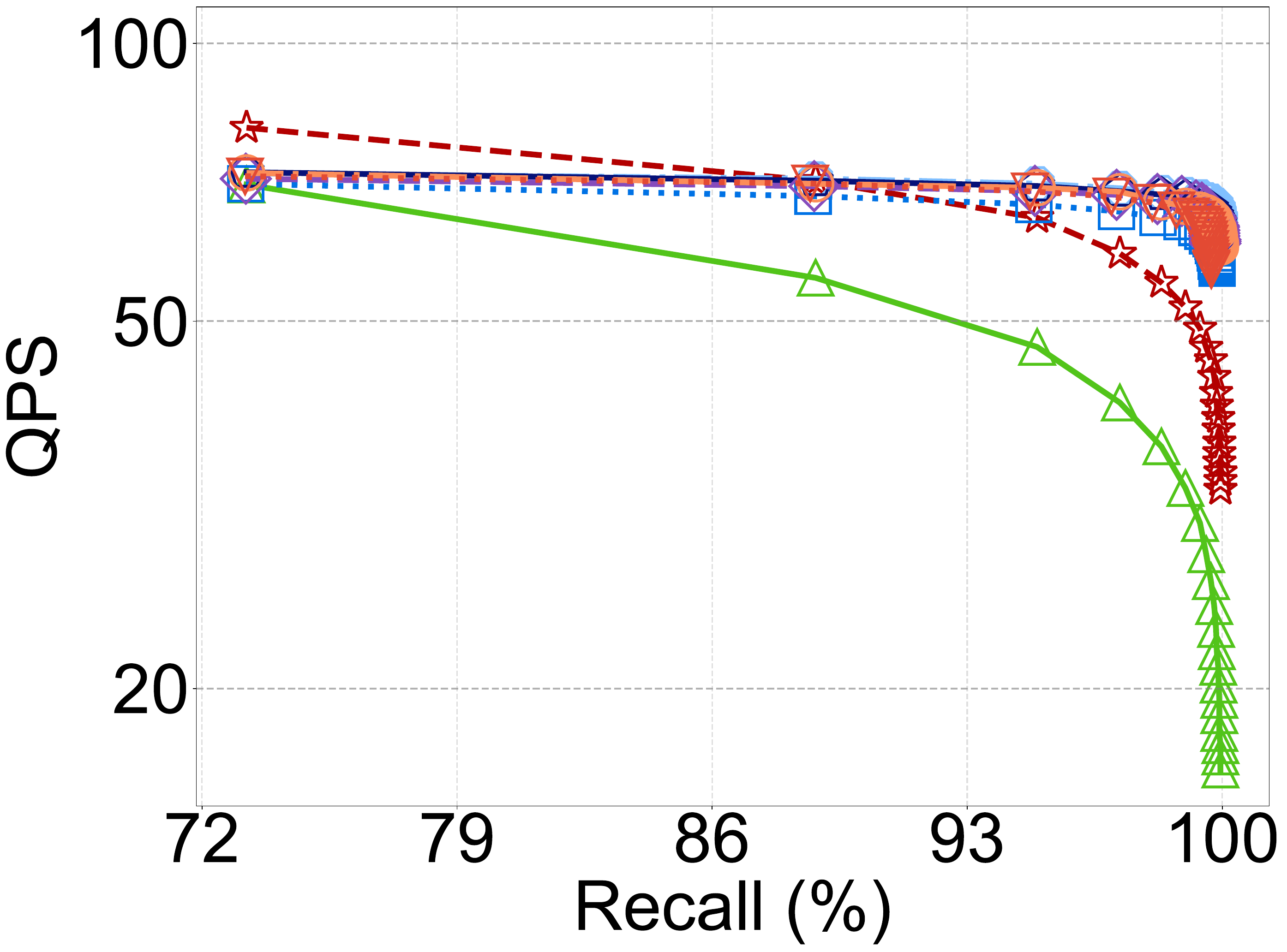}
        \vspace{-3.5ex}
        \caption{\Trevi (without SIMD)}
    \end{subfigure}

    \begin{subfigure}{0.24\textwidth}
        \centering
        \includegraphics[width=\textwidth]{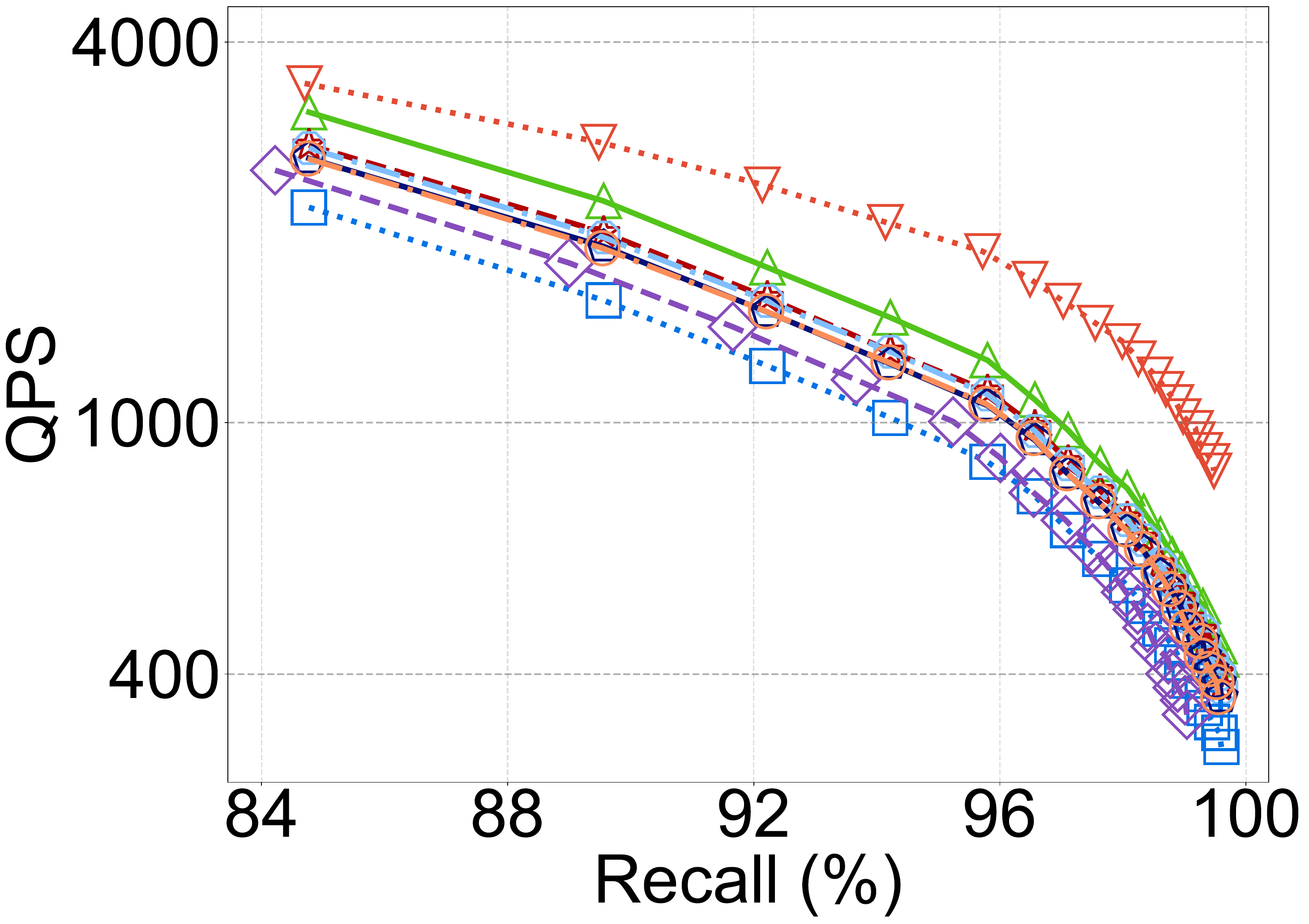}
        \vspace{-3.5ex}
        \caption{\Glove (with SIMD)}
    \end{subfigure}
    \begin{subfigure}{0.24\textwidth}
        \centering
        \includegraphics[width=\textwidth]{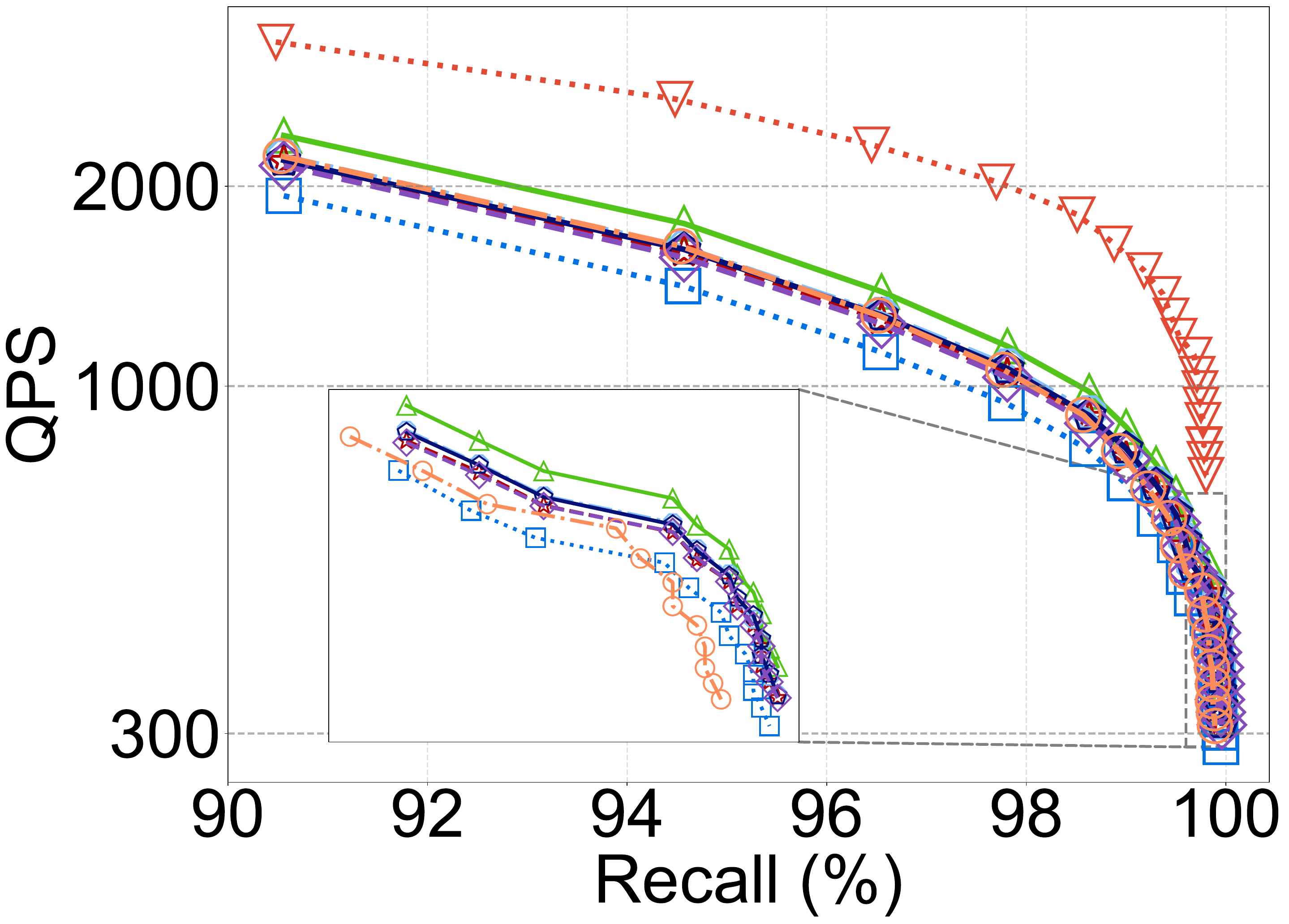}
        \vspace{-3.5ex}
        \caption{\Sift (with SIMD)}\label{fig:with-simd-sift}
    \end{subfigure}
    \begin{subfigure}{0.238\textwidth}
        \centering
        \includegraphics[width=\textwidth]{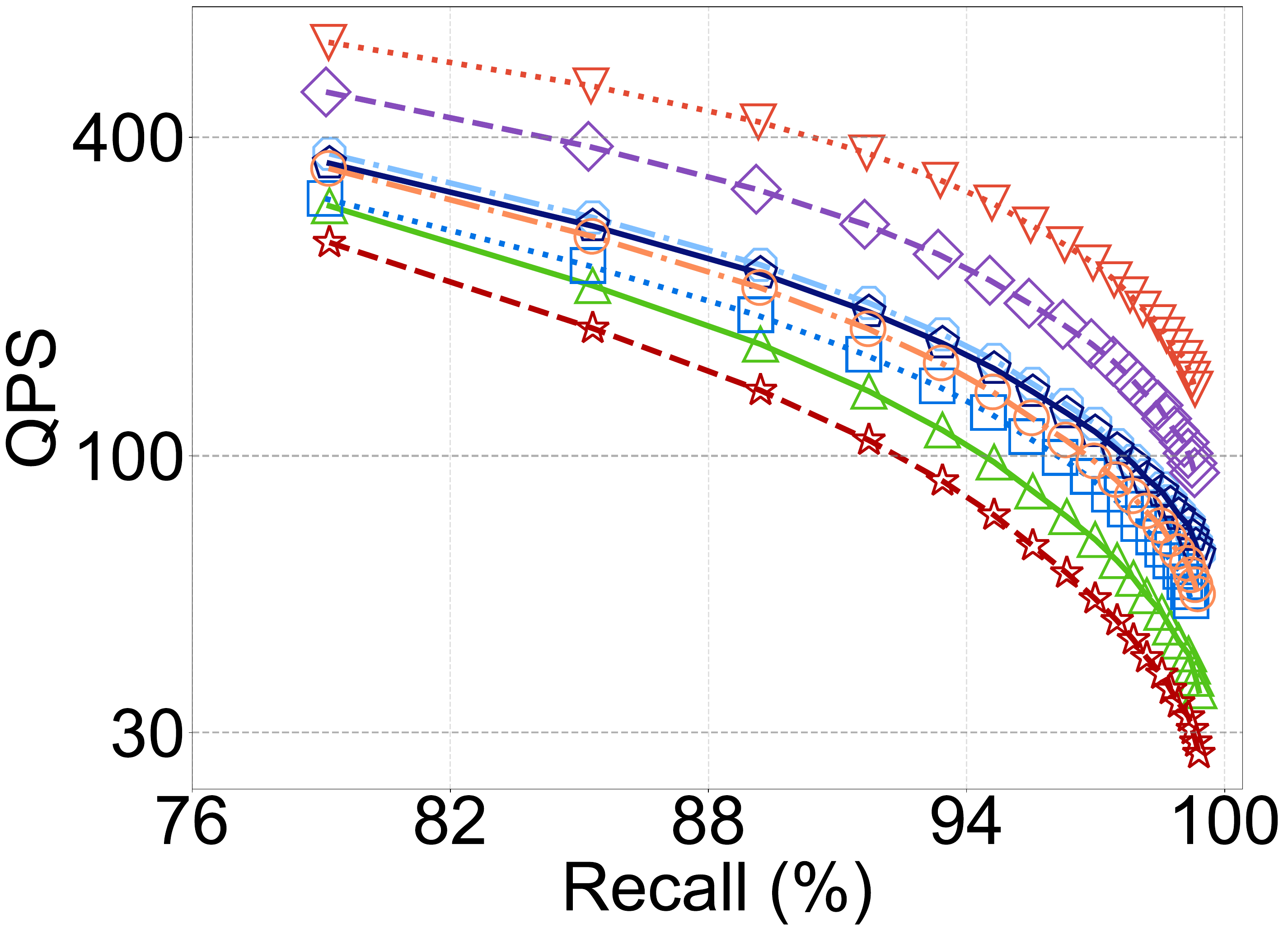}
        \vspace{-3.5ex}
        \caption{\Gist (with SIMD)}
    \end{subfigure}
    \begin{subfigure}{0.235\textwidth}
        \centering
        \includegraphics[width=\textwidth]{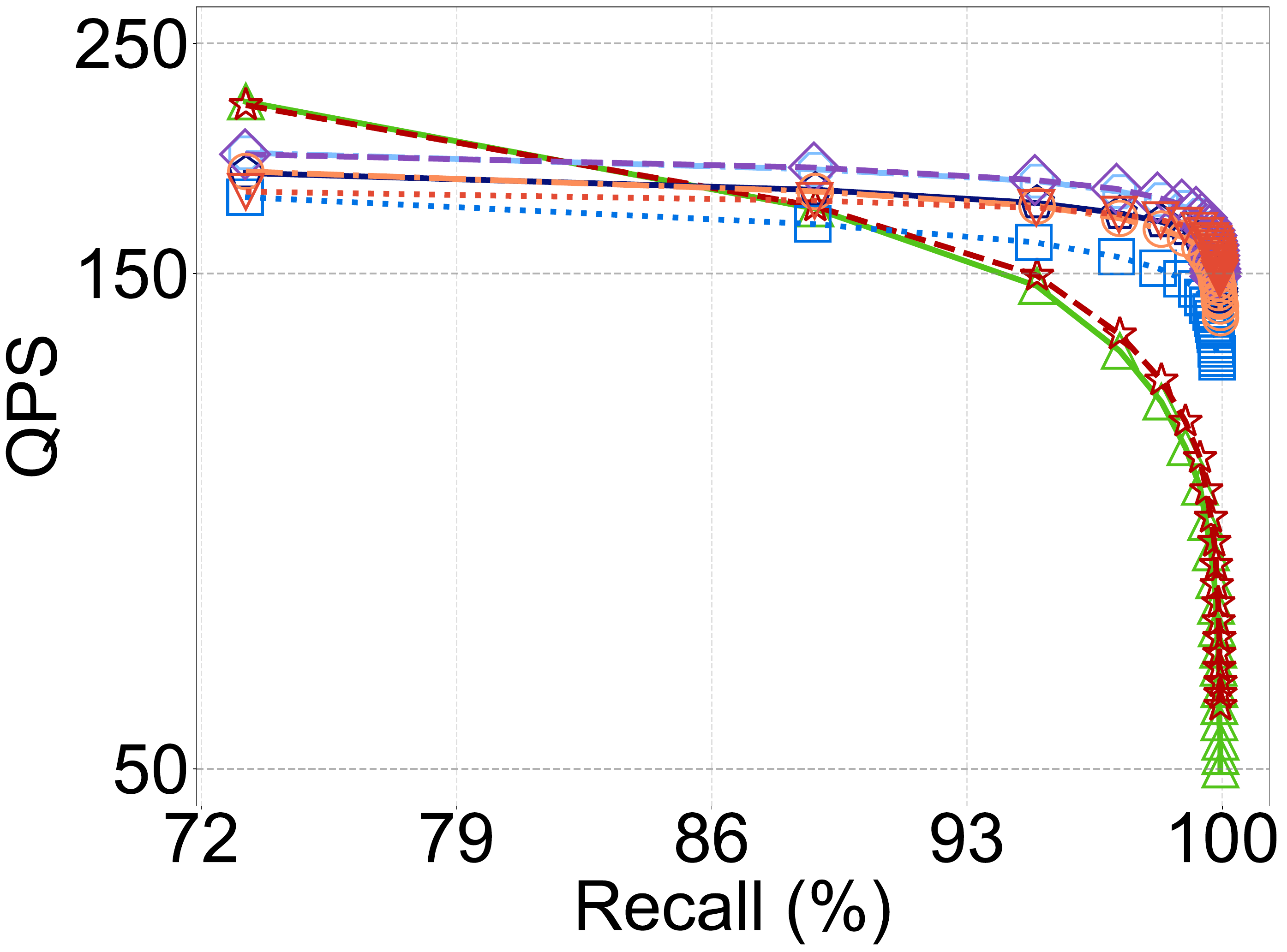}
        \vspace{-3.5ex}
        \caption{\Trevi (with SIMD)}
    \end{subfigure}

    \begin{subfigure}{0.24\textwidth}
        \centering
        \includegraphics[width=\textwidth]{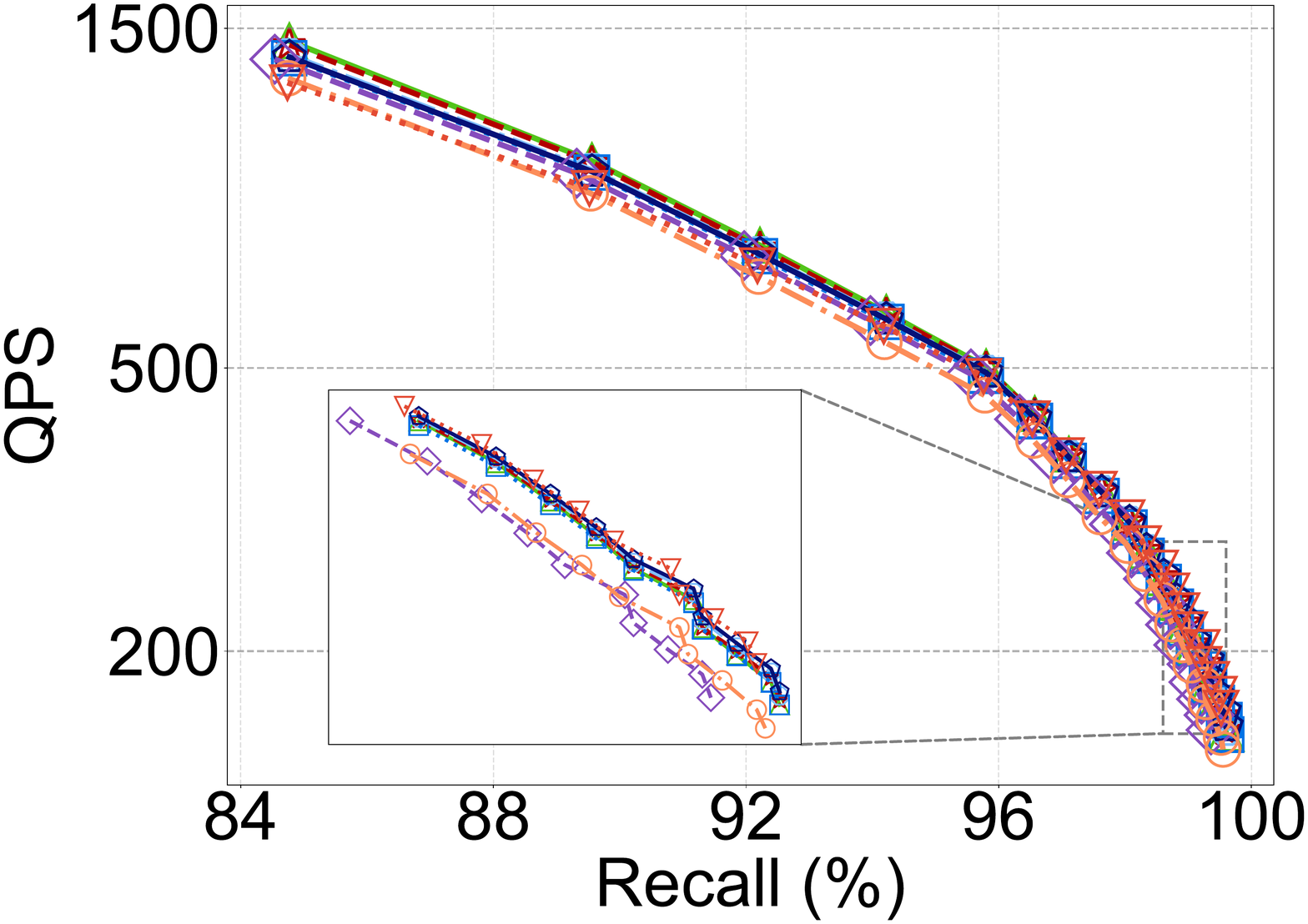}
        \vspace{-3.5ex}
        \caption{\Glove (with GPU)}
    \end{subfigure}
    \begin{subfigure}{0.24\textwidth}
        \centering
        \includegraphics[width=\textwidth]{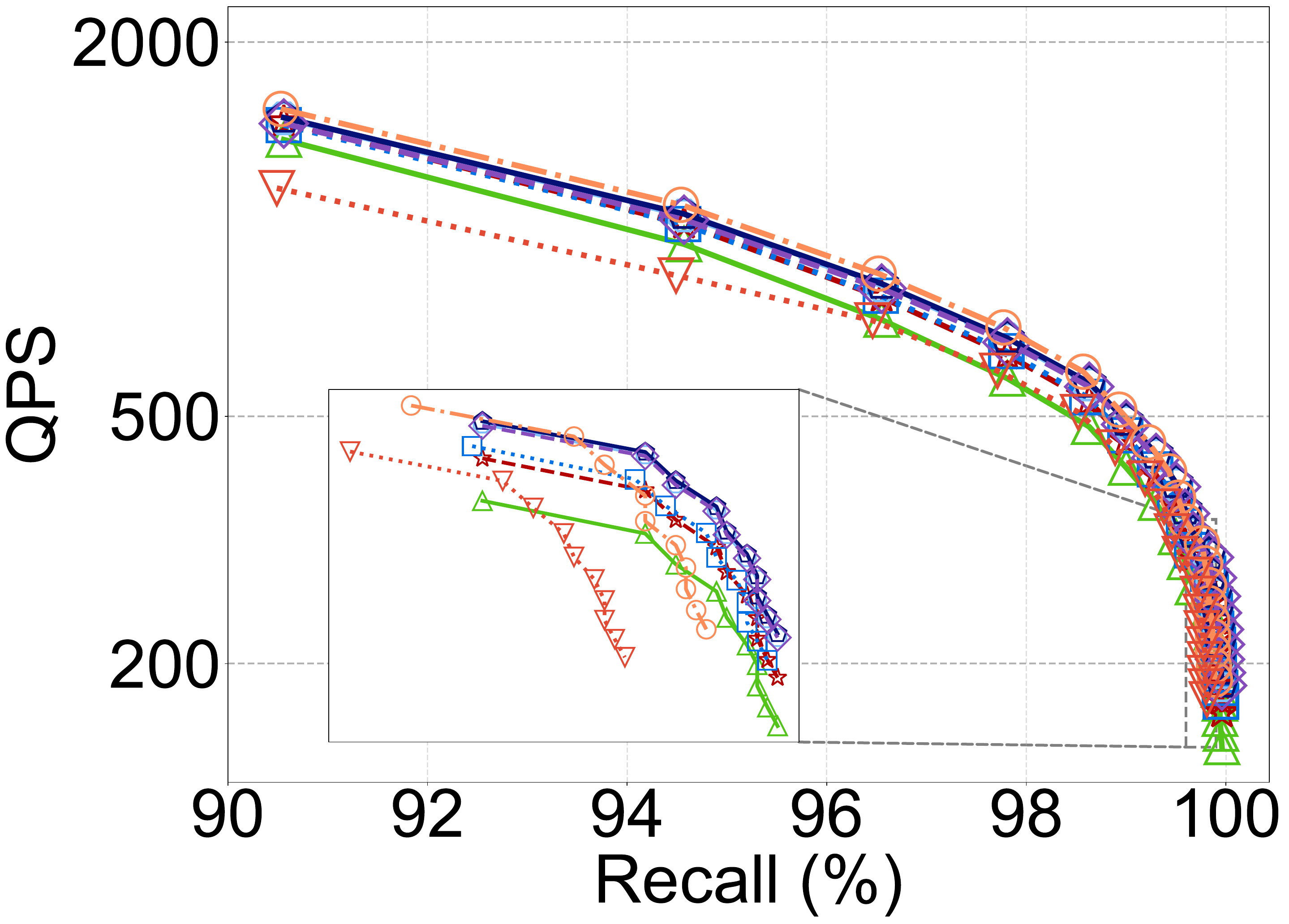}
        \vspace{-3.5ex}
        \caption{\Sift (with GPU)}
    \end{subfigure}
    \begin{subfigure}{0.238\textwidth}
        \centering
        \includegraphics[width=\textwidth]{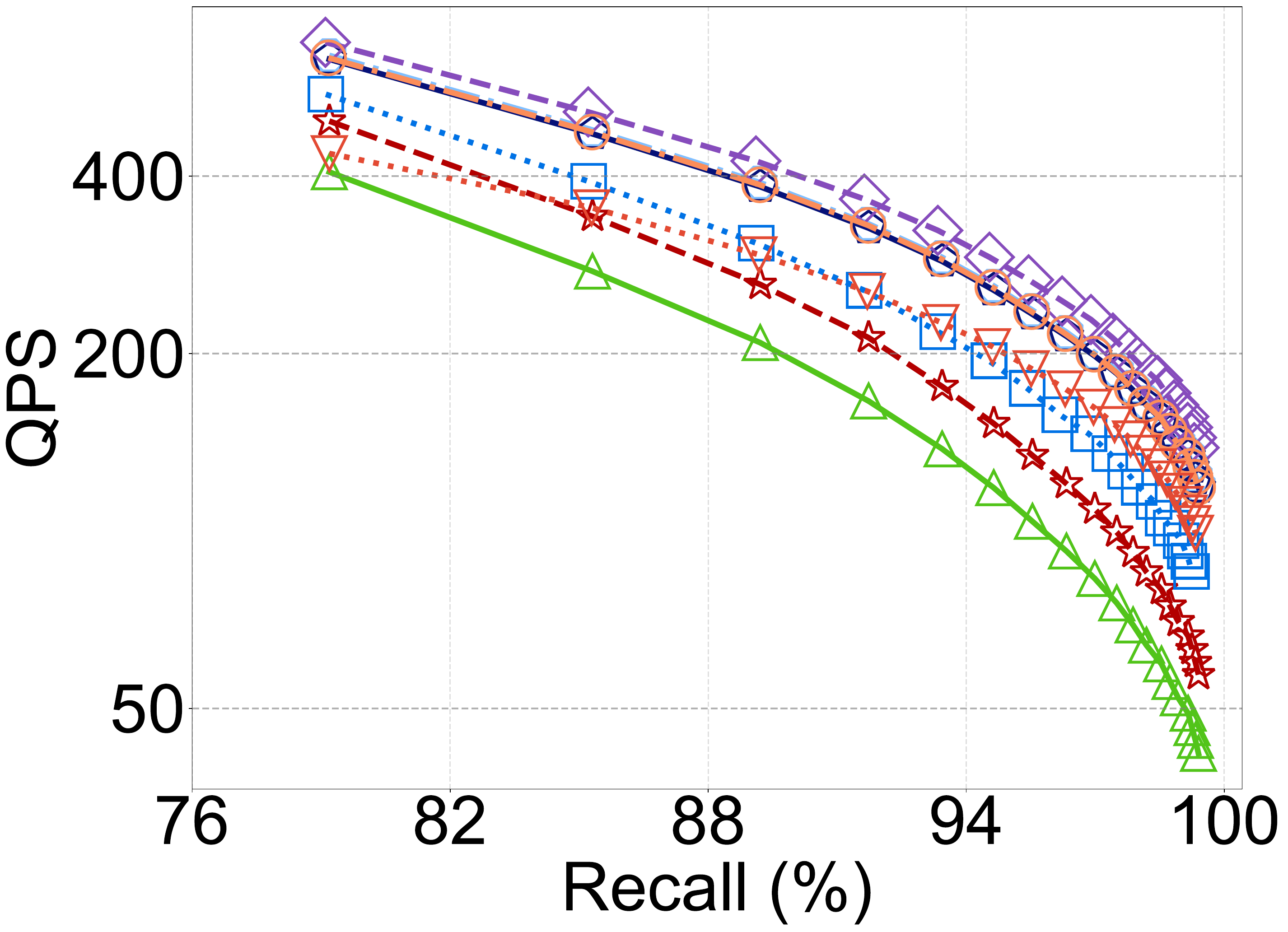}
        \vspace{-3.5ex}
        \caption{\Gist (with GPU)}
    \end{subfigure}
    \begin{subfigure}{0.235\textwidth}
        \centering
        \includegraphics[width=\textwidth]{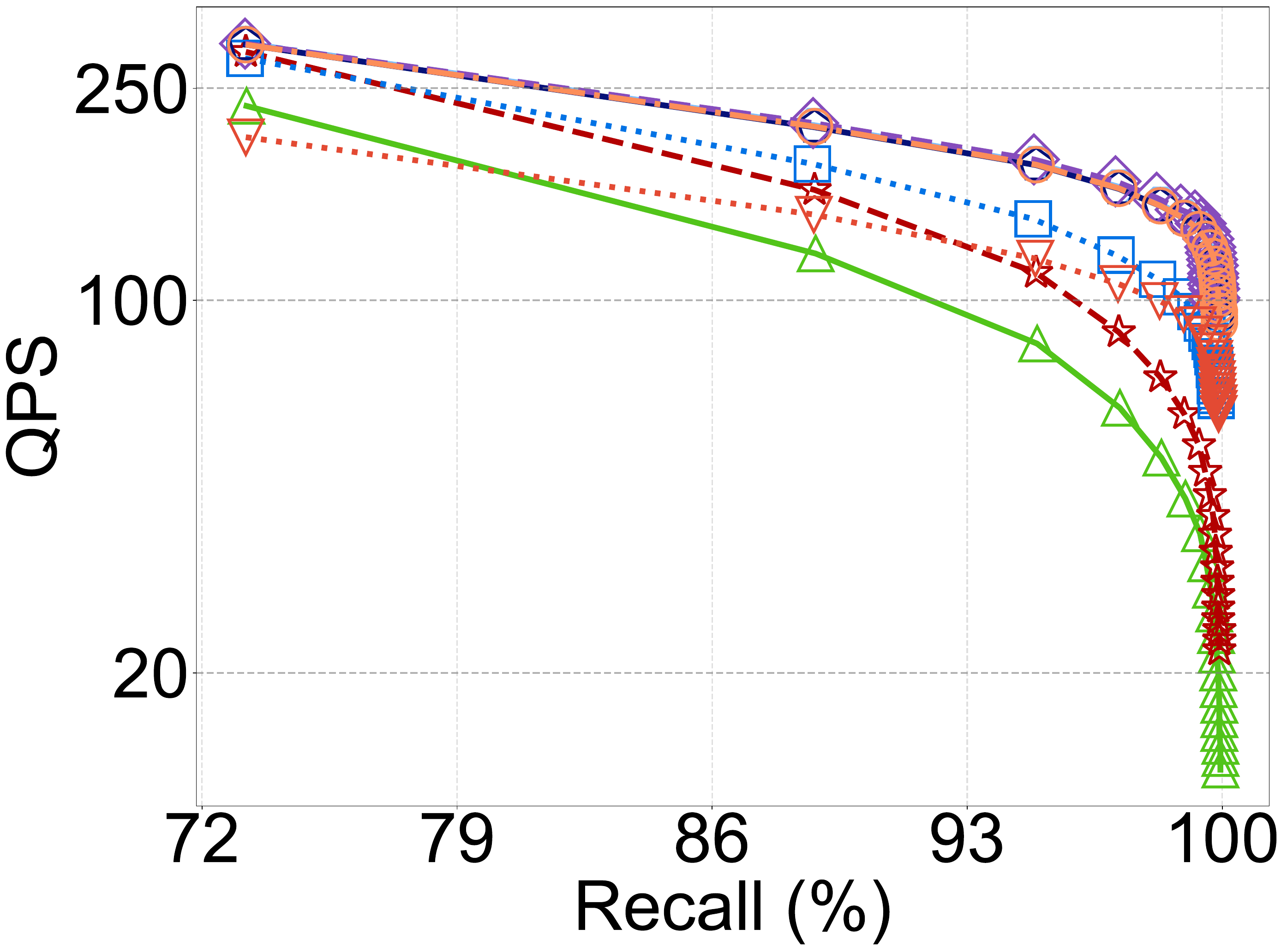}
        \vspace{-3.5ex}
        \caption{\Trevi (with GPU)}
    \end{subfigure}
    \vspace{-1.0ex}
    \caption{Comparisons of DCO methods on the IVF index across different hardware configurations}\label{fig:IVF-hardware}
    \vspace{-3.0ex}
\end{figure*}

\vspace{-1ex}
\subsection{Diversified Hardware Condition}
We evaluate DCO methods across different hardware platforms: GPU and CPU (with/without SIMD support). 
Consequently, we use the IVF index for this experiment, as HNSW often lacks native GPU support. 
%The dimensionality pruning ratio remain stable across these hardware conditions (see the last column of \figref{fig:IVF-hardware}), we focus primarily on QPS and recall. 

\fakeparagraph{Performance on CPU with/without SIMD Support}
As illustrated in the first two rows of \figref{fig:IVF-hardware}, the usage of SIMD  significantly alters the performance ranking of DCO methods in terms of QPS.
For example, on the \Sift dataset, \DDCopq is the least efficient method with SIMD disabled, but becomes the most efficient when SIMD is enabled.
This is because SIMD enables the parallel computation of multiple quantization distances by packing them into a single instruction.
The \Loop method \RPDScanning consistently maintains a high QPS on the \Trevi dataset,
where the gap to the optimal one is usually within 1\% for recall over 89\%.
% Another notable example is \DDCopq, which transitions from the least efficient method without SIMD to the most efficient with SIMD for recall below 98.6\%, on the \Sift dataset. 

We also observe that SIMD provides a substantial QPS boost for all methods.
For example, \FDScanning exhibits a QPS increase of up to 3.3$\times$ on the \Gist dataset when using SIMD. 
Given that modern AMD and Intel processors usually support SIMD instruction sets like SSE and AVX~\cite{guide2011intel}, this improvement is almost a ``free lunch''.
This justifies our benchmark to enable SIMD as the default setting. 
However, enabling SIMD narrows the efficiency advantage of DCO methods over \FDScanning, reducing the speedup from 2.0--4.4$\times$ to 1.2--3.8$\times$ on the IVF index.
\zheng{This is because SIMD accelerates the vector distance computation for both \FDScanning and other DCO methods, whereas the extra computational overhead of the latter erodes their relative gain.}
On HNSW, the speedup reduces from 1.9--3.1$\times$ to 1.1--2.1$\times$ (see full paper \cite{fullpaper}).

\fakeparagraph{Performance on GPU}
The third row of \figref{fig:IVF-hardware} plots the QPS-recall curves of DCO methods running on GPU.
Most methods exhibit similar performance on the \Glove dataset, while \DDCres and \DDCpca underperform the others.
For other datasets at recall above 73\%, \RPDScanning, \DADE, \DDCres, and \DDCpca achieve similar QPS, often outperforming the others.
\RPDScanning performs better than \ADSampling.
The results of \RPDScanning demonstrate that \Loop methods remain competitive with GPU acceleration.
Overall, \ADSampling, \DADE, \DDCres, \DDCpca, and \DDCopq improve QPS with the same recall over \FDScanning by up to 4.9, 7.0, 7.6, 6.9, and 4.7$\times$ respectively, across these datasets.

On the \Trevi dataset, most methods achieve a higher QPS than \FDScanning, a pattern that differs from the results in \figref{fig:time-accuracy}.
This performance inversion is driven by the IVF index's larger number of DCO operations, which dilutes the $O(D^2)$ pre-processing cost per query.
As a result, the time spent on online pre-processing query vectors is dwarfed by the total time of DCOs and thus does not create a bottleneck. 

\vspace{-0.5ex}
\fakeparagraph{Performance Comparisons: CPU vs GPU}
Compared to CPU without SIMD, DCO methods can improve their QPS by up to 3.0--4.3$\times$ on GPU.
This improvement benefits from vector-level parallelization, which allows concurrent DCOs to be performed for a batch of query vectors.
However, this advantage drops to roughly 1.1--2.2$\times$, when compared to CPU with SIMD enabled.
Based on these comparison results, we have two major observations: (1) GPU is preferable for evaluating the performance of DCOs on IVF for higher dimensional datasets, as it delivers the best performance, and (2) HNSW is better suited to CPU with SIMD, since it does not natively support GPU execution.

\stepcounter{takeaway}
\begin{tipbox}
\textbf{Takeaway \#\thetakeaway: Hardware configuration is a critical yet overlooked factor in DCO evaluation.} 
Hardware configuration can drastically alter performance rankings.
For instance, enabling SIMD on CPU elevates \DDCopq from the least to the most efficient method on the \Sift dataset.
\end{tipbox}

\begin{figure}[htbp]
	\centering
    \includegraphics[width=0.43\textwidth]{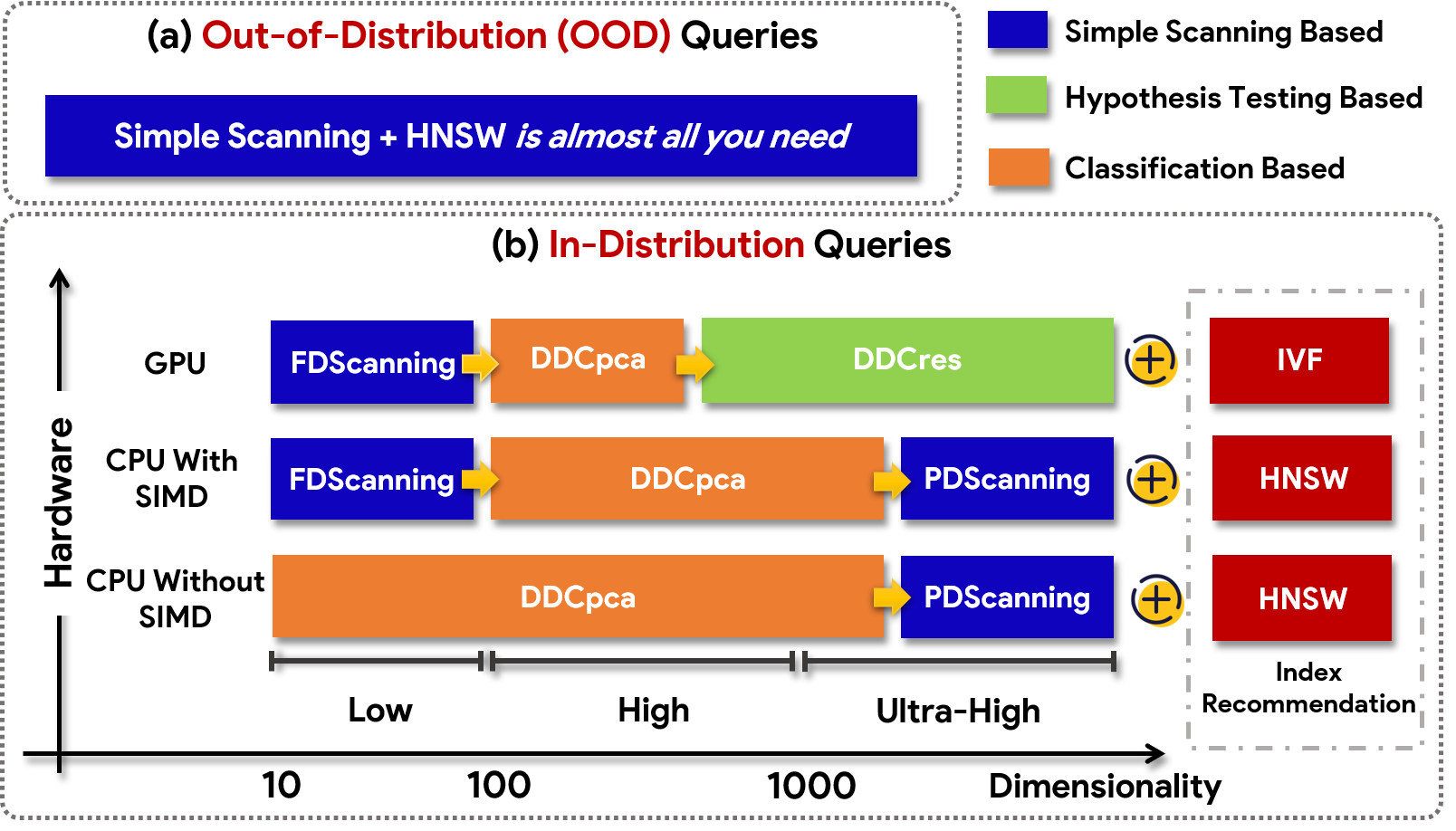}
    \vspace{-1ex}
	\caption{Recommendations of DCO methods}\label{fig:recommendations}
    \vspace{-1ex}
\end{figure}

\subsection{Summary: Recommendations, Merits, and Limitations}\label{sec:result-summary}

\fakeparagraph{Overall Recommendation}
Our experimental results lead to a clear set of recommendations for selecting the optimal DCO method in different scenarios, as shown in \figref{fig:recommendations}.
These recommendations are mainly determined by each method's average QPS ranking for achieving a recall rate over 95\%.

The optimal DCO algorithm selection depends on three primary factors: \textit{data dimensionality}, \textit{hardware configurations}, and \textit{query distribution}.
For example, on CPUs with SIMD support, the best choice for in-distribution queries shifts from \FDScanning to \DDCpca, and then to \PDScanning, as data dimensionality increases.
% Overall, \Loop methods (\FDScanning and \PDScanning) excel in OOD queries, low- and ultra-high-dimensional data.
Overall, \Loop methods, \FDScanning and \PDScanning, often outperform the others for OOD queries, and also perform well on low- and ultra-high-dimensional data for in-distribution queries. 
For in-distribution high-dimensional data, \DDCpca is often optimal with the HNSW index on CPUs, whereas \DDCres is the most efficient with the IVF index on GPUs.

Based on the above recommendations, we next analyze the main merits and limitations of existing DCO methods to guide their practical usage and future research.

\fakeparagraph{Merits}
Existing DCO methods offer several key advantages:
\begin{itemize}
    \item \textbf{Enhanced Query Throughput.} 
    These methods can substantially improve the query throughput of vector similarity search.
    For instance, on a CPU with SIMD enabled, \RPDScanning, \ADSampling, \DADE, \DDCres, \DDCpca, and \DDCopq improved QPS by up to 1.9, 1.4, 1.8, 1.9, 2.1, and 1.6$\times$ on HNSW, respectively.

    \item \textbf{Native Recall Preservation.} 
    Most DCO methods preserve the native recall of underlying vector indexes across various scenarios, despite their approximate nature.
    For instance, \Hypothesis methods exhibit a recall variation usually below 2\% in our experimental study.

    \item \textbf{Accelerate Index Building \& Data Insertion.} 
    Several methods enable faster index building and data insertion.

    \item \textbf{Extensibility to Other Distance Metrics.} 
    Although primarily designed for Euclidean distance, these algorithms can be adapted to other prevalent metrics while maintaining recall, such as inner product and cosine similarity.
\end{itemize}

% 0.1--0.4$\times$
\fakeparagraph{Limitations}
We also identify the following key limitations:
\begin{itemize}
    \item \textbf{No Silver Bullet for Universal Efficiency.}
    The efficiency advantages of these methods are not universal but highly dependent on data dimensionality. 
    On low- and ultra-high-dimensional datasets, most SOTA methods are slower than \Loop baselines, with a QPS reduction of up to 8\%--42\% and 77\%--79\% respectively.

    \item \textbf{Vulnerability to Out-of-Distribution (OOD) Queries.}
    Their performance is not robust to distribution shifts of query vectors.
    Experiments on multimodal datasets (\Laion and \TextImage) show that \Loop methods (\eg \FDScanning) can be more efficient than several SOTA methods, including \ADSampling, \DADE, \DDCres, \DDCpca, and \DDCopq.

    \item \textbf{Dependence on Hardware Environment.}
    The efficiency gains of DCO methods are contingent on hardware.
    When SIMD enabled, the QPS of all methods improves, but their advantages over \FDScanning narrow significantly, dropping from 2.0--4.4$\times$ to 1.2--3.8$\times$ on IVF, and from 1.9--3.1$\times$ to 1.1--2.1$\times$ on HNSW (see our full paper \cite{fullpaper}).

    \item \textbf{Restrictive Assumptions Limit General Applicability.}
    Most DCO methods rely on specific assumptions about data distributions, query workloads, or distance metrics. 
    For example, despite being a top performer in several scenarios, \DDCpca requires prior knowledge of the query workload and parameter $k$, sufficient training data, and is primarily designed for Euclidean and IP distances.
\end{itemize}

\section{Related Work}\label{sec:related}

We review the related work from the following two aspects: \textit{vector similarity search} and \textit{vector distance operation}.

\fakeparagraph{Vector Similarity Search}
Vector similarity search is broadly categorized into \textit{exact} and \textit{approximate} solutions.
While exact solutions guarantee correctness, they often suffer from high query latency on large-scale datasets due to the curse of dimensionality \cite{toth2017handbook}.
Consequently, recent work has focused primarily on approximate algorithms \cite{DBLP:journals/tkde/LiZSWLZL20,DBLP:journals/pvldb/Qin0X020,DBLP:conf/kdd/Qin000W21}. 
A common optimization technique in these algorithms is the usage of specialized indexes, which generally fall into three categories: \textit{graph-based indexes} (\eg HNSW \cite{DBLP:journals/pami/MalkovY20}, HVS \cite{DBLP:journals/pvldb/LuKXI21}, and SHG \cite{DBLP:journals/pvldb/GongZC25}), \textit{partition-based indexes} (\eg IVF \cite{DBLP:journals/pami/BabenkoL15} and Ada-IVF\cite{mohoney2024incremental}), and \textit{hash-based indexes} (\eg PM-LSH \cite{DBLP:journals/vldb/ZhengZWNLJ22}).

% \begin{itemize}
%     \item \textbf{Graph-based indexes} (\eg HNSW \cite{DBLP:journals/pami/MalkovY20}, HVS \cite{DBLP:journals/pvldb/LuKXI21}, and SHG \cite{DBLP:journals/pvldb/GongZC25}) organize vectors into a graph (or a hierarchy of graphs), where each node represents a vector and is connected by edges to sufficiently nearby vectors. 

%     \item \textbf{Partition-based indexes} (\eg IVF \cite{DBLP:journals/pami/BabenkoL15} and Ada-IVF\cite{mohoney2024incremental}) divide the vector space into multiple shards, enabling many shards to be pruned during query processing.
%     One of the most popular partitioning strategy is high-dimensional vector clustering \cite{DBLP:journals/tkde/FuCCWH24}. 
    
%     \item \textbf{Hash-based indexes}, such as QALSH \cite{huang2015query} and PM-LSH \cite{DBLP:journals/vldb/ZhengZWNLJ22}, employ locality-sensitive hashing (LSH) \cite{DBLP:conf/stoc/IndykM98} to map similar vectors into the same hash bucket, thereby narrowing the search space. 
% \end{itemize}
In practice, graph and partition-based indexes are the most widely adopted in production vector databases \cite{DBLP:conf/sigmod/WangYGJXLWGLXYY21,weaviate} and search engines \cite{douze2024faiss,qdrant}.
Typical examples include HNSW \cite{DBLP:journals/pami/MalkovY20} and IVF \cite{DBLP:journals/pami/BabenkoL15}, respectively. 
Recent studies also explore hybrid models, such as APG-LSH \cite{DBLP:journals/pvldb/ZhaoTHZZ23}, which integrates both graph and LSH to leverage their complementary strengths.
While existing experimental studies and surveys \cite{DBLP:journals/debu/TianYZ0Z023,pan2024survey,DBLP:journals/tkde/LiZSWLZL20,DBLP:journals/pvldb/WangXY021,DBLP:journals/csur/ChenGSLZMJ23} have primarily focused on index structures, our work shifts the focus to its foundation operation (see below).

% IVF \cite{DBLP:journals/pami/BabenkoL15}, divide the vector space into regions, enabling efficient pruning of irrelevant partitions during search. IVF is also commonly used in production vector databases\cite{DBLP:conf/sigmod/WangYGJXLWGLXYY21,weaviate,qdrant} and scales effectively to datasets with tens of millions of vectors or more\cite{douze2024faiss}.

%Vector similarity search in indexing-based solutions typically adopts a filter-and-refine framework. During query processing, a large number of \textit{vector distance operations} are performed. Besides the standard distance computation, several other operations are commonly used:

\fakeparagraph{Vector Distance Operation}
Query processing in these indexing-based solutions involves numerous \textit{vector distance operations}, such as vector distance computation, comparison \cite{DBLP:journals/pacmmod/GaoL23, DBLP:journals/pvldb/DengCZWZZ24, yang2025effective, DBLP:journals/debu/0007X0H0P024, DBLP:conf/isca/LiJTZ025}, quantization \cite{DBLP:journals/tkde/FuCCWH24,jegou2010product}, and fusion \cite{wang2023efficient, ait2025rwalks}.
% Beyond the standard \textit{distance computation operation}, several other operations are commonly used:
% \begin{itemize}
%     \item \textbf{Comparison.} 
%     Distance Comparison Operation (DCO) compares the distance between a data vector and a query vector against a given threshold.

%     \item \textbf{Quantization.}
%     Quantization reduces the dimensionality of a high-dimensional vector and enables faster (approximate) distance computation.

%     \item \textbf{Fusion.}
%     This operation integrates vector embeddings and their associated attributes into a single distance function \cite{wang2023efficient, ait2025rwalks}.
%     It is often applied within a hybrid data model that combines vector and relational data.
% \end{itemize}

Among these operations, DCO is our primary concern. 
Although straightforward, the DCO methods, \FDScanning and \PDScanning, are commonly employed in vector databases and search engines \cite{muja2009fast, douze2024faiss, DBLP:journals/pami/MalkovY20, DBLP:conf/sigmod/WangYGJXLWGLXYY21, yang2020pase, LibMetricSpace}.
Moreover, \PDScanning has also been applied in approximate nearest neighbor search in near-data processing architecture \cite{DBLP:conf/isca/LiJTZ025}.
\RPDScanning \cite{DBLP:journals/debu/0007X0H0P024} applies PCA in \PDScanning to prioritize dimensions with greater impact on distances.
While Gao \etal \cite{DBLP:journals/pacmmod/GaoL23} first introduced hypothesis testing to support DCOs, Deng \etal \cite{DBLP:journals/pvldb/DengCZWZZ24} and Yang \etal \cite{yang2025effective} later built upon this technique to propose more effective solutions.
Furthermore, Yang \etal \cite{yang2025effective} formulated DCO as a classification problem, and developed two \Class methods: \DDCpca and \DDCopq.

\section{Conclusion and Future Direction}\label{sec:conclusion}

Distance comparison operation (DCO) has emerged as a promising method for accelerating vector similarity search.
While prior work has introduced several effective DCO algorithms, their evaluations have often been conducted in restricted settings, failing to fully reveal their metrics and expose their limitations under broader yet realistic conditions.
To bridge this gap, we have conducted a comprehensive benchmark on eight state-of-the-art (SOTA) methods across ten diverse datasets.
Our evaluation demonstrates that \textit{no current method represents a silver bullet}: for out-of-distribution (OOD) queries, or on vector datasets with relatively low or ultra-high dimensionality, SOTA methods can even increase query latency compared to naive baselines.

Motivated by these observations, we identify the following research directions for future work:

(1) Developing specialized optimization methods tailored to relatively low- or ultra-high-dimensional datasets.

(2) Improving robustness against OOD queries, which are  prevalent in real-world applications like multimodal retrieval.

(3) Incorporating hardware configuration (\eg GPU and SIMD) as a primary design consideration.

(4) Extending the scope of DCOs beyond Euclidean metrics and other limiting assumptions to enhance generality.

\clearpage
\section*{AI-Generated Content Acknowledgement}
During the preparation of this work, we used DeepSeek-R1 in order to assist with proofreading and correction of grammatical errors and typos throughout the entire manuscript. This AI tool was applied to main sections of the article at a superficial, non-substantive level to improve readability. The authors reviewed and edited all AI-suggested changes and take full responsibility for the content of the manuscript.
\balance
\bibliographystyle{IEEEtran}
% \bibliography{dco}

\begin{thebibliography}{10}
\providecommand{\url}[1]{#1}
\csname url@samestyle\endcsname
\providecommand{\newblock}{\relax}
\providecommand{\bibinfo}[2]{#2}
\providecommand{\BIBentrySTDinterwordspacing}{\spaceskip=0pt\relax}
\providecommand{\BIBentryALTinterwordstretchfactor}{4}
\providecommand{\BIBentryALTinterwordspacing}{\spaceskip=\fontdimen2\font plus
\BIBentryALTinterwordstretchfactor\fontdimen3\font minus
  \fontdimen4\font\relax}
\providecommand{\BIBforeignlanguage}[2]{{%
\expandafter\ifx\csname l@#1\endcsname\relax
\typeout{** WARNING: IEEEtran.bst: No hyphenation pattern has been}%
\typeout{** loaded for the language `#1'. Using the pattern for}%
\typeout{** the default language instead.}%
\else
\language=\csname l@#1\endcsname
\fi
#2}}
\providecommand{\BIBdecl}{\relax}
\BIBdecl

\bibitem{DBLP:journals/pacmmod/GaoL23}
J.~Gao and C.~Long, ``High-dimensional approximate nearest neighbor search:
  with reliable and efficient distance comparison operations,'' \emph{Proc.
  {ACM} Manag. Data}, vol.~1, no.~2, pp. 137:1--137:27, 2023.

\bibitem{DBLP:journals/pvldb/DengCZWZZ24}
L.~Deng, P.~Chen, X.~Zeng, T.~Wang, Y.~Zhao, and K.~Zheng, ``Efficient
  data-aware distance comparison operations for high-dimensional approximate
  nearest neighbor search,'' \emph{{PVLDB}}, vol.~18, no.~3, pp. 812--821,
  2024.

\bibitem{yang2025effective}
M.~Yang, W.~Li, J.~Jin, X.~Zhong, X.~Wang, Z.~Shen, W.~Jia, and W.~Wang,
  ``Effective and general distance computation for approximate nearest neighbor
  search,'' in \emph{{ICDE}}, 2025, pp. 1098--1110.

\bibitem{DBLP:journals/debu/0007X0H0P024}
Z.~Wang, H.~Xiong, Q.~Wang, Z.~He, P.~Wang, T.~Palpanas, and W.~Wang,
  ``Dimensionality-reduction techniques for approximate nearest neighbor
  search: {A} survey and evaluation,'' \emph{{IEEE} Data Eng. Bull.}, vol.~48,
  no.~3, pp. 63--80, 2024.

\bibitem{DBLP:conf/isca/LiJTZ025}
Y.~Li, Y.~Jin, B.~Tian, H.~Zhang, and M.~Gao, ``{ANSMET:} approximate nearest
  neighbor search with near-memory processing and hybrid early termination,''
  in \emph{{ISCA}}, 2025, pp. 1093--1107.

\bibitem{DBLP:journals/tkde/LiSW24}
L.~Li, W.~Sun, and B.~Wu, ``Dforest: {A} minimal dimensionality-aware indexing
  for high-dimensional exact similarity search,'' \emph{{IEEE} Trans. Knowl.
  Data Eng.}, vol.~36, no.~10, pp. 5092--5105, 2024.

\bibitem{DBLP:conf/sigmod/WangYGJXLWGLXYY21}
J.~Wang, X.~Yi, R.~Guo, H.~Jin, P.~Xu, S.~Li, X.~Wang, X.~Guo, C.~Li, X.~Xu,
  K.~Yu, Y.~Yuan, Y.~Zou, J.~Long, Y.~Cai, Z.~Li, Z.~Zhang, Y.~Mo, J.~Gu,
  R.~Jiang, Y.~Wei, and C.~Xie, ``Milvus: {A} purpose-built vector data
  management system,'' in \emph{{SIGMOD}}, 2021, pp. 2614--2627.

\bibitem{weaviate}
\BIBentryALTinterwordspacing
``{Weaviate}.'' [Online]. Available: \url{https://github.com/weaviate/weaviate}
\BIBentrySTDinterwordspacing

\bibitem{qdrant}
\BIBentryALTinterwordspacing
``{Qdrant}.'' [Online]. Available: \url{https://github.com/qdrant/qdrant}
\BIBentrySTDinterwordspacing

\bibitem{code}
\BIBentryALTinterwordspacing
``Source code and datasets in the benchmark.'' [Online]. Available:
  \url{https://github.com/zzlin237/dco-benchmark.git}
\BIBentrySTDinterwordspacing

\bibitem{sun2025gaussdb}
G.~Li, J.~Sun, J.~Pan, J.~Wang, Y.~Xie, R.~Liu, and W.~Nie, ``Gaussdb-vector:
  {A} large-scale persistent real-time vector database for {LLM}
  applications,'' \emph{{PVLDB}}, vol.~18, no.~12, pp. 4951--4963, 2025.

\bibitem{han2023comprehensive}
Y.~Han, C.~Liu, and P.~Wang, ``A comprehensive survey on vector database:
  Storage and retrieval technique, challenge,'' \emph{CoRR}, vol.
  abs/2310.11703, 2023.

\bibitem{casella2024statistical}
G.~Casella and R.~Berger, \emph{Statistical inference}.\hskip 1em plus 0.5em
  minus 0.4em\relax Chapman and Hall/CRC, 2024.

\bibitem{james2013introduction}
G.~James, D.~Witten, T.~Hastie, and R.~Tibshirani, \emph{An introduction to
  statistical learning: with applications in R}.\hskip 1em plus 0.5em minus
  0.4em\relax Springer, 2013, vol. 103.

\bibitem{szeliski2022computer}
R.~Szeliski, \emph{Computer vision: algorithms and applications}.\hskip 1em
  plus 0.5em minus 0.4em\relax Springer Nature, 2022.

\bibitem{manning1999foundations}
C.~Manning and H.~Schutze, \emph{Foundations of statistical natural language
  processing}.\hskip 1em plus 0.5em minus 0.4em\relax MIT press, 1999.

\bibitem{min2023recent}
B.~Min, H.~Ross, E.~Sulem, A.~P.~B. Veyseh, T.~H. Nguyen, O.~Sainz, E.~Agirre,
  I.~Heintz, and D.~Roth, ``Recent advances in natural language processing via
  large pre-trained language models: {A} survey,'' \emph{{ACM} Comput. Surv.},
  vol.~56, no.~2, pp. 30:1--30:40, 2024.

\bibitem{pecher2024survey}
B.~Pecher, I.~Srba, and M.~Bielikov{\'{a}}, ``A survey on stability of learning
  with limited labelled data and its sensitivity to the effects of
  randomness,'' \emph{{ACM} Comput. Surv.}, vol.~57, no.~1, pp. 19:1--19:40,
  2025.

\bibitem{toth2017handbook}
C.~D. Toth, J.~O'Rourke, and J.~E. Goodman, \emph{Handbook of discrete and
  computational geometry}.\hskip 1em plus 0.5em minus 0.4em\relax CRC press,
  2017.

\bibitem{DBLP:journals/pvldb/WangXY021}
M.~Wang, X.~Xu, Q.~Yue, and Y.~Wang, ``A comprehensive survey and experimental
  comparison of graph-based approximate nearest neighbor search,''
  \emph{{PVLDB}}, vol.~14, no.~11, pp. 1964--1978, 2021.

\bibitem{pan2024survey}
J.~J. Pan, J.~Wang, and G.~Li, ``Survey of vector database management
  systems,'' \emph{{VLDB} J.}, vol.~33, no.~5, pp. 1591--1615, 2024.

\bibitem{johnson2019billion}
J.~Johnson, M.~Douze, and H.~J{\'{e}}gou, ``Billion-scale similarity search
  with gpus,'' \emph{{IEEE} Trans. Big Data}, vol.~7, no.~3, pp. 535--547,
  2021.

\bibitem{douze2024faiss}
M.~Douze, A.~Guzhva, C.~Deng, J.~Johnson, G.~Szilvasy, P.~Mazar{\'{e}},
  M.~Lomeli, L.~Hosseini, and H.~J{\'{e}}gou, ``The faiss library,''
  \emph{CoRR}, vol. abs/2401.08281, 2024.

\bibitem{DBLP:conf/www/ChenCJYDH23}
P.~H. Chen, W.~Chang, J.~Jiang, H.~Yu, I.~S. Dhillon, and C.~Hsieh, ``{FINGER:}
  fast inference for graph-based approximate nearest neighbor search,'' in
  \emph{{WWW}}, 2023, pp. 3225--3235.

\bibitem{muja2009fast}
M.~Muja and D.~G. Lowe, ``Fast approximate nearest neighbors with automatic
  algorithm configuration,'' in \emph{{VISAPP}}, 2009, pp. 331--340.

\bibitem{abdi2010principal}
H.~Abdi and L.~J. Williams, ``Principal component analysis,'' \emph{WIRES
  COMPUT STAT}, vol.~2, no.~4, pp. 433--459, 2010.

\bibitem{DBLP:journals/pami/MalkovY20}
Y.~A. Malkov and D.~A. Yashunin, ``Efficient and robust approximate nearest
  neighbor search using hierarchical navigable small world graphs,''
  \emph{{IEEE} Trans. Pattern Anal. Mach. Intell.}, vol.~42, no.~4, pp.
  824--836, 2020.

\bibitem{DBLP:journals/pami/BabenkoL15}
A.~Babenko and V.~S. Lempitsky, ``The inverted multi-index,'' \emph{{IEEE}
  Trans. Pattern Anal. Mach. Intell.}, vol.~37, no.~6, pp. 1247--1260, 2015.

\bibitem{bigann}
\BIBentryALTinterwordspacing
``{Big {ANN} Benchmark},'' 2024. [Online]. Available:
  \url{https://big-ann-benchmarks.com/}
\BIBentrySTDinterwordspacing

\bibitem{glove}
\BIBentryALTinterwordspacing
``{GloVe: Global Vectors for Word Representation}.'' [Online]. Available:
  \url{https://nlp.stanford.edu/projects/glove/}
\BIBentrySTDinterwordspacing

\bibitem{DBLP:journals/corr/abs-2111-02114}
C.~Schuhmann, R.~Vencu, R.~Beaumont, R.~Kaczmarczyk, C.~Mullis, A.~Katta,
  T.~Coombes, J.~Jitsev, and A.~Komatsuzaki, ``{LAION-400M:} open dataset of
  clip-filtered 400 million image-text pairs,'' \emph{CoRR}, vol.
  abs/2111.02114, 2021.

\bibitem{nguyen2016ms}
T.~Nguyen, M.~Rosenberg, X.~Song, J.~Gao, S.~Tiwary, R.~Majumder, and L.~Deng,
  ``{MS} {MARCO:} {A} human generated machine reading comprehension dataset,''
  in \emph{{CoCo@NIPS}}, vol. 1773, 2016.

\bibitem{openai2022improved}
``New and improved embedding model,''
  \url{https://openai.com/index/new-and-improved-embedding-model/}.

\bibitem{fullpaper}
\BIBentryALTinterwordspacing
``Distance comparison operations are not silver bullets in vector similarity
  search: An experimental study on their merits and limits (full paper).''
  [Online]. Available: \url{https://github.com/zzlin237/dco-benchmark.git}
\BIBentrySTDinterwordspacing

\bibitem{jayaram2019diskann}
S.~Jayaram~Subramanya, F.~Devvrit, H.~V. Simhadri, R.~Krishnawamy, and
  R.~Kadekodi, ``Diskann: Fast accurate billion-point nearest neighbor search
  on a single node,'' \emph{{NeurIPS}}, vol.~32, 2019.

\bibitem{DBLP:conf/nips/ZhangWCCZMHDMWP23}
H.~Zhang, Y.~Wang, Q.~Chen, R.~Chang, T.~Zhang, Z.~Miao, Y.~Hou, Y.~Ding,
  X.~Miao, H.~Wang, B.~Pang, Y.~Zhan, H.~Sun, W.~Deng, Q.~Zhang, F.~Yang,
  X.~Xie, M.~Yang, and B.~Cui, ``Model-enhanced vector index,'' in
  \emph{{NeurIPS}}, 2023.

\bibitem{guide2011intel}
P.~Guide, ``Intel{\textregistered} 64 and ia-32 architectures software
  developer’s manual,'' \emph{Volume 3B: system programming guide, Part},
  vol.~2, no.~11, pp. 0--40, 2011.

\bibitem{DBLP:journals/tkde/LiZSWLZL20}
W.~Li, Y.~Zhang, Y.~Sun, W.~Wang, M.~Li, W.~Zhang, and X.~Lin, ``Approximate
  nearest neighbor search on high dimensional data - experiments, analyses, and
  improvement,'' \emph{{IEEE} Trans. Knowl. Data Eng.}, vol.~32, no.~8, pp.
  1475--1488, 2020.

\bibitem{DBLP:journals/pvldb/Qin0X020}
J.~Qin, W.~Wang, C.~Xiao, and Y.~Zhang, ``Similarity query processing for
  high-dimensional data,'' \emph{{PVLDB}}, vol.~13, no.~12, pp. 3437--3440,
  2020.

\bibitem{DBLP:conf/kdd/Qin000W21}
J.~Qin, W.~Wang, C.~Xiao, Y.~Zhang, and Y.~Wang, ``High-dimensional similarity
  query processing for data science,'' in \emph{{KDD}}, 2021, pp. 4062--4063.

\bibitem{DBLP:journals/pvldb/LuKXI21}
K.~Lu, M.~Kudo, C.~Xiao, and Y.~Ishikawa, ``{HVS:} hierarchical graph structure
  based on voronoi diagrams for solving approximate nearest neighbor search,''
  \emph{{PVLDB}}, vol.~15, no.~2, pp. 246--258, 2021.

\bibitem{DBLP:journals/pvldb/GongZC25}
Z.~Gong, Y.~Zeng, and L.~Chen, ``Accelerating approximate nearest neighbor
  search in hierarchical graphs: Efficient level navigation with shortcuts,''
  \emph{{PVLDB}}, vol.~18, no.~10, pp. 3518--3530, 2025.

\bibitem{mohoney2024incremental}
J.~Mohoney, A.~Pacaci, S.~R. Chowdhury, U.~F. Minhas, J.~Pound, C.~Renggli,
  N.~Reyhani, I.~F. Ilyas, T.~Rekatsinas, and S.~Venkataraman, ``Incremental
  ivf index maintenance for streaming vector search,'' \emph{arXiv preprint
  arXiv:2411.00970}, 2024.

\bibitem{DBLP:journals/vldb/ZhengZWNLJ22}
B.~Zheng, X.~Zhao, L.~Weng, Q.~V.~H. Nguyen, H.~Liu, and C.~S. Jensen,
  ``{PM-LSH:} a fast and accurate in-memory framework for high-dimensional
  approximate {NN} and closest pair search,'' \emph{{VLDB} J.}, vol.~31, no.~6,
  pp. 1339--1363, 2022.

\bibitem{DBLP:journals/pvldb/ZhaoTHZZ23}
X.~Zhao, Y.~Tian, K.~Huang, B.~Zheng, and X.~Zhou, ``Towards efficient index
  construction and approximate nearest neighbor search in high-dimensional
  spaces,'' \emph{{PVLDB}}, vol.~16, no.~8, pp. 1979--1991, 2023.

\bibitem{DBLP:journals/debu/TianYZ0Z023}
Y.~Tian, Z.~Yue, R.~Zhang, X.~Zhao, B.~Zheng, and X.~Zhou, ``Approximate
  nearest neighbor search in high dimensional vector databases: Current
  research and future directions,'' \emph{{IEEE} Data Eng. Bull.}, vol.~47,
  no.~3, pp. 39--54, 2023.

\bibitem{DBLP:journals/csur/ChenGSLZMJ23}
L.~Chen, Y.~Gao, X.~Song, Z.~Li, Y.~Zhu, X.~Miao, and C.~S. Jensen, ``Indexing
  metric spaces for exact similarity search,'' \emph{{ACM} Comput. Surv.},
  vol.~55, no.~6, pp. 128:1--128:39, 2023.

\bibitem{DBLP:journals/tkde/FuCCWH24}
Y.~Fu, C.~Chen, X.~Chen, W.~Wong, and B.~He, ``Optimizing the number of
  clusters for billion-scale quantization-based nearest neighbor search,''
  \emph{{IEEE} Trans. Knowl. Data Eng.}, vol.~36, no.~11, pp. 6786--6800, 2024.

\bibitem{jegou2010product}
H.~J{\'{e}}gou, M.~Douze, and C.~Schmid, ``Product quantization for nearest
  neighbor search,'' \emph{{IEEE} Trans. Pattern Anal. Mach. Intell.}, vol.~33,
  no.~1, pp. 117--128, 2011.

\bibitem{wang2023efficient}
M.~Wang, L.~Lv, X.~Xu, Y.~Wang, Q.~Yue, and J.~Ni, ``An efficient and robust
  framework for approximate nearest neighbor search with attribute
  constraint,'' in \emph{{NeurIPS}}, 2023.

\bibitem{ait2025rwalks}
A.~A. Aomar, K.~Echihabi, M.~Arnaboldi, I.~Alagiannis, D.~Hilloulin, and
  M.~Cherkaoui, ``Rwalks: Random walks as attribute diffusers for filtered
  vector search,'' \emph{Proc. {ACM} Manag. Data}, vol.~3, no.~3, pp.
  212:1--212:26, 2025.

\bibitem{yang2020pase}
W.~Yang, T.~Li, G.~Fang, and H.~Wei, ``{PASE:} postgresql
  ultra-high-dimensional approximate nearest neighbor search extension,'' in
  \emph{{SIGMOD}}, 2020, pp. 2241--2253.

\bibitem{LibMetricSpace}
\BIBentryALTinterwordspacing
K.~Figueroa, G.~Navarro, and E.~Chavez, ``The metric spaces library maintained
  by the sisap initiative,'' 2017. [Online]. Available:
  \url{https://github.com/kaarinita/metricSpaces}
\BIBentrySTDinterwordspacing

\bibitem{hnswlib}
\BIBentryALTinterwordspacing
``Hnswlib: fast approximate nearest neighbor search,'' 2025. [Online].
  Available: \url{https://github.com/nmslib/hnswlib}
\BIBentrySTDinterwordspacing

\end{thebibliography}

% Generated by IEEEtran.bst, version: 1.12 (2007/01/11)

\clearpage

\appendices
\section{Detailed Implementation}
\zheng{We implemented all DCO algorithms from scratch within a unified framework, adhering to their open-source implementations~\cite{DBLP:journals/pacmmod/GaoL23, DBLP:journals/pvldb/DengCZWZZ24, yang2025effective}. This section will elaborate on the relevant implementation details.

\fakeparagraph{Unified Framework}
Every DCO algorithm was deployed within the same vector similarity search framework with the same index. During online query processing, all compared methods are implemented in C++, while Python is only used for pre-processing (\ie PCA and training),
which is in line with existing work on DCOs~\cite{DBLP:journals/pacmmod/GaoL23, DBLP:journals/pvldb/DengCZWZZ24, yang2025effective}.

\fakeparagraph{Identical Index Configuration}
In terms of indexes, we fix a consistent underlying data layout for all methods, where vectors in the IVF index are stored contiguously within each partition while those in the HSNW index follow the original insertion sequence, and none of the DCO methods are allowed to modify this predefined layout. 

\fakeparagraph{Batch Queries}
Each query within a batch is processed sequentially and independently. 
We adopt this established evaluation protocol to maintain consistency with the baseline methods for a fair comparison.

\fakeparagraph{Hardware Optimization}
For HNSW, we utilize the SIMD-accelerated vector distance computation routines from the Hnswlib library~\cite{hnswlib}, where a single SIMD instruction processes multiple dimension pairs simultaneously. On the GPU, we pre-load the IVF index into device memory and adopt fine-grained parallelism: a single CUDA kernel is launched per partition, with each thread processing one candidate vector. This configuration is identical across all DCO methods to ensure consistent execution patterns.}

\fakeparagraph{Consistent Implementation Languages} 
The online query processing of all compared methods is implemented in C++, while Python is only used for pre-processing (\ie PCA and training), which is in line with existing work on DCOs~\cite{DBLP:journals/pacmmod/GaoL23, DBLP:journals/pvldb/DengCZWZZ24, yang2025effective}.

\section{Query Processing via HNSW on CPUs with SIMD Disabled}\label{app:HNSW-NO-SIMD}

This experiment evaluates the performance improvement from applying DCOs to vector similarity search on HNSW with SIMD disabled.

\begin{figure}[h]
    \centering
    \includegraphics[width=0.75\linewidth]{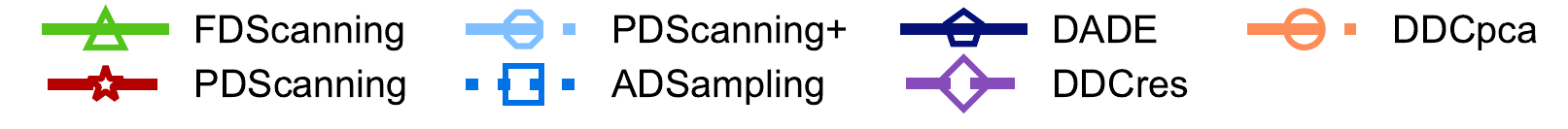}
    \begin{subfigure}{0.48\linewidth}
        \centering
        \includegraphics[width=\textwidth]{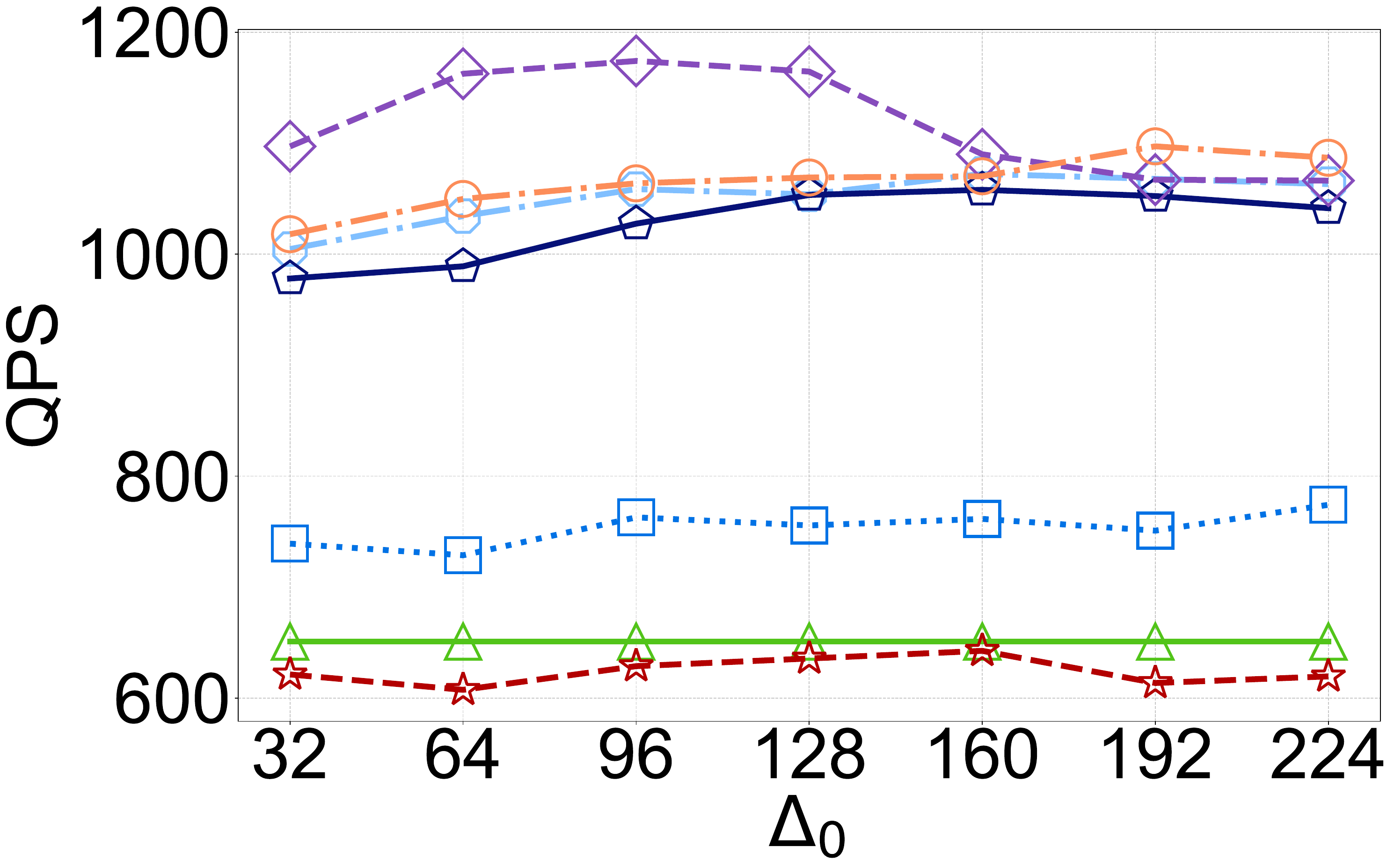}\vspace{-1.0ex}
        \caption{HNSW}\label{fig:initial-dimension-hnsw}
    \end{subfigure}%
    \begin{subfigure}{0.48\linewidth}
        \centering
        \includegraphics[width=\textwidth]{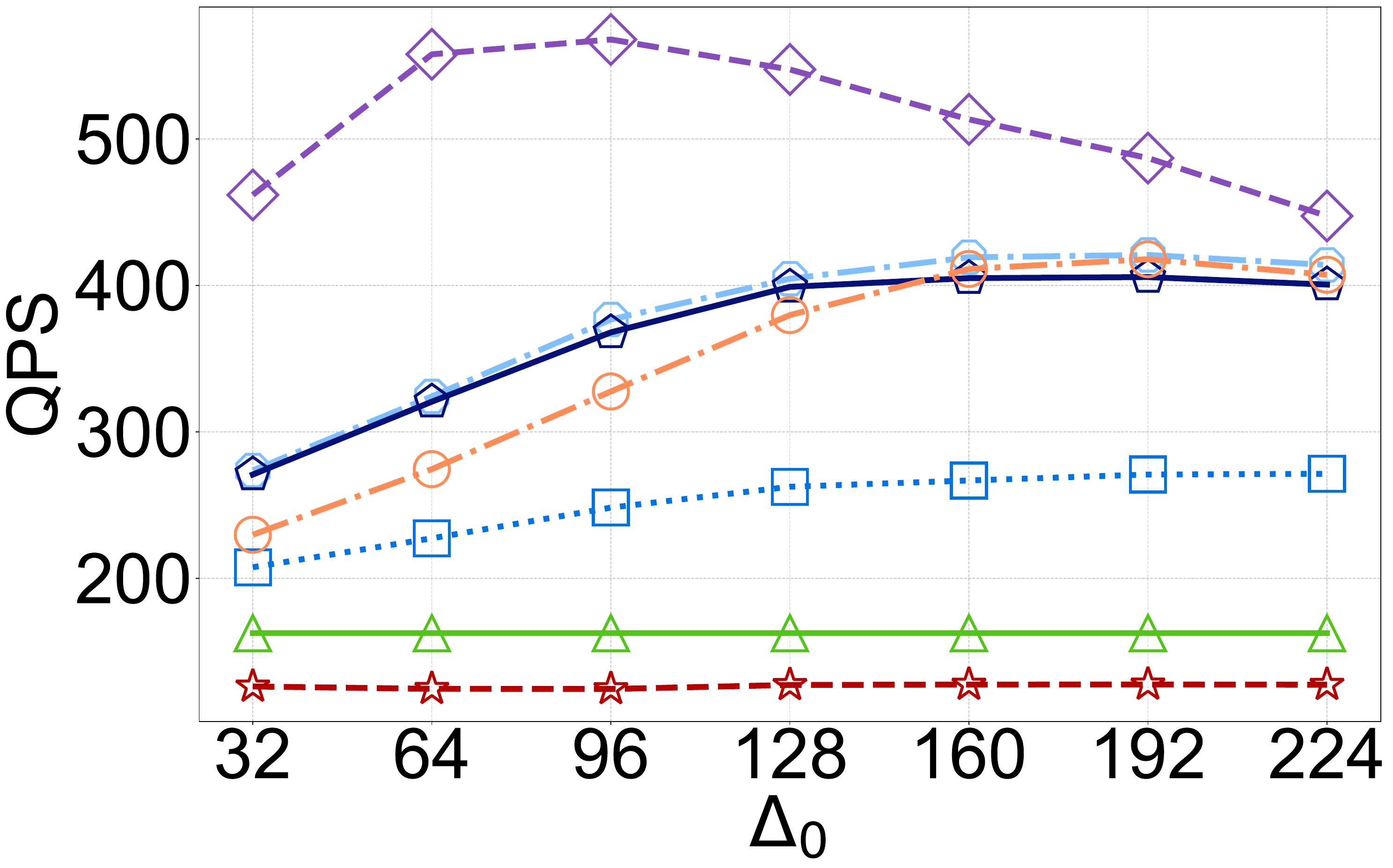}\vspace{-1.0ex}
        \caption{IVF}\label{fig:initial-dimension-ivf}
    \end{subfigure}
    \caption{Impact of initial step size $\Delta_0$}\label{fig:initial-dimension}
\end{figure}

\begin{figure}[h]
    \centering
    \includegraphics[width=0.75\linewidth]{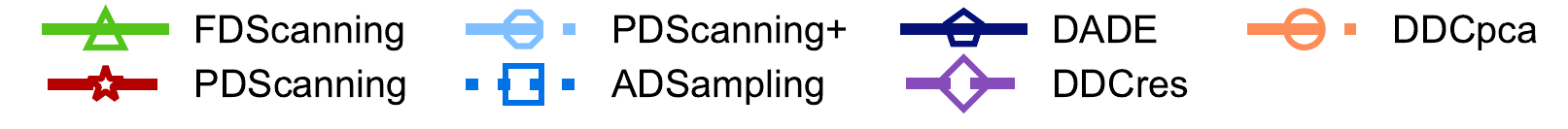}
    \begin{subfigure}{0.48\linewidth}
        \centering
        \includegraphics[width=\textwidth]{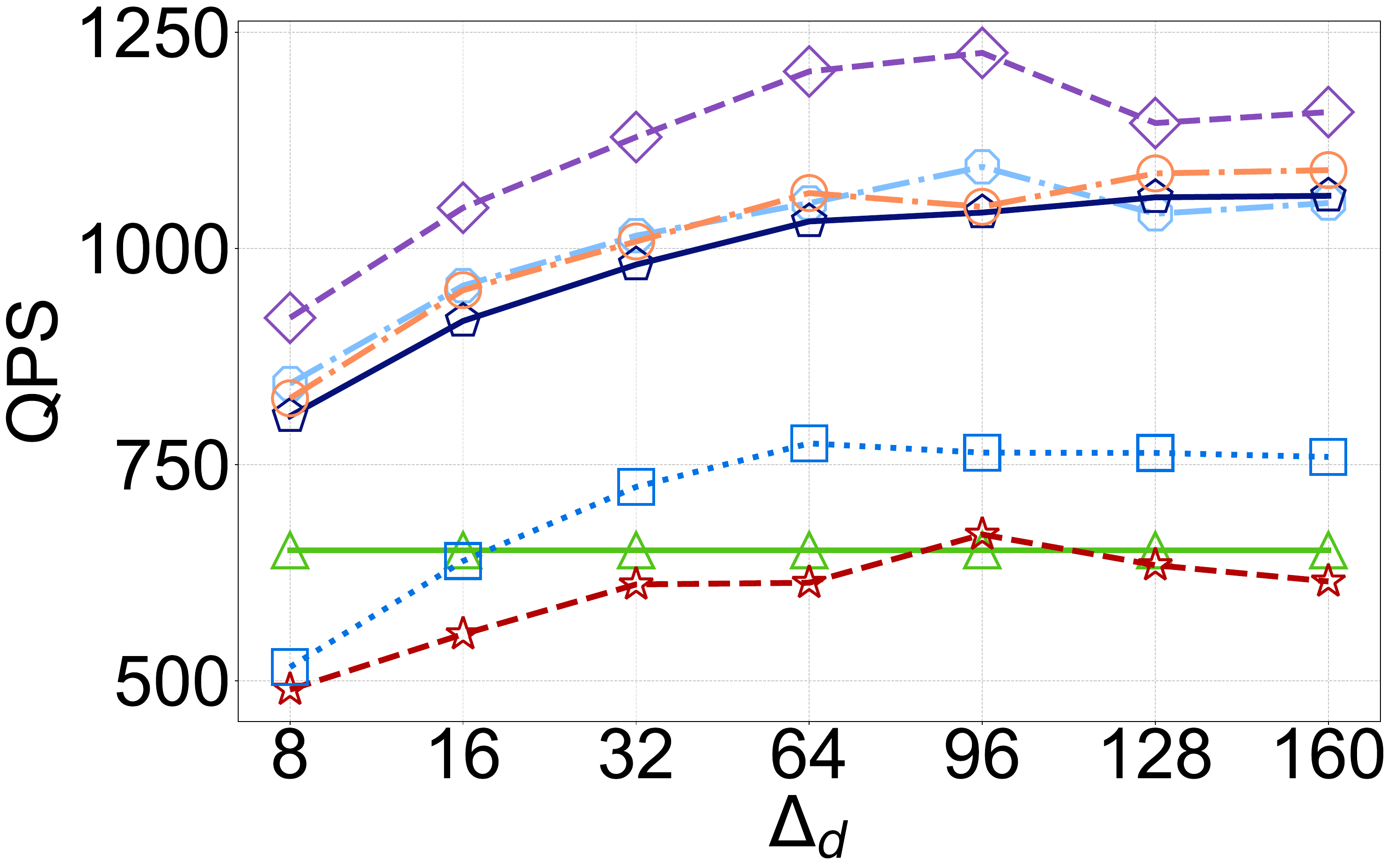}\vspace{-1.0ex}
        \caption{HNSW}\label{fig:sampling-step-hnsw}
    \end{subfigure}%
    \begin{subfigure}{0.48\linewidth}
        \centering
        \includegraphics[width=\textwidth]{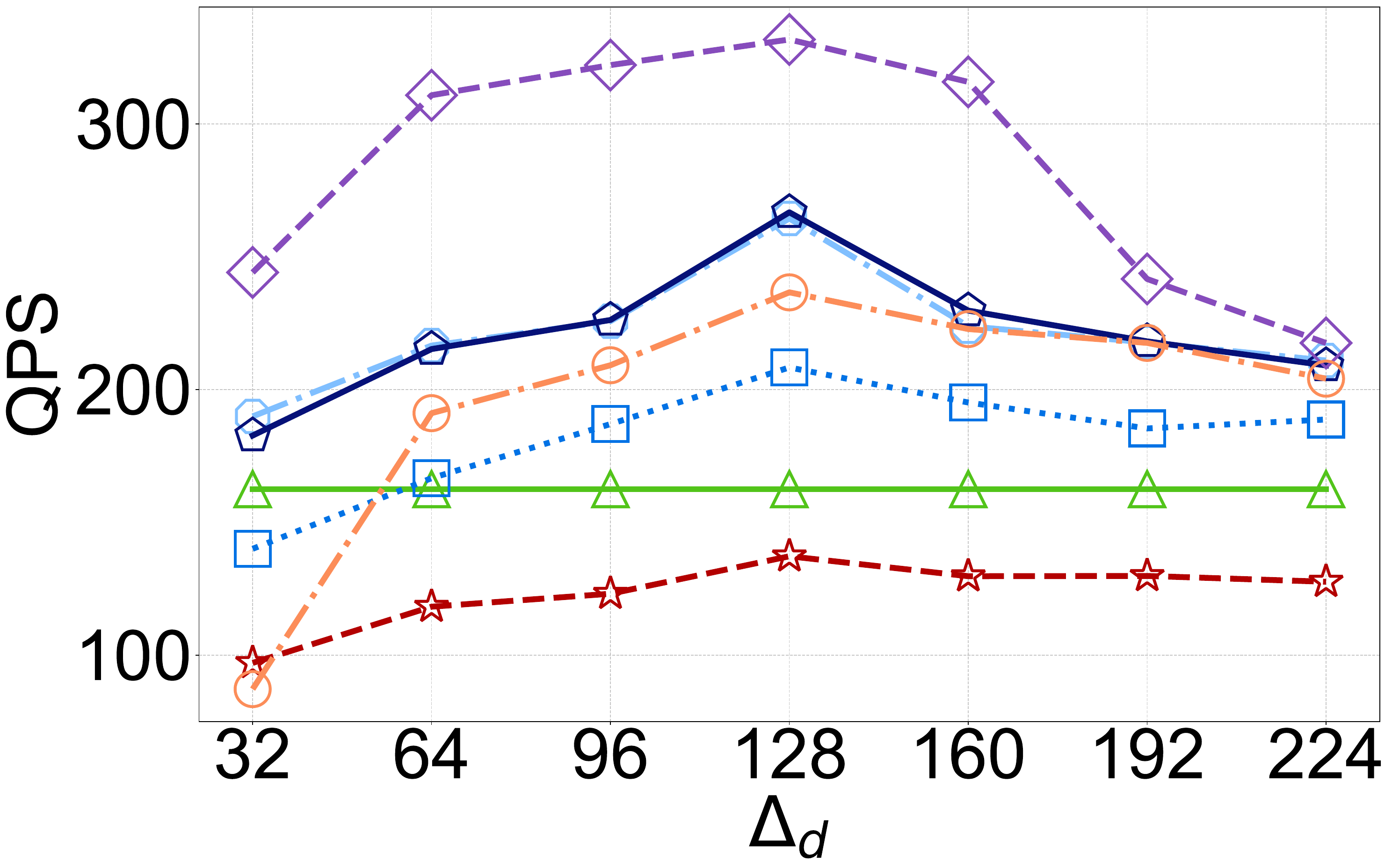}\vspace{-1.0ex}
        \caption{IVF}\label{fig:sampling-step-ivf}
    \end{subfigure}
    \caption{Impact of incremental step size $\Delta_d$}\label{fig:sampling-step}
\end{figure}

\fakeparagraph{Overall Query Performance without SIMD} Compared to the result in \figref{fig:time-accuracy}, the rankings of different DCO methods have undergone dramatic changes when SIMD is disabled. Specifically, \FDScanning no longer achieves the best performance on the \Deep dataset. Meanwhile, compared to \FDScanning, \Hypothesis methods improve QPS by up to 1.9--2.5$\times$, while the \Class methods achieve 2.1--3.1$\times$ improvements. These advantages over \FDScanning are larger than those observed when SIMD is enabled. This phenomenon can be attributed to the fact that SIMD primarily accelerates the distance computation itself, while the overhead of the termination decision logic, \ie determining whether to terminate computation remains unchanged. Consequently, the relative benefit of DCOs is diluted when SIMD is active, as the non-computational components dominate a larger fraction of the total time cost.

We also observe cases where the baseline \PDScanning outperforms other methods.  
As shown in \figref{fig:HNSW-NO_SIMD-trevi-20} and \figref{fig:HNSW-NO_SIMD-trevi-100}, \PDScanning achieves the highest QPS before the recall reaches 98\%. This demonstrates that the online pre-processing time $O(N^2)$ is also a bottleneck for the SOTA methods for some ultra-high-dimensional datasets on HNSW when SIMD is disabled.

\stepcounter{takeaway}
\vspace{-0.5ex}
\begin{tipbox}
\textbf{Takeaway: Enabling SIMD narrows the efficiency advantage of DCO methods over \FDScanning,} reducing the speedup from 1.9--3.1$\times$ to 1.1--2.1$\times$ on HNSW.
\end{tipbox}

\begin{figure*}[t]
    \centering
    \includegraphics[width=0.9\textwidth]{figure/query_performance/legend.pdf}
    \begin{subfigure}{0.24\textwidth}
        \centering
        \includegraphics[width=\textwidth]{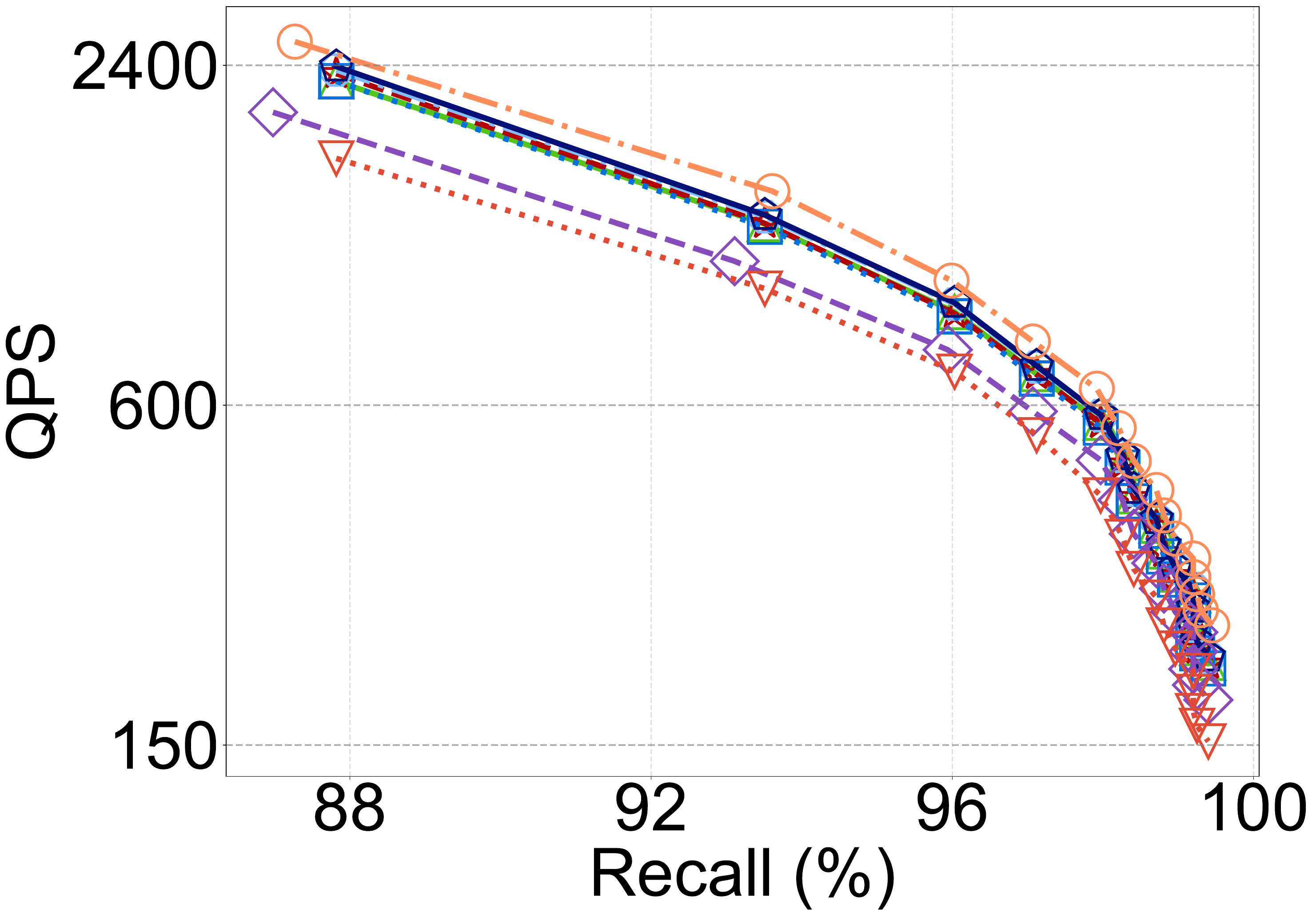}\vspace{-1.0ex}
        \caption{\Deep ($k=20$)}
    \end{subfigure}
    \begin{subfigure}{0.24\textwidth}
        \centering
        \includegraphics[width=\textwidth]{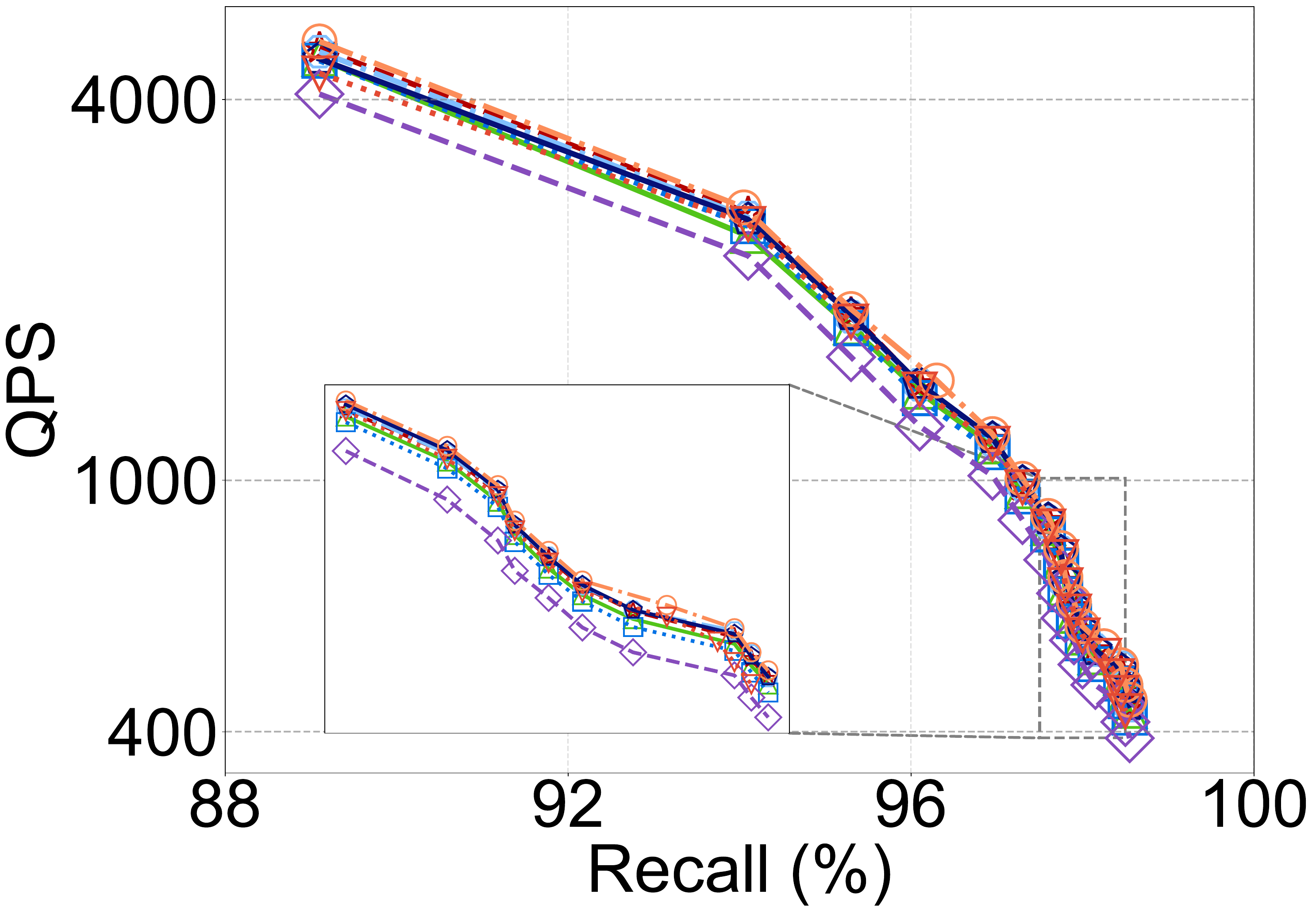}\vspace{-1.0ex}
        \caption{\Glove ($k=20$)}
    \end{subfigure}
    \begin{subfigure}{0.24\textwidth}
        \centering
        \includegraphics[width=\textwidth]{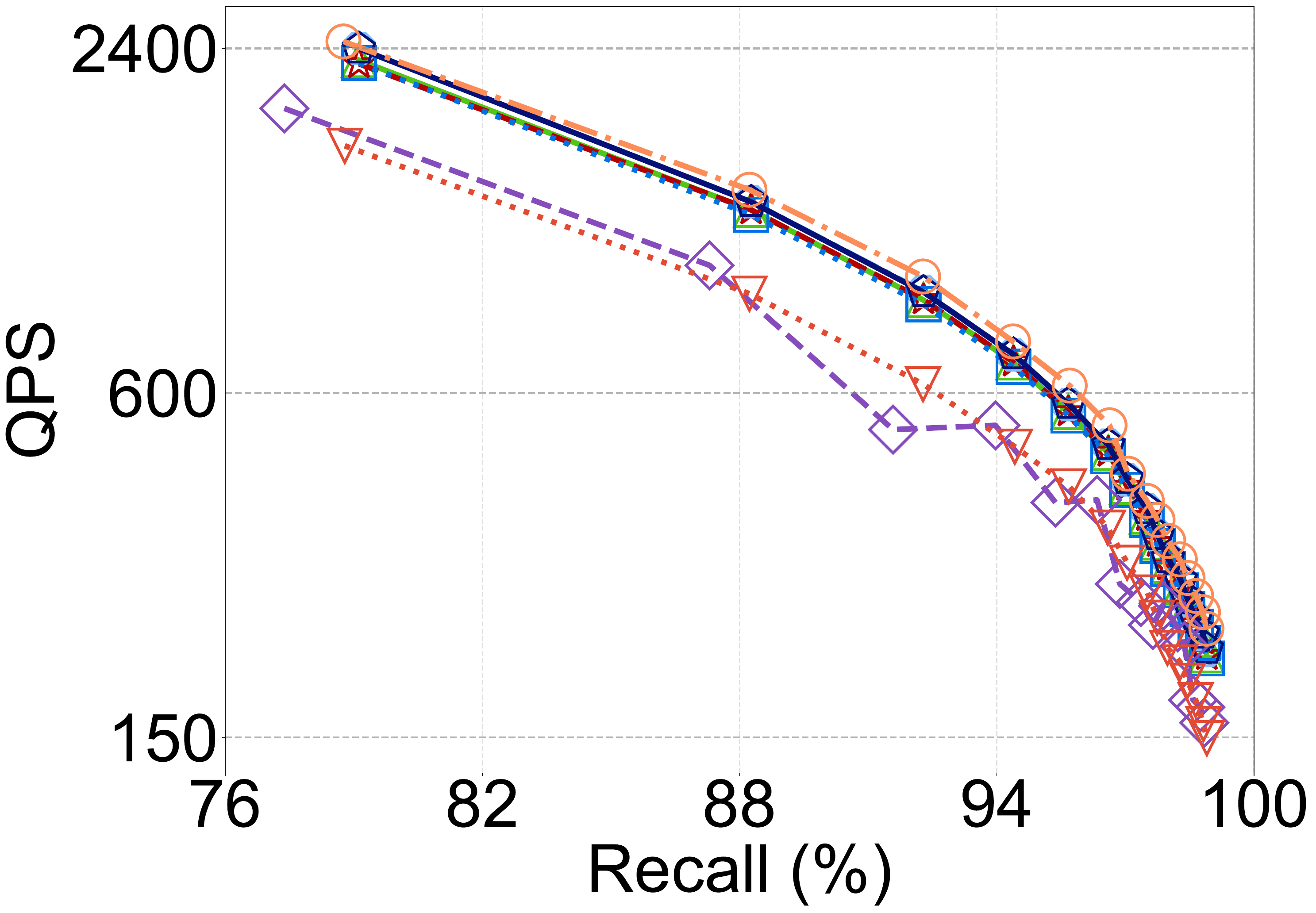}\vspace{-1.0ex}
        \caption{\Deep ($k=100$)}
    \end{subfigure}
    \begin{subfigure}{0.24\textwidth}
        \centering
        \includegraphics[width=\textwidth]{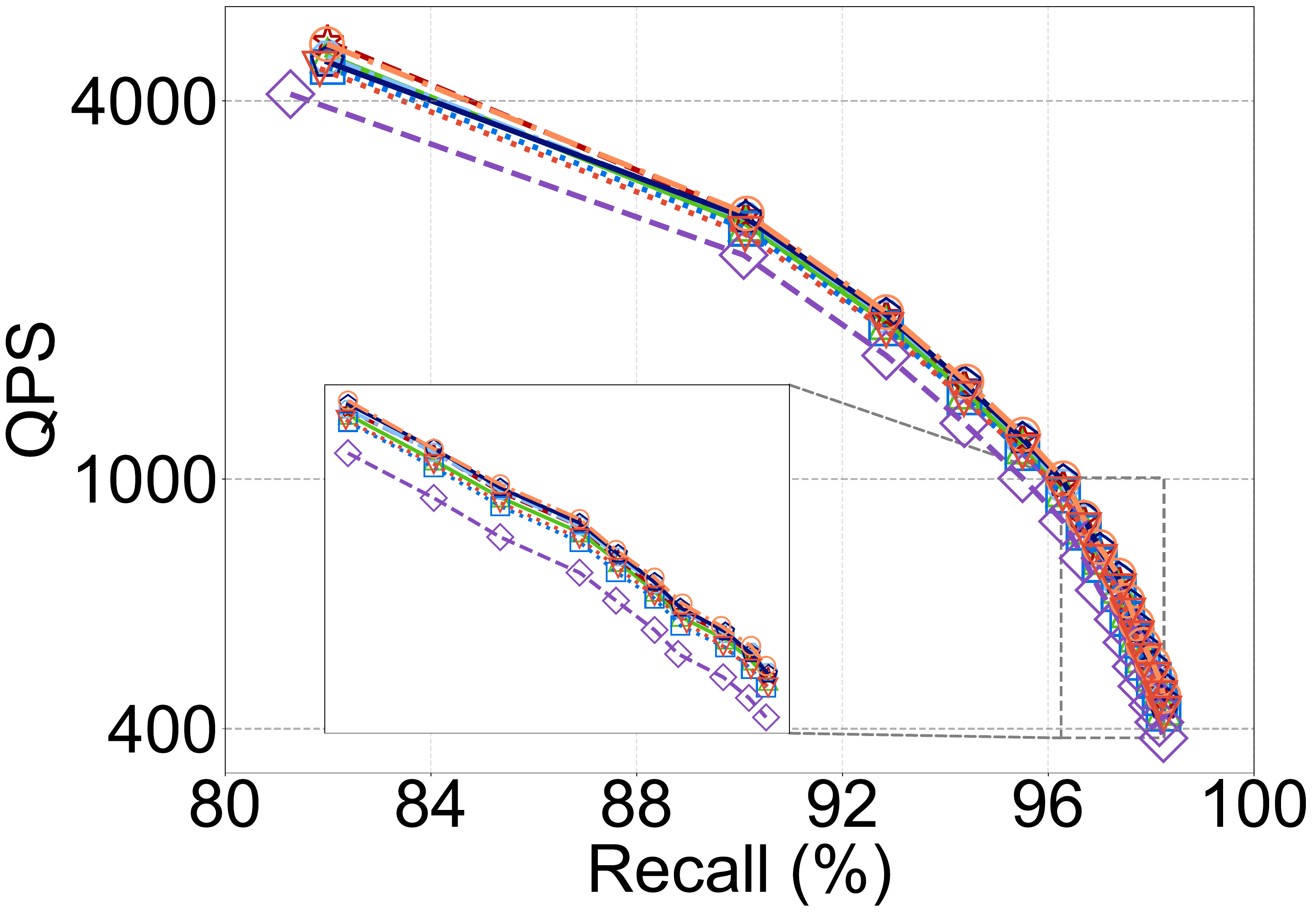}\vspace{-1.0ex}
        \caption{\Glove ($k=100$)}
    \end{subfigure}

    \begin{subfigure}{0.24\textwidth}
        \centering
        \includegraphics[width=\textwidth]{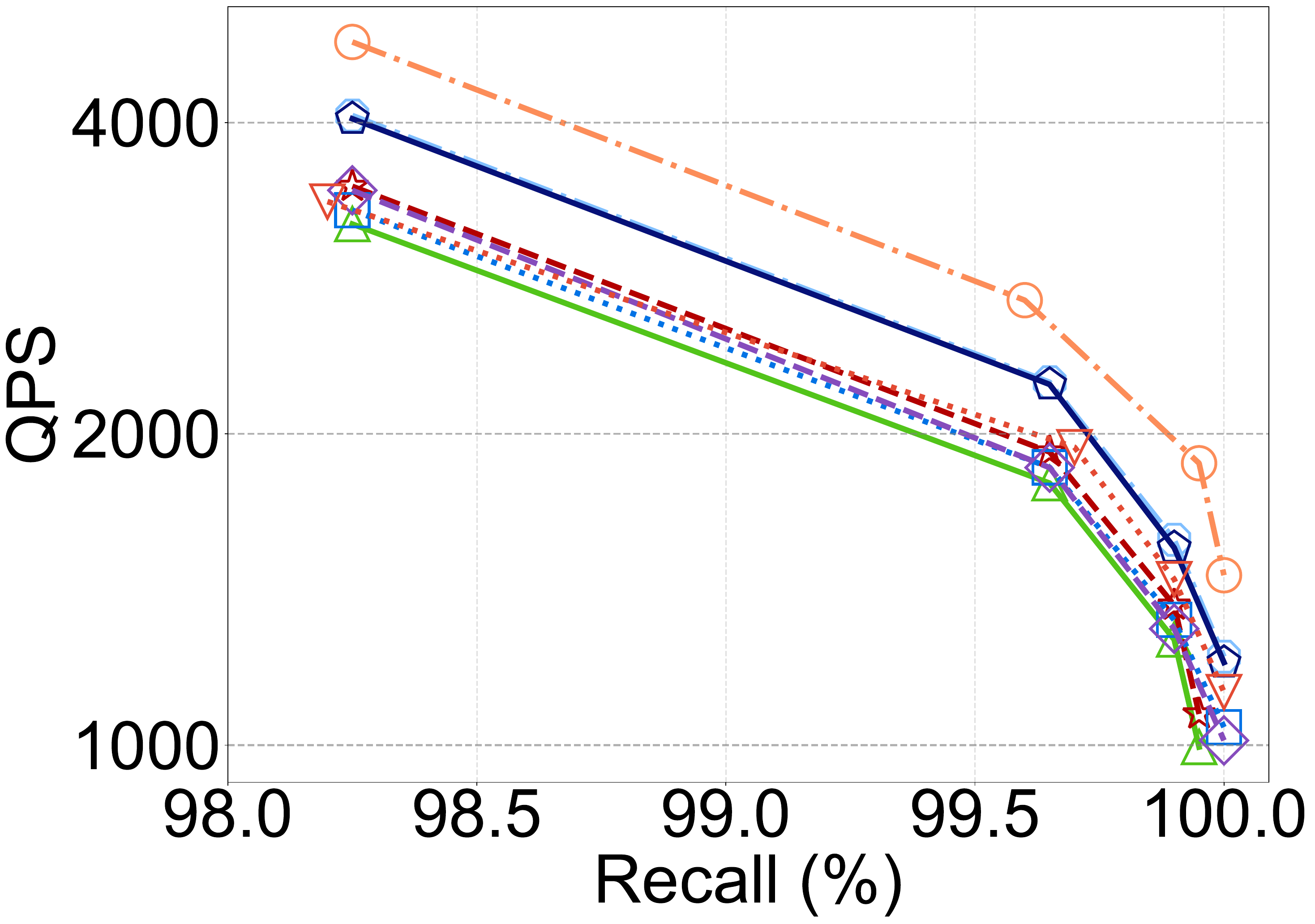}\vspace{-1.0ex}
        \caption{\Sift ($k=20$)}
    \end{subfigure}
    \begin{subfigure}{0.24\textwidth}
        \centering
        \includegraphics[width=\textwidth]{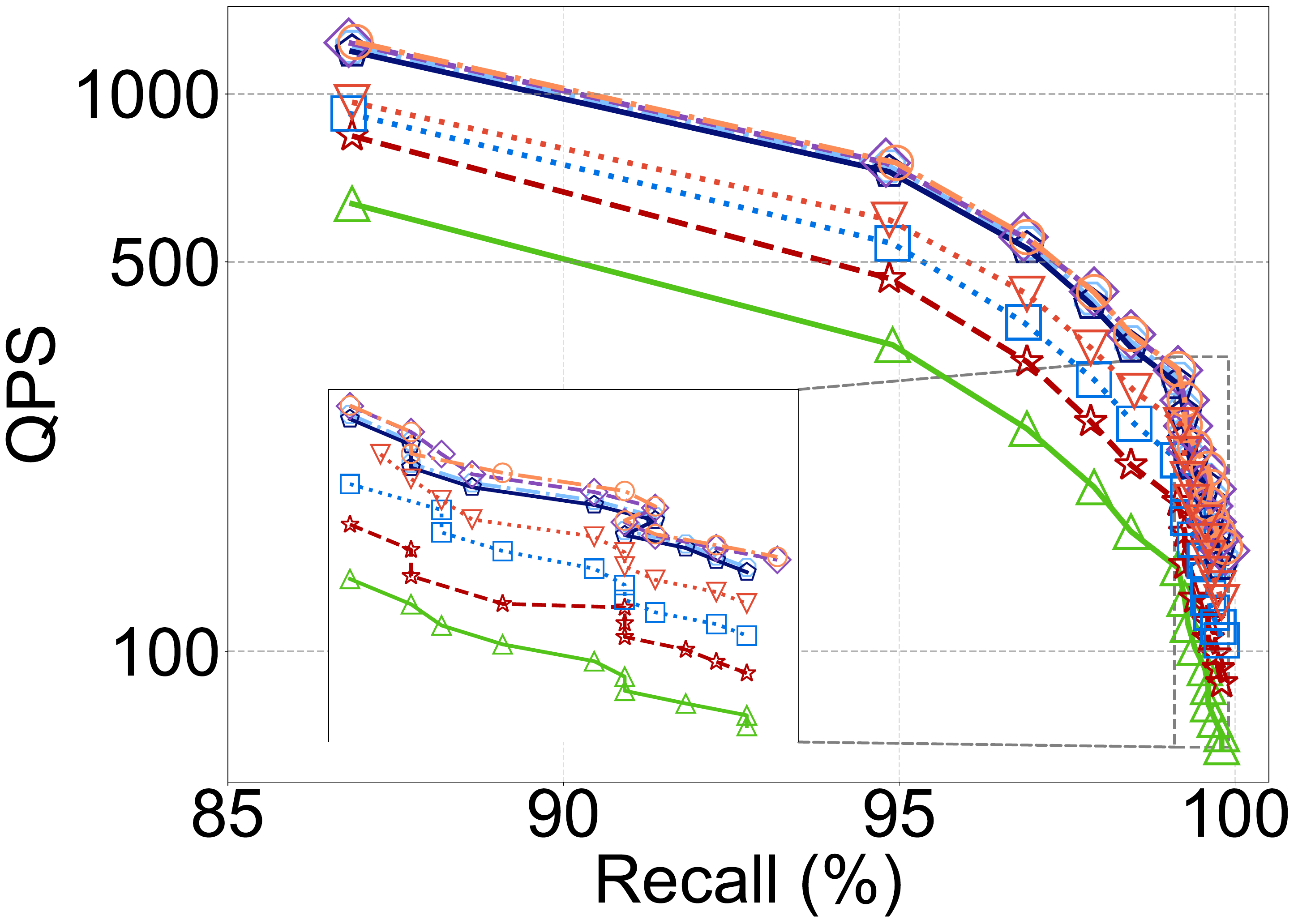}\vspace{-1.0ex}
        \caption{\Gist ($k=20$)}
    \end{subfigure}
    \begin{subfigure}{0.24\textwidth}
        \centering
        \includegraphics[width=\textwidth]{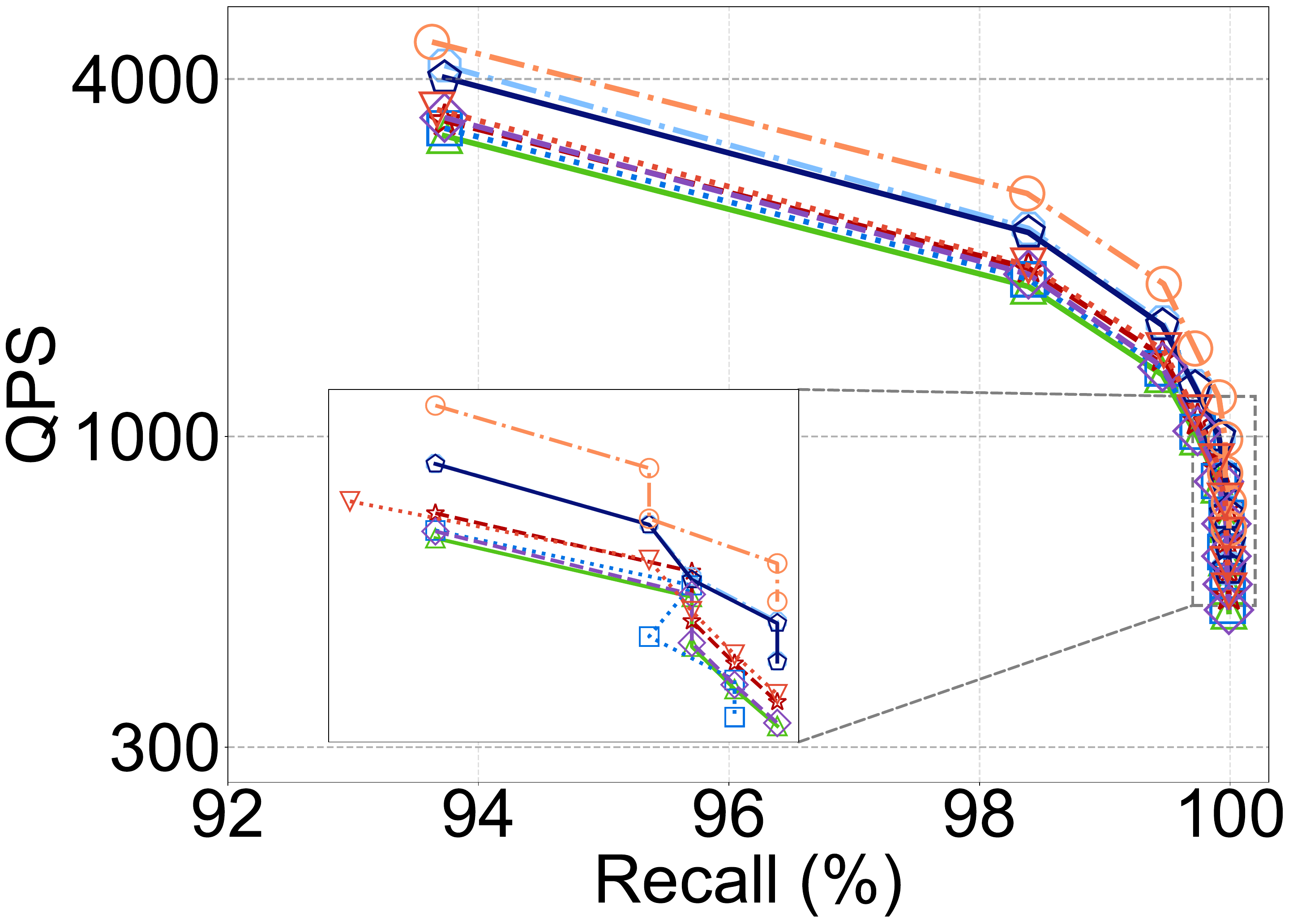}\vspace{-1.0ex}
        \caption{\Sift ($k=100$)}
    \end{subfigure}
    \begin{subfigure}{0.24\textwidth}
        \centering
        \includegraphics[width=\textwidth]{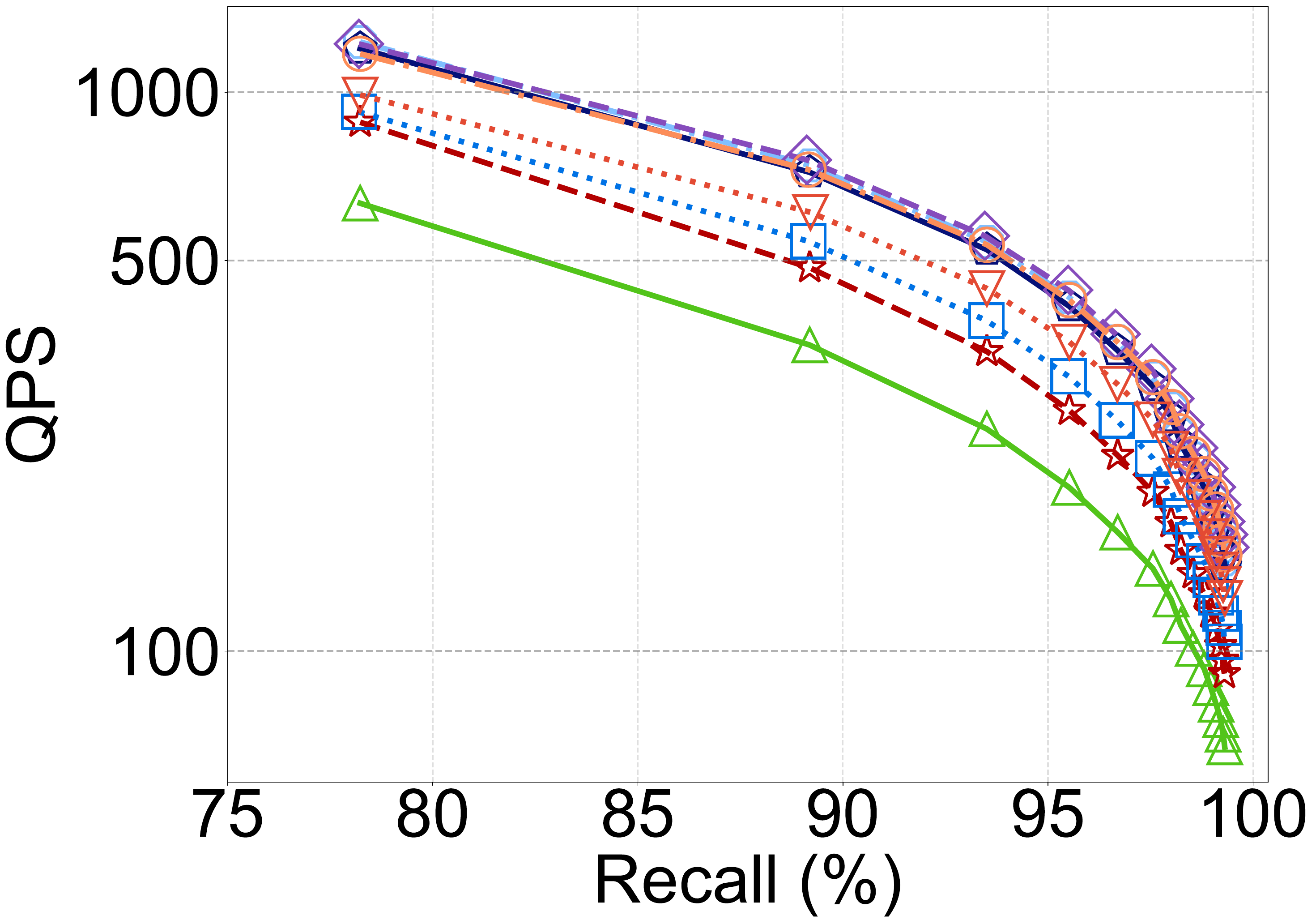}\vspace{-1.0ex}
        \caption{\Gist ($k=100$)}
    \end{subfigure}

    \begin{subfigure}{0.24\textwidth}
        \centering
        \includegraphics[width=\textwidth]{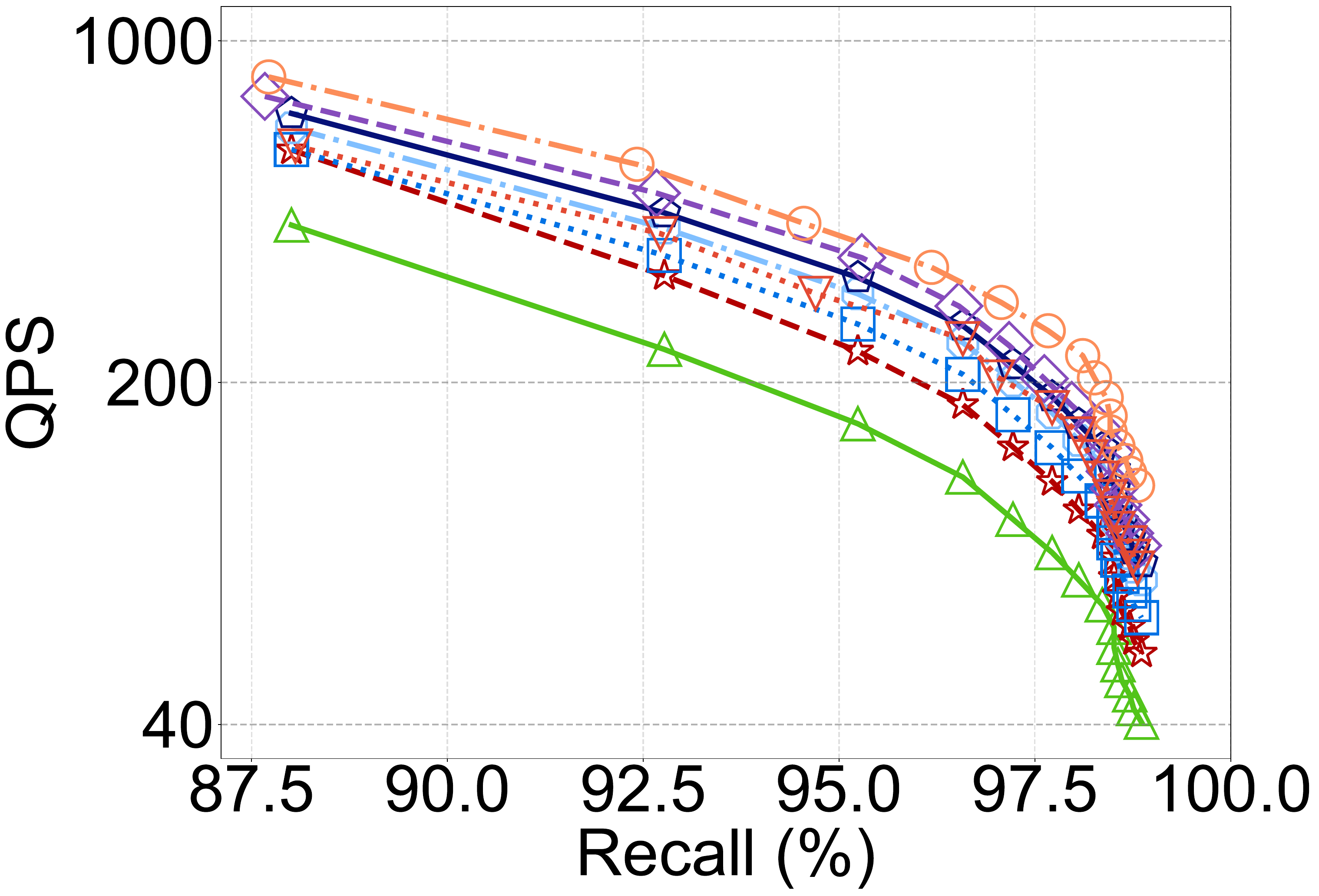}\vspace{-1.0ex}
        \caption{\Openai ($k=20$)}
    \end{subfigure}
    \begin{subfigure}{0.23\textwidth}
        \centering
        \includegraphics[width=\textwidth]{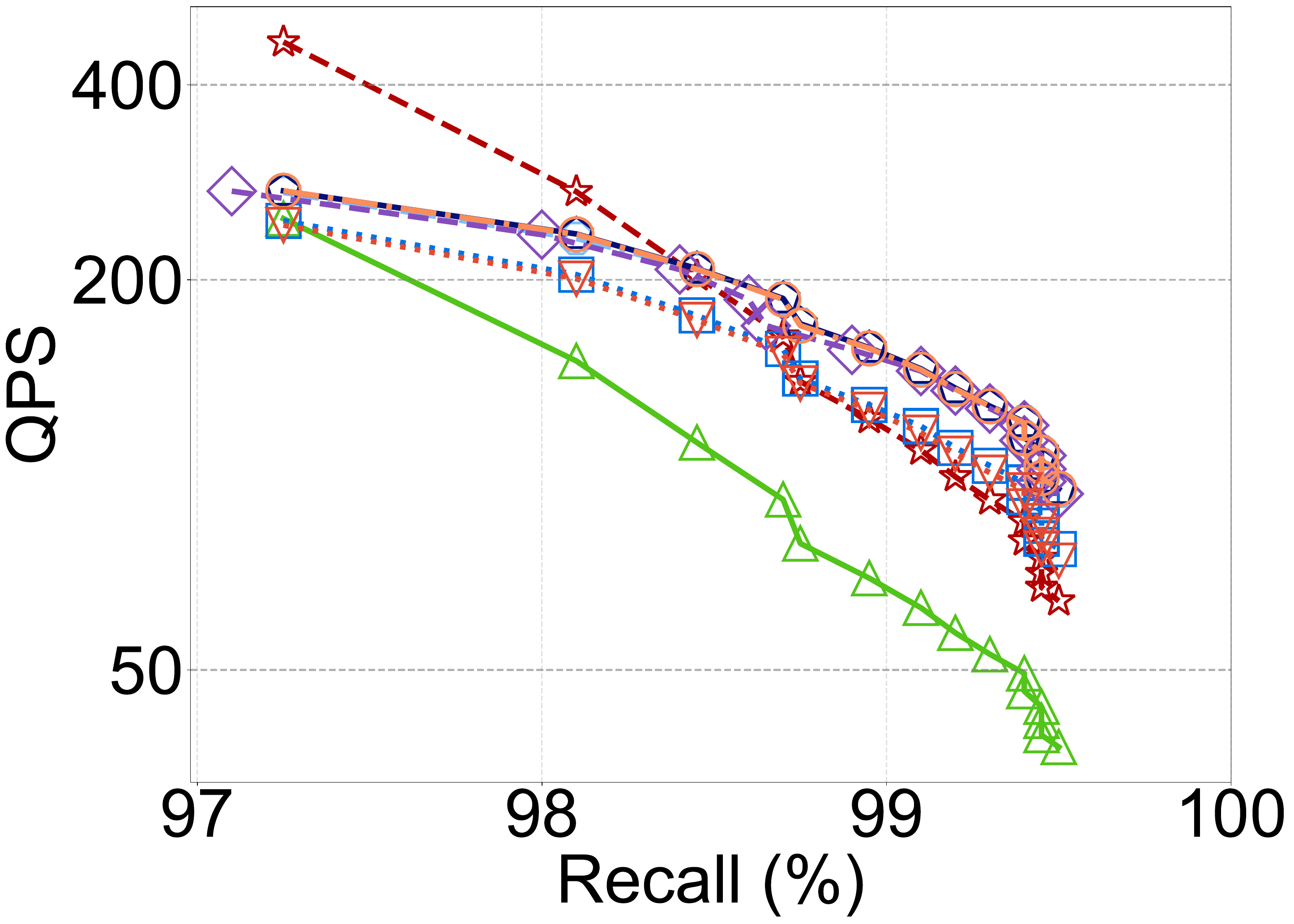}\vspace{-1.0ex}
        \caption{\Trevi ($k=20$)}\label{fig:HNSW-NO_SIMD-trevi-20}
    \end{subfigure}
    \begin{subfigure}{0.24\textwidth}
        \centering
        \includegraphics[width=\textwidth]{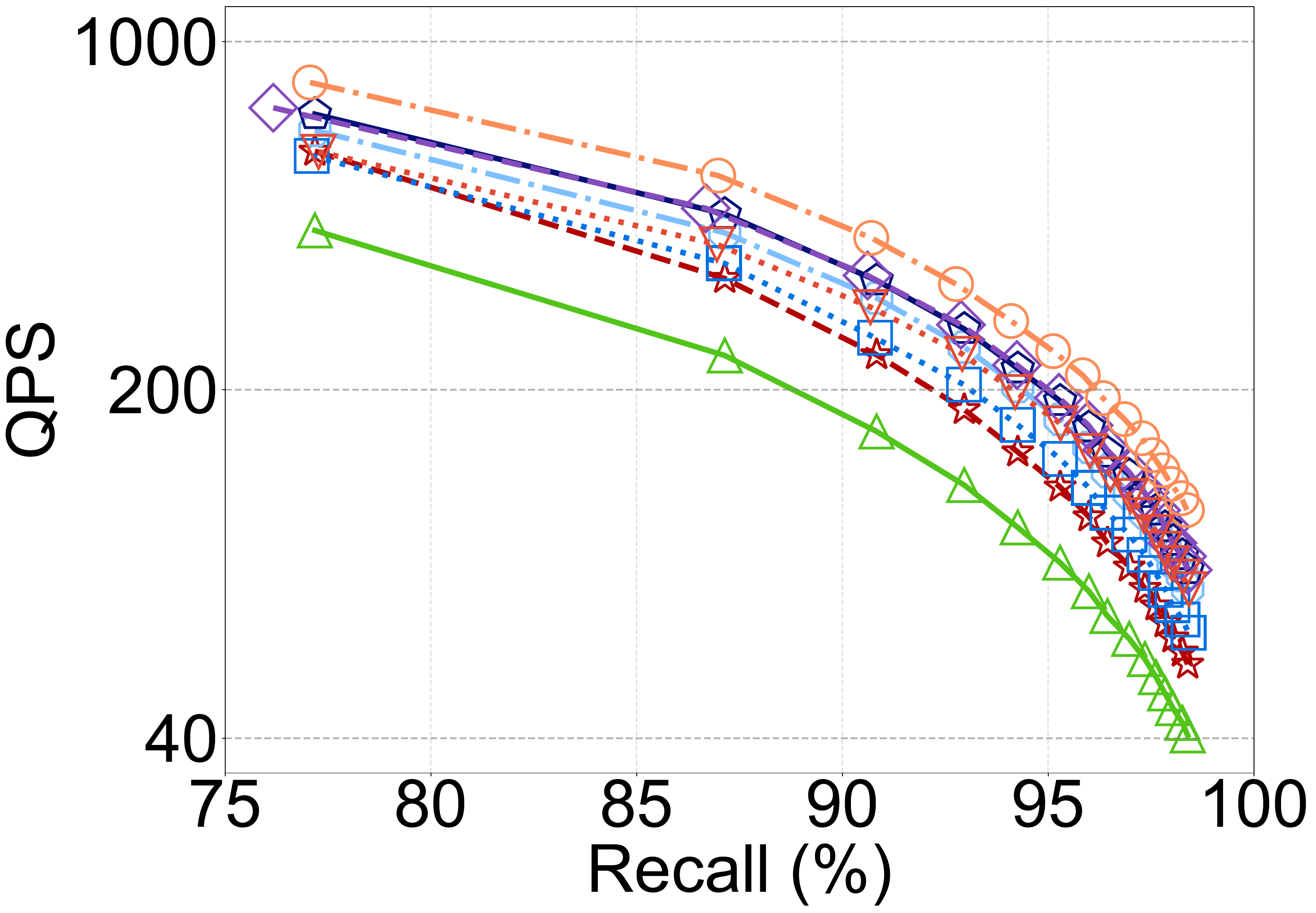}\vspace{-1.0ex}
        \caption{\Openai ($k=100$)}\label{fig:HNSW-NO_SIMD-trevi-100}
    \end{subfigure}
    \begin{subfigure}{0.23\textwidth}
        \centering
        \includegraphics[width=\textwidth]{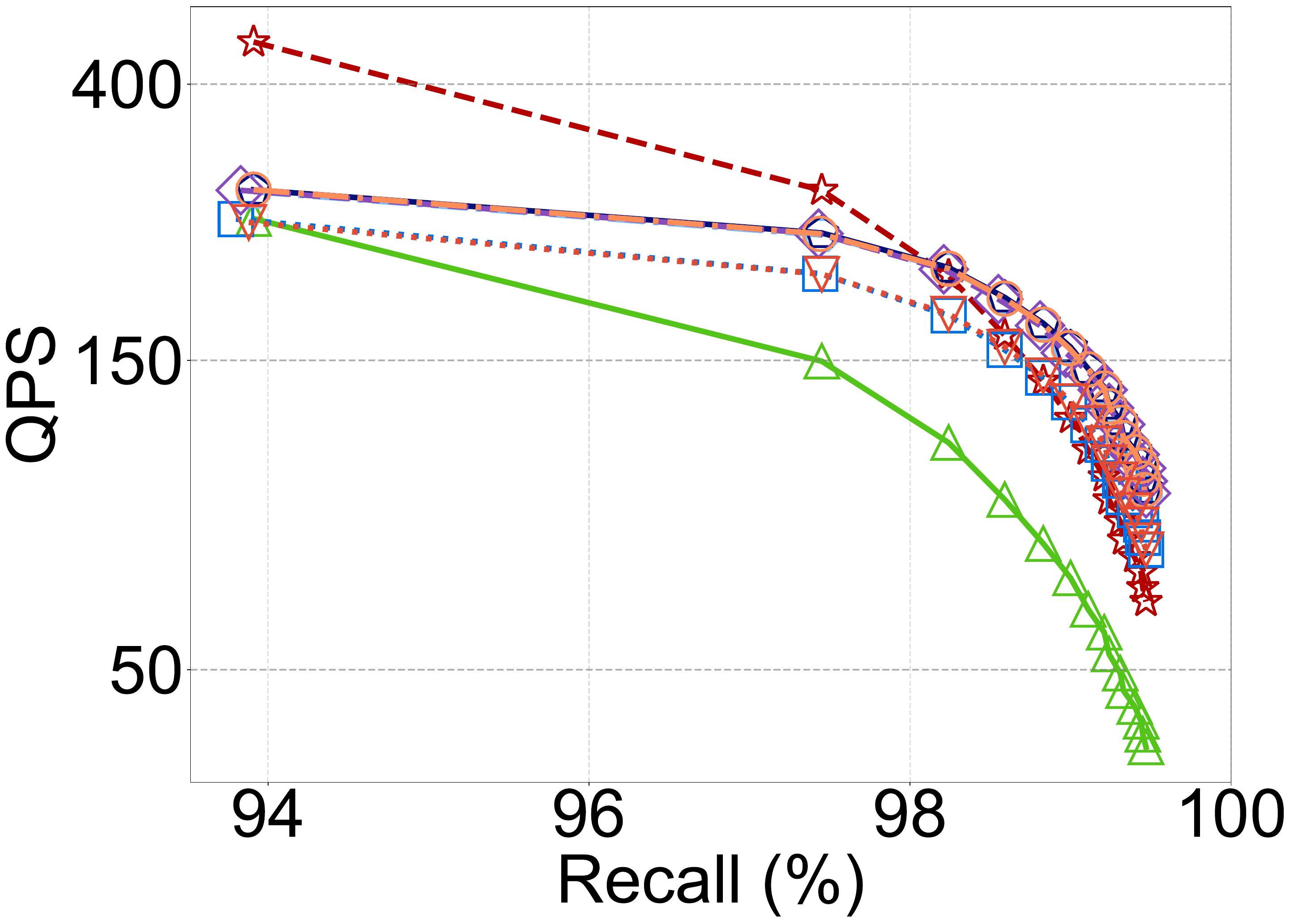}\vspace{-1.0ex}
        \caption{\Trevi ($k=100$)}
    \end{subfigure}
    \caption{Query performance when DCOs are applied to vector similarity search on CPUs with SIMD disabled}
    \label{fig:without-SIMD}
\end{figure*}

\section{Query Processing via HNSW with AVX2 SIMD Instruction Set}

\zheng{We evaluated the query performance of DCO methods using the AVX2 SIMD instruction set. As shown in \figref{fig:avx}, the overall performance trends are similar to those obtained with the SSE SIMD instruction set. For instance, on the low-dimensional \Glove dataset, most DCO methods fail to outperform \FDScanning. 
On the ultra-high-dimensional \Trevi dataset, both \FDScanning and \PDScanning maintain a performance lead until the recall rate exceeds 98\%. 

Since AVX2 is a more advanced instruction set than SSE, DCO methods (\eg \FDScanning) typically run faster with AVX2 than with SSE. 
As a result, the relative speedup provided by DCOs diminishes when AVX2 is employed. 
Notably, with AVX2, \FDScanning can surpass more recent SOTA methods. 
For example, \FDScanning outperforms \ADSampling on the \Gist dataset in \figref{fig:avx}.}

\begin{figure*}[t]
    \centering
    \includegraphics[width=0.9\textwidth]{figure/query_performance/legend.pdf}
    \begin{subfigure}{0.24\textwidth}
        \centering
        \includegraphics[width=\textwidth]{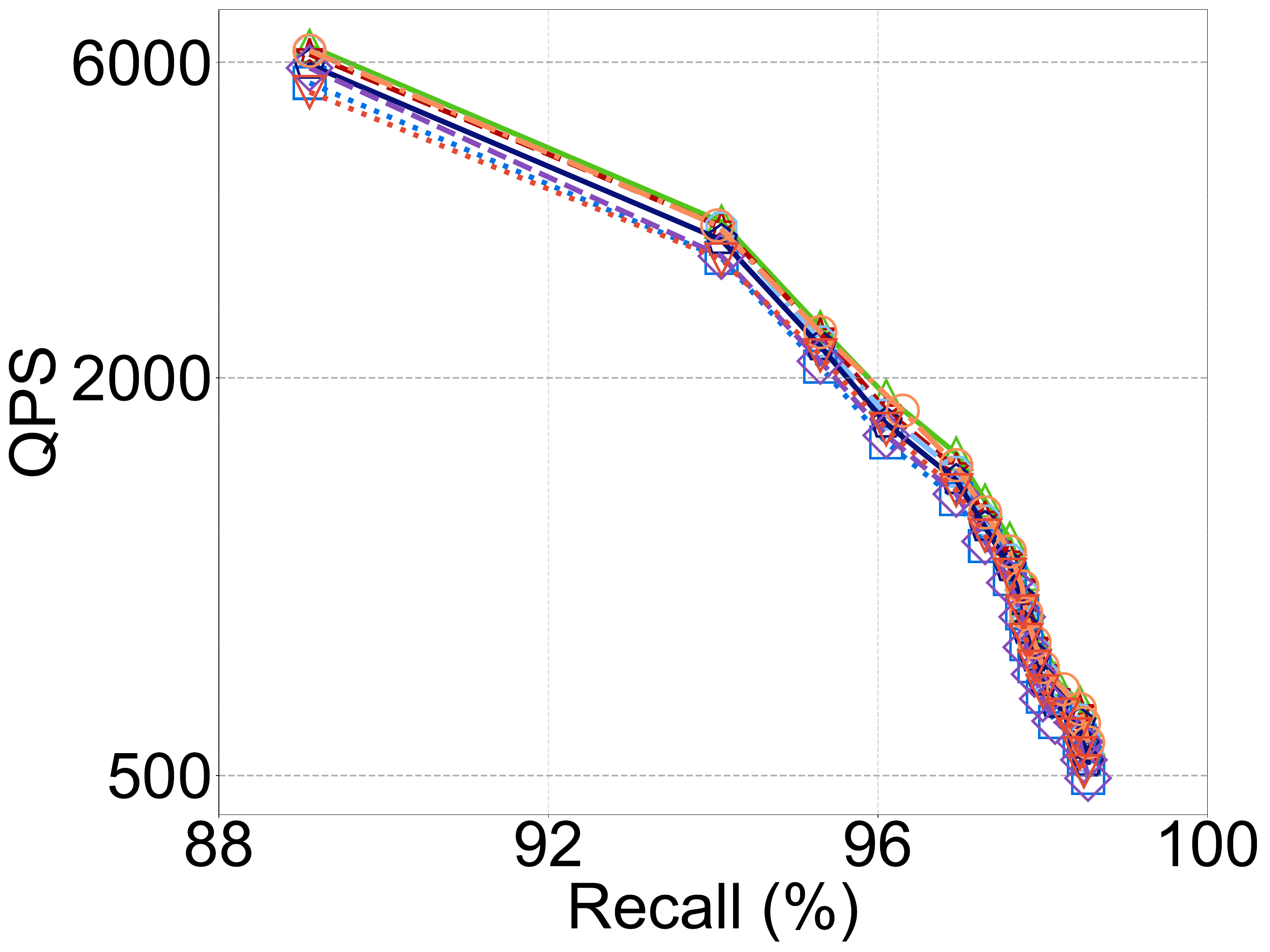}\vspace{-1.0ex}
        \caption{\Glove ($k=20$)}
    \end{subfigure}
    \begin{subfigure}{0.24\textwidth}
        \centering
        \includegraphics[width=\textwidth]{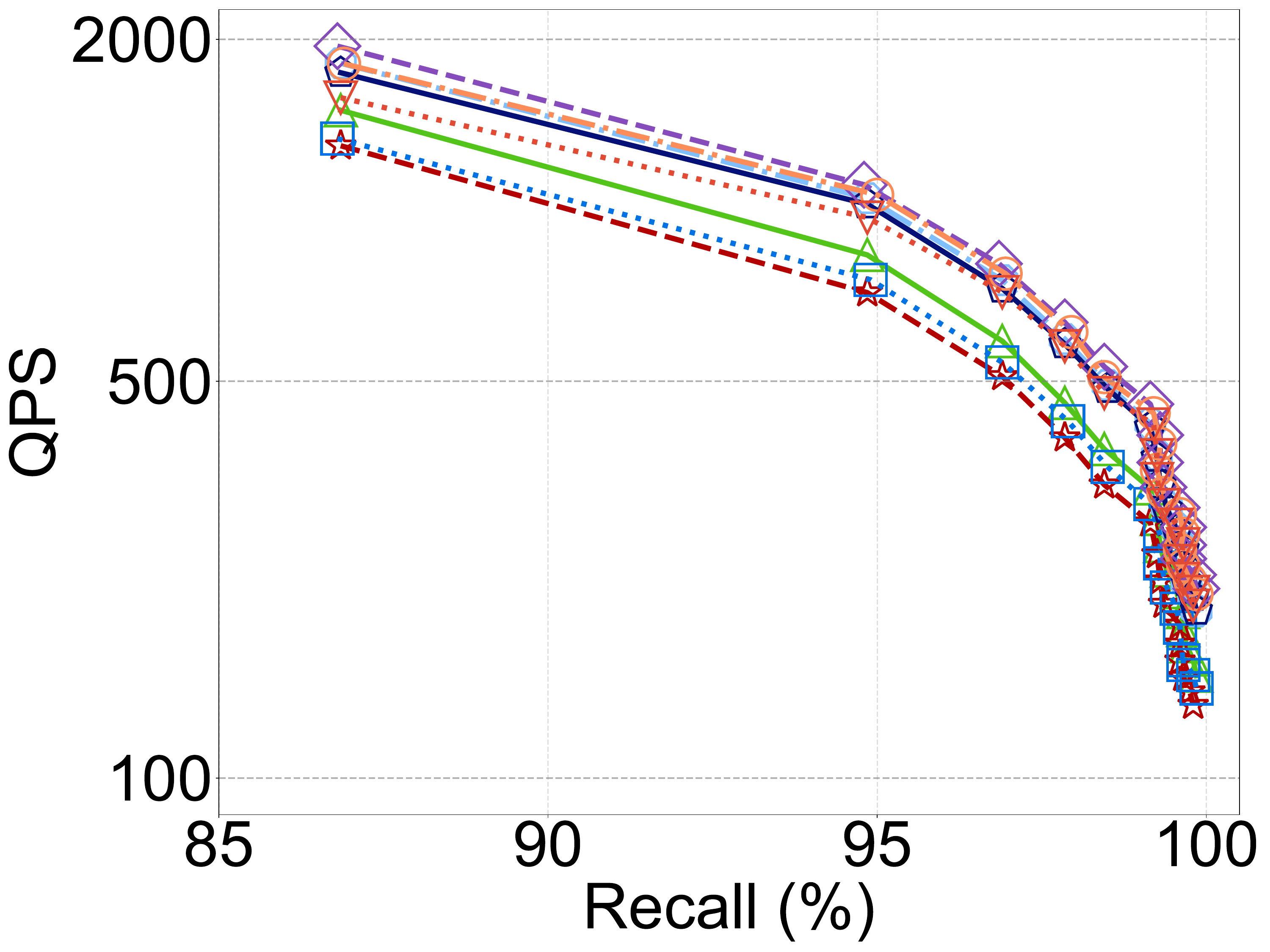}\vspace{-1.0ex}
        \caption{\Gist ($k=20$)}
    \end{subfigure}
    \begin{subfigure}{0.24\textwidth}
        \centering
        \includegraphics[width=\textwidth]{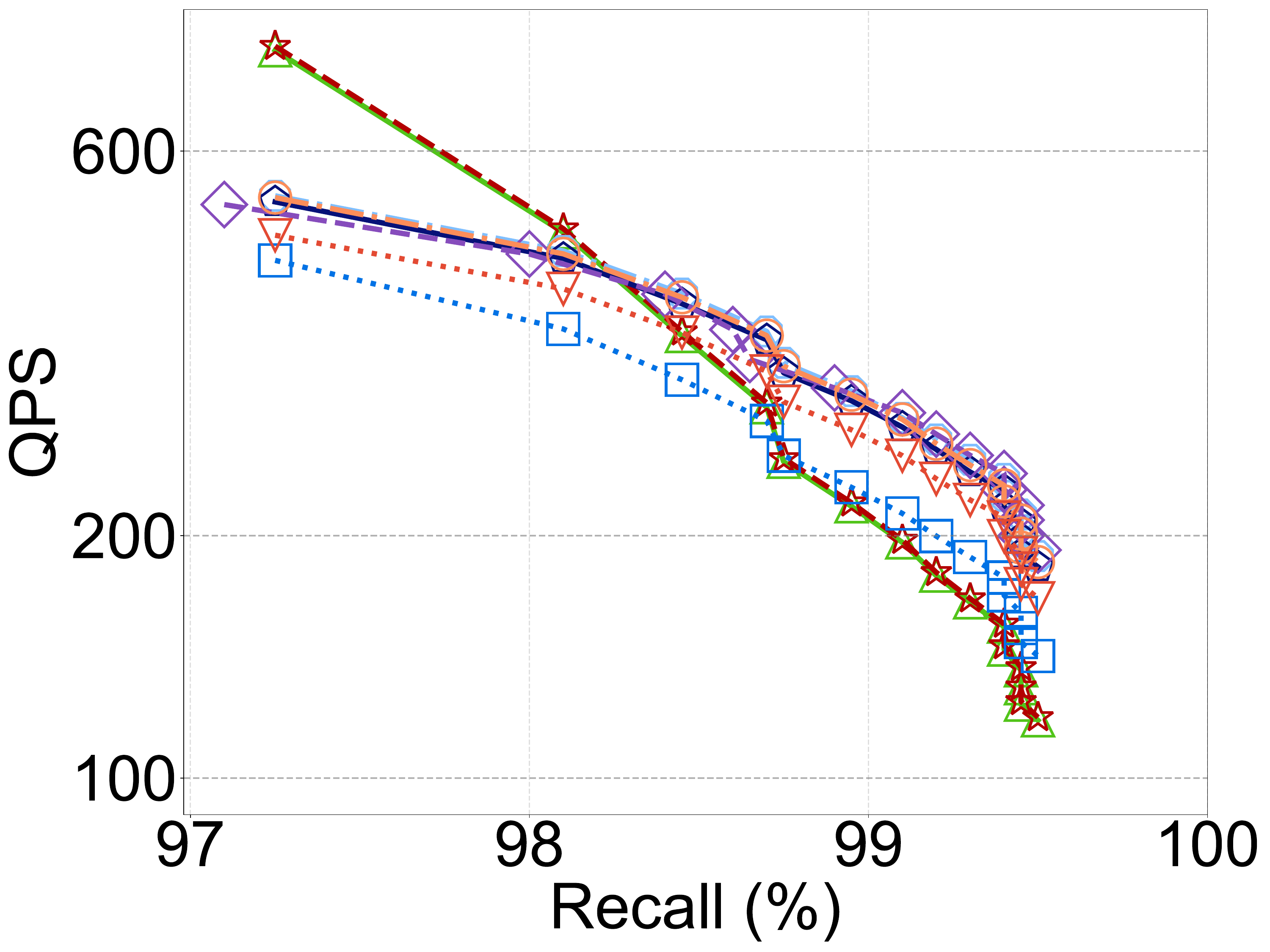}\vspace{-1.0ex}
        \caption{\Trevi ($k=20$)}
    \end{subfigure}
    \\
    \begin{subfigure}{0.24\textwidth}
        \centering
        \includegraphics[width=\textwidth]{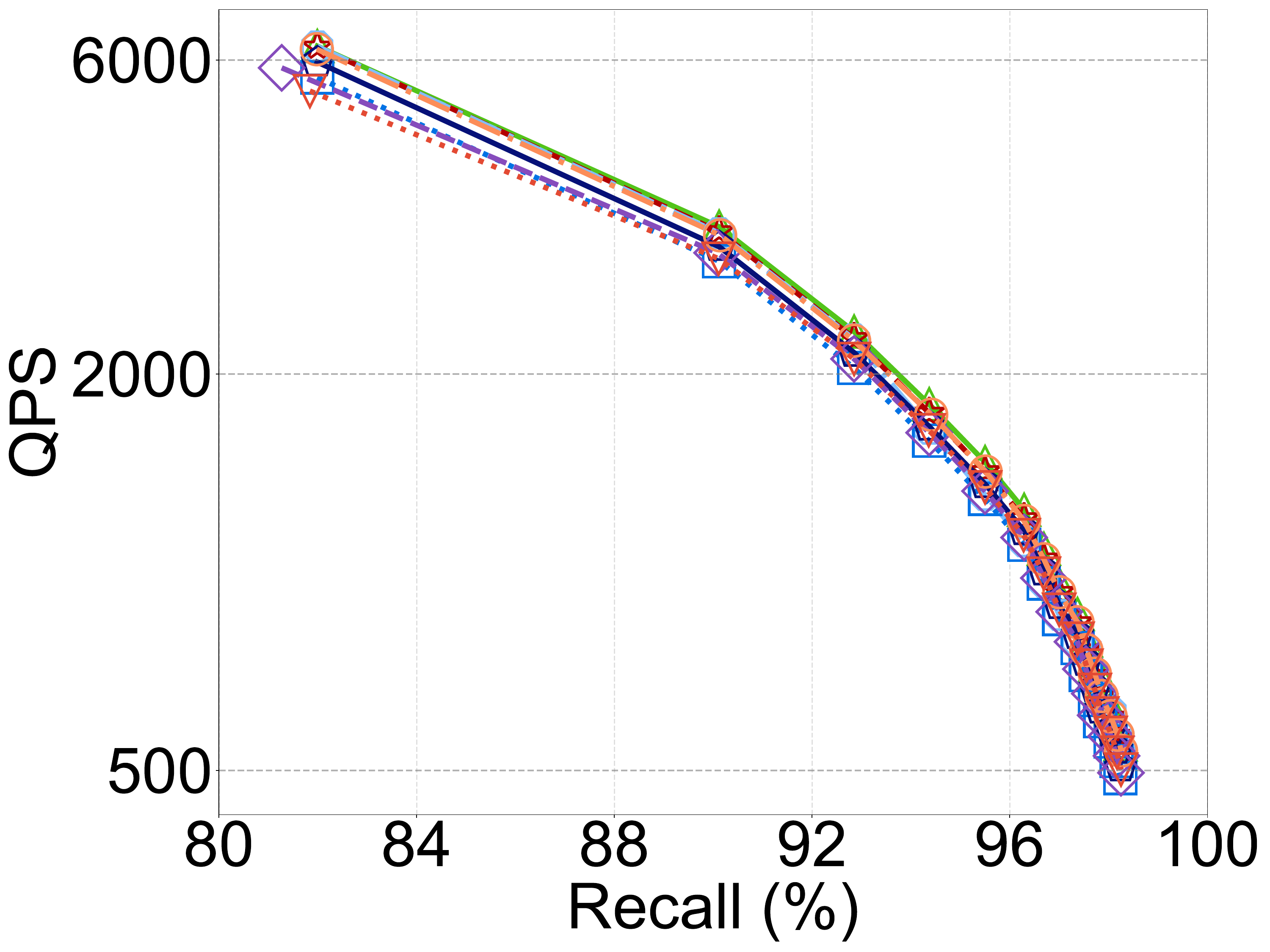}\vspace{-1.0ex}
        \caption{\Glove ($k=100$)}
    \end{subfigure}
    \begin{subfigure}{0.24\textwidth}
        \centering
        \includegraphics[width=\textwidth]{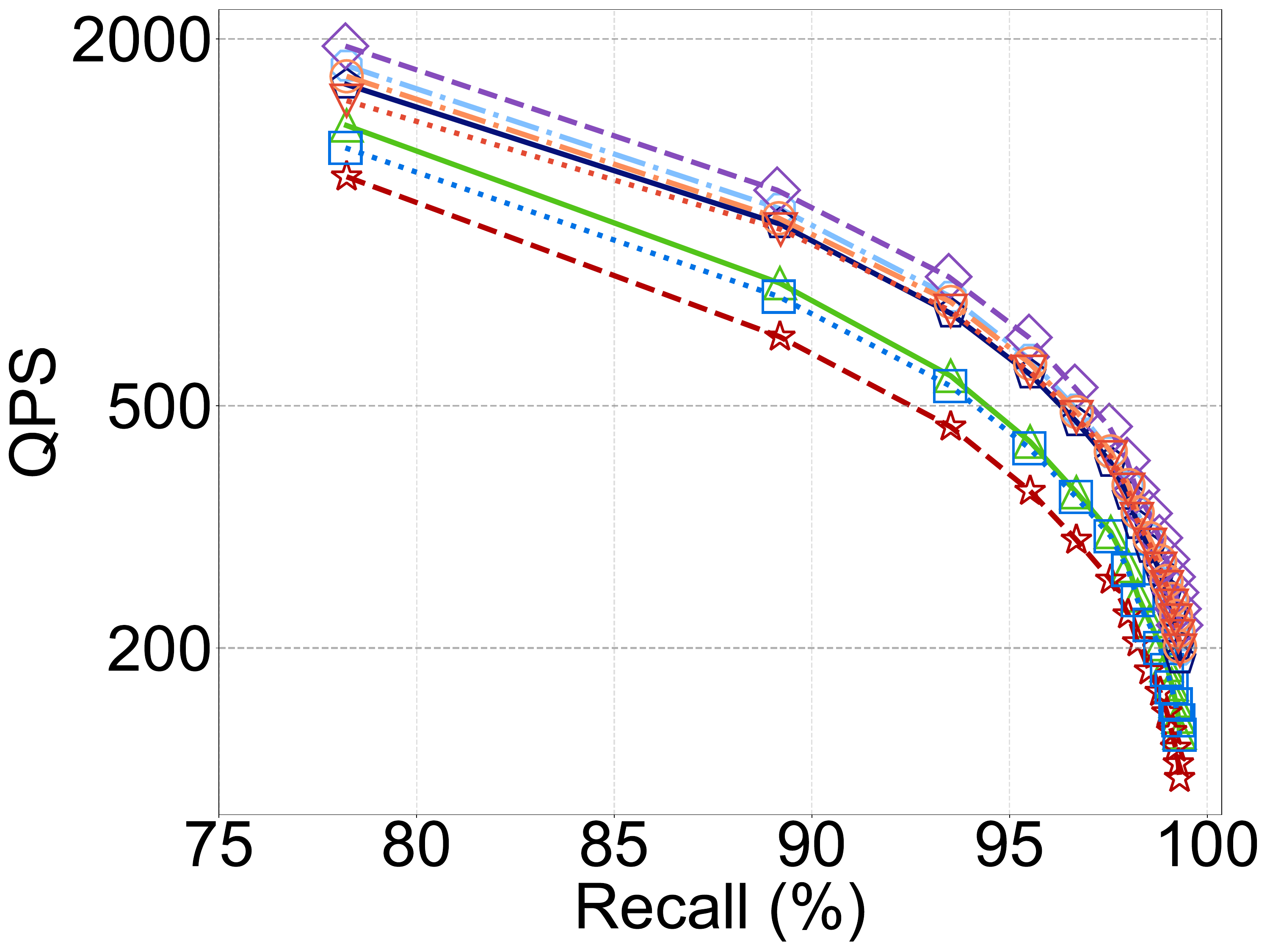}\vspace{-1.0ex}
        \caption{\Gist ($k=100$)}
    \end{subfigure}
    \begin{subfigure}{0.24\textwidth}
        \centering
        \includegraphics[width=\textwidth]{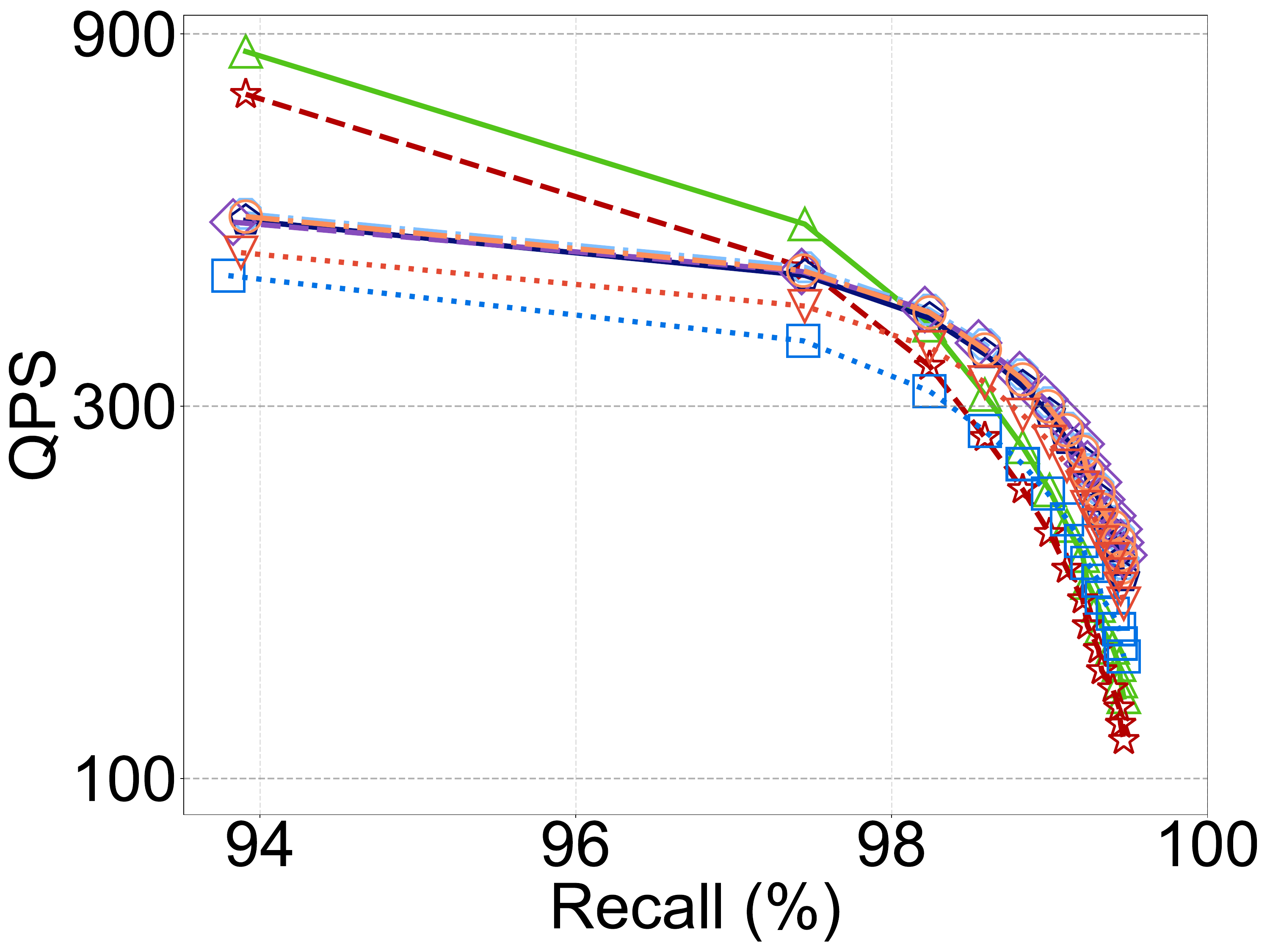}\vspace{-1.0ex}
        \caption{\Trevi ($k=100$)}
    \end{subfigure}
    \caption{Performance comparison on AVX2}\label{fig:avx}
\end{figure*}

\section{Out-of-Distribution (OOD) Queries on CPUs with SIMD Disabled}\label{app:IVF-OOD}
To further evaluate the robustness of DCOs under out-of-distribution (OOD) queries, we also conduct this experiment with SIMD disabled.

As shown in \figref{fig:laion-id-no-simd} and \figref{fig:textimage-id-no-simd}, \DDCpca achieves the highest QPS on both datasets for in-distribution queries.  
However, it becomes the least efficient method under OOD queries, performing up to 1.6$\times$ and 1.8$\times$ slower than \FDScanning and \PDScanning, respectively.  In this setting, aside from \DDCopq, which achieves the highest QPS on the \Laion dataset, the loop-based baselines \PDScanning and \RPDScanning achieve the highest QPS on both datasets for recall over 97\%. This aligns with our earlier finding: the SOTA methods often fail to maintain efficiency gains on OOD queries.

\begin{figure*}[t]
    \centering
    \includegraphics[width=0.9\textwidth]{figure/query_performance/legend.pdf}
    \begin{subfigure}{0.24\textwidth}
        \centering
        \includegraphics[width=\textwidth]{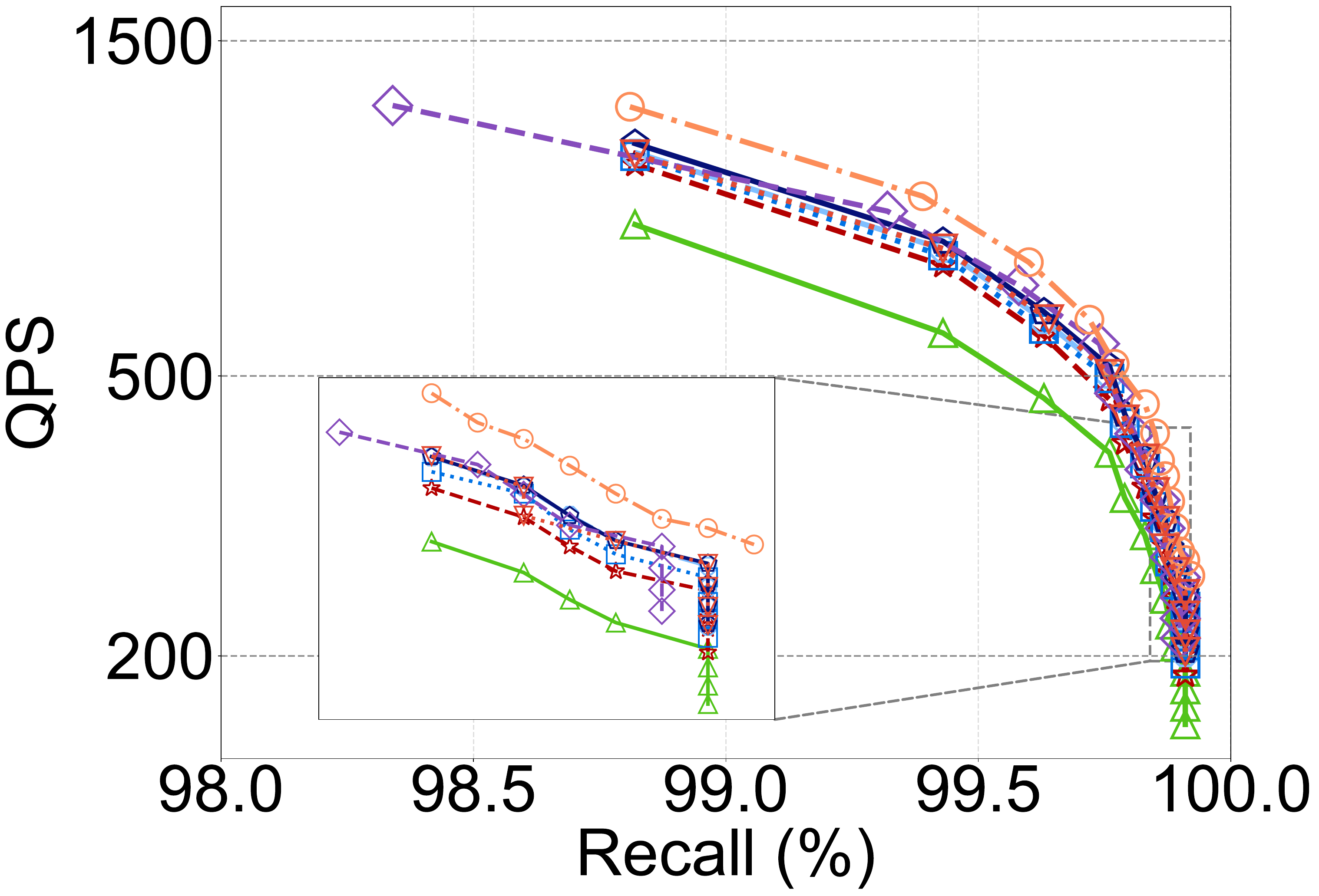}\vspace{-1.0ex}
        \caption{\Laion (in-distribution)}\label{fig:laion-id-no-simd}
    \end{subfigure}
    \begin{subfigure}{0.24\textwidth}
        \centering
        \includegraphics[width=\textwidth]{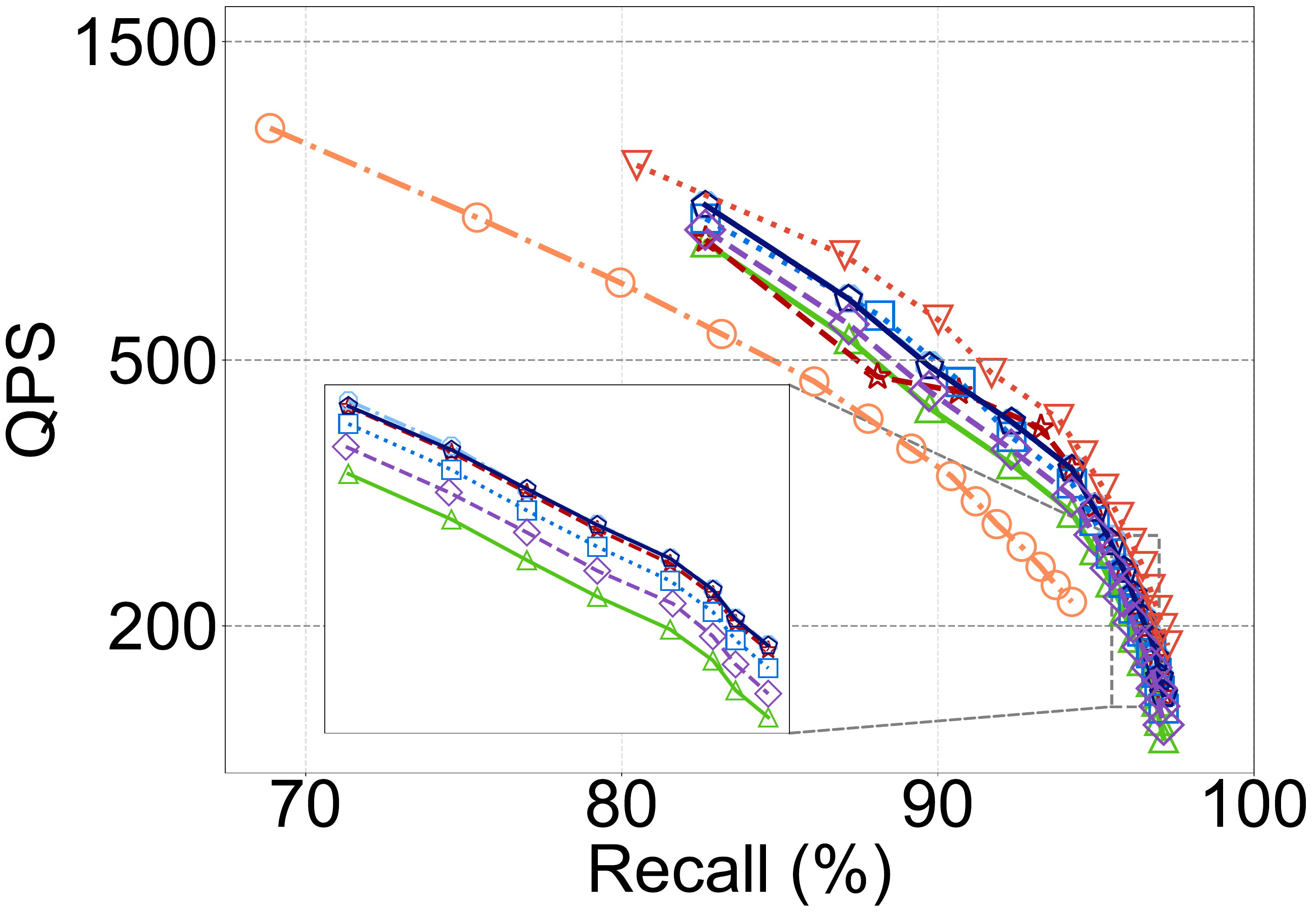}\vspace{-1.0ex}
        \caption{\Laion (OOD)}\label{fig:laion-ood-no-simd}
    \end{subfigure}
    \begin{subfigure}{0.24\textwidth}
        \centering
        \includegraphics[width=\textwidth]{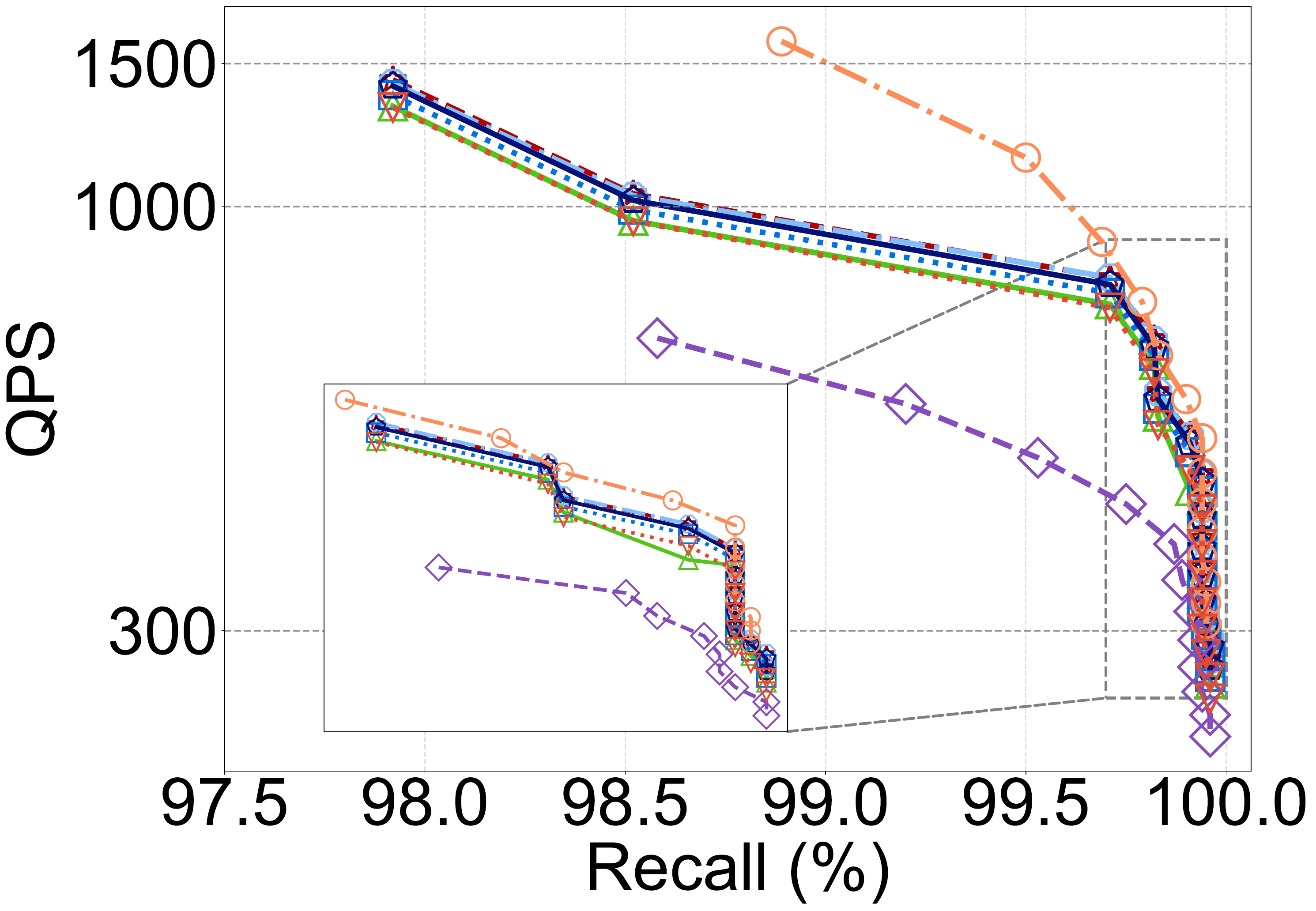}\vspace{-1.0ex}
        \caption{\TextImage (in-distribution)}\label{fig:textimage-id-no-simd}
    \end{subfigure}
    \begin{subfigure}{0.24\textwidth}
        \centering
        \includegraphics[width=\textwidth]{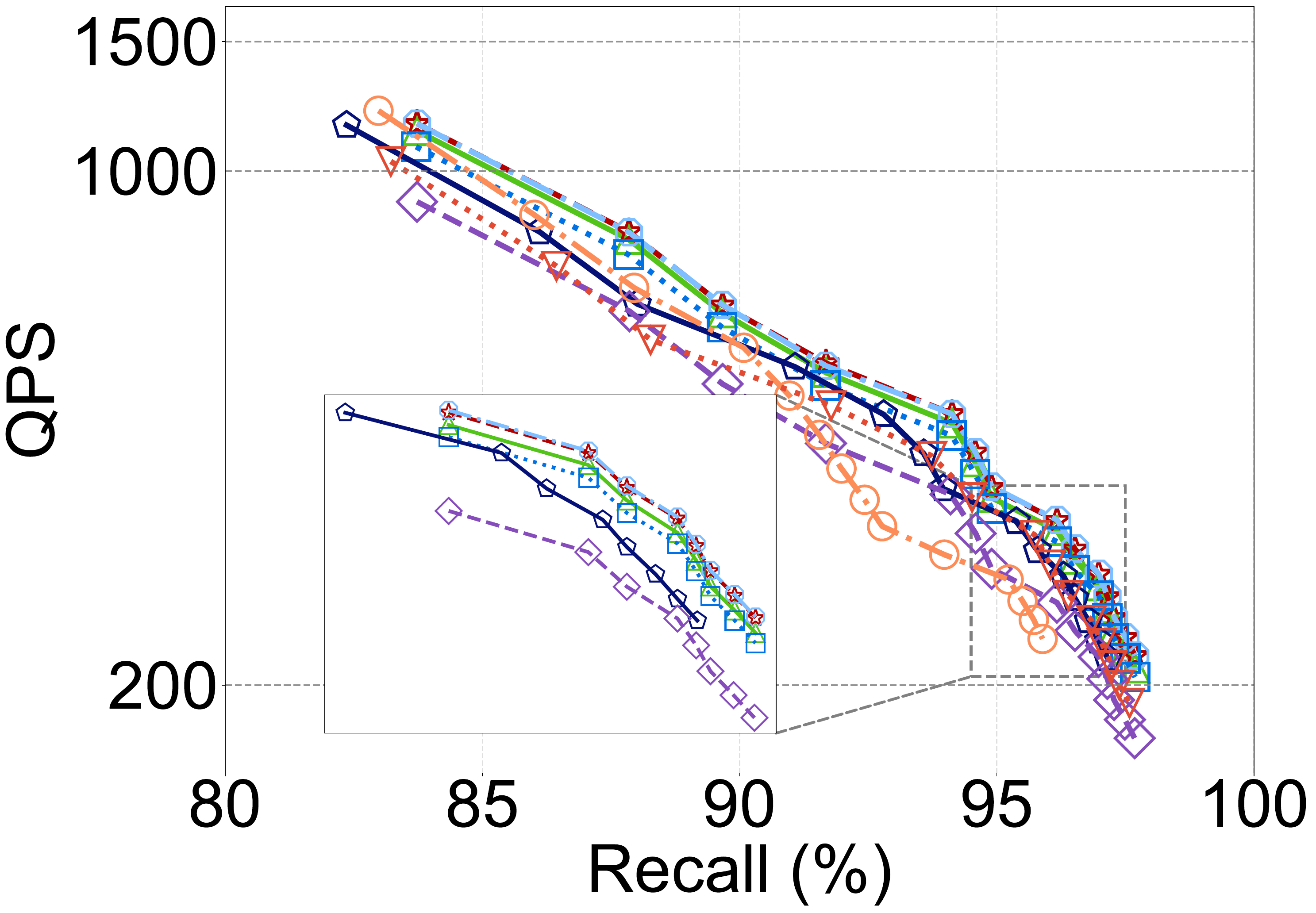}\vspace{-1.0ex}
        \caption{\TextImage (OOD)}\label{fig:textimage-ood-no-simd}
    \end{subfigure}
    \caption{Results on in-distribution and OOD queries without SIMD }\label{fig:out-of-distribution-no-simd}
\end{figure*}

\section{Experiment with Cosine Similarity}
\zheng{This experiment evaluates the extension of DCO methods to cosine similarity. To conduct this experiment, we select three datasets of varying dimensionality: \Glove, \Gist, and \Trevi.}

\begin{figure*}[t]
    \centering
    \includegraphics[width=0.9\textwidth]{figure/query_performance/legend.pdf}
    \begin{subfigure}{0.24\textwidth}
        \centering
        \includegraphics[width=\textwidth]{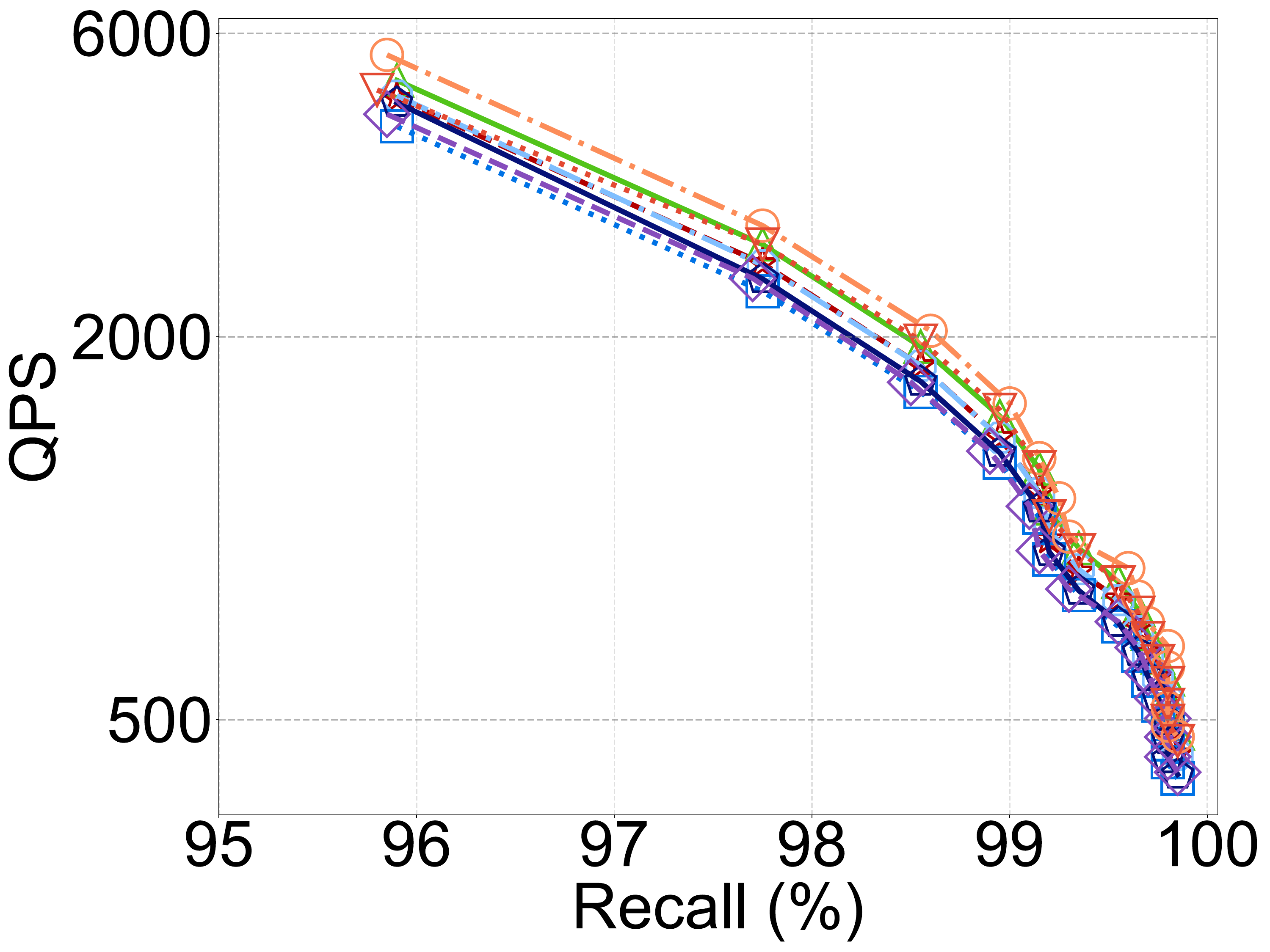}\vspace{-1.0ex}
        \caption{\Glove, $k=20$}
    \end{subfigure}
    \begin{subfigure}{0.24\textwidth}
        \centering
        \includegraphics[width=\textwidth]{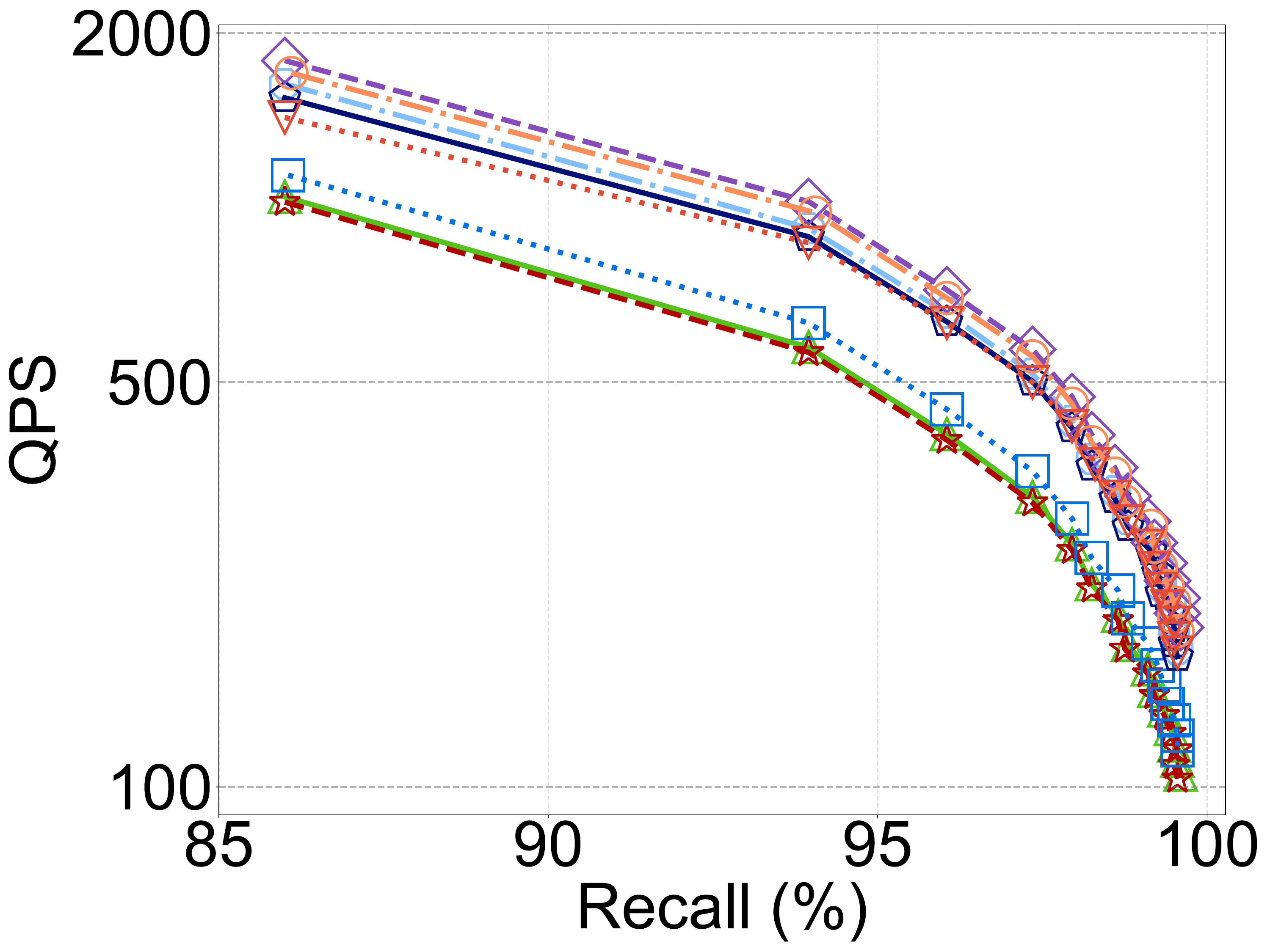}\vspace{-1.0ex}
        \caption{\Gist, $k=20$}
    \end{subfigure}
    \begin{subfigure}{0.24\textwidth}
        \centering
        \includegraphics[width=\textwidth]{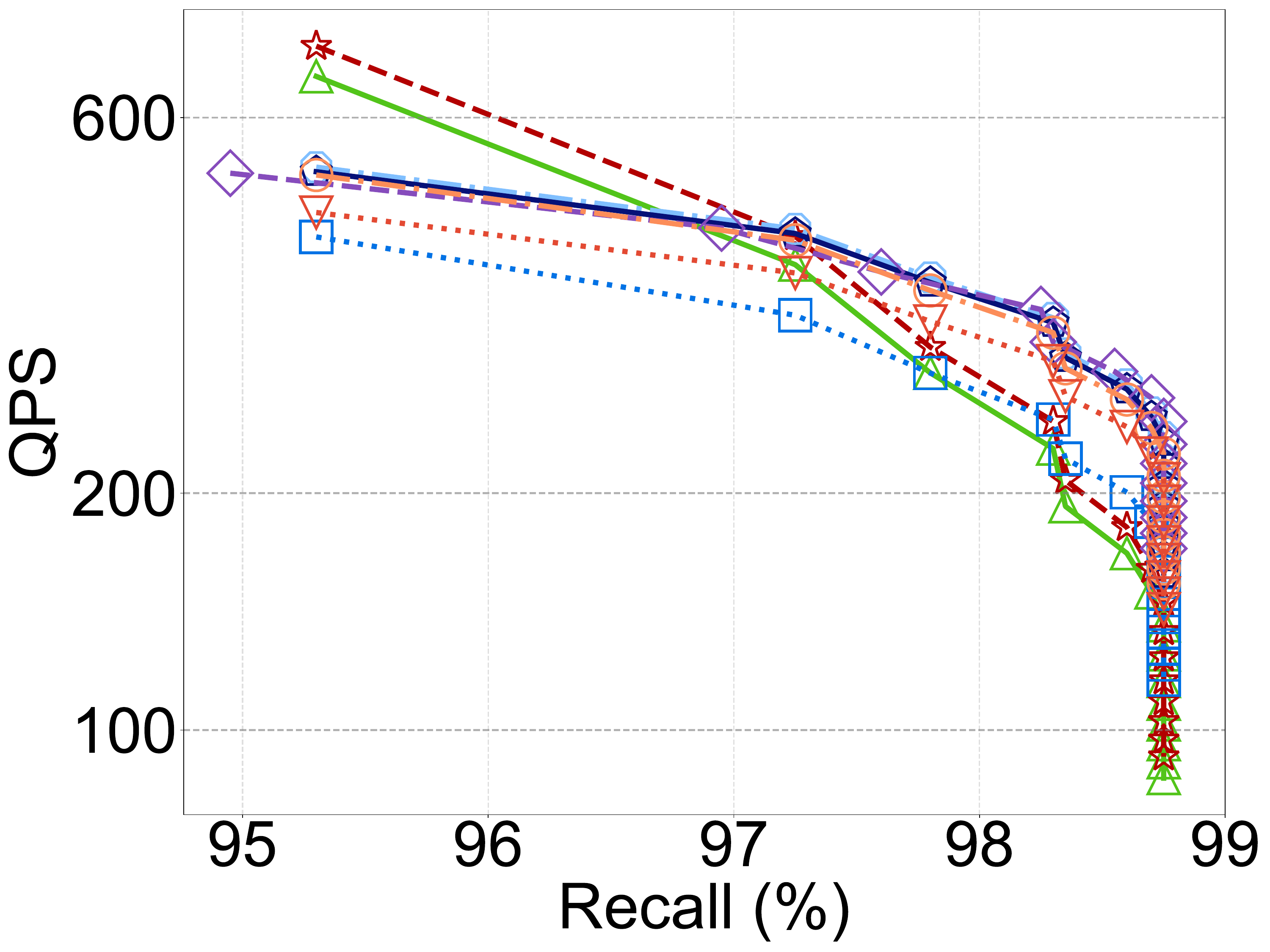}\vspace{-1.0ex}
        \caption{\Trevi, $k=20$}
    \end{subfigure}
    \\
    \begin{subfigure}{0.24\textwidth}
        \centering
        \includegraphics[width=\textwidth]{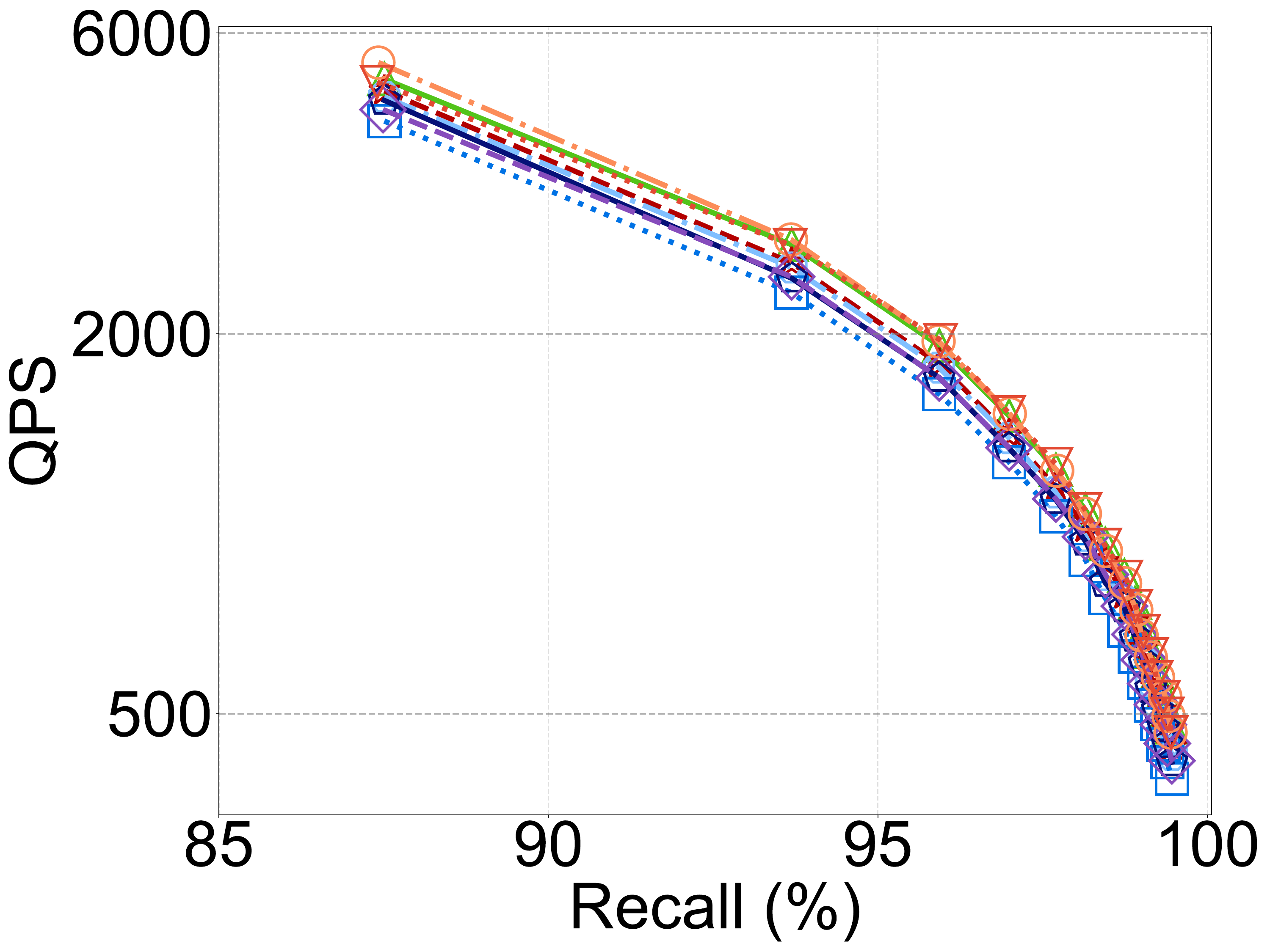}\vspace{-1.0ex}
        \caption{\Glove, $k=100$}
    \end{subfigure}
    \begin{subfigure}{0.24\textwidth}
        \centering
        \includegraphics[width=\textwidth]{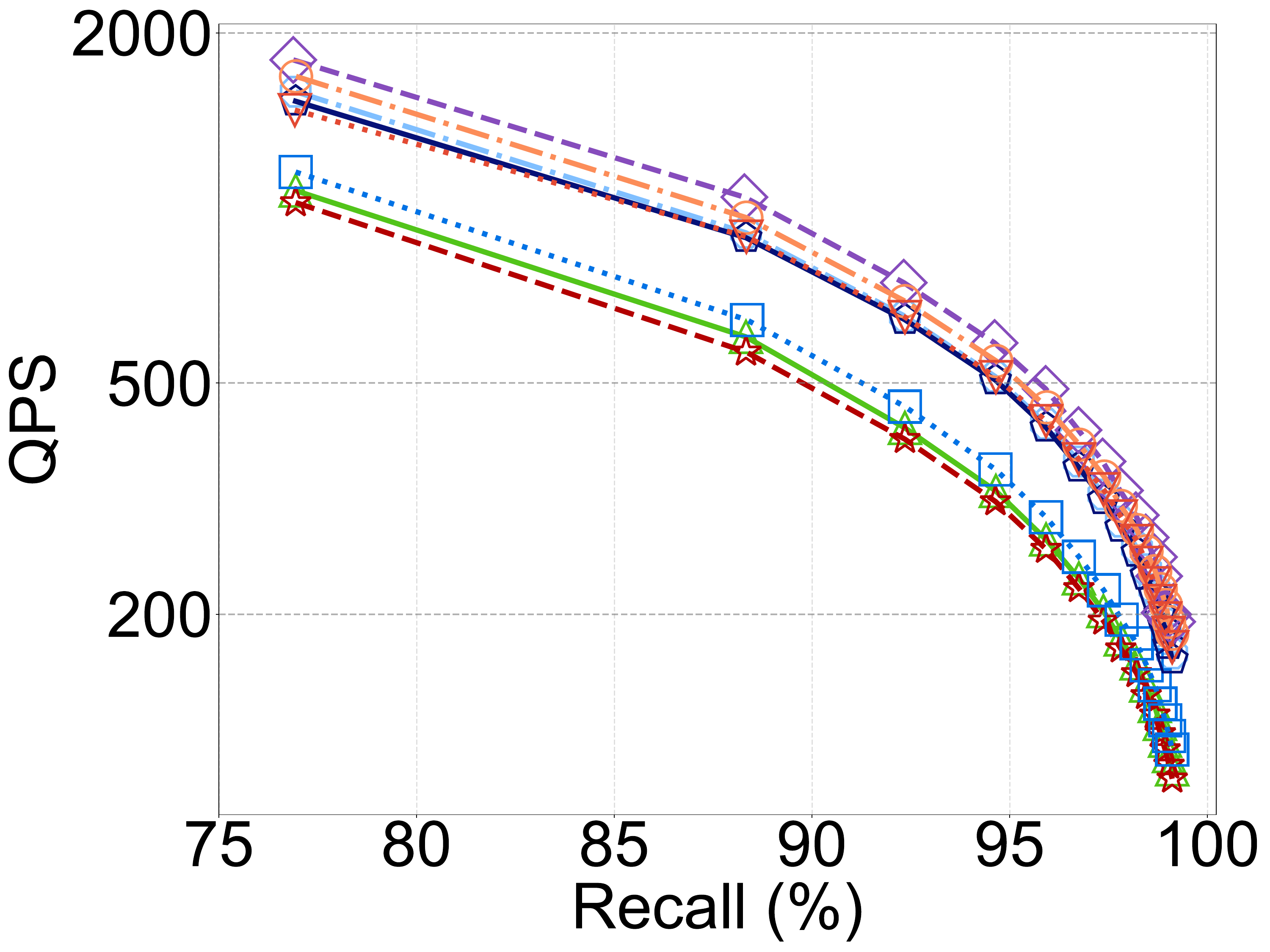}\vspace{-1.0ex}
        \caption{\Gist, $k=100$}
    \end{subfigure}
    \begin{subfigure}{0.24\textwidth}
        \centering
        \includegraphics[width=\textwidth]{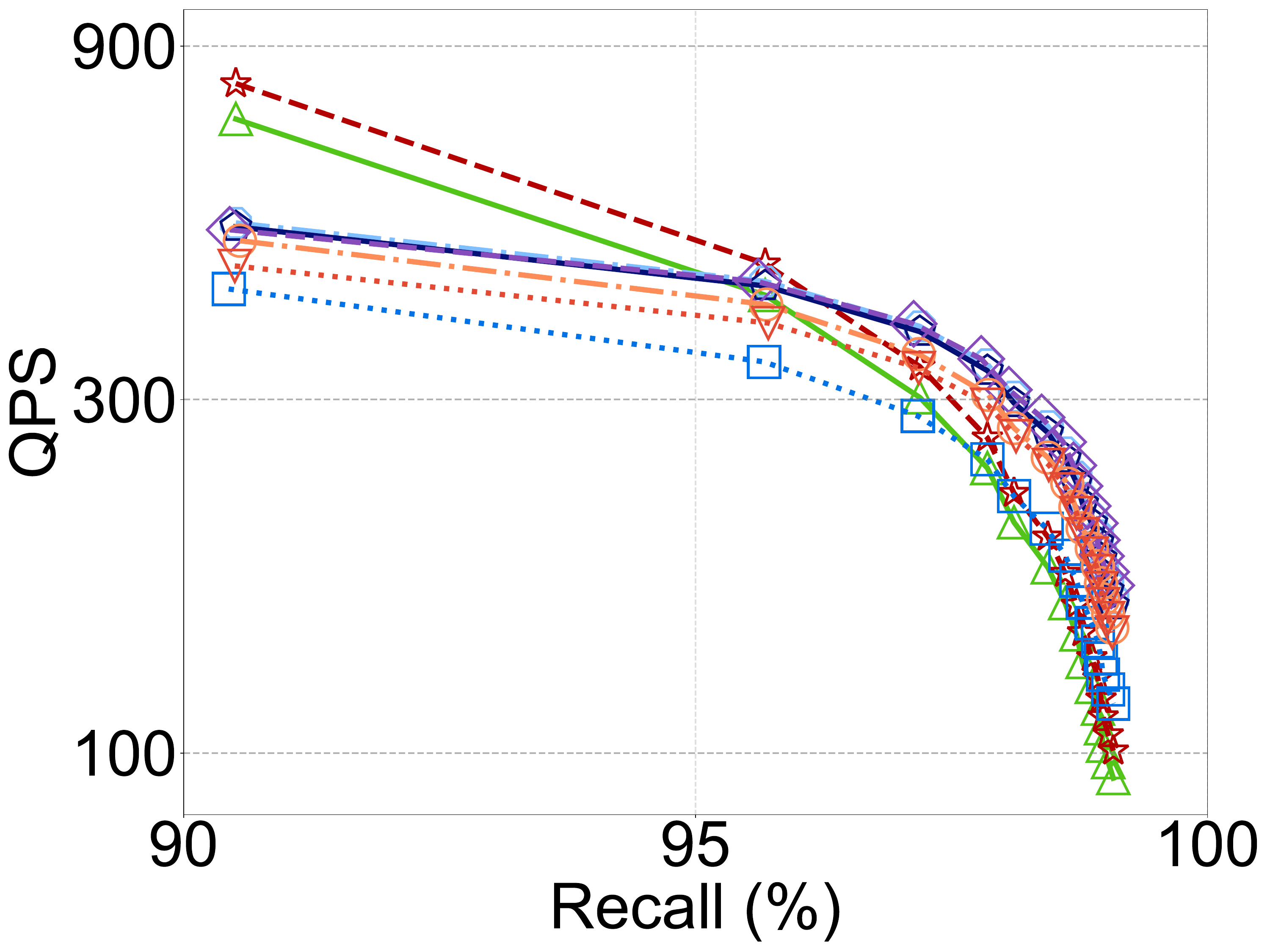}\vspace{-1.0ex}
        \caption{\Trevi, $k=100$}
    \end{subfigure}
    \caption{Search performance using cosine similarity }\label{fig:cosine-similarity}
\end{figure*}

\zheng{As illustrated in \figref{fig:cosine-similarity}, the overall performance trend for cosine similarity is consistent with that observed for Euclidean distance and inner product.
On the low-dimensional datasets (\eg \Glove), the efficiency advantage of the SOTA DCO methods is less pronounced, and they are even outperformed by \FDScanning in some cases.
For the ultra-high-dimensional dataset (\eg \Trevi), the performance bottleneck caused by online pre-processing time persists consistently. This performance pattern stems primarily from two key factors: (1) most DCO extensions for cosine similarity are directly derived from their Euclidean distance equivalents, and (2) cosine similarity is mathematically equivalent to the inner product for normalized vectors.}

\section{Experiments of Diverse Hardware Configurations on \Msmacro Dataset}\label{app:IVF-hardware-extend}

We further evaluate DCO methods across different hardware platforms on the \Msmacro dataset, which has the highest dimensionality among all evaluated datasets.

\fakeparagraph{Performance on CPU with and without SIMD Support}
As shown in \figref{fig:msmacro-without-SIMD} and \figref{fig:msmacro-with-SIMD}, 
\PDScanning outperforms the SOTA methods due to the larger online pre-processing time before the recall reaches 89\%. As the dimensionality further increases, \DDCres attains the highest QPS, regardless of whether SIMD is enabled. Specifically, it improves QPS by up to $1.7\times$ and $2.9\times$ over \FDScanning on CPUs with and without SIMD, respectively.

\fakeparagraph{Performance on GPU} As shown in \figref{fig:msmacro-gpu}, \DDCres consistently achieves the highest QPS. Notably, the bottleneck caused by online pre-processing time, observed on CPU, does not manifest on GPU.  
This can be attributed to the GPU’s superior efficiency in matrix multiplication and parallel computation. The SOTA methods demonstrate strong performance on the ultra-high-dimensional \Msmacro dataset, improving QPS by up to 1.4--5.6$\times$ over \FDScanning.

\begin{figure*}[t]
    \centering
    \begin{subfigure}{0.24\textwidth}
        \centering
        \includegraphics[width=\textwidth]{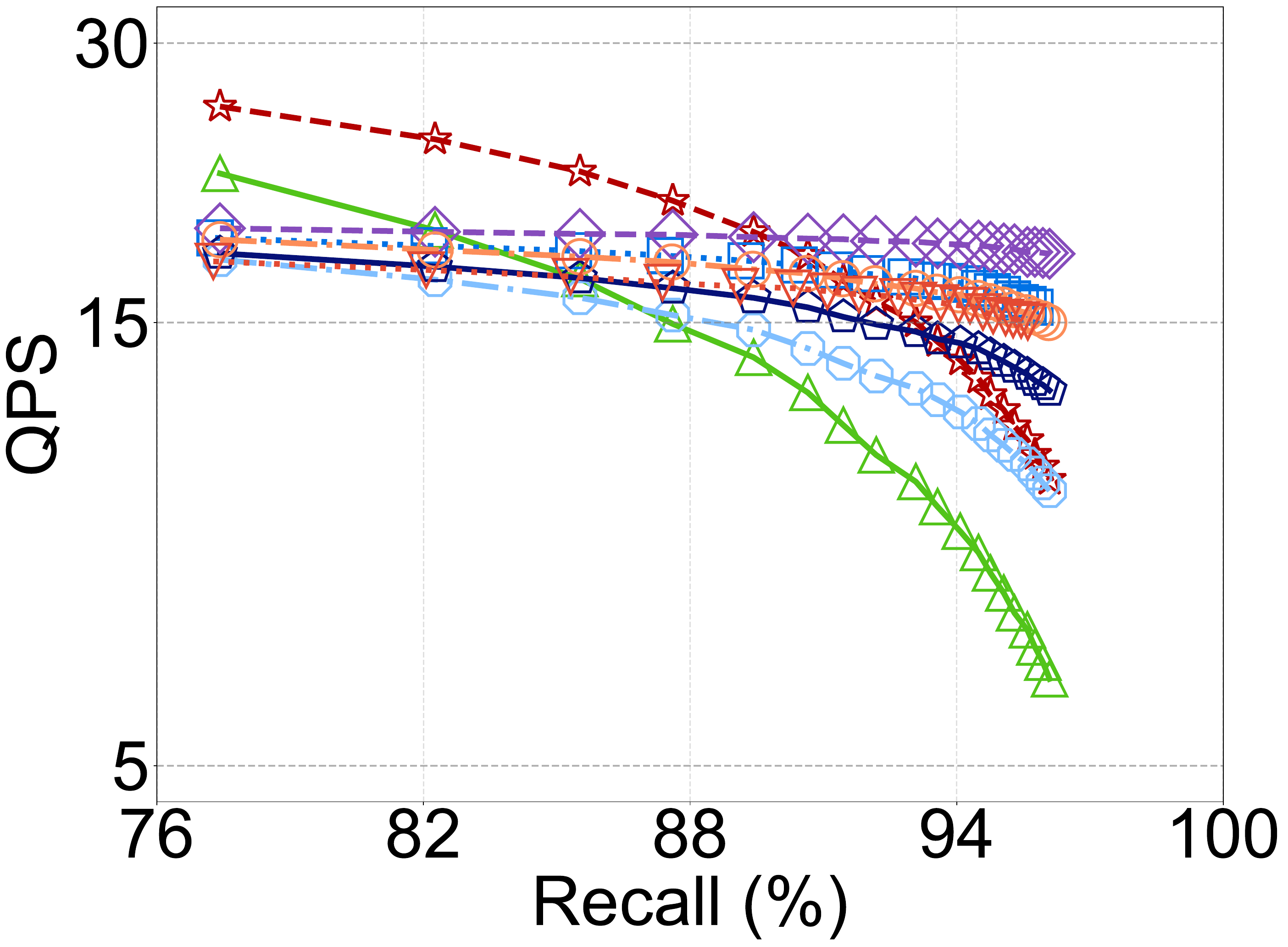}
        \caption{\Msmacro (without SIMD)}\label{fig:msmacro-without-SIMD}
    \end{subfigure}
    \begin{subfigure}{0.25\textwidth}
        \centering
        \includegraphics[width=\textwidth]{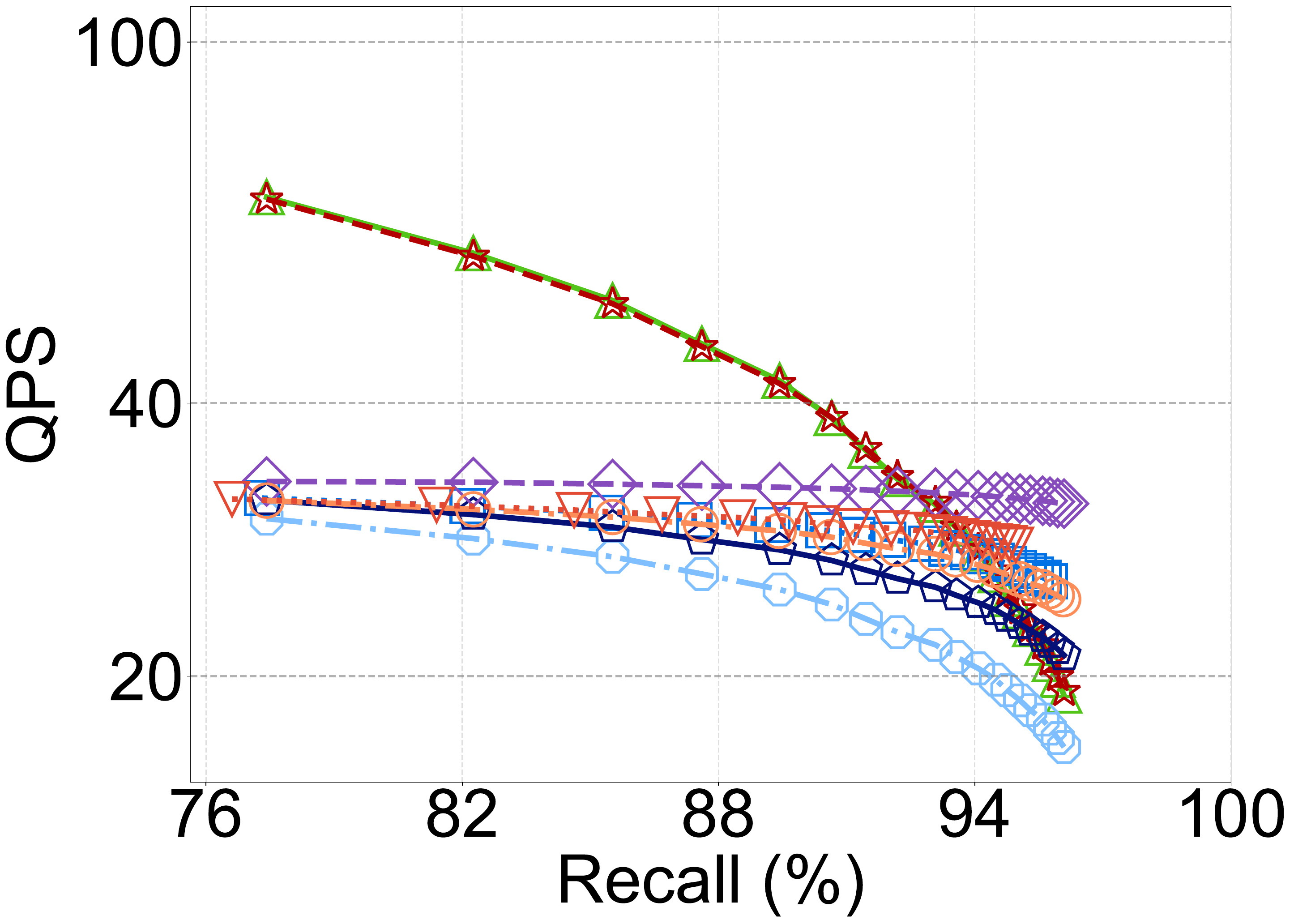}
        \caption{\Msmacro (with SIMD)}\label{fig:msmacro-with-SIMD}
    \end{subfigure}
    \begin{subfigure}{0.24\textwidth}
        \centering
        \includegraphics[width=\textwidth]{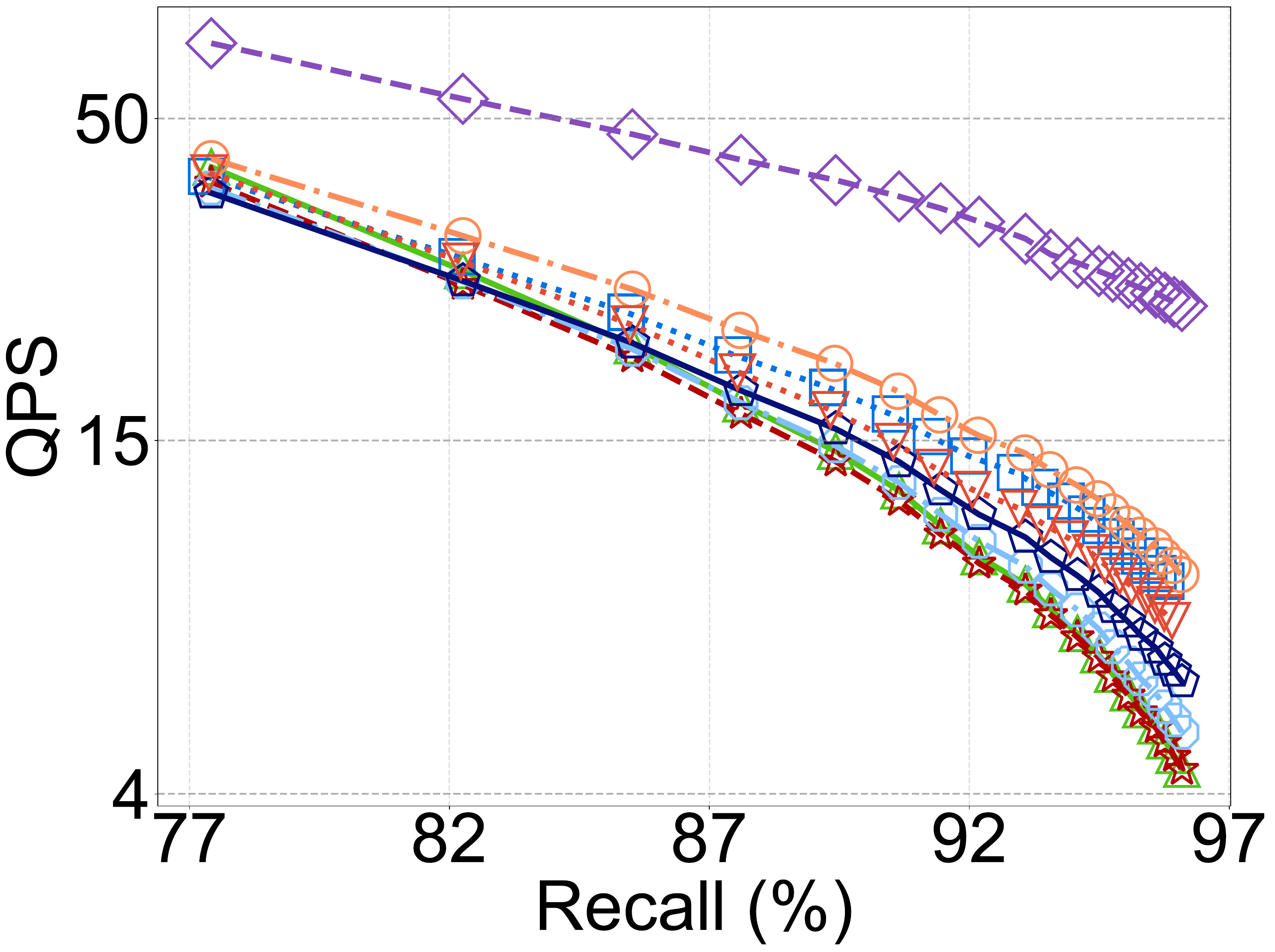}
        \caption{\Msmacro (with GPU)}\label{fig:msmacro-gpu}
    \end{subfigure}
    \caption{Comparisons of DCO methods using the IVF index on \Msmacro dataset across different hardware environments}\label{fig:extend-hardware}
\end{figure*}

\begin{figure*}[t]
    \centering

    \begin{subfigure}{0.24\textwidth}
        \centering
        \includegraphics[width=\textwidth]{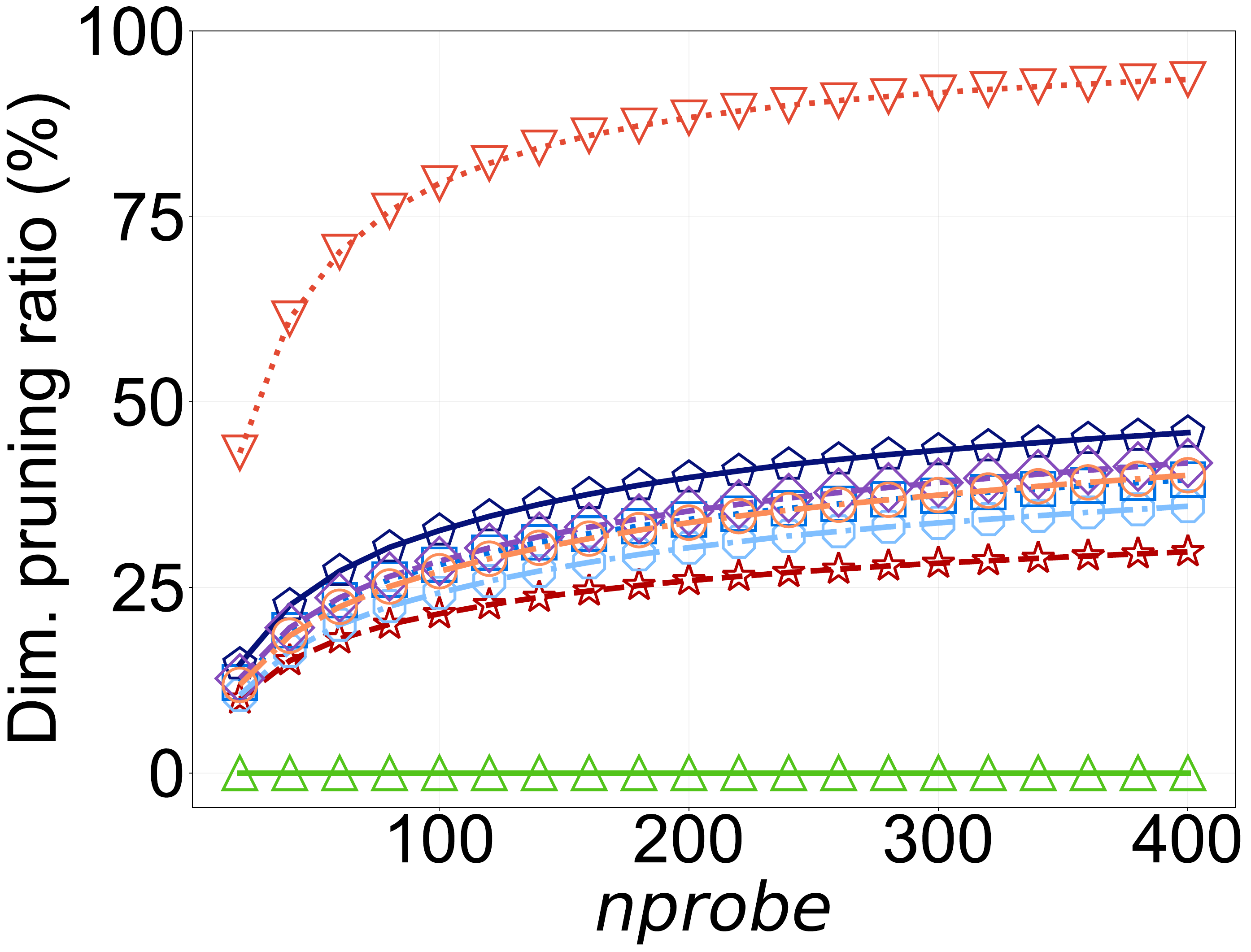}
        \caption{\Glove}\label{fig:pruning-ivf-glove}
    \end{subfigure}
    \begin{subfigure}{0.24\textwidth}
        \centering
        \includegraphics[width=\textwidth]{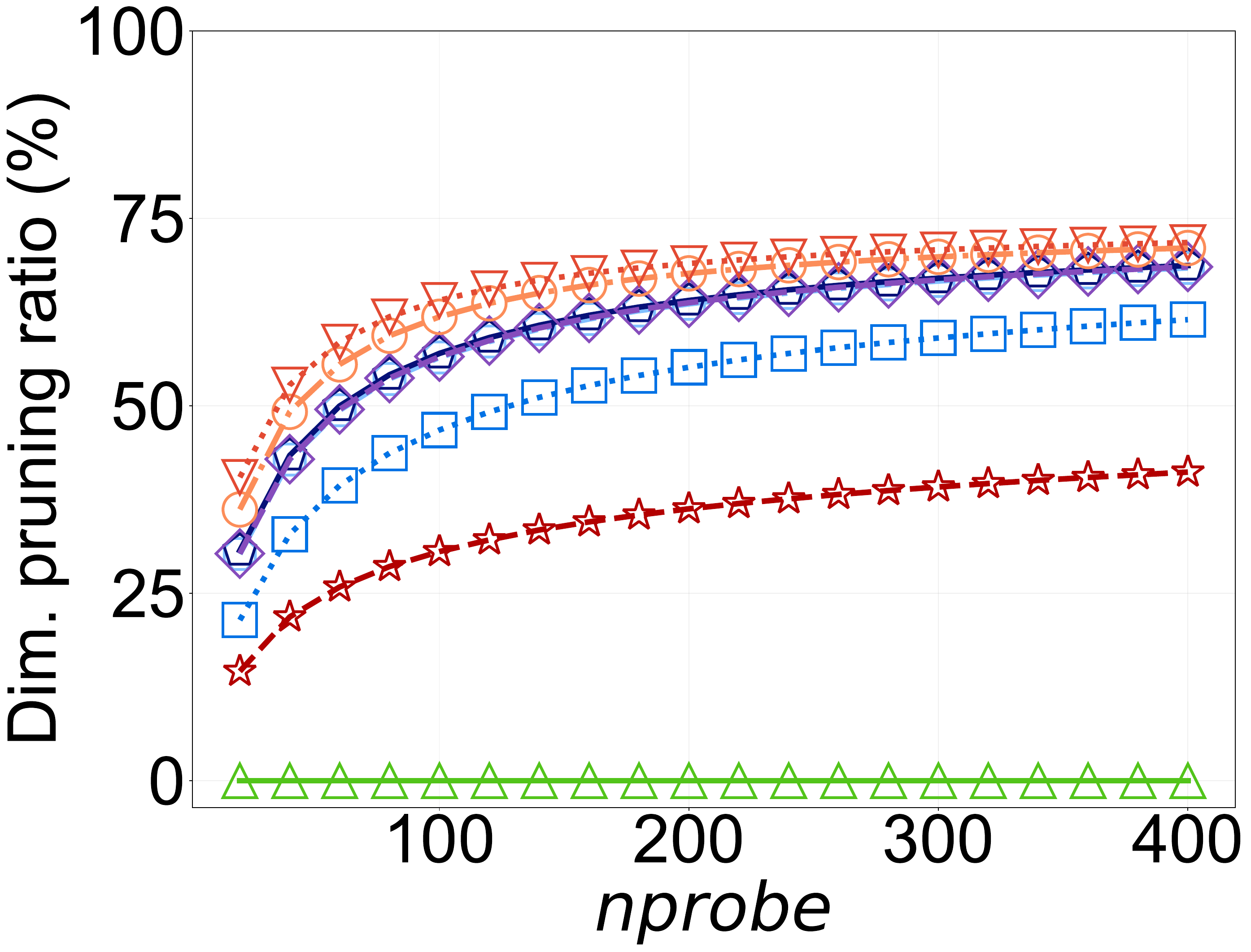}
        \caption{\Sift}\label{fig:pruning-ivf-sift}
    \end{subfigure}
    \begin{subfigure}{0.24\textwidth}
        \centering
        \includegraphics[width=\textwidth]{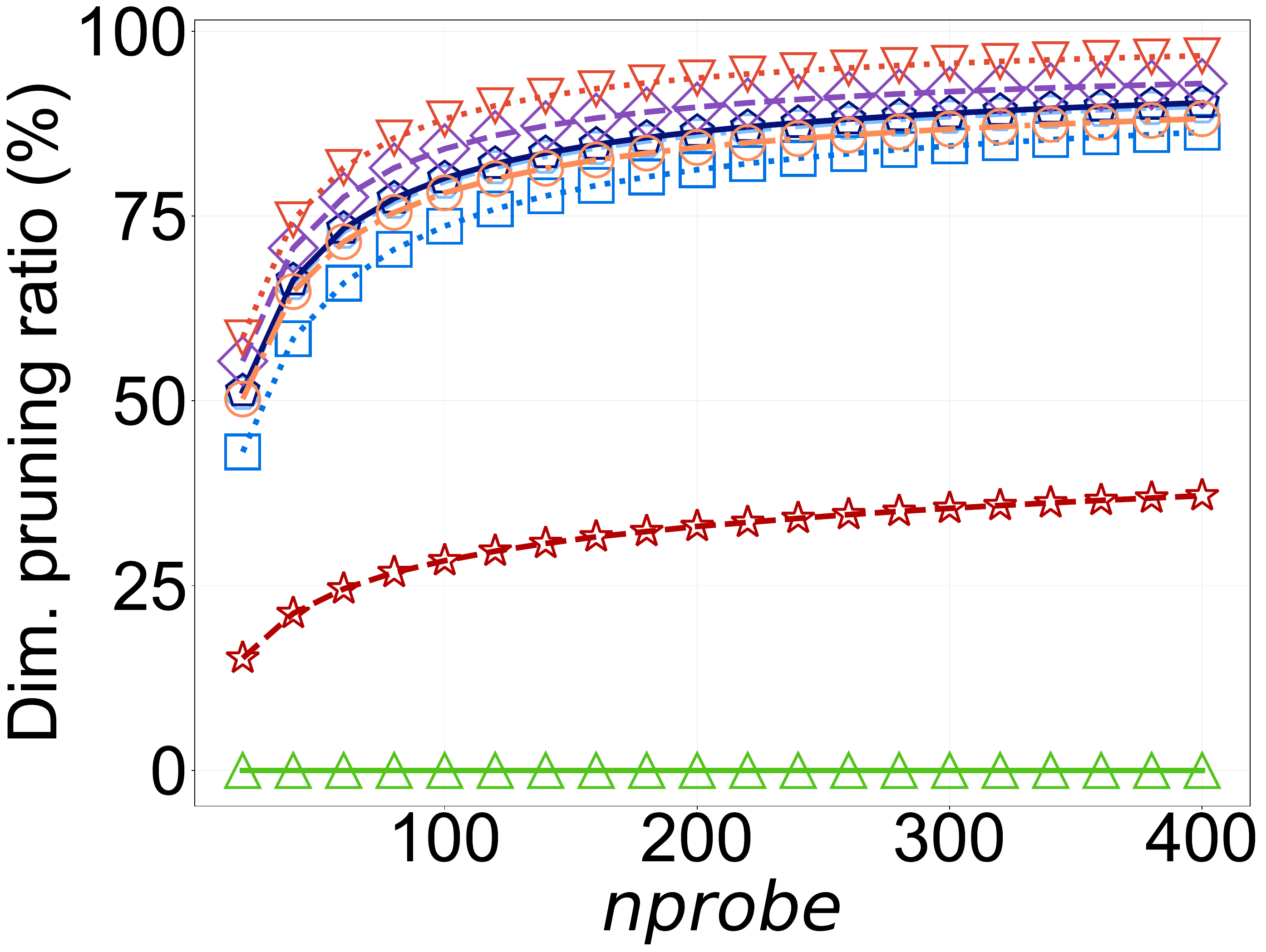}
        \caption{\Gist}\label{fig:pruning-ivf-gist}
    \end{subfigure}
    \begin{subfigure}{0.24\textwidth}
        \centering
        \includegraphics[width=\textwidth]{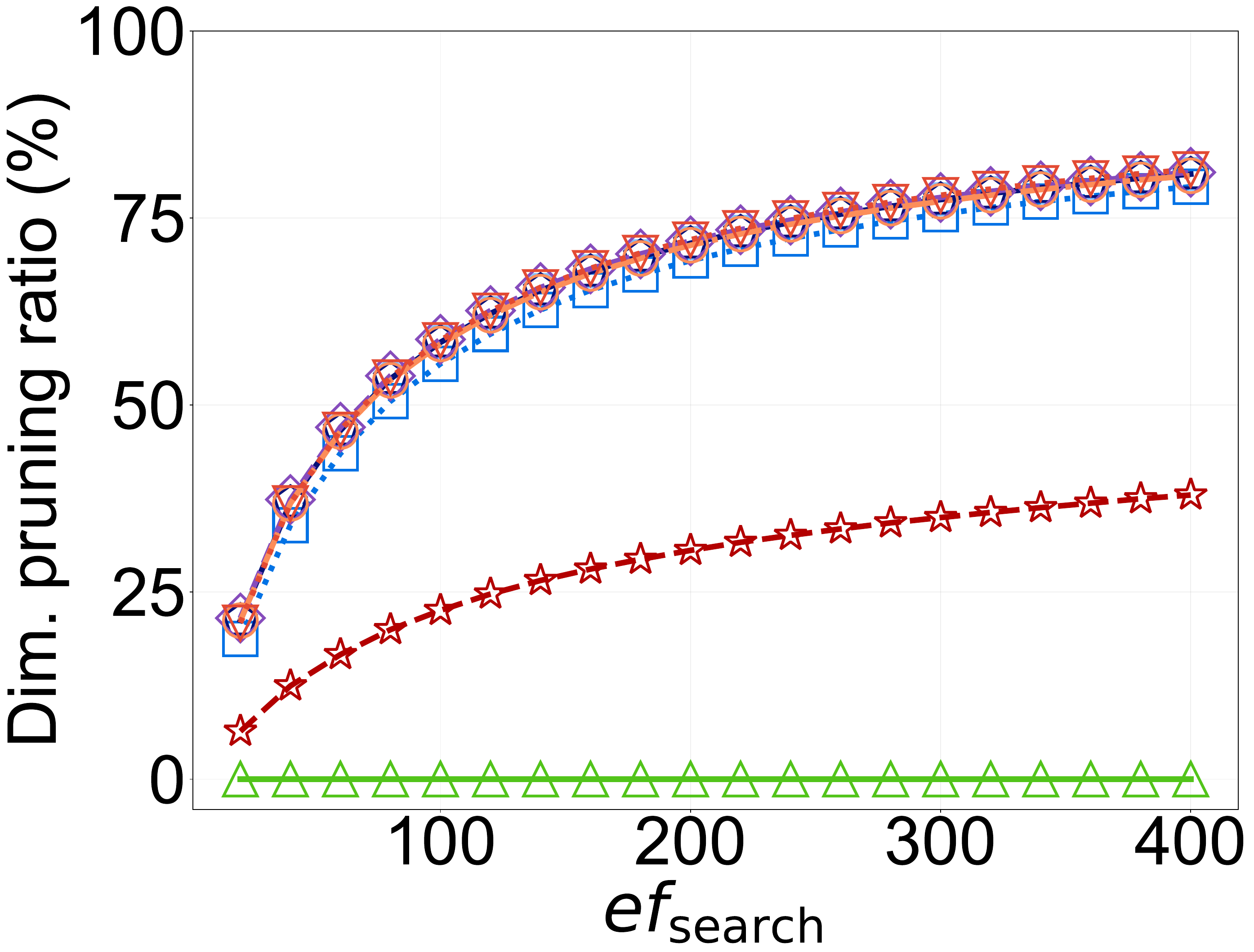}
        \caption{\Trevi}\label{fig:pruning-ivf-trevi}
    \end{subfigure}
    
    \begin{subfigure}{0.24\textwidth}
        \centering
        \includegraphics[width=\textwidth]{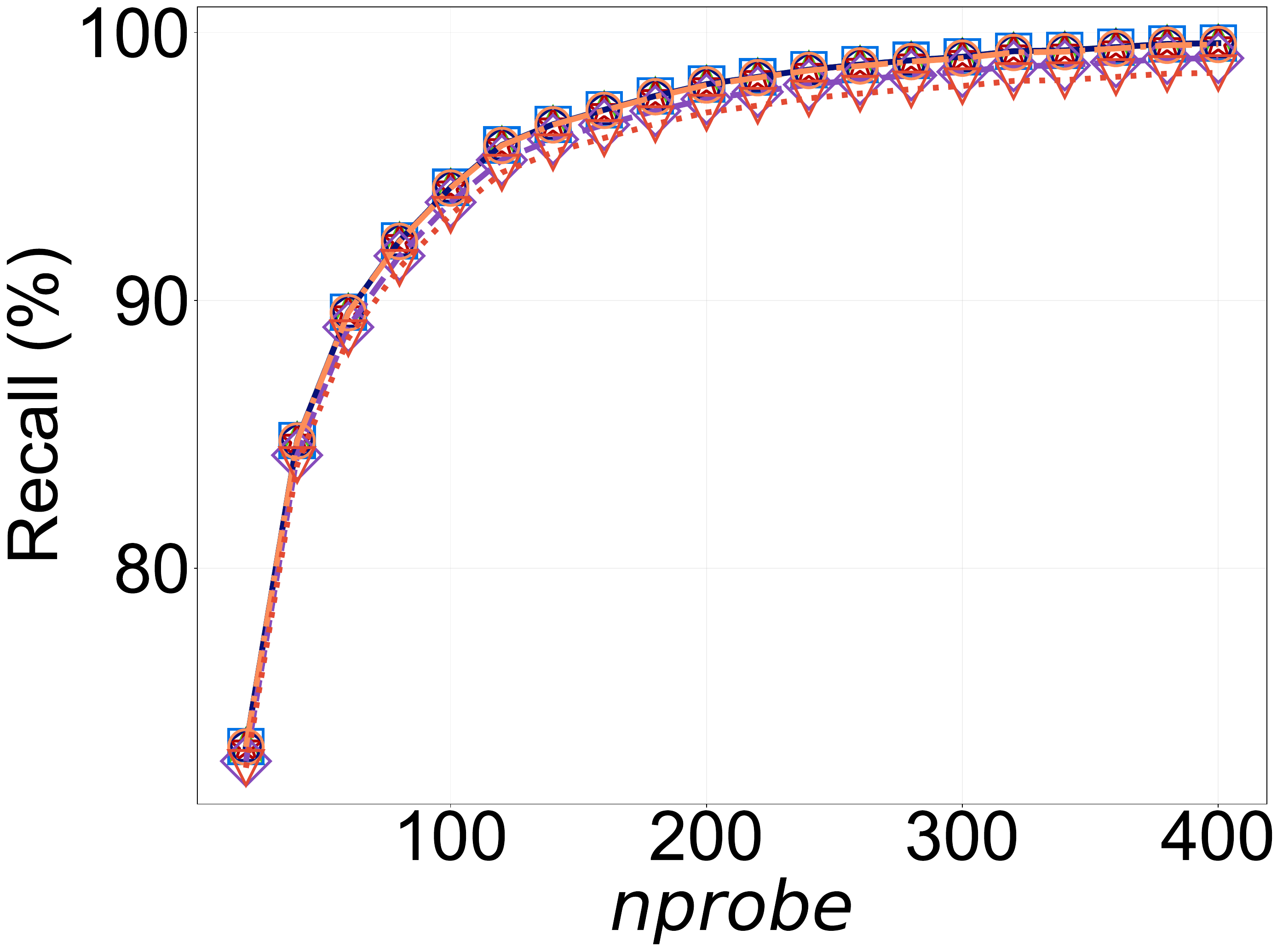}
        \caption{\Glove}\label{fig:pruning-ivf-glove-recall}
    \end{subfigure}
    \begin{subfigure}{0.24\textwidth}
        \centering
        \includegraphics[width=\textwidth]{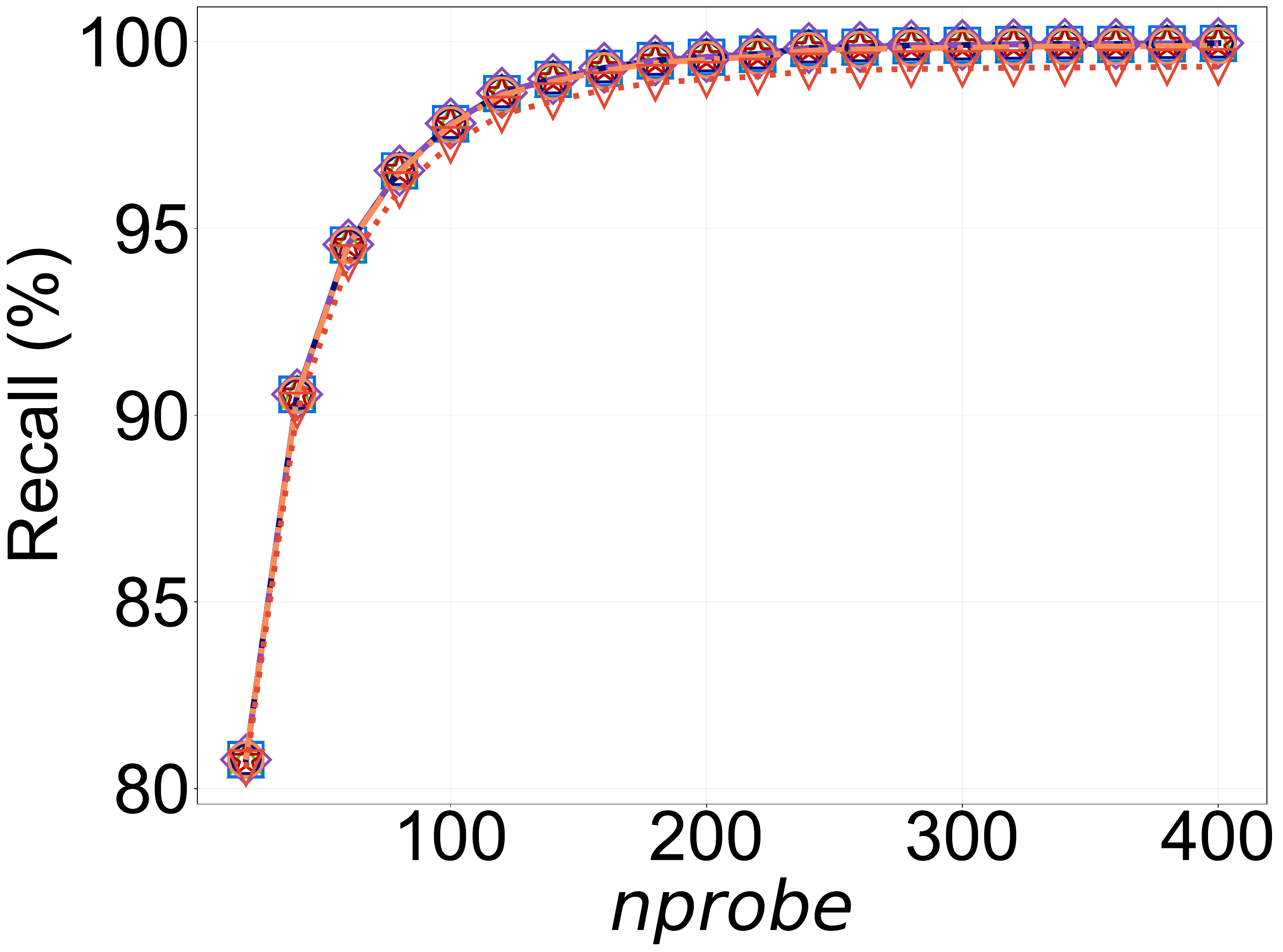}
        \caption{\Sift}\label{fig:pruning-ivf-sift-recall}
    \end{subfigure}
    \begin{subfigure}{0.24\textwidth}
        \centering
        \includegraphics[width=\textwidth]{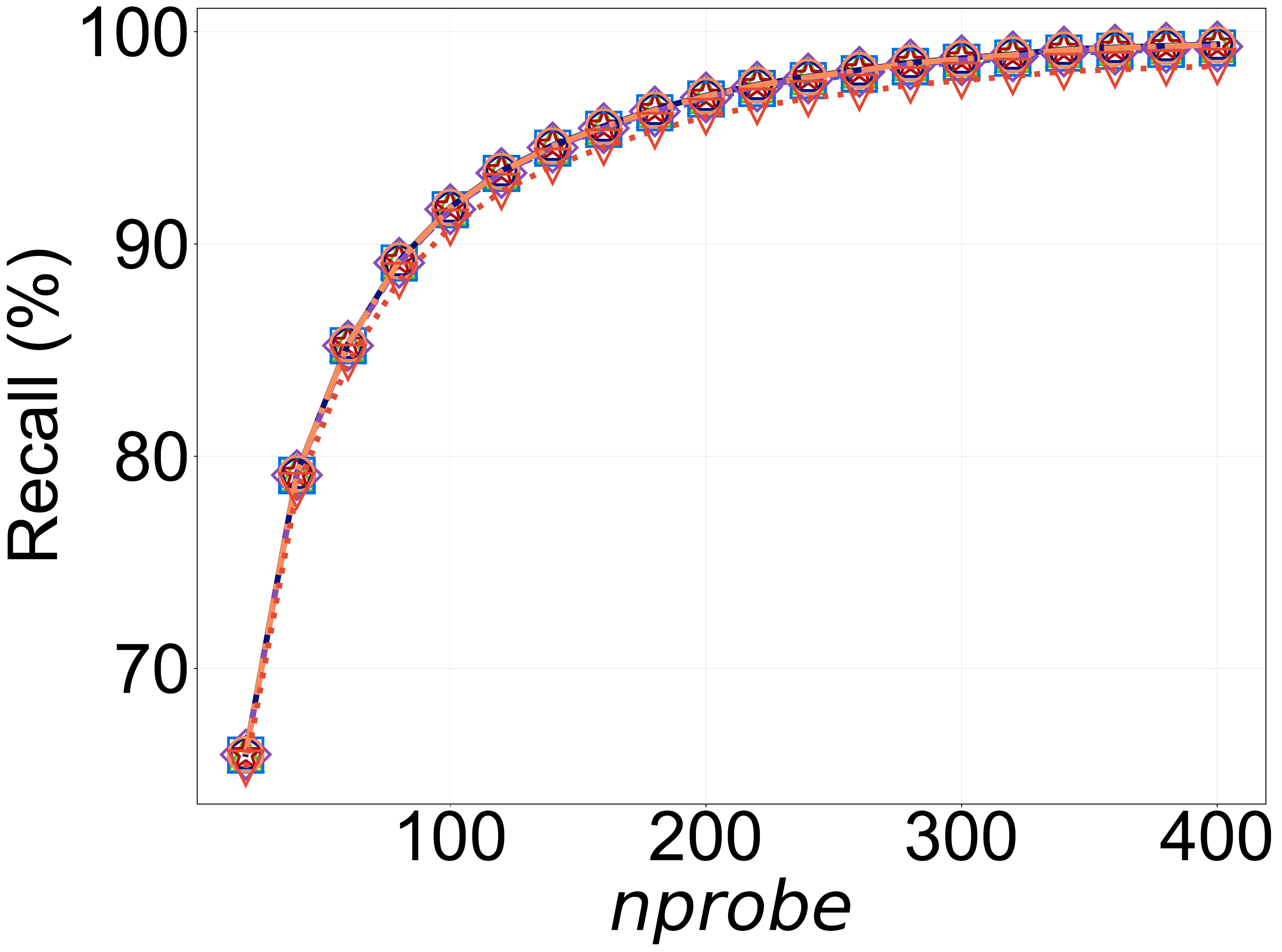}
        \caption{\Gist}\label{fig:pruning-ivf-gist-recall}
    \end{subfigure}
    \begin{subfigure}{0.24\textwidth}
        \centering
        \includegraphics[width=\textwidth]{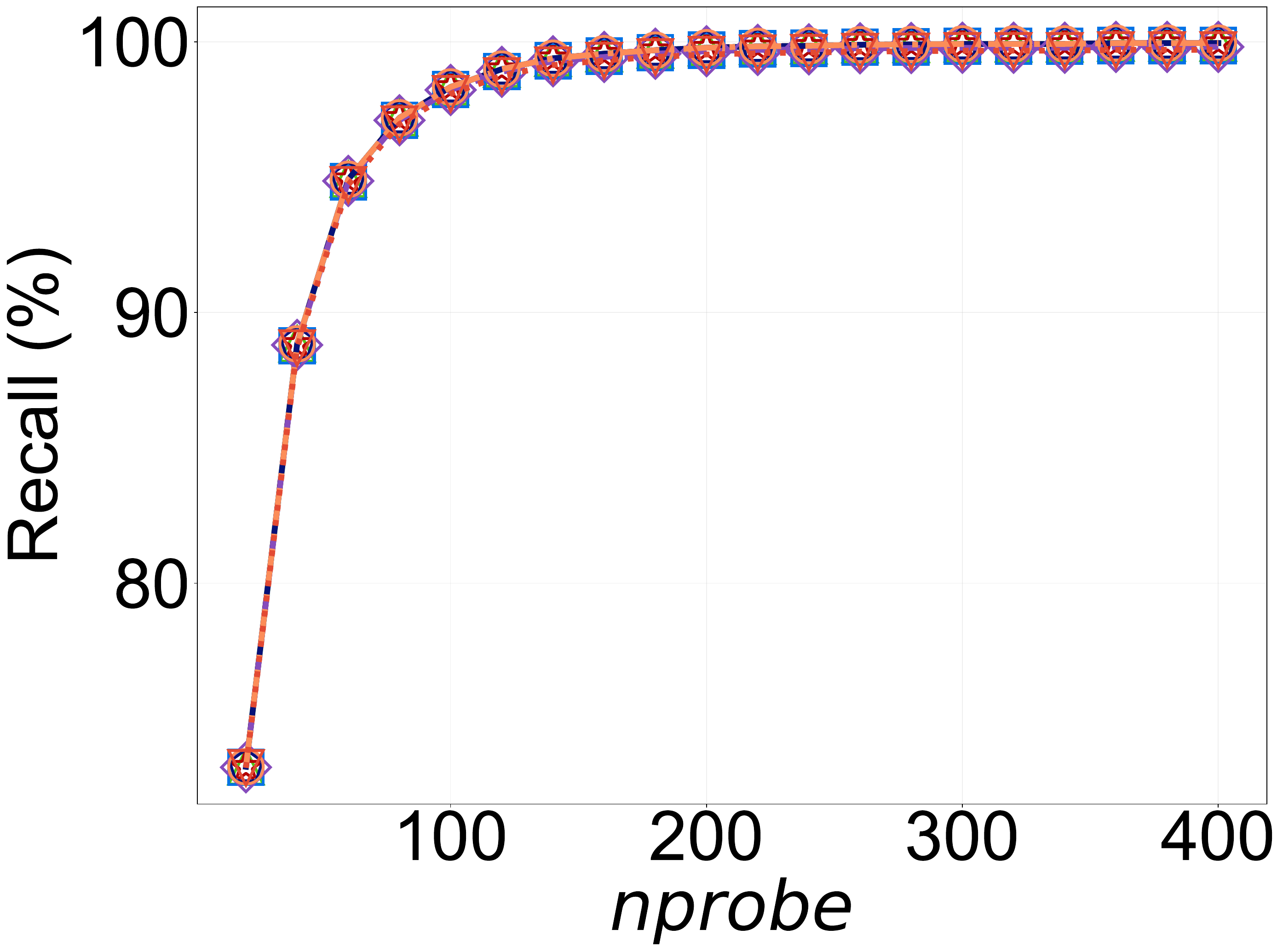}
        \caption{\Trevi}\label{fig:pruning-ivf-trevi-recall}
    \end{subfigure}
    \caption{Dimension pruning via DCO and its impact on recall on IVF}\label{fig:dco-pruning-ivf}
\end{figure*}

\section{Pruning Capability of DCO under the IVF Index}

We evaluate and analyze the capability of DCO on IVF. \figref{fig:dco-pruning-ivf} illustrates the dimension pruning ratios of DCOs and their recall across three datasets of varying dimensionality. 

\fakeparagraph{Pruning Capability of DCO on IVF} The results also demonstrate that the pruning effectiveness of DCOs is dimension-dependent: higher pruning ratios are consistently observed on high-dimensional and ultra-high-dimensional datasets. On these datasets, \DDCopq typically achieves the highest pruning ratio. However, it is worth noting that, in addition to dimension scanning, \DDCopq incurs additional overhead from computing quantized distances. Consequently, despite its strong pruning capability, \DDCopq exhibits poorer query performance on IVF when SIMD is disabled. When SIMD is enabled, thereby significantly accelerating quantized distance computation on IVF, the performance of \DDCopq aligns much more closely with its pruning capability. For other DCO methods, a higher pruning capability generally translates to better acceleration when SIMD is disabled. Moreover, all methods maintain comparable recall, demonstrating that the minor approximation errors introduced by DCOs do not compromise query accuracy on IVF.

\fakeparagraph{Pruning Capability Comparisons: HNSW vs IVF} 
%Additional results on dimension pruning ratios for HNSW are presented in \figref{fig:dco-pruning-hnsw}. 
Combined with the findings in \figref{fig:dco-pruning}, we observe that the pruning capability of DCOs remains relatively stable on HNSW, whereas it increases with $nprobe$ in IVF. 
This behavior can be attributed to the characteristics of candidate selection in each index structure.  
In HNSW, the retrieved candidates are typically close to the query vector.  
In contrast, IVF retrieves many candidates that are far from the query vector; for these, DCO methods can quickly determine irrelevance using only a few scanned dimensions. As a result, the pruning ratio in IVF rises with increasing $nprobe$.

\section{Parameter Study}
This experiment investigates two primary parameters of DCOs: 
the \textit{initial step size} $\Delta_0$ (\ie the initial number of dimensions to scan),
and the \textit{incremental step size} $\Delta_d$ used for each subsequent scan.
We systematically evaluate their effects on both HNSW and IVF indexes using the \Gist dataset.
Our study primarily reports QPS, as recall remains largely unaffected.
Notice that, prior work \cite{jayaram2019diskann, DBLP:journals/pvldb/WangXY021, DBLP:conf/nips/ZhangWCCZMHDMWP23} usually sets these parameters as $\Delta_d = \Delta_0 = 32$.

\fakeparagraph{Impact of Initial Step Size $\Delta_0$}
\figref{fig:initial-dimension} presents the QPS of each method on both HNSW and IVF indexes.
In \figref{fig:initial-dimension-hnsw}, the QPS of \Loop methods remains relatively stable, whereas \Hypothesis and \Class methods achieve a notable QPS increase by tuning $\Delta_0$ appropriately.
A similar trend is observed on the IVF index.
\ADSampling, \DADE, \DDCres, and \DDCpca improve QPS by up to 1.3, 1.5, 1.2, and 1.8$\times$ respectively, in \figref{fig:initial-dimension-ivf}.

The efficiency gain of tuning $\Delta_0$ is more pronounced on IVF than on HNSW.
This is because IVF stores each data partition contiguously in the main memory, a data layout that is cache-friendly.
In contrast, the graph-based index HNSW lacks such a compact data layout.
Consequently, even \RPDScanning achieves a considerable QPS improvement by adjusting $\Delta_0$ in \figref{fig:initial-dimension-ivf}.

\fakeparagraph{Impact of Incremental Step Size $\Delta_d$}
\figref{fig:sampling-step} reports the QPS when varying $\Delta_d$.
Most algorithms are highly sensitive to this parameter, with \FDScanning being the only exception. 
The optimal $\Delta_d$ is 64, 96 or 160 for HNSW, and 64 for IVF.
Compared to the default $\Delta_d = 32$ in prior work \cite{jayaram2019diskann, DBLP:journals/pvldb/WangXY021, DBLP:conf/nips/ZhangWCCZMHDMWP23}, \ADSampling, \DADE, and \DDCres achieve speedups of up to 1.1, 1.2, and 1.1$\times$.

Overall, jointly using the optimal parameter combinations, \ADSampling, \DADE, and \DDCres can boost QPS by up to 1.5, 1.8, and 1.8$\times$ compared to the settings in prior work.
\stepcounter{takeaway}
\vspace{-0.5ex}
\begin{tipbox}
\textbf{Takeaway: While tuning $\Delta_0$ and $\Delta_d$ can substantially boost QPS, their optimal values are determined by datasets, indexes, and specific DCO methods.}  
\end{tipbox}

\section{Detailed Implementation}
\zheng{We implemented all DCO algorithms from scratch within a unified framework, adhering to their open-source implementations~\cite{DBLP:journals/pacmmod/GaoL23, DBLP:journals/pvldb/DengCZWZZ24, yang2025effective}. This section will elaborate on the relevant implementation details.

\fakeparagraph{Unified Framework}
Every DCO algorithm was deployed within the same vector similarity search framework with the same index. During online query processing, all compared methods are implemented in C++, while Python is only used for pre-processing (\ie PCA and training),
which is in line with existing work on DCOs~\cite{DBLP:journals/pacmmod/GaoL23, DBLP:journals/pvldb/DengCZWZZ24, yang2025effective}.

\fakeparagraph{Identical Index Configuration}
In terms of indexes, we fix a consistent underlying data layout for all methods, where vectors in the IVF index are stored contiguously within each partition while those in the HSNW index follow the original insertion sequence, and none of the DCO methods are allowed to modify this predefined layout. 

\fakeparagraph{Batch Queries}
Each query within a batch is processed sequentially and independently. 
We adopt this established evaluation protocol to maintain consistency with the baseline methods for a fair comparison.

\fakeparagraph{Hardware Optimization}
For HNSW, we utilize the SIMD-accelerated vector distance computation routines from the Hnswlib library~\cite{hnswlib}, where a single SIMD instruction processes multiple dimension pairs simultaneously. On the GPU, we pre-load the IVF index into device memory and adopt fine-grained parallelism: a single CUDA kernel is launched per partition, with each thread processing one candidate vector. This configuration is identical across all DCO methods to ensure consistent execution patterns.}

\fakeparagraph{Consistent Implementation Languages} 
The online query processing of all compared methods is implemented in C++, while Python is only used for pre-processing (\ie PCA and training), which is in line with existing work on DCOs~\cite{DBLP:journals/pacmmod/GaoL23, DBLP:journals/pvldb/DengCZWZZ24, yang2025effective}.

\section{Query Processing via HNSW on CPUs with SIMD Disabled}\label{app:HNSW-NO-SIMD}

This experiment evaluates the performance improvement from applying DCOs to vector similarity search on HNSW with SIMD disabled.

\begin{figure}[h]
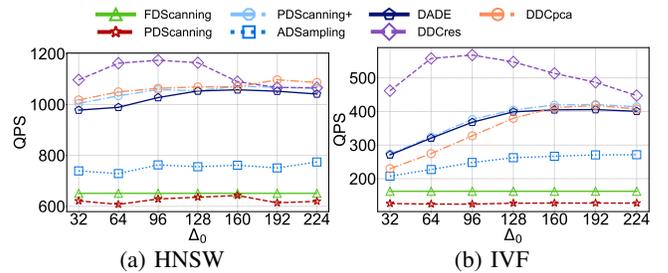

    \centering
    \includegraphics[width=0.75\linewidth]{figure/initial_dimension/legend.pdf}
    \begin{subfigure}{0.48\linewidth}
        \centering
        \includegraphics[width=\textwidth]{figure/initial_dimension/initdimlineHNSW.pdf}\vspace{-1.0ex}
        \caption{HNSW}\label{fig:initial-dimension-hnsw}
    \end{subfigure}%
    \begin{subfigure}{0.48\linewidth}
        \centering
        \includegraphics[width=\textwidth]{figure/initial_dimension/initdimlineIVF.pdf}\vspace{-1.0ex}
        \caption{IVF}\label{fig:initial-dimension-ivf}
    \end{subfigure}
    \caption{Impact of initial step size $\Delta_0$}\label{fig:initial-dimension}
\end{figure}

\begin{figure}[h]
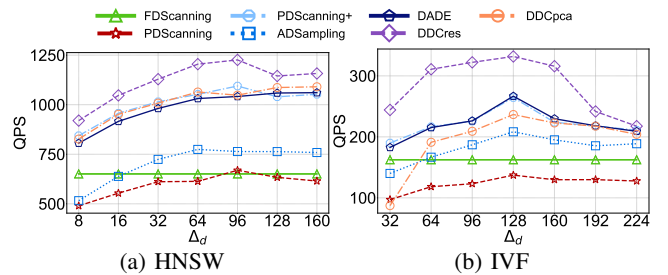

    \centering
    \includegraphics[width=0.75\linewidth]{figure/sampling_dimensions/legend.pdf}
    \begin{subfigure}{0.48\linewidth}
        \centering
        \includegraphics[width=\textwidth]{figure/sampling_dimensions/increase_step_line-HNSW.pdf}\vspace{-1.0ex}
        \caption{HNSW}\label{fig:sampling-step-hnsw}
    \end{subfigure}%
    \begin{subfigure}{0.48\linewidth}
        \centering
        \includegraphics[width=\textwidth]{figure/sampling_dimensions/increase_step_line-IVF.pdf}\vspace{-1.0ex}
        \caption{IVF}\label{fig:sampling-step-ivf}
    \end{subfigure}
    \caption{Impact of incremental step size $\Delta_d$}\label{fig:sampling-step}
\end{figure}

\fakeparagraph{Overall Query Performance without SIMD} Compared to the result in \figref{fig:time-accuracy}, the rankings of different DCO methods have undergone dramatic changes when SIMD is disabled. Specifically, \FDScanning no longer achieves the best performance on the \Deep dataset. Meanwhile, compared to \FDScanning, \Hypothesis methods improve QPS by up to 1.9--2.5$\times$, while the \Class methods achieve 2.1--3.1$\times$ improvements. These advantages over \FDScanning are larger than those observed when SIMD is enabled. This phenomenon can be attributed to the fact that SIMD primarily accelerates the distance computation itself, while the overhead of the termination decision logic, \ie determining whether to terminate computation remains unchanged. Consequently, the relative benefit of DCOs is diluted when SIMD is active, as the non-computational components dominate a larger fraction of the total time cost.

We also observe cases where the baseline \PDScanning outperforms other methods.  
As shown in \figref{fig:HNSW-NO_SIMD-trevi-20} and \figref{fig:HNSW-NO_SIMD-trevi-100}, \PDScanning achieves the highest QPS before the recall reaches 98\%. This demonstrates that the online pre-processing time $O(N^2)$ is also a bottleneck for the SOTA methods for some ultra-high-dimensional datasets on HNSW when SIMD is disabled.

\stepcounter{takeaway}
\vspace{-0.5ex}
\begin{tipbox}
\textbf{Takeaway: Enabling SIMD narrows the efficiency advantage of DCO methods over \FDScanning,} reducing the speedup from 1.9--3.1$\times$ to 1.1--2.1$\times$ on HNSW.
\end{tipbox}

\begin{figure*}[t]
    \centering
    \includegraphics[width=0.9\textwidth]{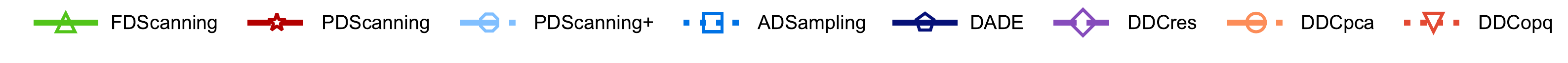}
    \begin{subfigure}{0.24\textwidth}
        \centering
        \includegraphics[width=\textwidth]{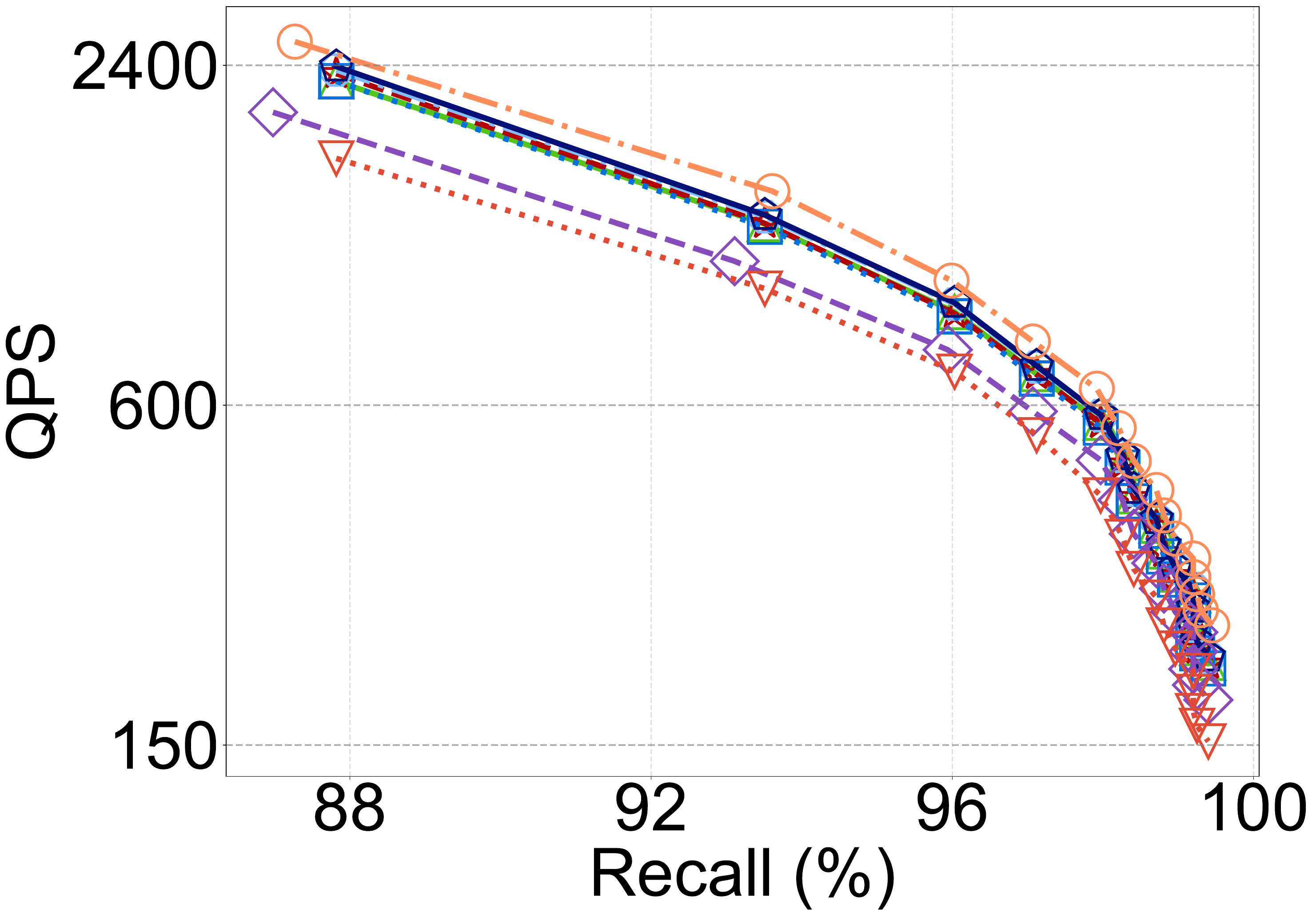}\vspace{-1.0ex}
        \caption{\Deep ($k=20$)}
    \end{subfigure}
    \begin{subfigure}{0.24\textwidth}
        \centering
        \includegraphics[width=\textwidth]{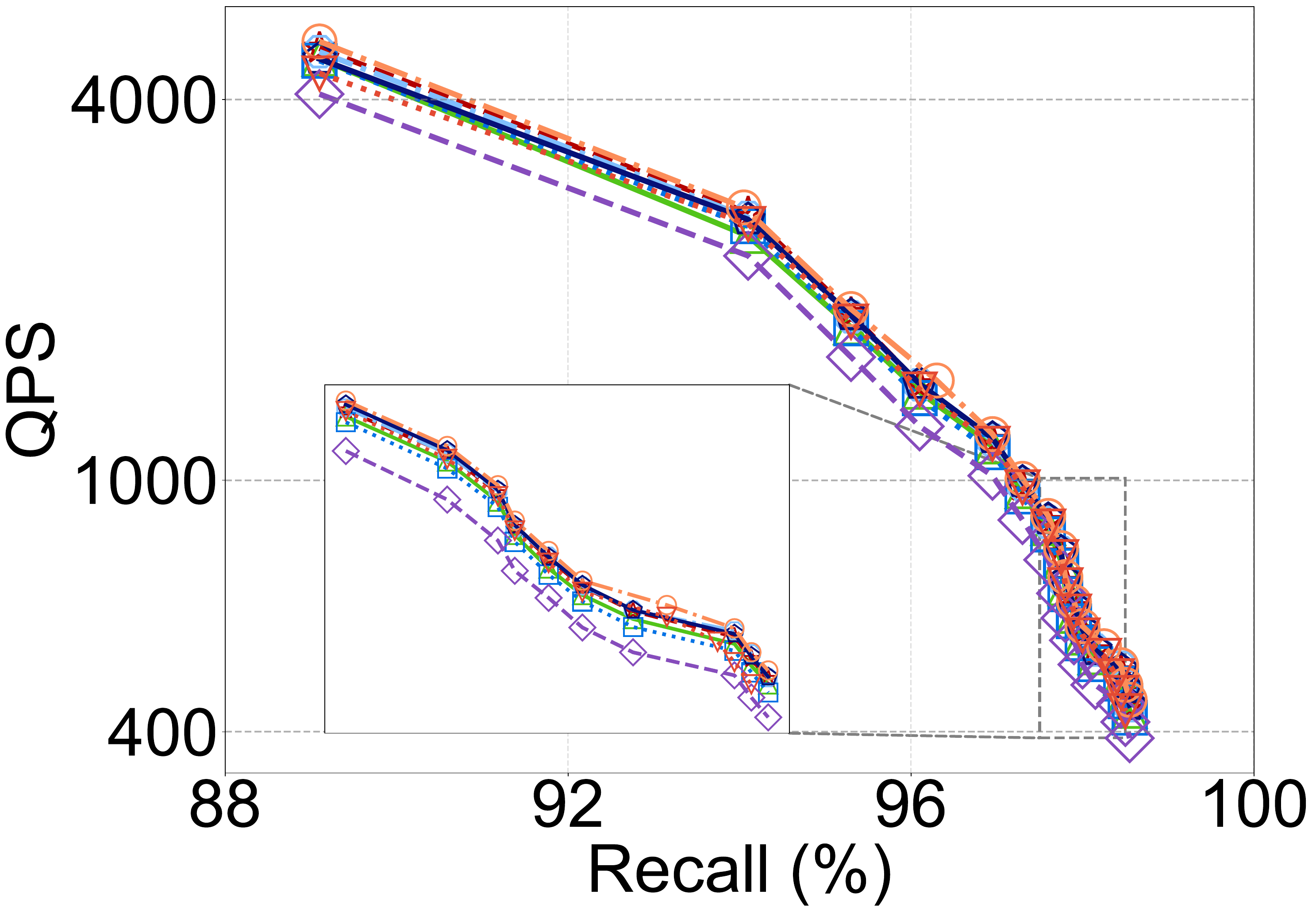}\vspace{-1.0ex}
        \caption{\Glove ($k=20$)}
    \end{subfigure}
    \begin{subfigure}{0.24\textwidth}
        \centering
        \includegraphics[width=\textwidth]{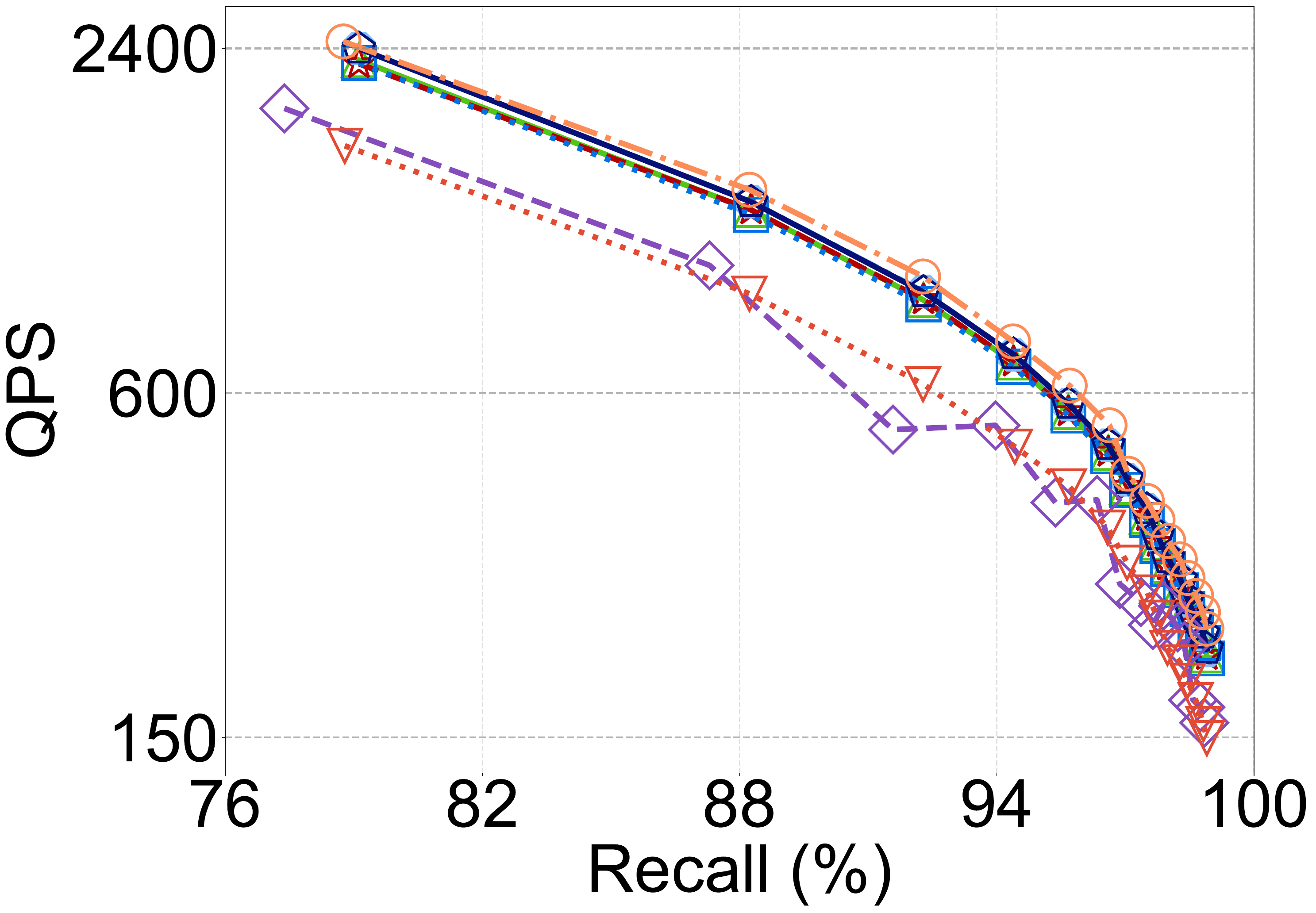}\vspace{-1.0ex}
        \caption{\Deep ($k=100$)}
    \end{subfigure}
    \begin{subfigure}{0.24\textwidth}
        \centering
        \includegraphics[width=\textwidth]{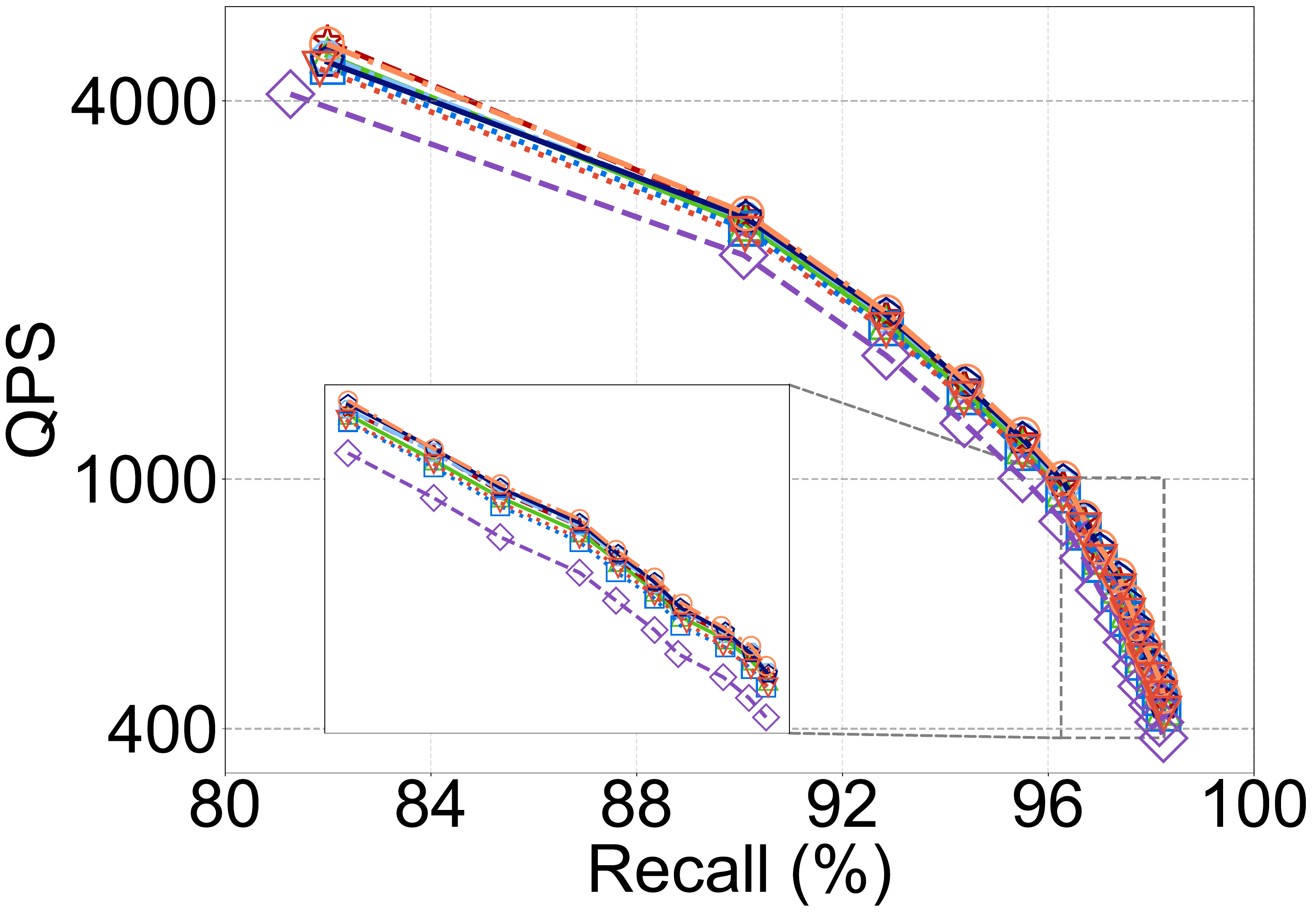}\vspace{-1.0ex}
        \caption{\Glove ($k=100$)}
    \end{subfigure}

    \begin{subfigure}{0.24\textwidth}
        \centering
        \includegraphics[width=\textwidth]{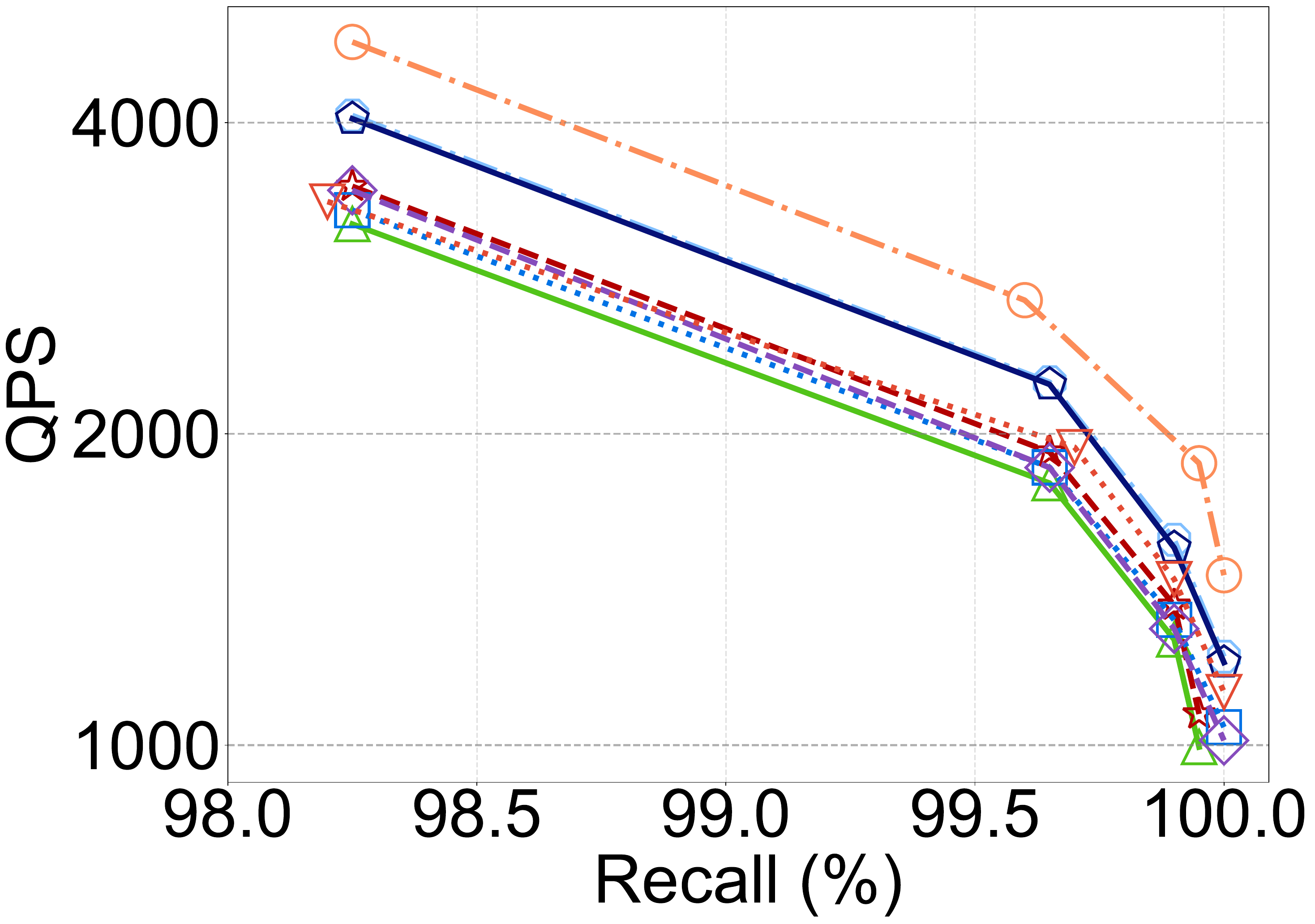}\vspace{-1.0ex}
        \caption{\Sift ($k=20$)}
    \end{subfigure}
    \begin{subfigure}{0.24\textwidth}
        \centering
        \includegraphics[width=\textwidth]{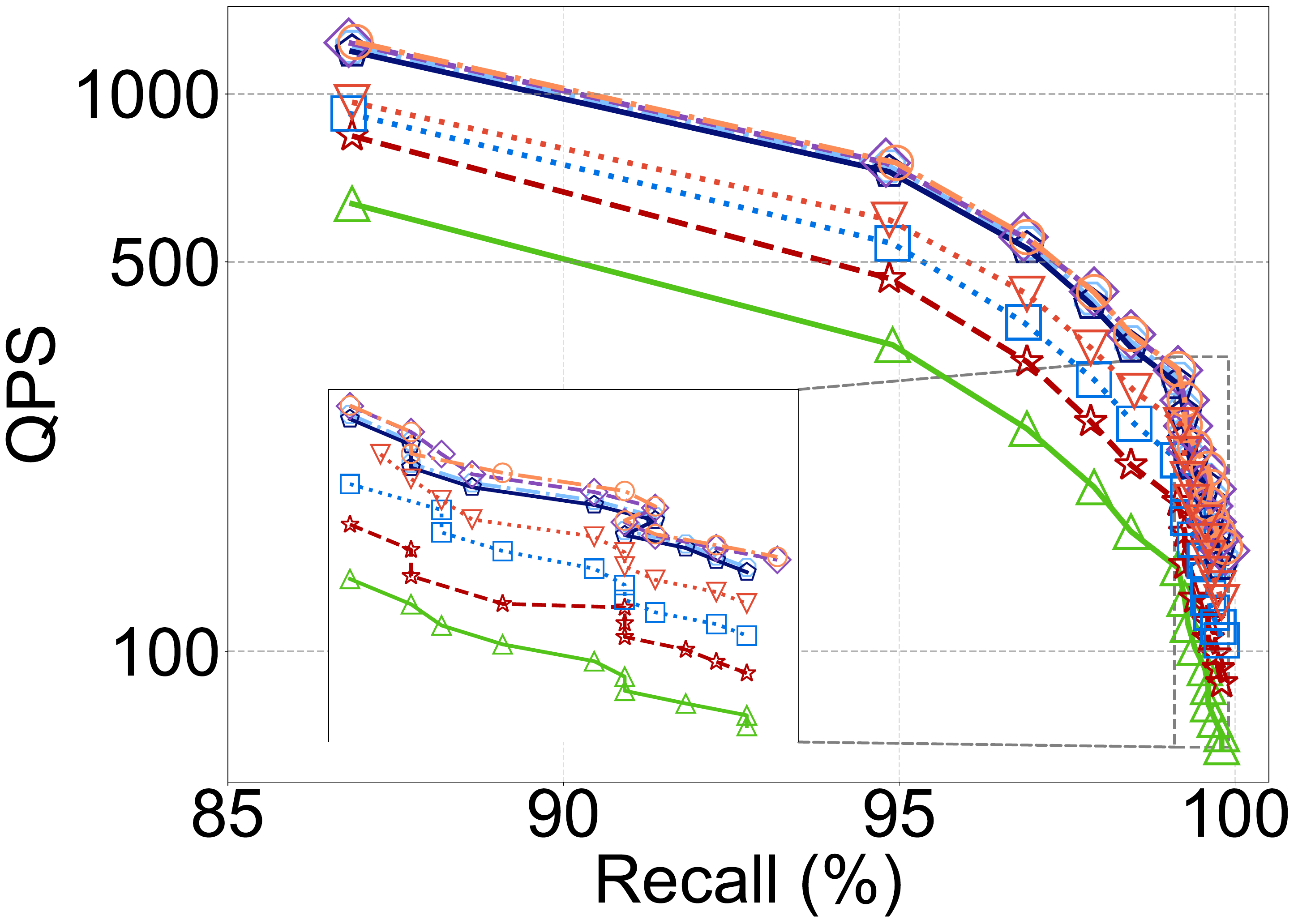}\vspace{-1.0ex}
        \caption{\Gist ($k=20$)}
    \end{subfigure}
    \begin{subfigure}{0.24\textwidth}
        \centering
        \includegraphics[width=\textwidth]{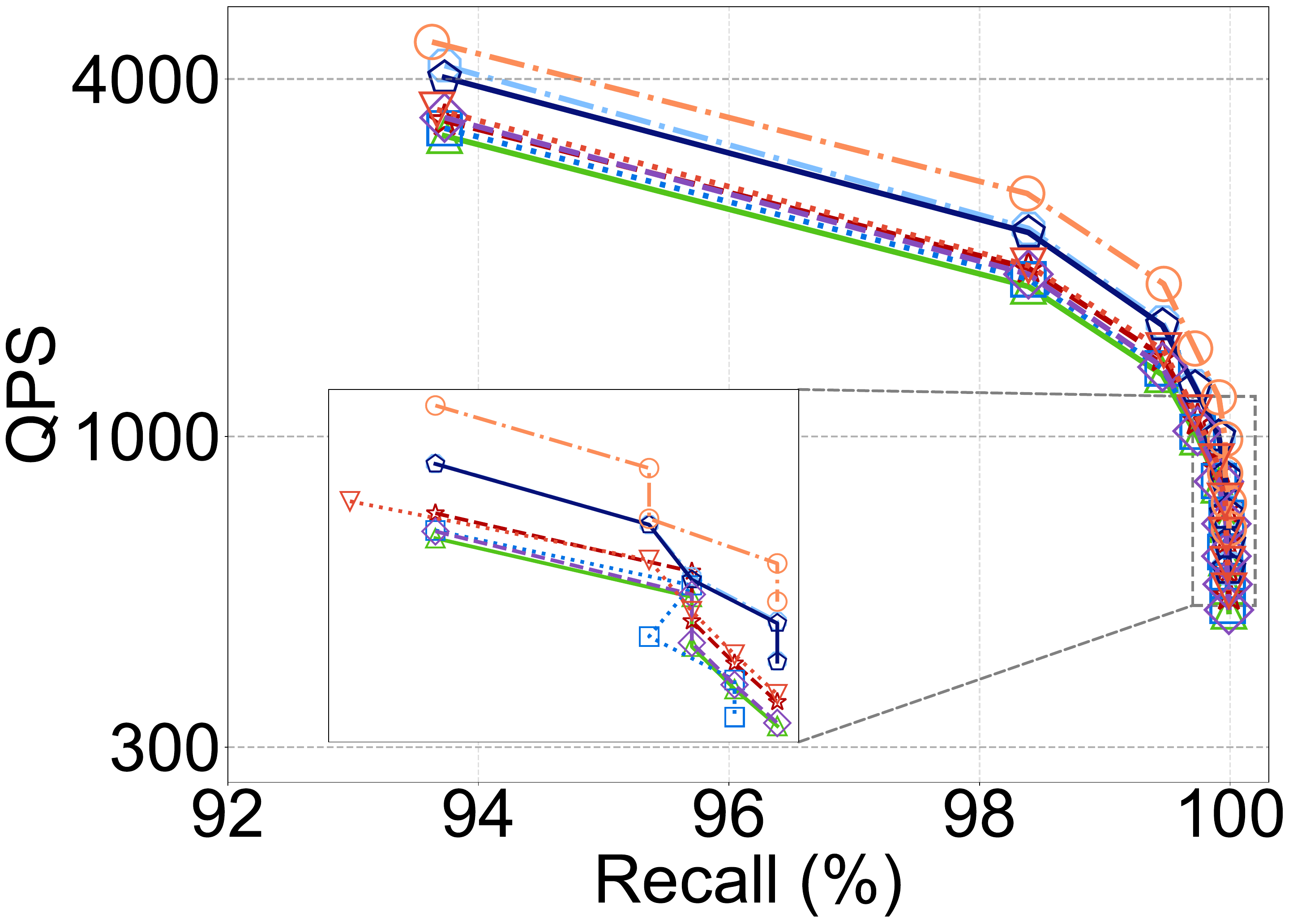}\vspace{-1.0ex}
        \caption{\Sift ($k=100$)}
    \end{subfigure}
    \begin{subfigure}{0.24\textwidth}
        \centering
        \includegraphics[width=\textwidth]{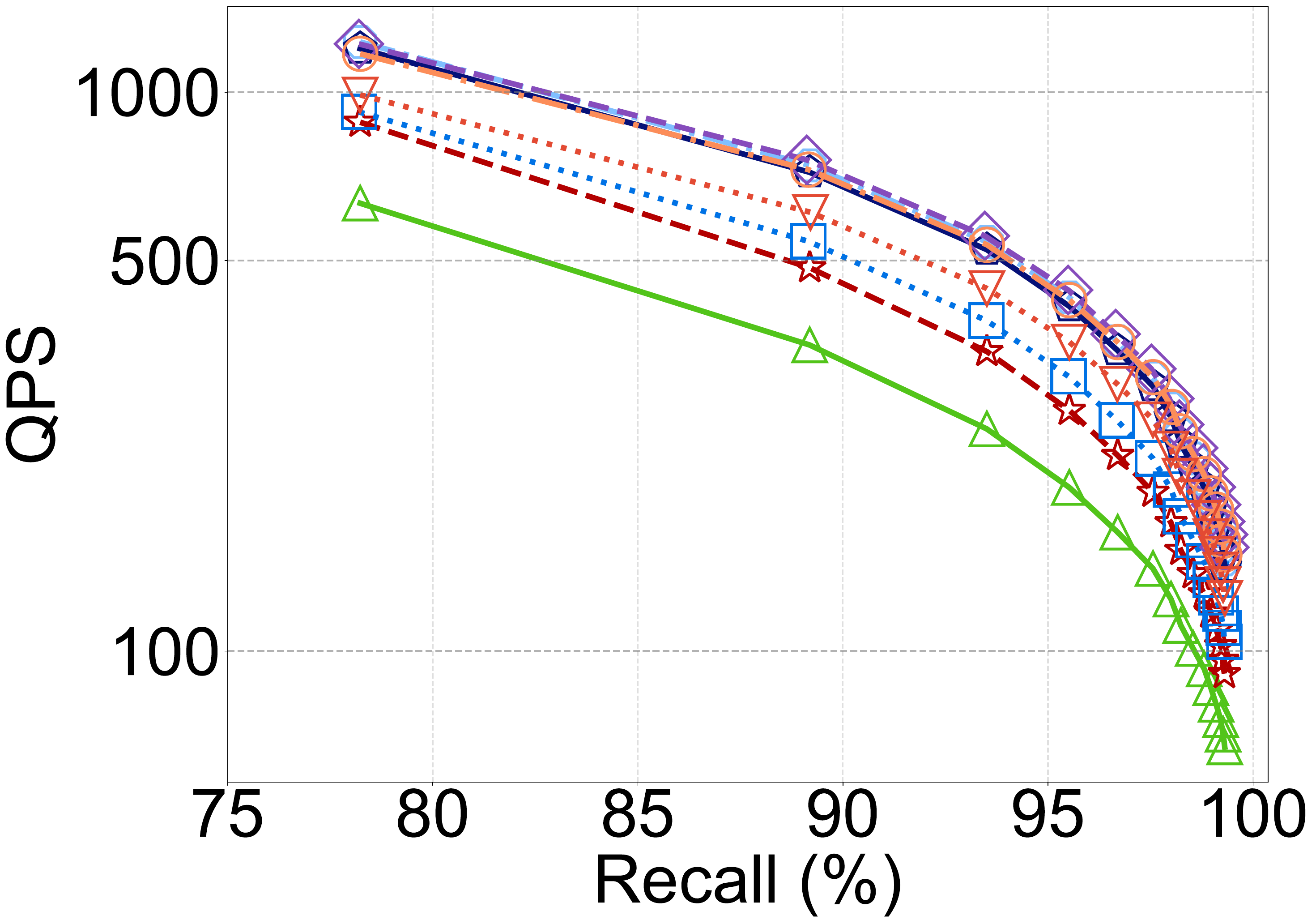}\vspace{-1.0ex}
        \caption{\Gist ($k=100$)}
    \end{subfigure}

    \begin{subfigure}{0.24\textwidth}
        \centering
        \includegraphics[width=\textwidth]{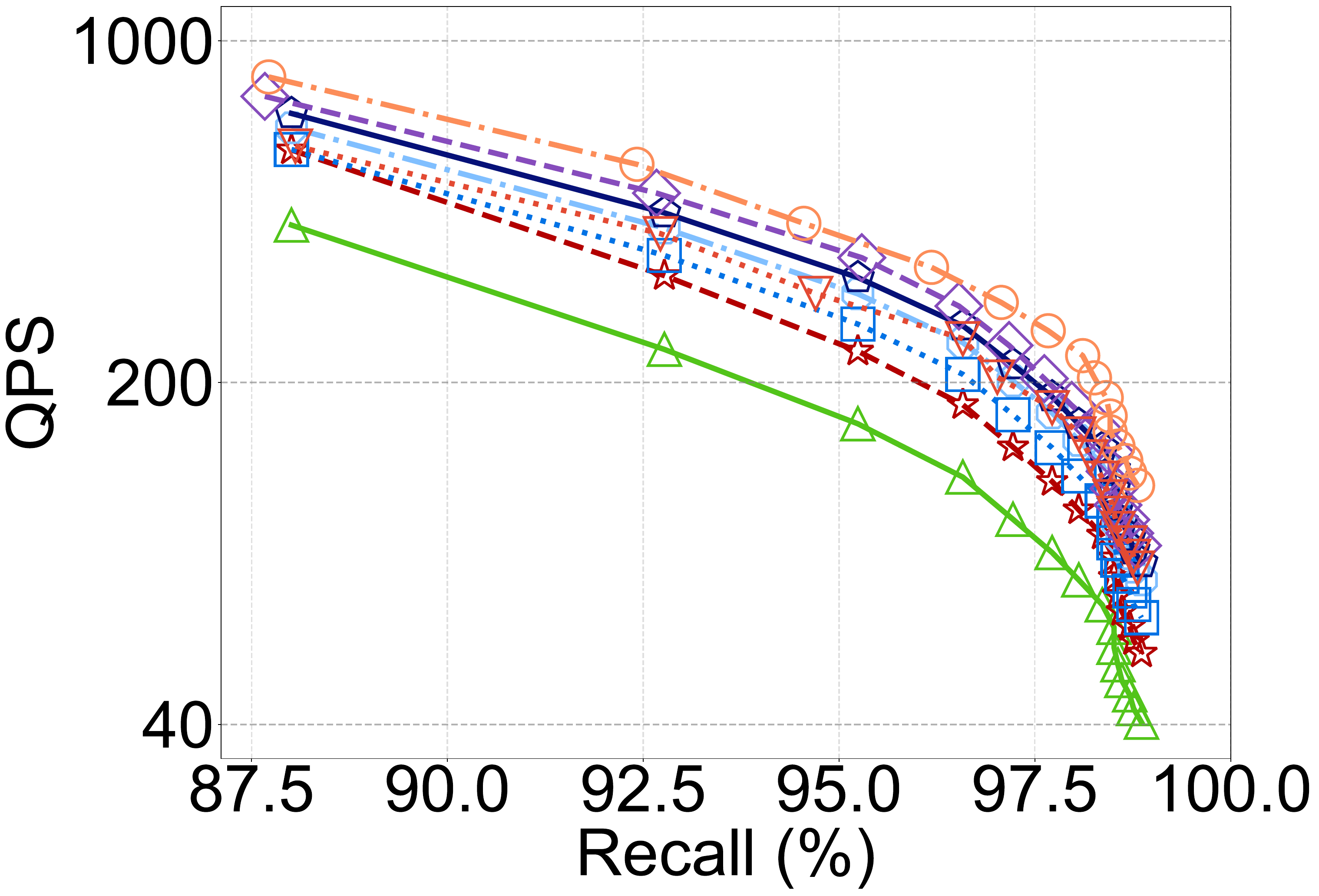}\vspace{-1.0ex}
        \caption{\Openai ($k=20$)}
    \end{subfigure}
    \begin{subfigure}{0.23\textwidth}
        \centering
        \includegraphics[width=\textwidth]{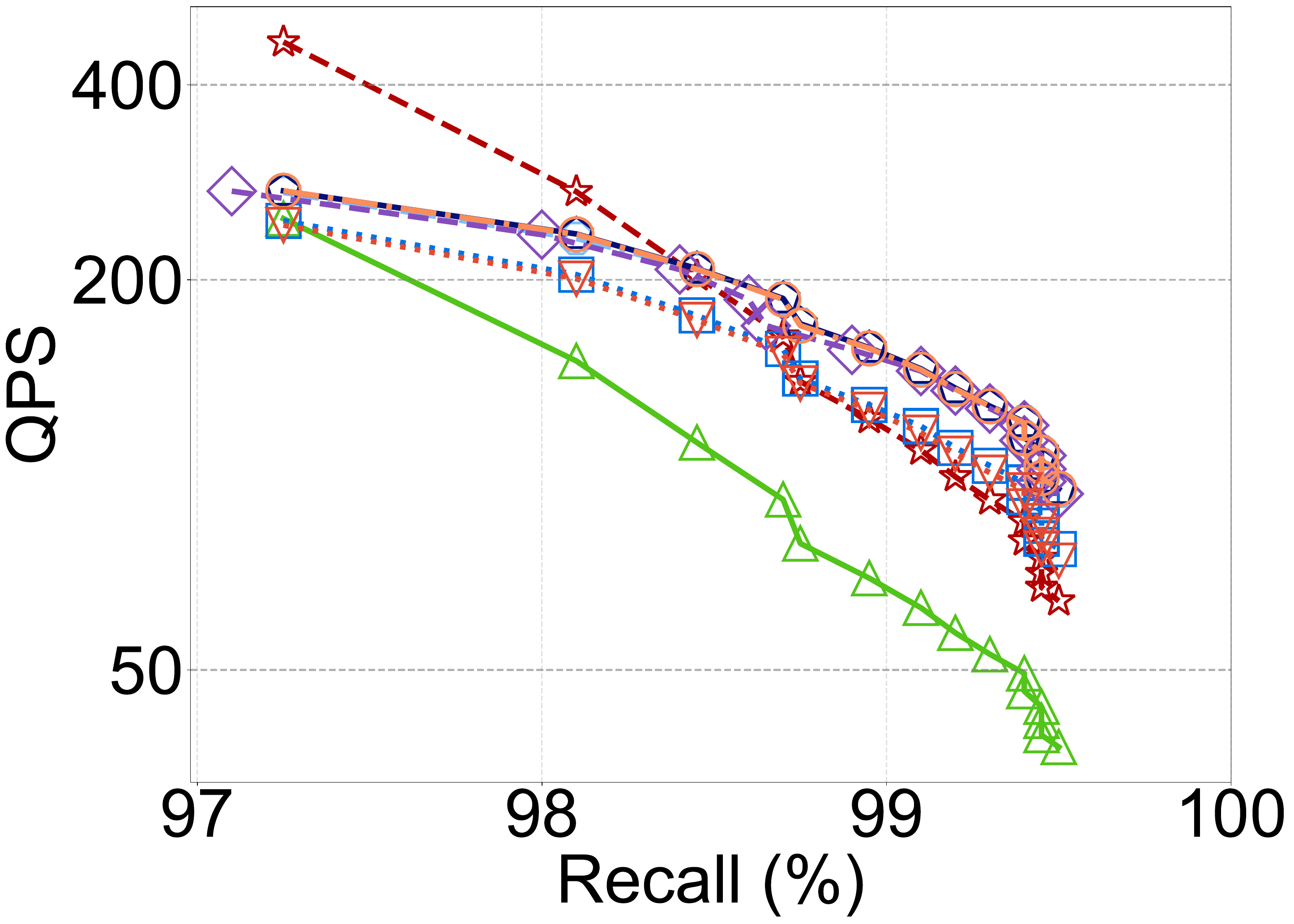}\vspace{-1.0ex}
        \caption{\Trevi ($k=20$)}\label{fig:HNSW-NO_SIMD-trevi-20}
    \end{subfigure}
    \begin{subfigure}{0.24\textwidth}
        \centering
        \includegraphics[width=\textwidth]{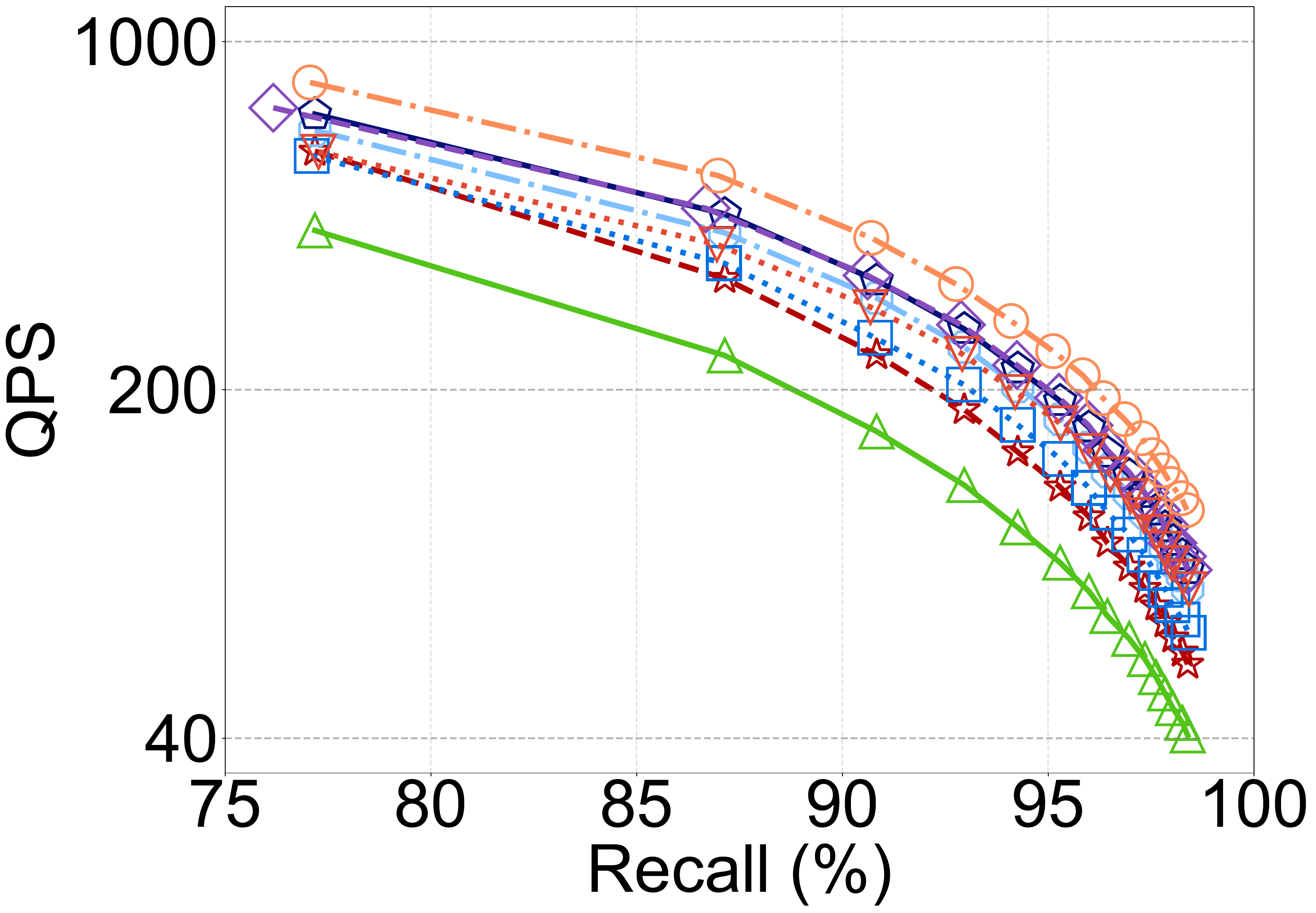}\vspace{-1.0ex}
        \caption{\Openai ($k=100$)}\label{fig:HNSW-NO_SIMD-trevi-100}
    \end{subfigure}
    \begin{subfigure}{0.23\textwidth}
        \centering
        \includegraphics[width=\textwidth]{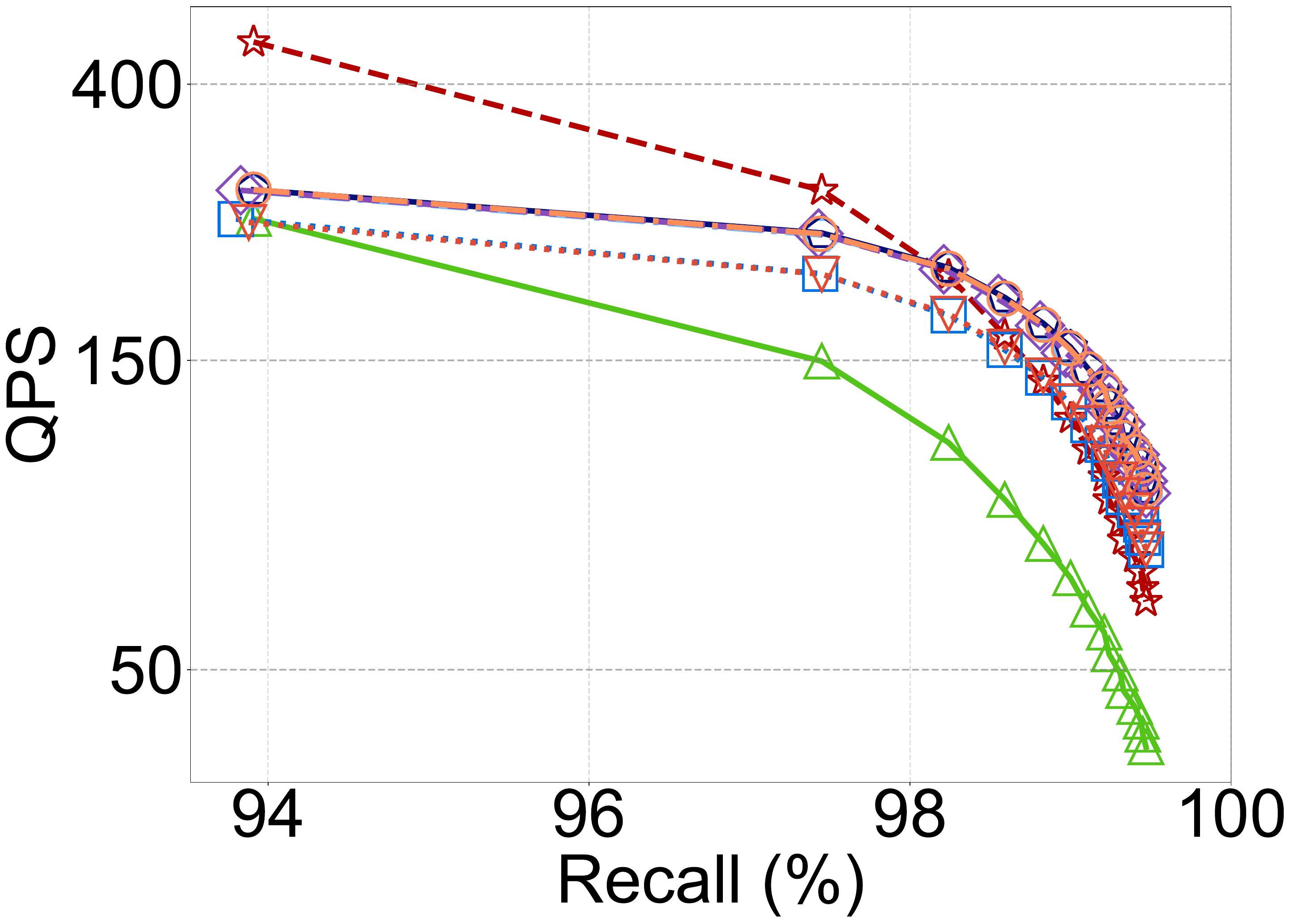}\vspace{-1.0ex}
        \caption{\Trevi ($k=100$)}
    \end{subfigure}
    \caption{Query performance when DCOs are applied to vector similarity search on CPUs with SIMD disabled}
    \label{fig:without-SIMD}
\end{figure*}

\section{Query Processing via HNSW with AVX2 SIMD Instruction Set}

\zheng{We evaluated the query performance of DCO methods using the AVX2 SIMD instruction set. As shown in \figref{fig:avx}, the overall performance trends are similar to those obtained with the SSE SIMD instruction set. For instance, on the low-dimensional \Glove dataset, most DCO methods fail to outperform \FDScanning. 
On the ultra-high-dimensional \Trevi dataset, both \FDScanning and \PDScanning maintain a performance lead until the recall rate exceeds 98\%. 

Since AVX2 is a more advanced instruction set than SSE, DCO methods (\eg \FDScanning) typically run faster with AVX2 than with SSE. 
As a result, the relative speedup provided by DCOs diminishes when AVX2 is employed. 
Notably, with AVX2, \FDScanning can surpass more recent SOTA methods. 
For example, \FDScanning outperforms \ADSampling on the \Gist dataset in \figref{fig:avx}.}

\begin{figure*}[t]
    \centering
    \includegraphics[width=0.9\textwidth]{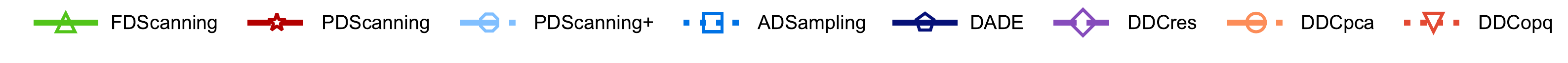}
    \begin{subfigure}{0.24\textwidth}
        \centering
        \includegraphics[width=\textwidth]{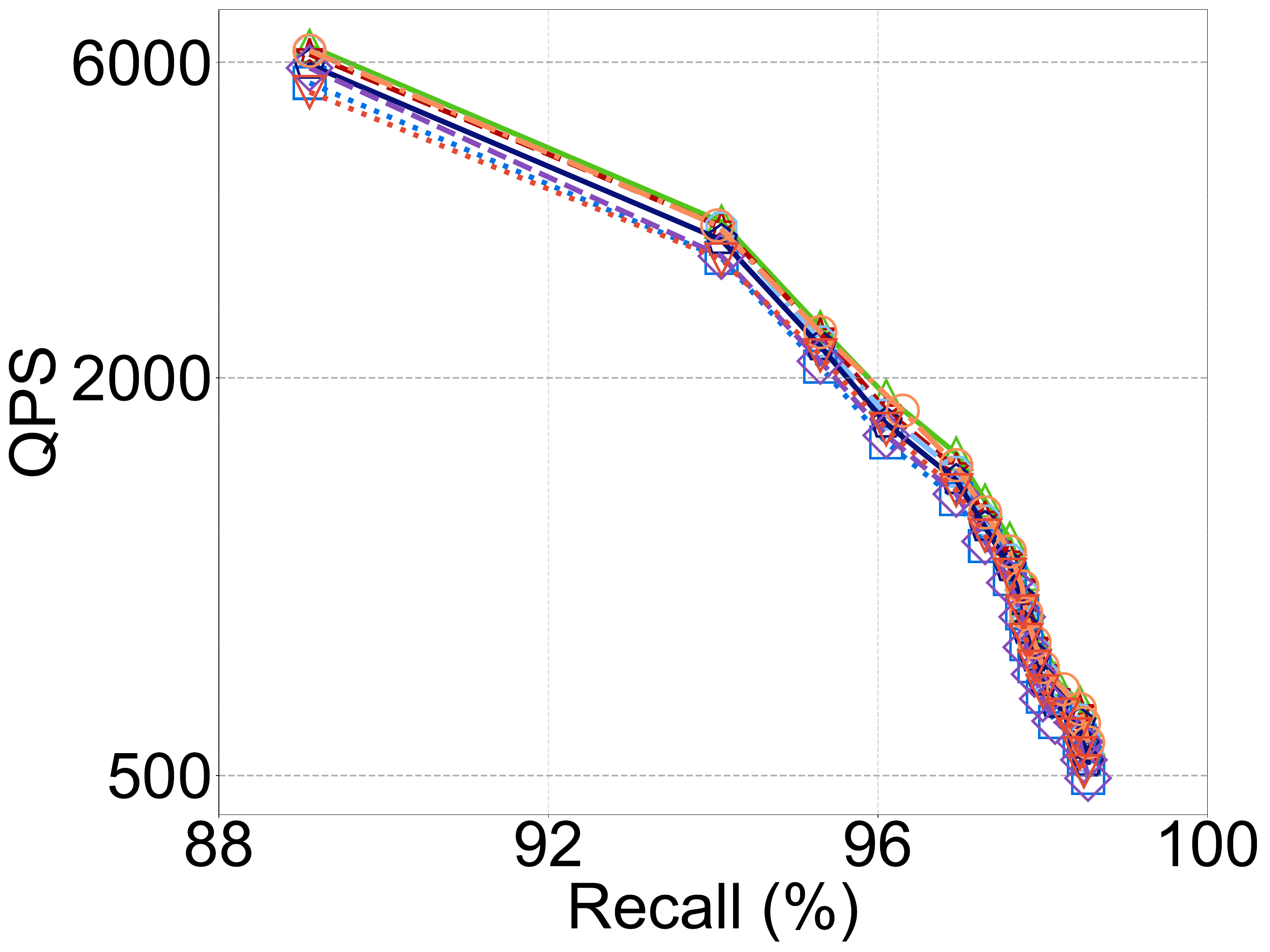}\vspace{-1.0ex}
        \caption{\Glove, $k=20$}\label{fig:cosine-glove}
    \end{subfigure}
    \begin{subfigure}{0.24\textwidth}
        \centering
        \includegraphics[width=\textwidth]{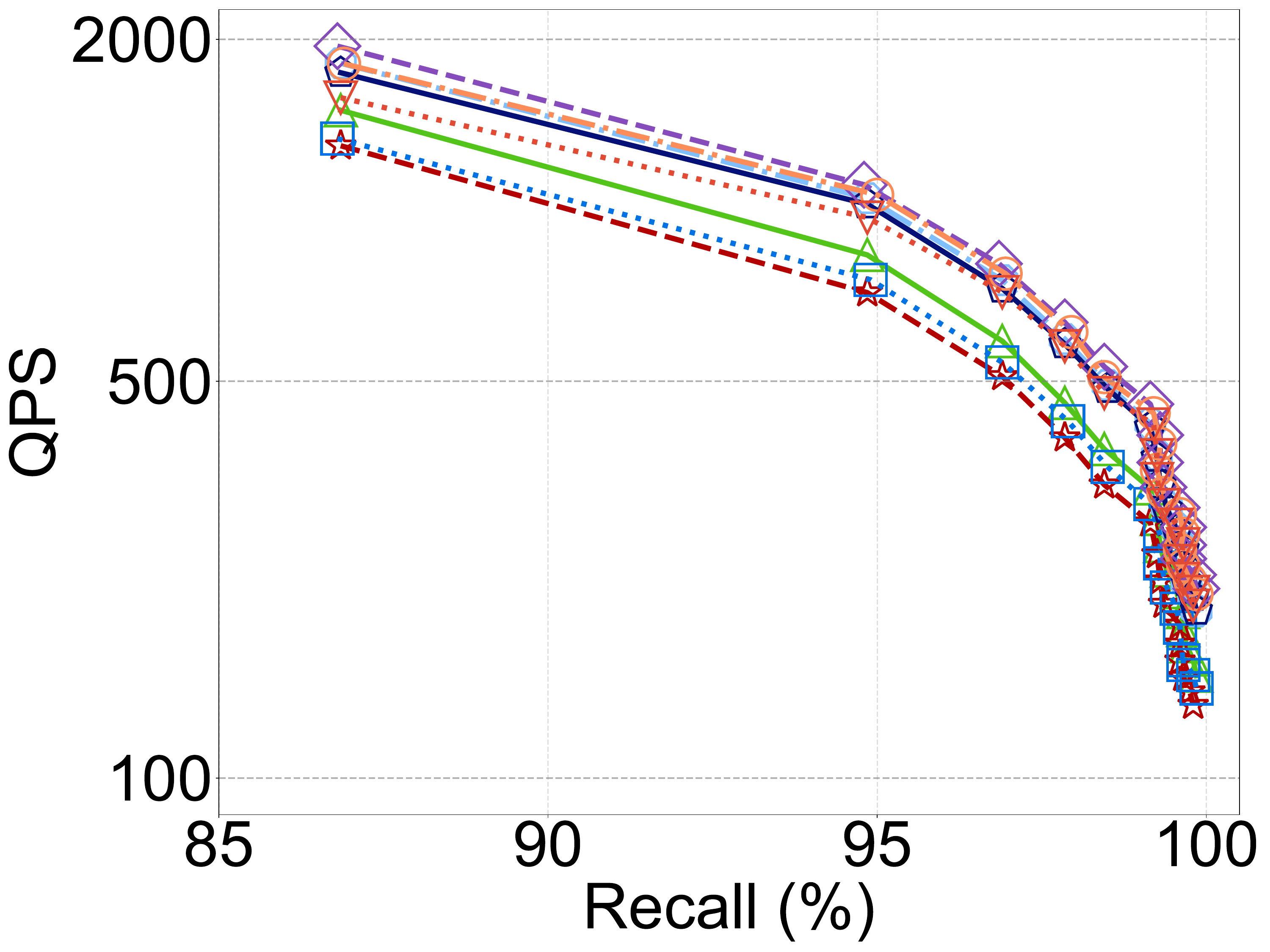}\vspace{-1.0ex}
        \caption{\Gist, $k=20$}\label{fig:cosine-gist}
    \end{subfigure}
    \begin{subfigure}{0.24\textwidth}
        \centering
        \includegraphics[width=\textwidth]{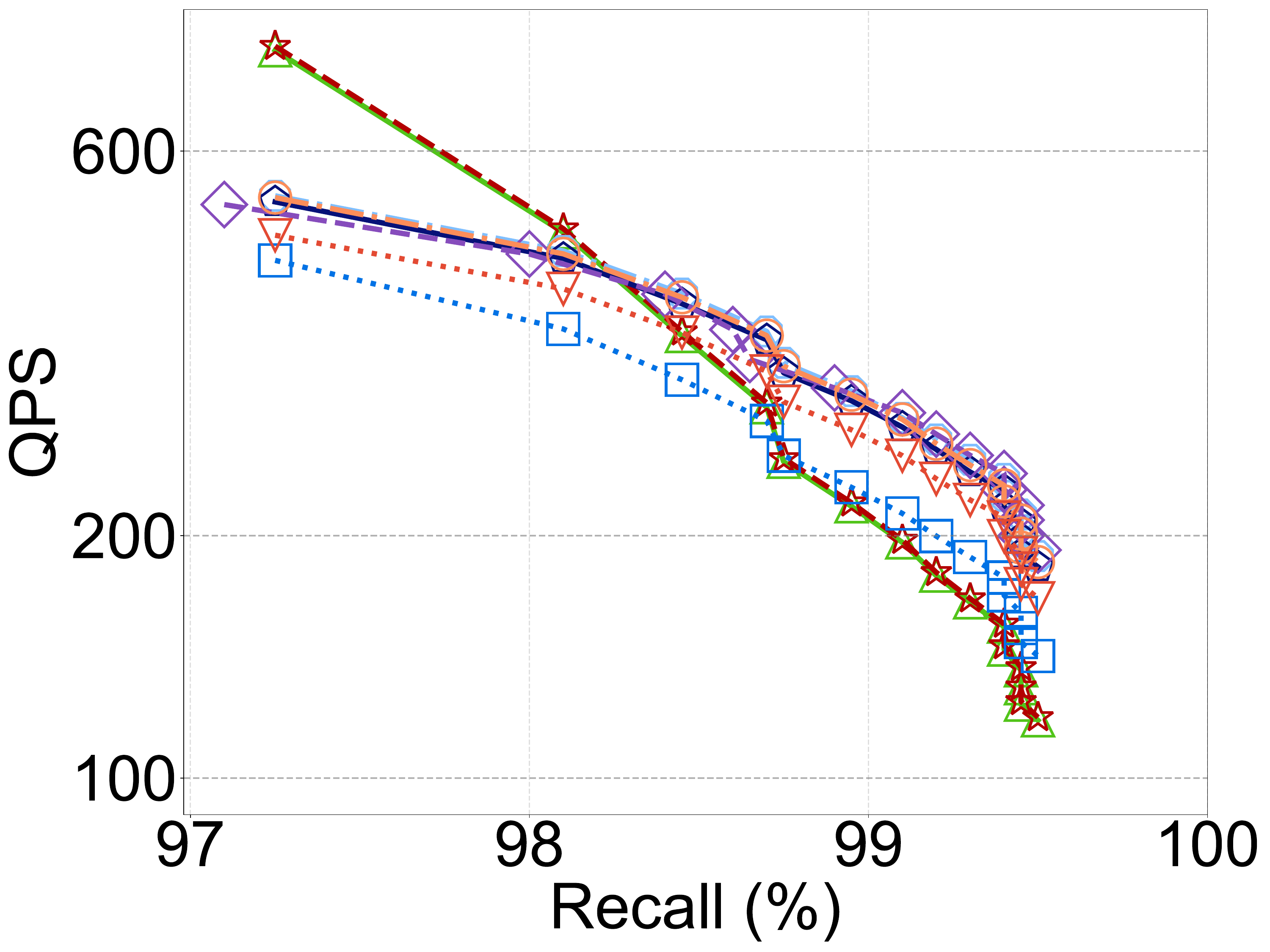}\vspace{-1.0ex}
        \caption{\Trevi, $k=20$}\label{fig:cosine-trevi}
    \end{subfigure}
    \\
    \begin{subfigure}{0.24\textwidth}
        \centering
        \includegraphics[width=\textwidth]{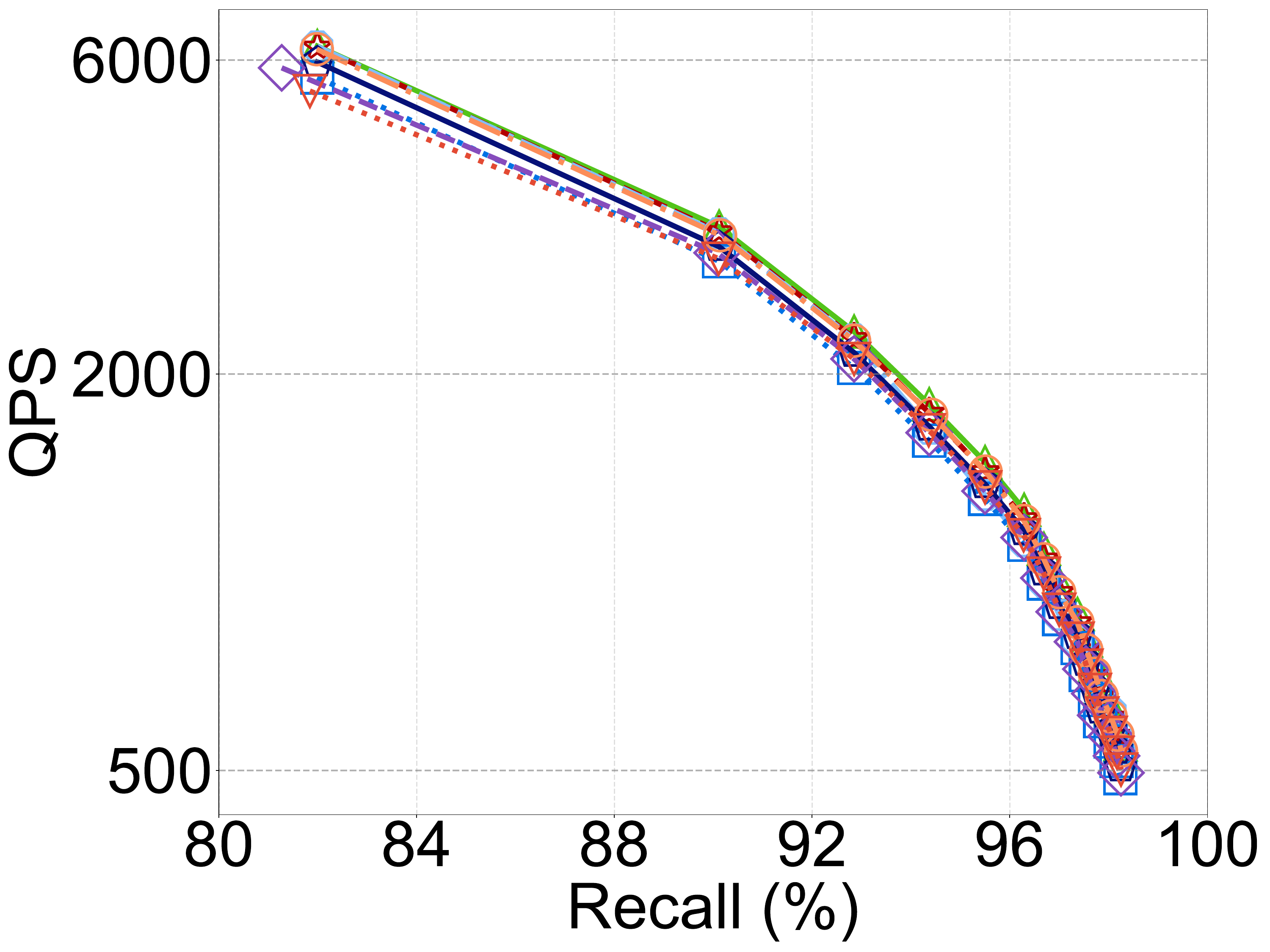}\vspace{-1.0ex}
        \caption{\Glove, $k=100$}\label{fig:cosine-glove}
    \end{subfigure}
    \begin{subfigure}{0.24\textwidth}
        \centering
        \includegraphics[width=\textwidth]{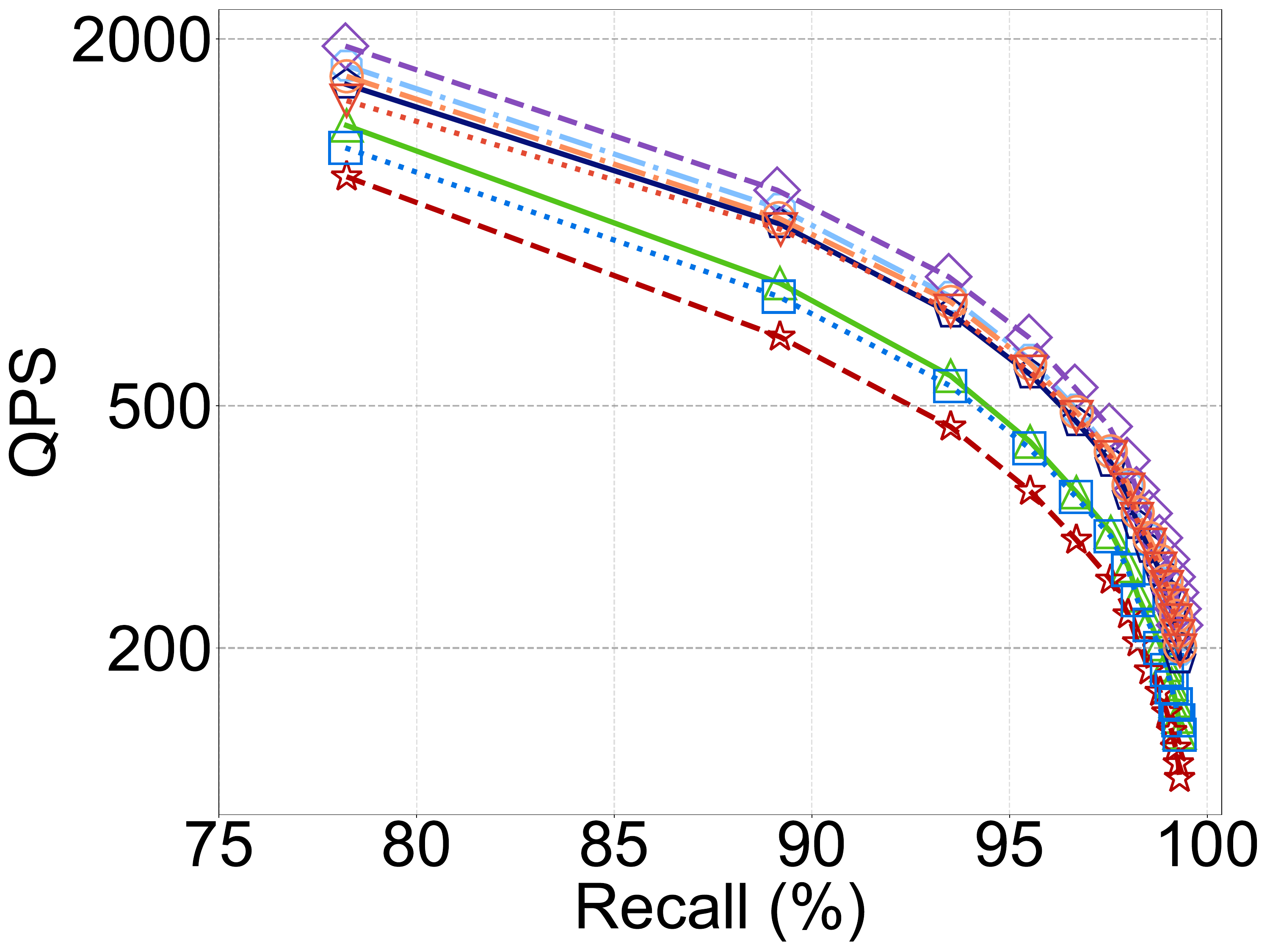}\vspace{-1.0ex}
        \caption{\Gist, $k=100$}\label{fig:cosine-gist}
    \end{subfigure}
    \begin{subfigure}{0.24\textwidth}
        \centering
        \includegraphics[width=\textwidth]{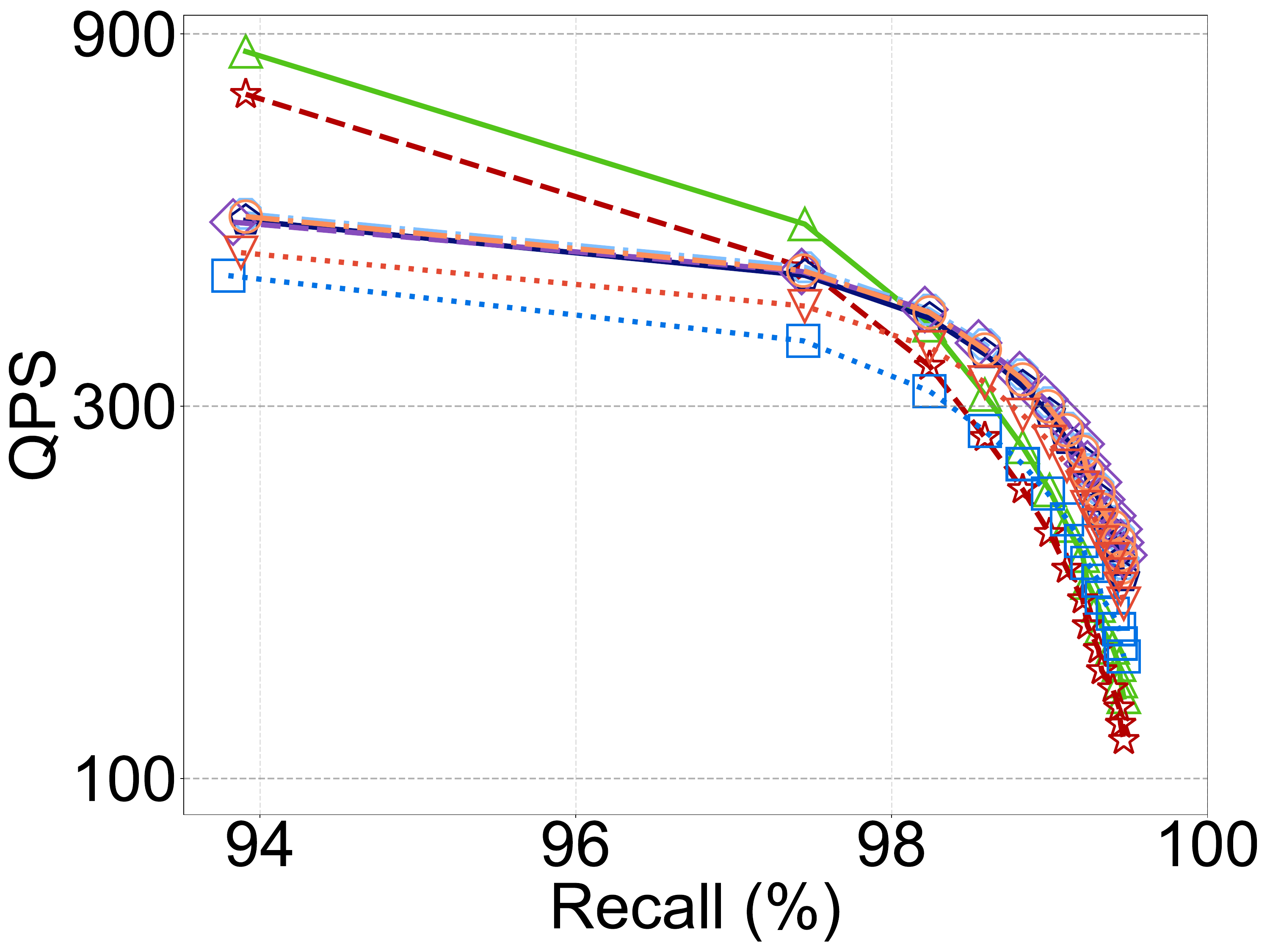}\vspace{-1.0ex}
        \caption{\Trevi, $k=100$}\label{fig:cosine-trevi}
    \end{subfigure}
    \caption{Performance comparison on AVX2}\label{fig:avx}
\end{figure*}

\section{Out-of-Distribution (OOD) Queries on CPUs with SIMD Disabled}\label{app:IVF-OOD}
To further evaluate the robustness of DCOs under out-of-distribution (OOD) queries, we also conduct this experiment with SIMD disabled.

As shown in \figref{fig:laion-id-no-simd} and \figref{fig:textimage-id-no-simd}, \DDCpca achieves the highest QPS on both datasets for in-distribution queries.  
However, it becomes the least efficient method under OOD queries, performing up to 1.6$\times$ and 1.8$\times$ slower than \FDScanning and \PDScanning, respectively.  In this setting, aside from \DDCopq, which achieves the highest QPS on the \Laion dataset, the loop-based baselines \PDScanning and \RPDScanning achieve the highest QPS on both datasets for recall over 97\%. This aligns with our earlier finding: the SOTA methods often fail to maintain efficiency gains on OOD queries.

\begin{figure*}[t]
    \centering
    \includegraphics[width=0.9\textwidth]{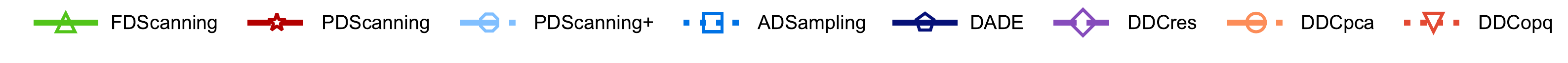}
    \begin{subfigure}{0.24\textwidth}
        \centering
        \includegraphics[width=\textwidth]{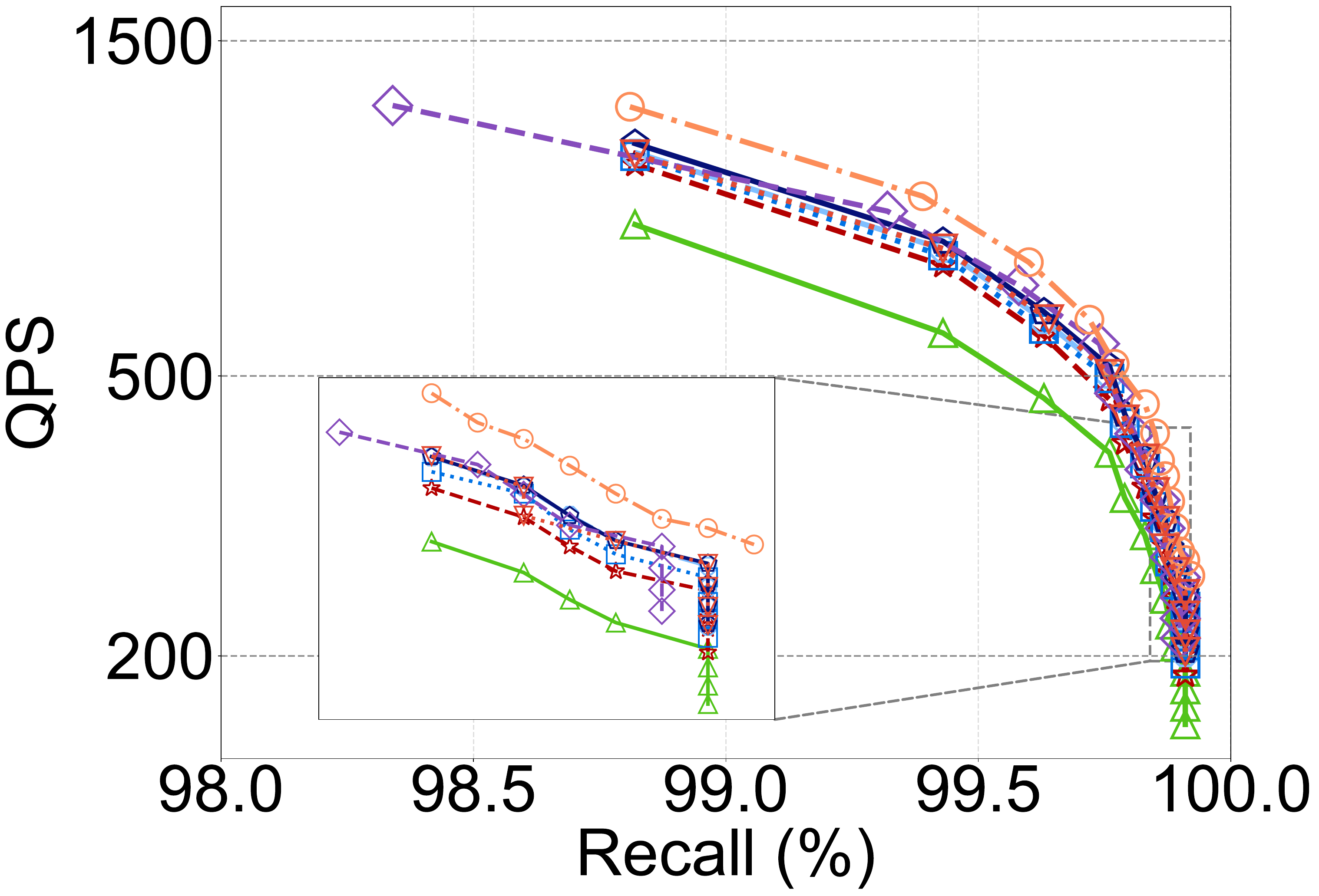}\vspace{-1.0ex}
        \caption{\Laion (in-distribution)}\label{fig:laion-id-no-simd}
    \end{subfigure}
    \begin{subfigure}{0.24\textwidth}
        \centering
        \includegraphics[width=\textwidth]{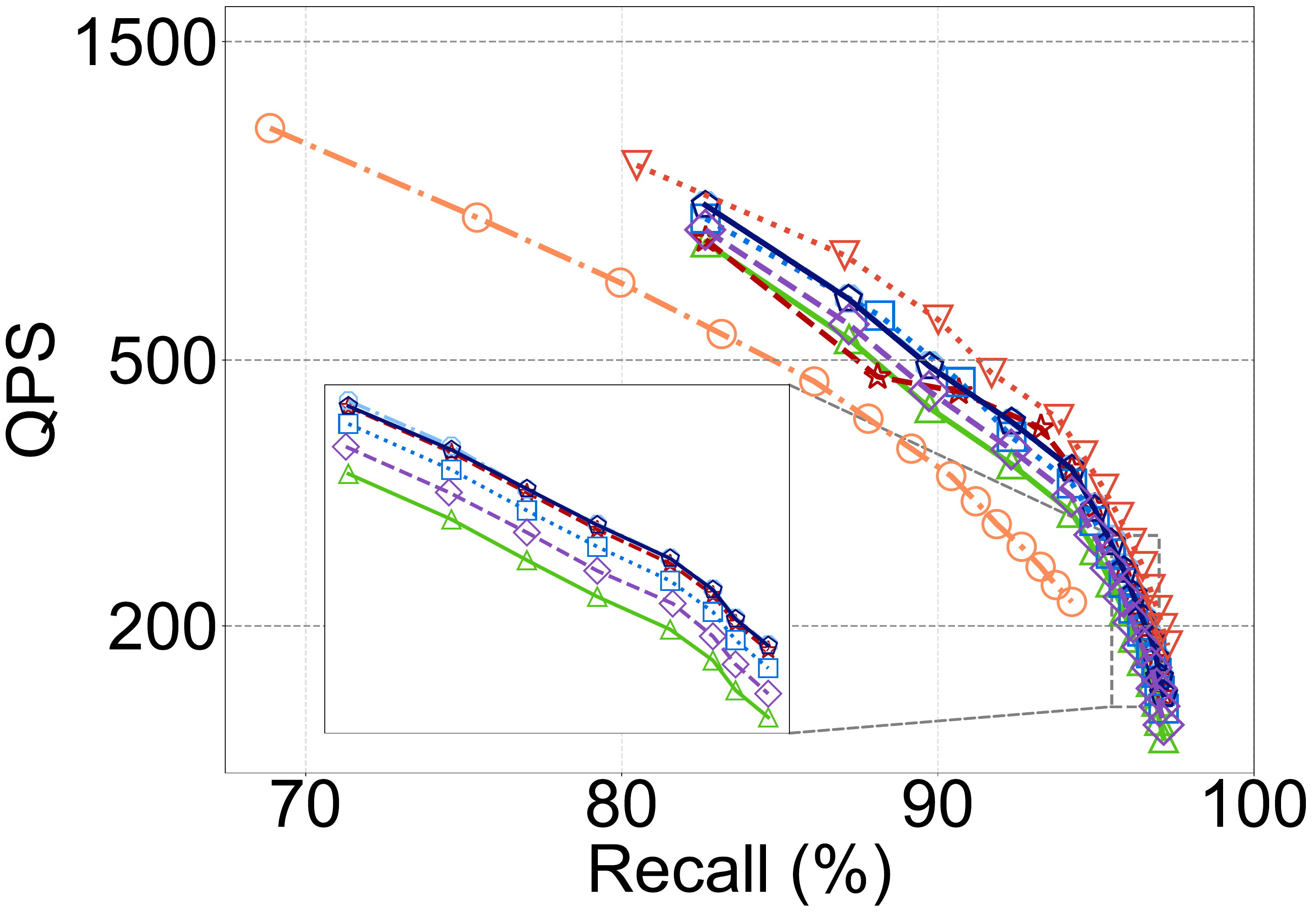}\vspace{-1.0ex}
        \caption{\Laion (OOD)}\label{fig:laion-ood-no-simd}
    \end{subfigure}
    \begin{subfigure}{0.24\textwidth}
        \centering
        \includegraphics[width=\textwidth]{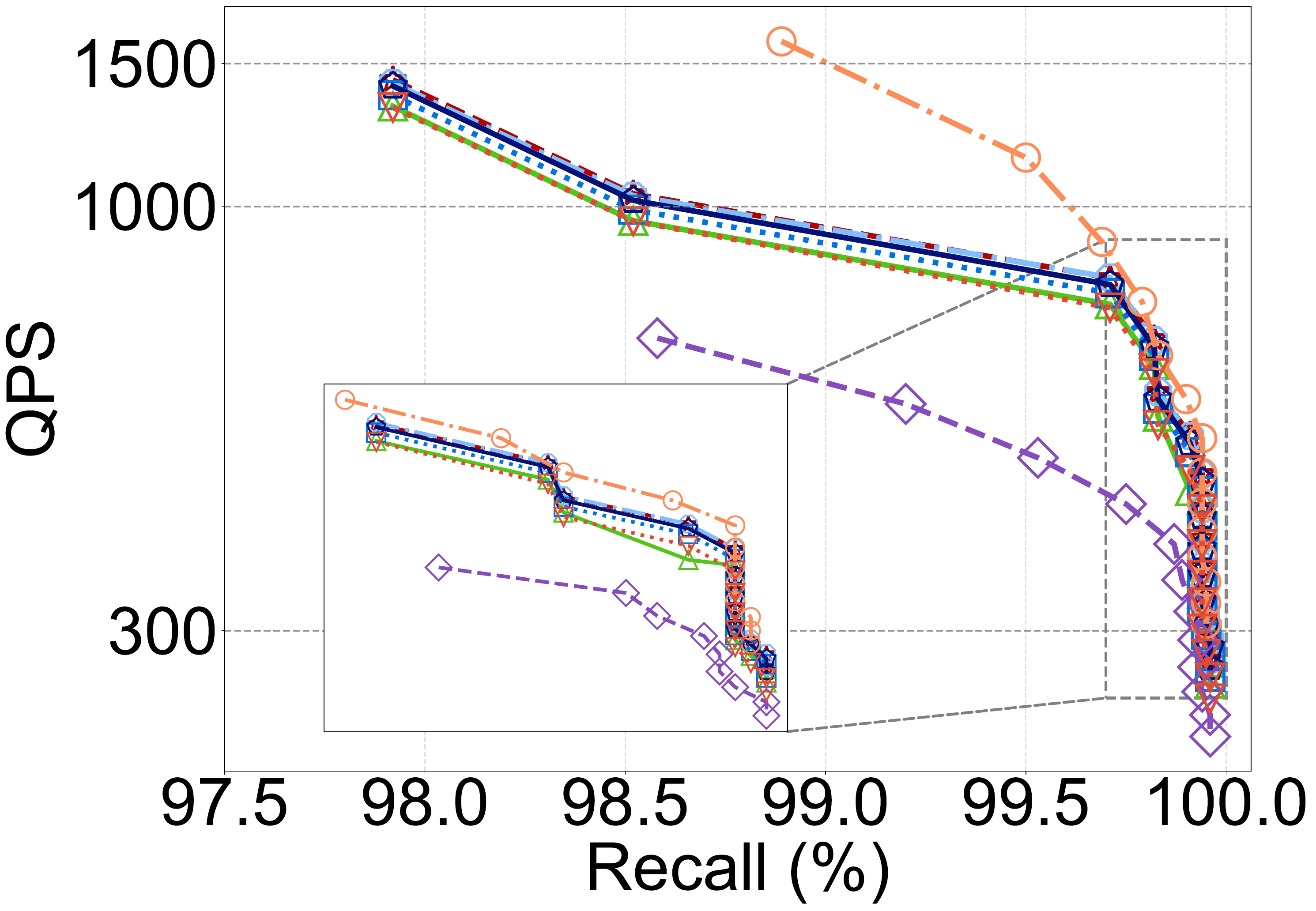}\vspace{-1.0ex}
        \caption{\TextImage (in-distribution)}\label{fig:textimage-id-no-simd}
    \end{subfigure}
    \begin{subfigure}{0.24\textwidth}
        \centering
        \includegraphics[width=\textwidth]{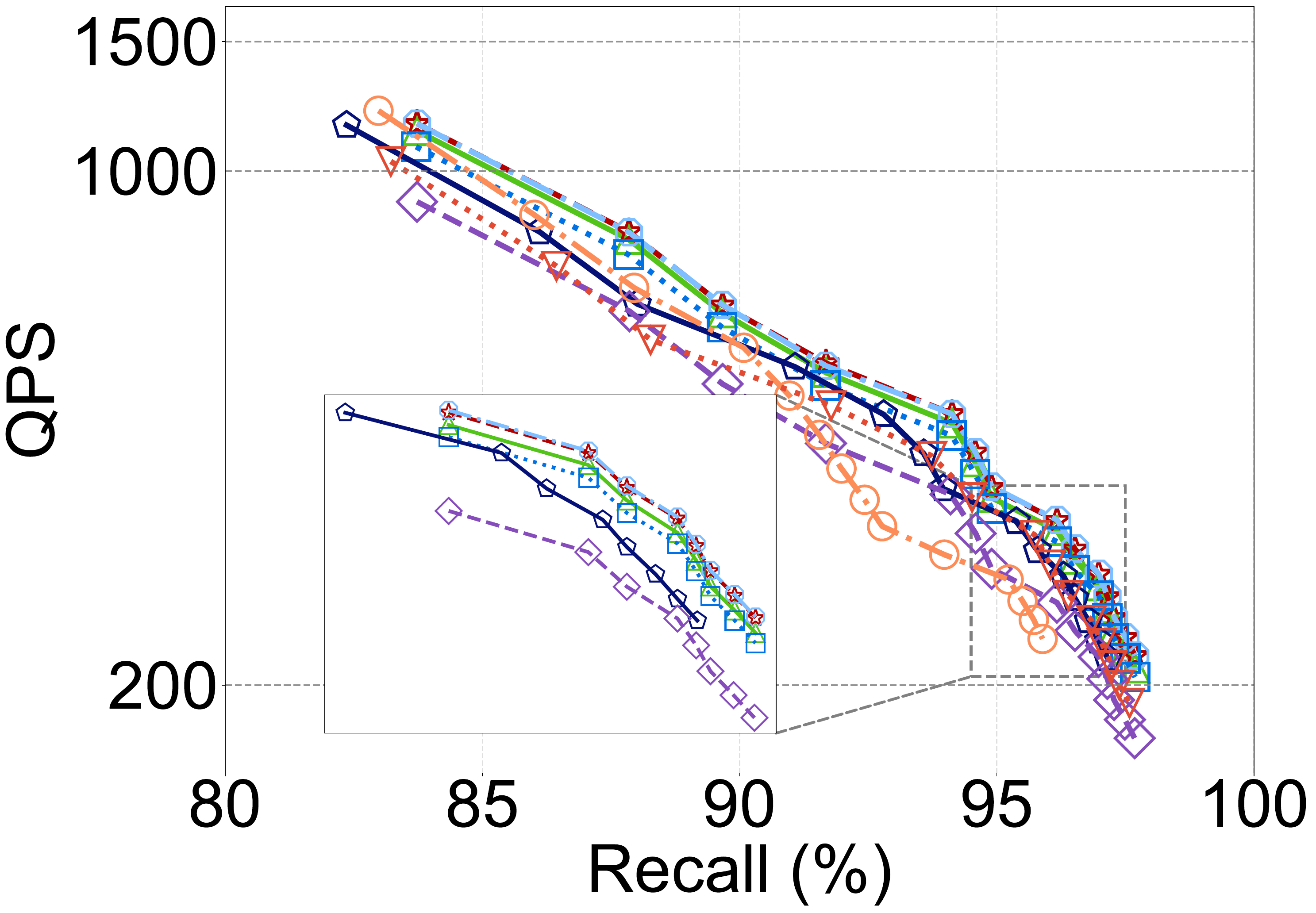}\vspace{-1.0ex}
        \caption{\TextImage (OOD)}\label{fig:textimage-ood-no-simd}
    \end{subfigure}
    \caption{Results on in-distribution and OOD queries without SIMD }\label{fig:out-of-distribution-no-simd}
\end{figure*}

\section{Experiment with Cosine Similarity}
\zheng{This experiment evaluates the extension of DCO methods to cosine similarity. To conduct this experiment, we select three datasets of varying dimensionality: \Glove, \Gist, and \Trevi.}

\begin{figure*}[t]
    \centering
    \includegraphics[width=0.9\textwidth]{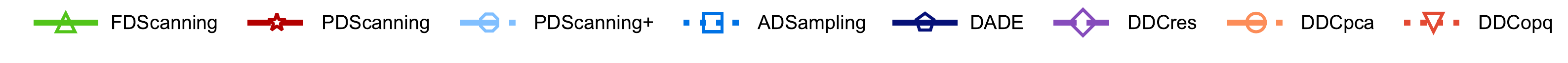}
    \begin{subfigure}{0.24\textwidth}
        \centering
        \includegraphics[width=\textwidth]{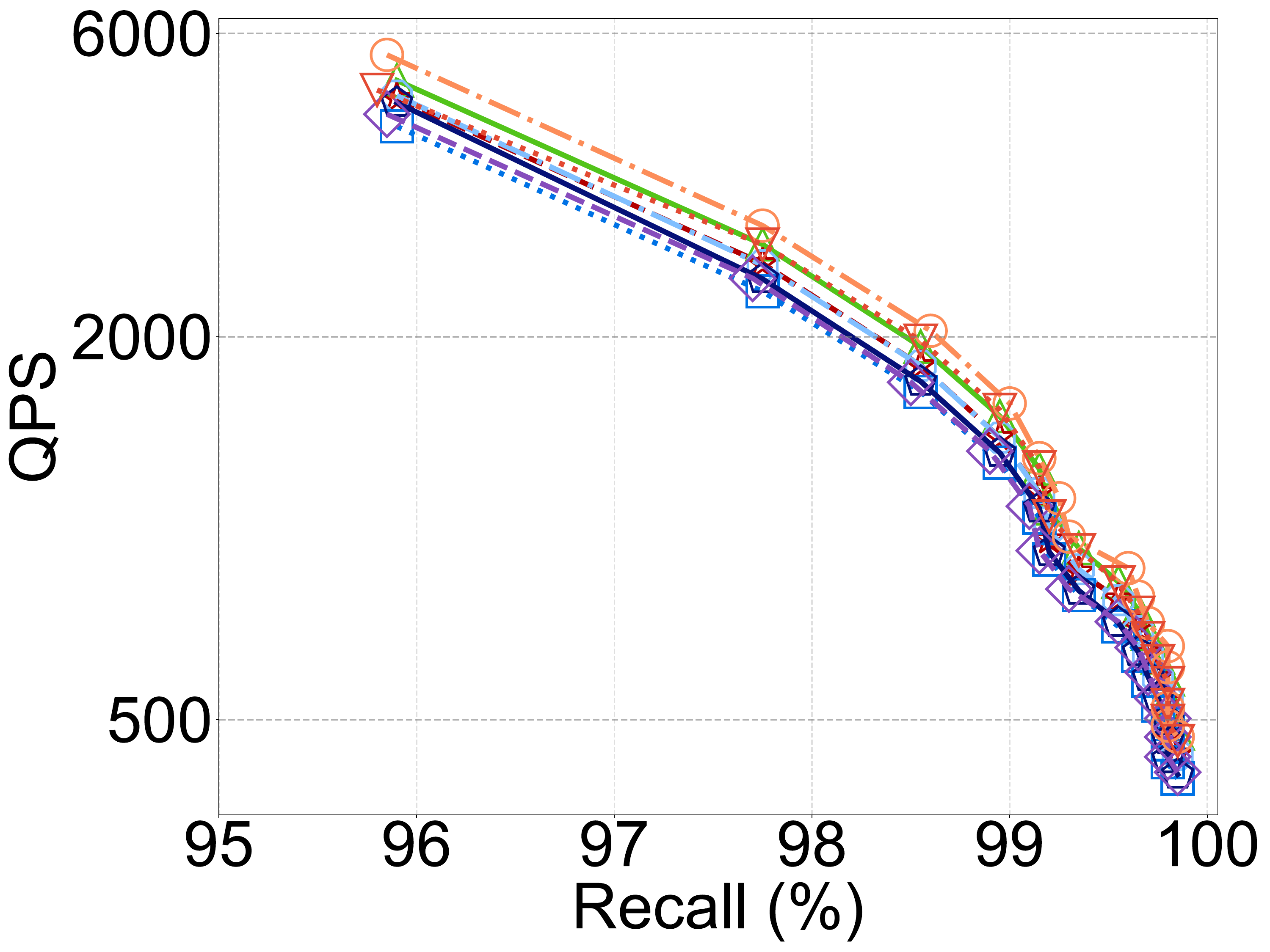}\vspace{-1.0ex}
        \caption{\Glove, $k=20$}\label{fig:cosine-glove}
    \end{subfigure}
    \begin{subfigure}{0.24\textwidth}
        \centering
        \includegraphics[width=\textwidth]{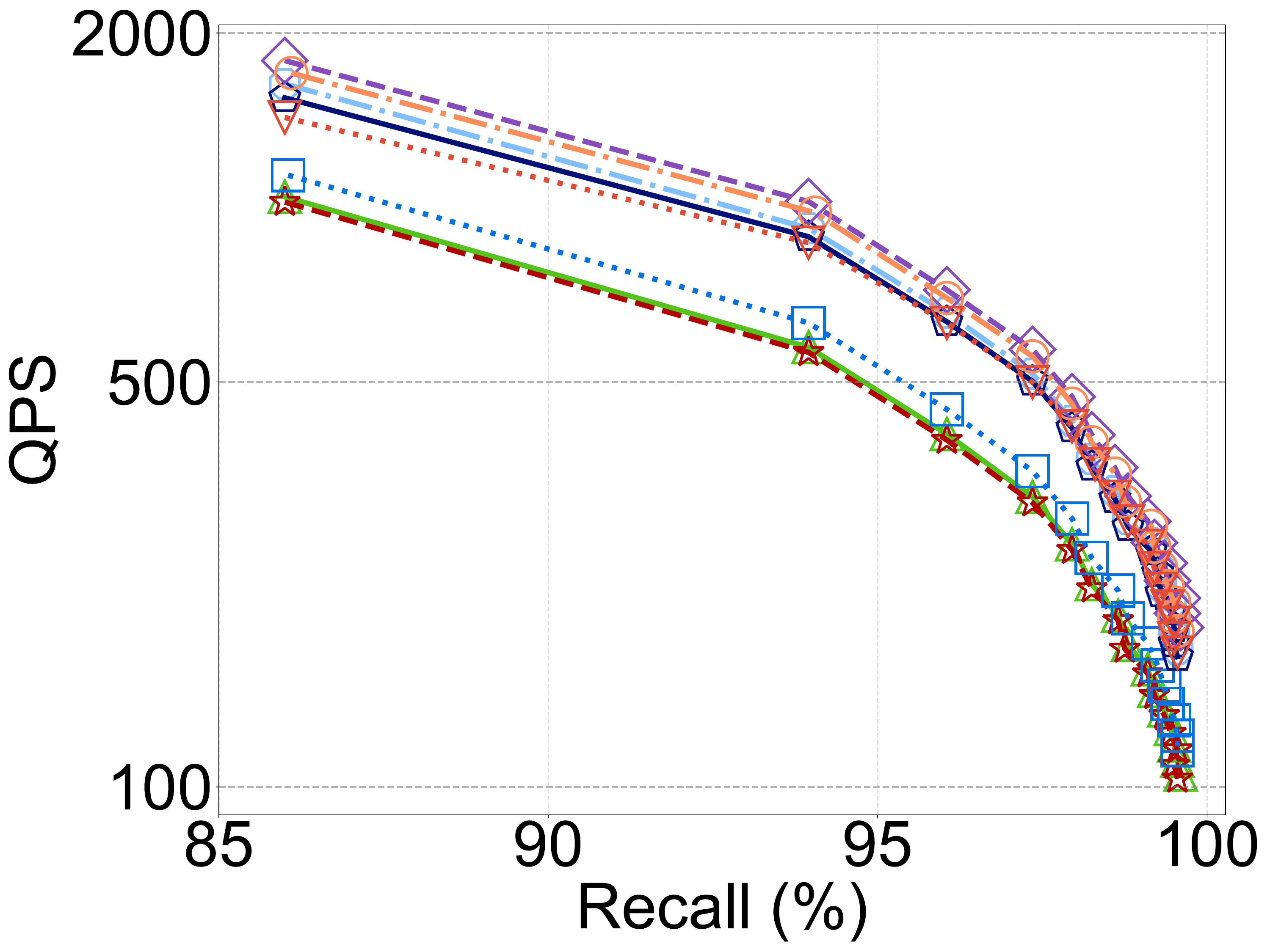}\vspace{-1.0ex}
        \caption{\Gist, $k=20$}\label{fig:cosine-gist}
    \end{subfigure}
    \begin{subfigure}{0.24\textwidth}
        \centering
        \includegraphics[width=\textwidth]{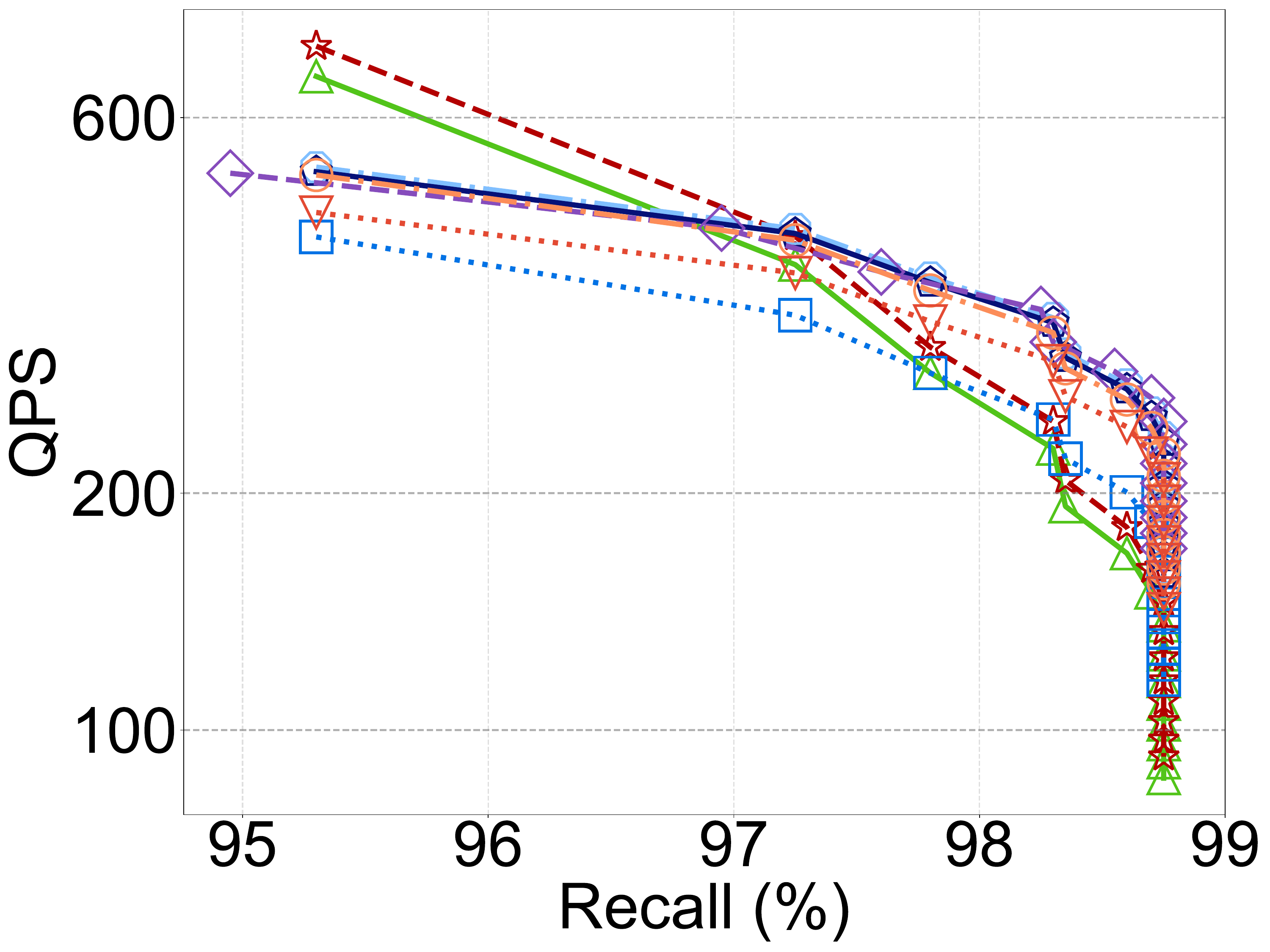}\vspace{-1.0ex}
        \caption{\Trevi, $k=20$}\label{fig:cosine-trevi}
    \end{subfigure}
    \\
    \begin{subfigure}{0.24\textwidth}
        \centering
        \includegraphics[width=\textwidth]{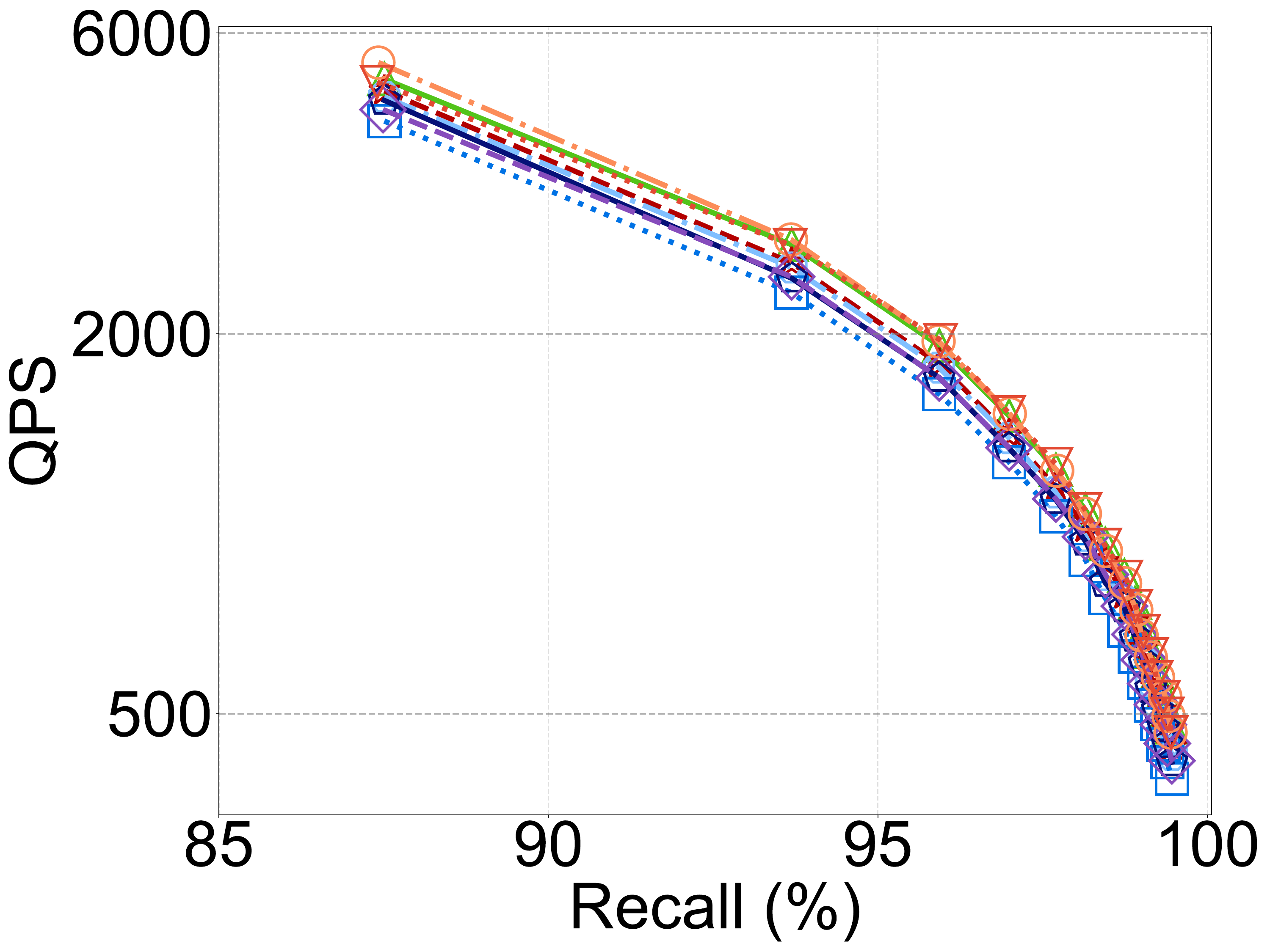}\vspace{-1.0ex}
        \caption{\Glove, $k=100$}\label{fig:cosine-glove}
    \end{subfigure}
    \begin{subfigure}{0.24\textwidth}
        \centering
        \includegraphics[width=\textwidth]{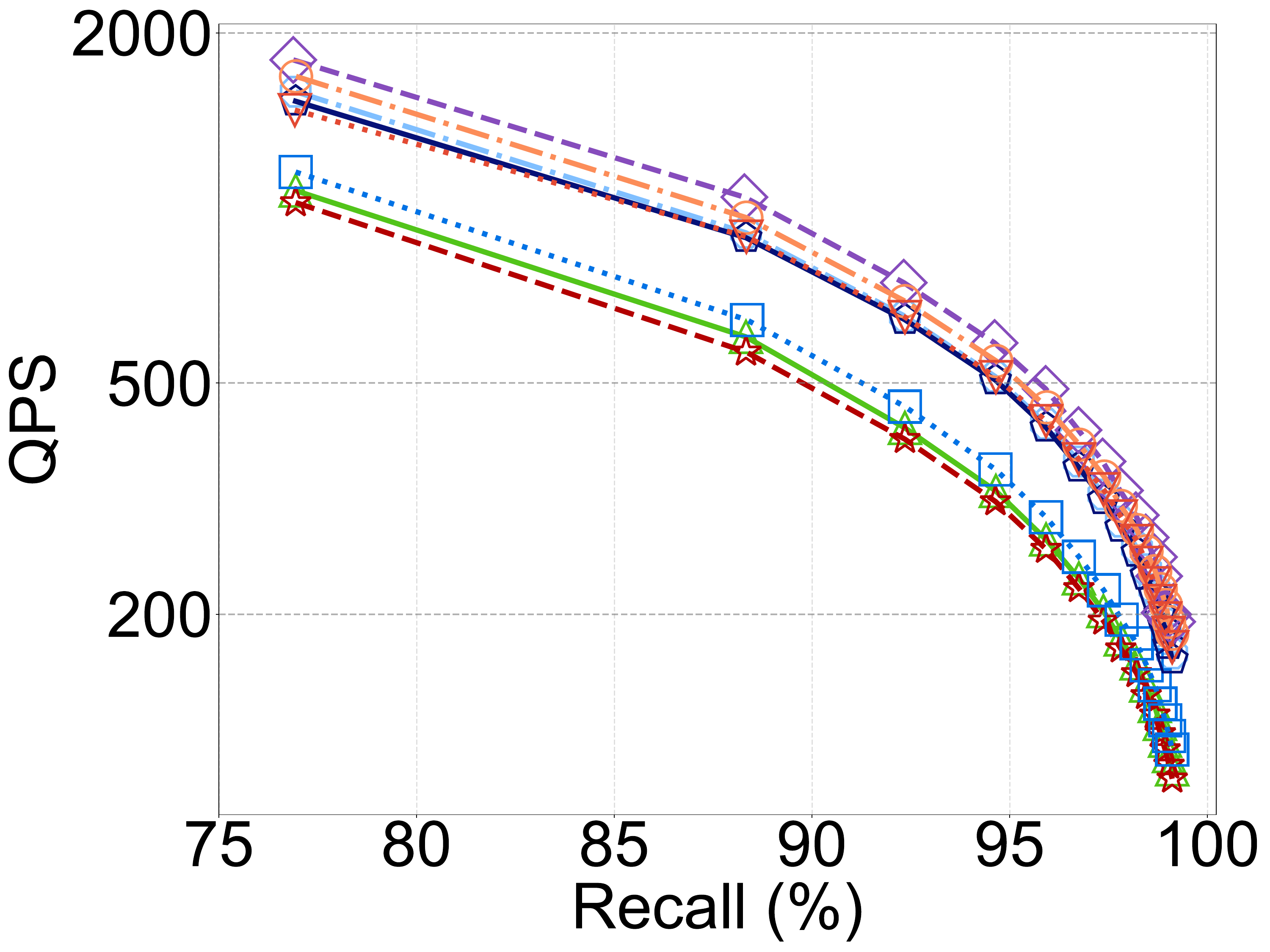}\vspace{-1.0ex}
        \caption{\Gist, $k=100$}\label{fig:cosine-gist}
    \end{subfigure}
    \begin{subfigure}{0.24\textwidth}
        \centering
        \includegraphics[width=\textwidth]{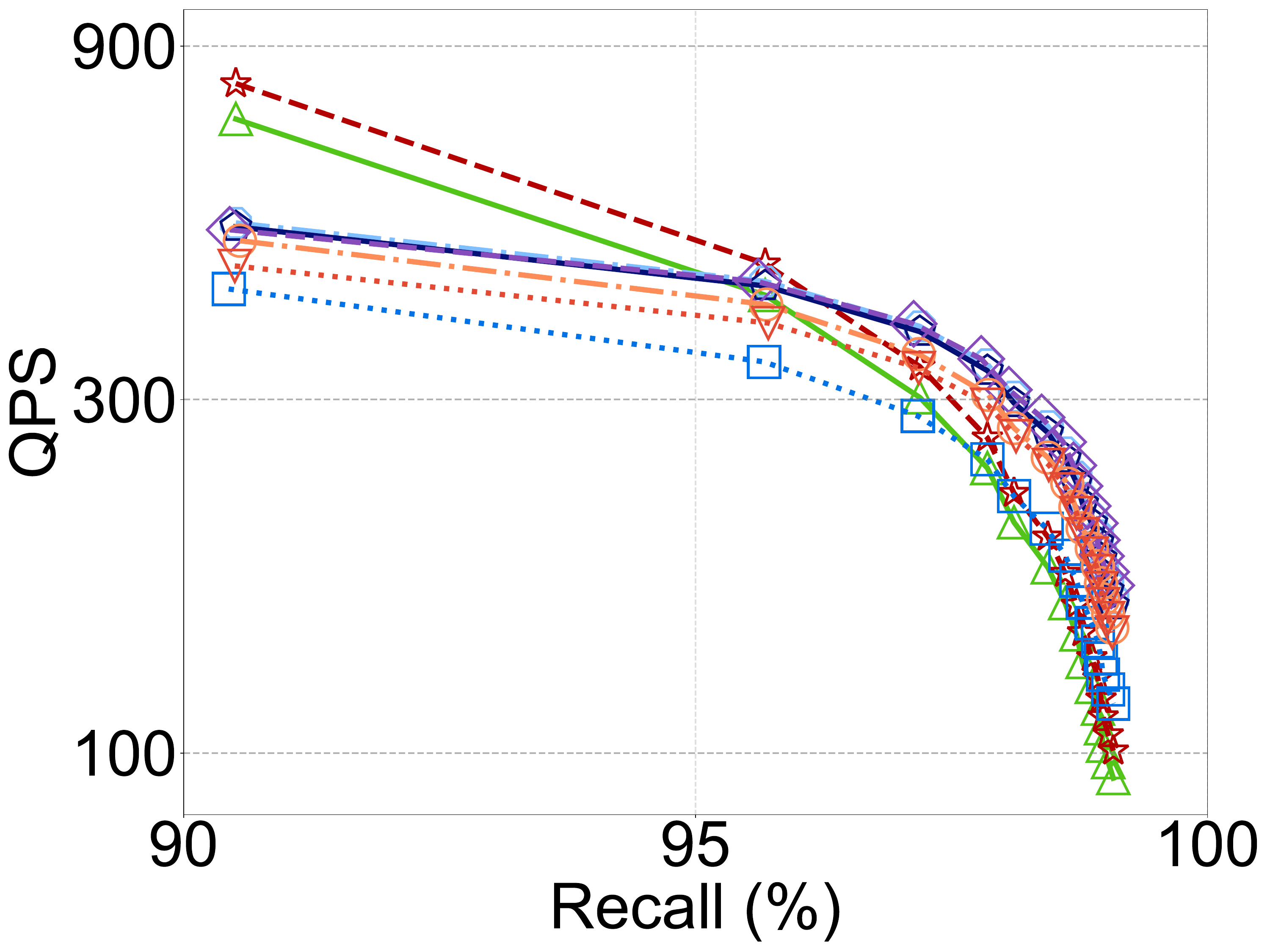}\vspace{-1.0ex}
        \caption{\Trevi, $k=100$}\label{fig:cosine-trevi}
    \end{subfigure}
    \caption{Search performance using cosine similarity }\label{fig:cosine-similarity}
\end{figure*}

\zheng{As illustrated in \figref{fig:cosine-similarity}, the overall performance trend for cosine similarity is consistent with that observed for Euclidean distance and inner product.
On the low-dimensional datasets (\eg \Glove), the efficiency advantage of the SOTA DCO methods is less pronounced, and they are even outperformed by \FDScanning in some cases.
For the ultra-high-dimensional dataset (\eg \Trevi), the performance bottleneck caused by online pre-processing time persists consistently. This performance pattern stems primarily from two key factors: (1) most DCO extensions for cosine similarity are directly derived from their Euclidean distance equivalents, and (2) cosine similarity is mathematically equivalent to the inner product for normalized vectors.}

\section{Experiments of Diverse Hardware Configurations on \Msmacro Dataset}\label{app:IVF-hardware-extend}

We further evaluate DCO methods across different hardware platforms on the \Msmacro dataset, which has the highest dimensionality among all evaluated datasets.

\fakeparagraph{Performance on CPU with and without SIMD Support}
As shown in \figref{fig:msmacro-without-SIMD} and \figref{fig:msmacro-with-SIMD}, 
\PDScanning outperforms the SOTA methods due to the larger online pre-processing time before the recall reaches 89\%. As the dimensionality further increases, \DDCres attains the highest QPS, regardless of whether SIMD is enabled. Specifically, it improves QPS by up to $1.7\times$ and $2.9\times$ over \FDScanning on CPUs with and without SIMD, respectively.

\fakeparagraph{Performance on GPU} As shown in \figref{fig:msmacro-gpu}, \DDCres consistently achieves the highest QPS. Notably, the bottleneck caused by online pre-processing time, observed on CPU, does not manifest on GPU.  
This can be attributed to the GPU’s superior efficiency in matrix multiplication and parallel computation. The SOTA methods demonstrate strong performance on the ultra-high-dimensional \Msmacro dataset, improving QPS by up to 1.4--5.6$\times$ over \FDScanning.

\begin{figure*}[t]
    \centering
    \begin{subfigure}{0.24\textwidth}
        \centering
        \includegraphics[width=\textwidth]{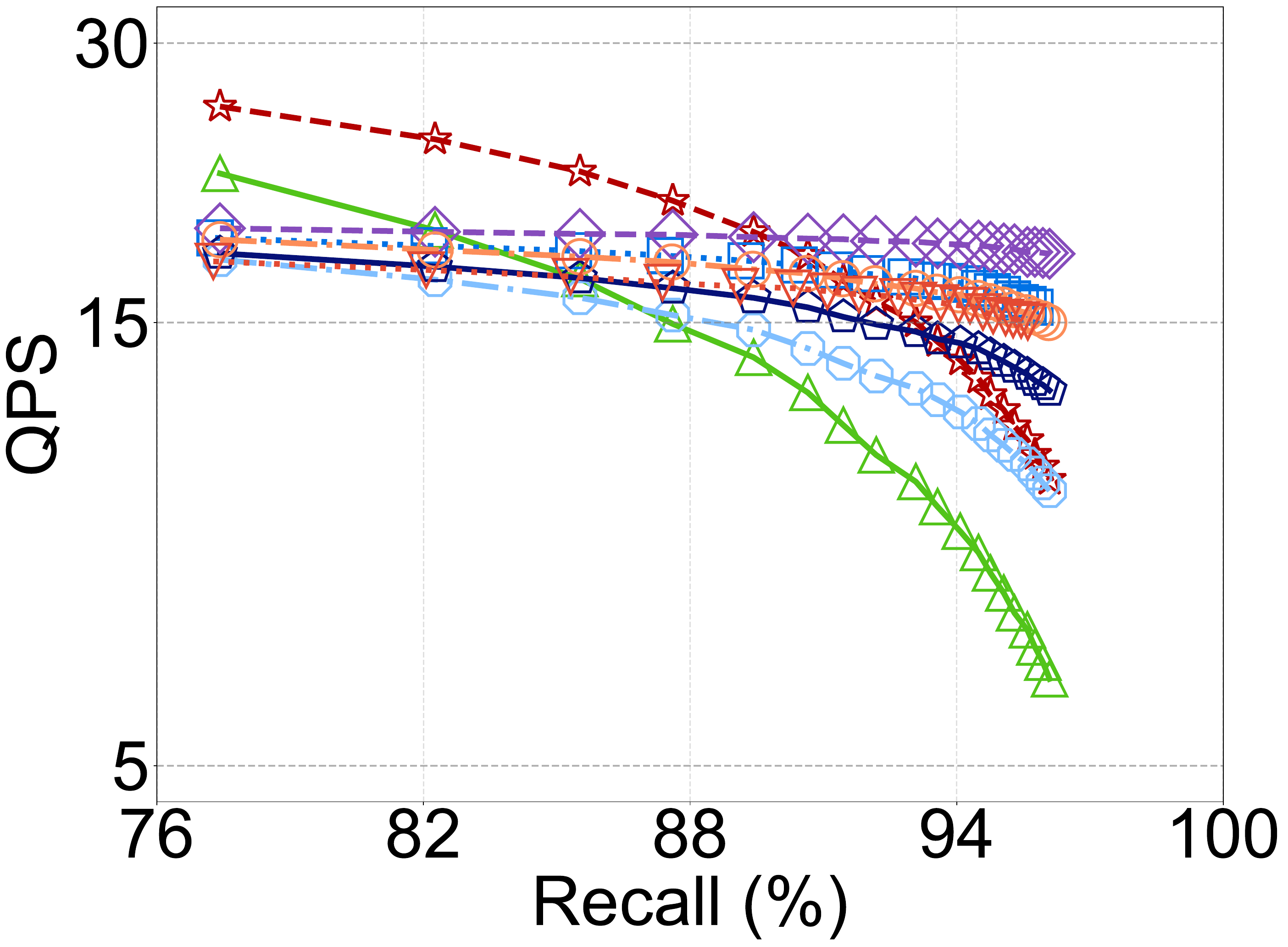}
        \caption{\Msmacro (without SIMD)}\label{fig:msmacro-without-SIMD}
    \end{subfigure}
    \begin{subfigure}{0.25\textwidth}
        \centering
        \includegraphics[width=\textwidth]{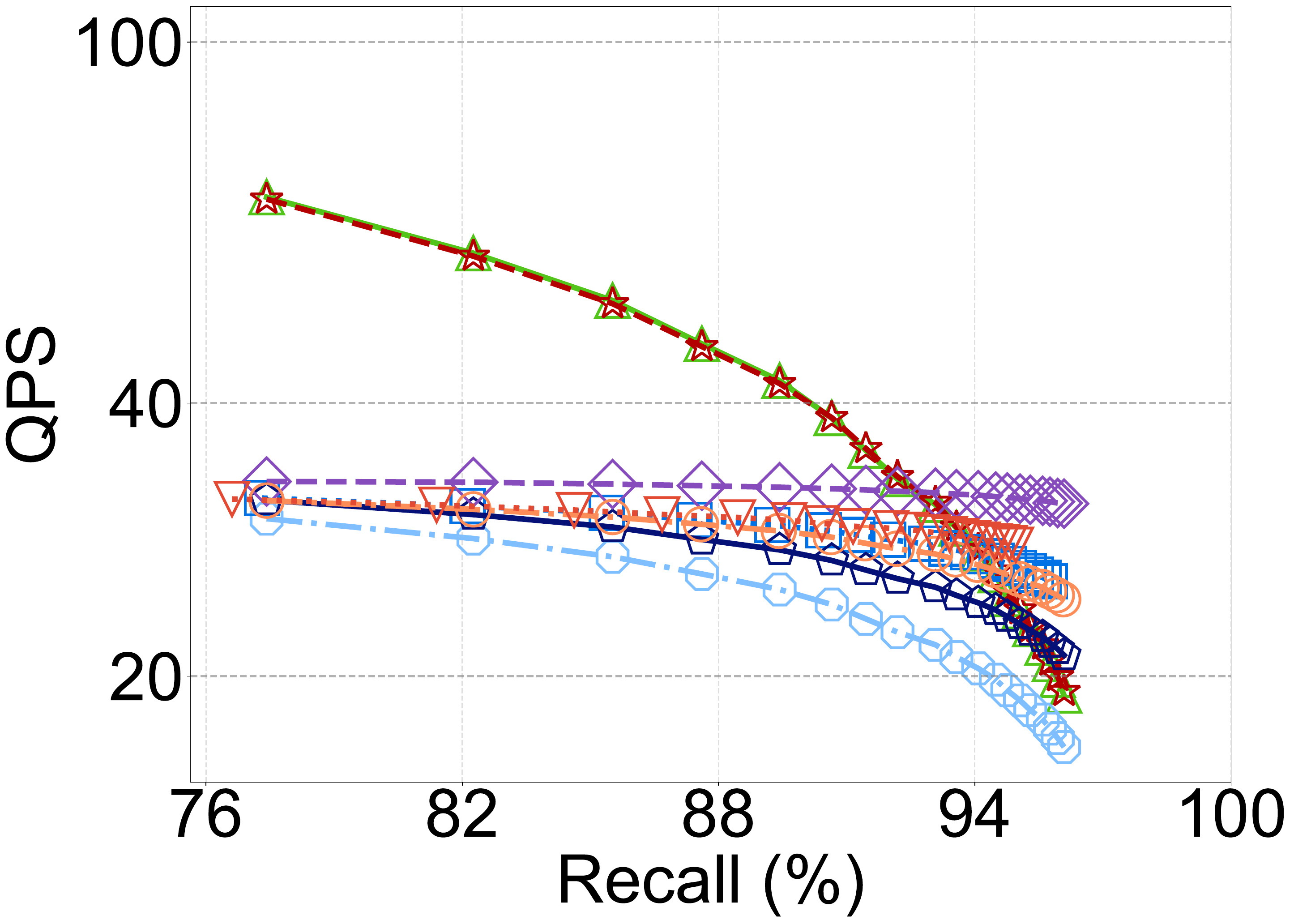}
        \caption{\Msmacro (with SIMD)}\label{fig:msmacro-with-SIMD}
    \end{subfigure}
    \begin{subfigure}{0.24\textwidth}
        \centering
        \includegraphics[width=\textwidth]{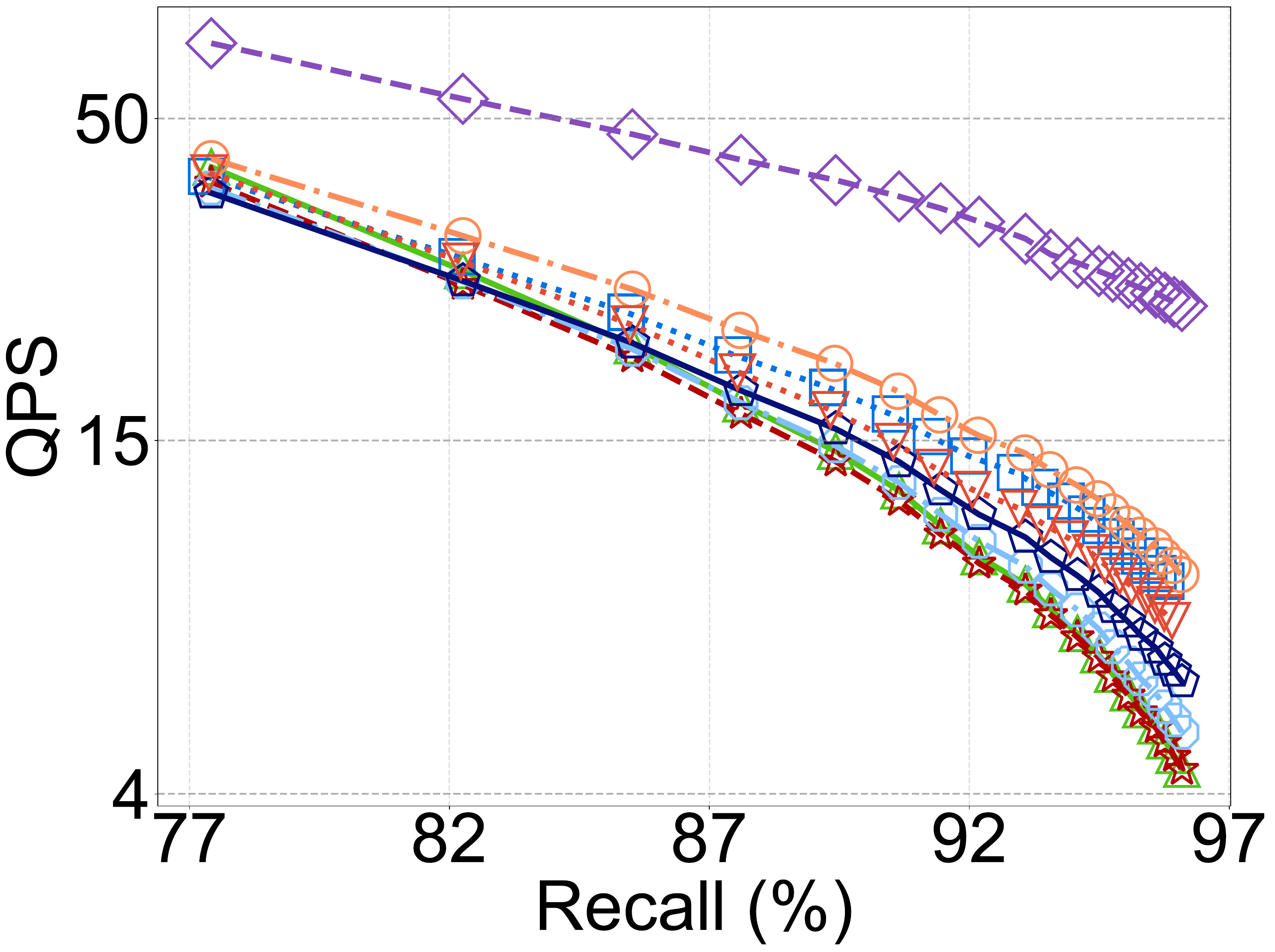}
        \caption{\Msmacro (with GPU)}\label{fig:msmacro-gpu}
    \end{subfigure}
    \caption{Comparisons of DCO methods using the IVF index on \Msmacro dataset across different hardware environments}\label{fig:extend-hardware}
\end{figure*}

\begin{figure*}[t]
    \centering

    \begin{subfigure}{0.24\textwidth}
        \centering
        \includegraphics[width=\textwidth]{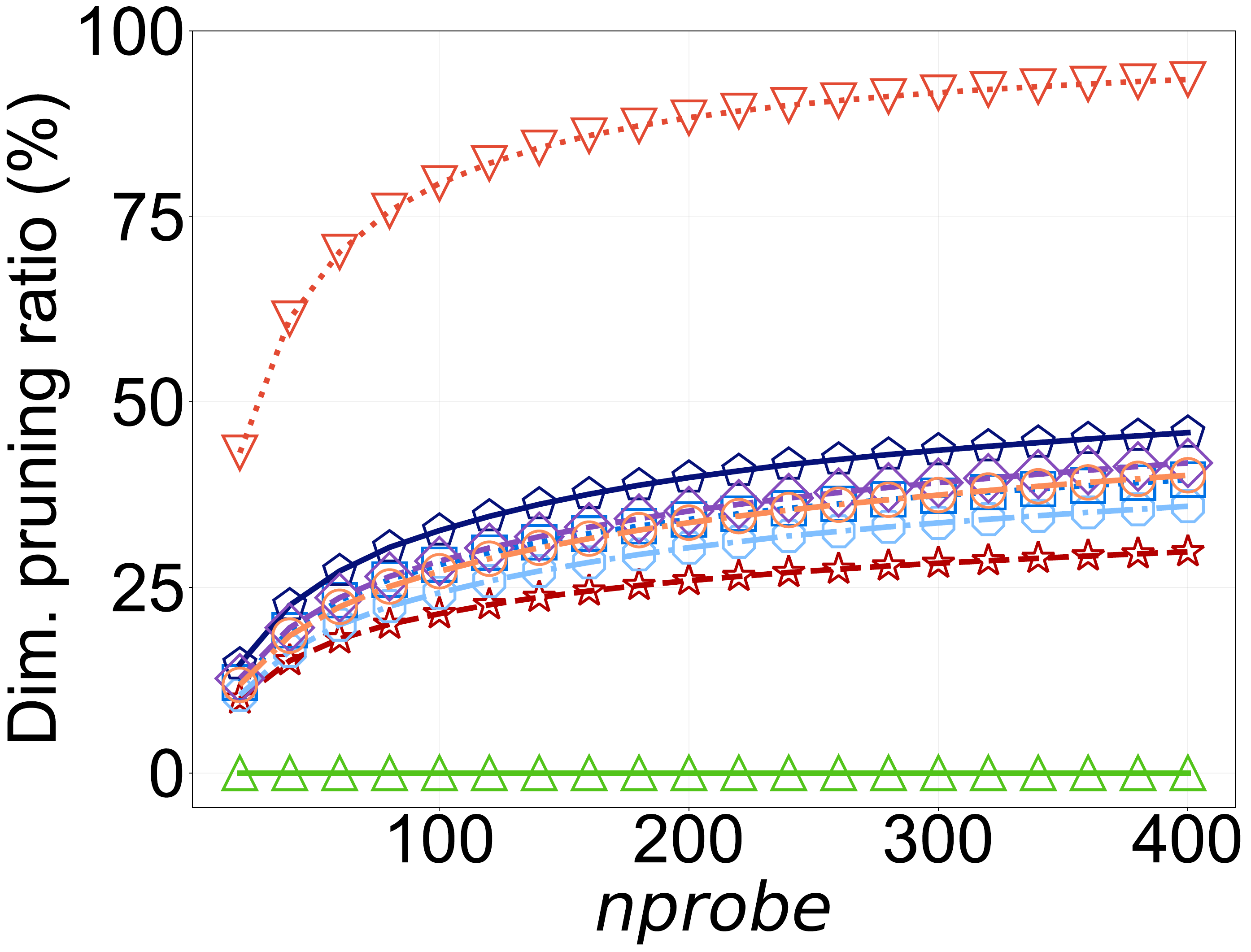}
        \caption{\Glove}\label{fig:pruning-ivf-glove}
    \end{subfigure}
    \begin{subfigure}{0.24\textwidth}
        \centering
        \includegraphics[width=\textwidth]{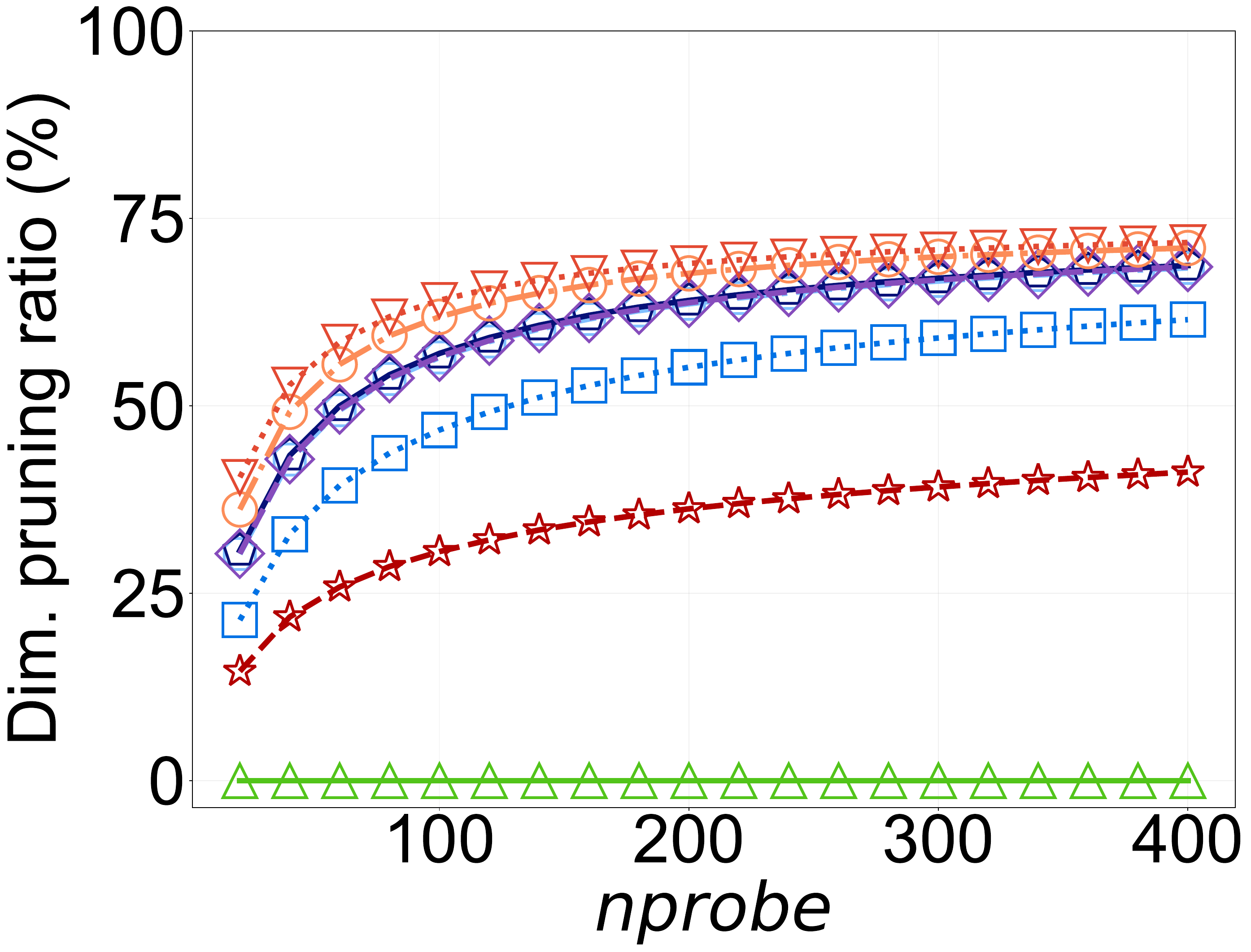}
        \caption{\Sift}\label{fig:pruning-ivf-sift}
    \end{subfigure}
    \begin{subfigure}{0.24\textwidth}
        \centering
        \includegraphics[width=\textwidth]{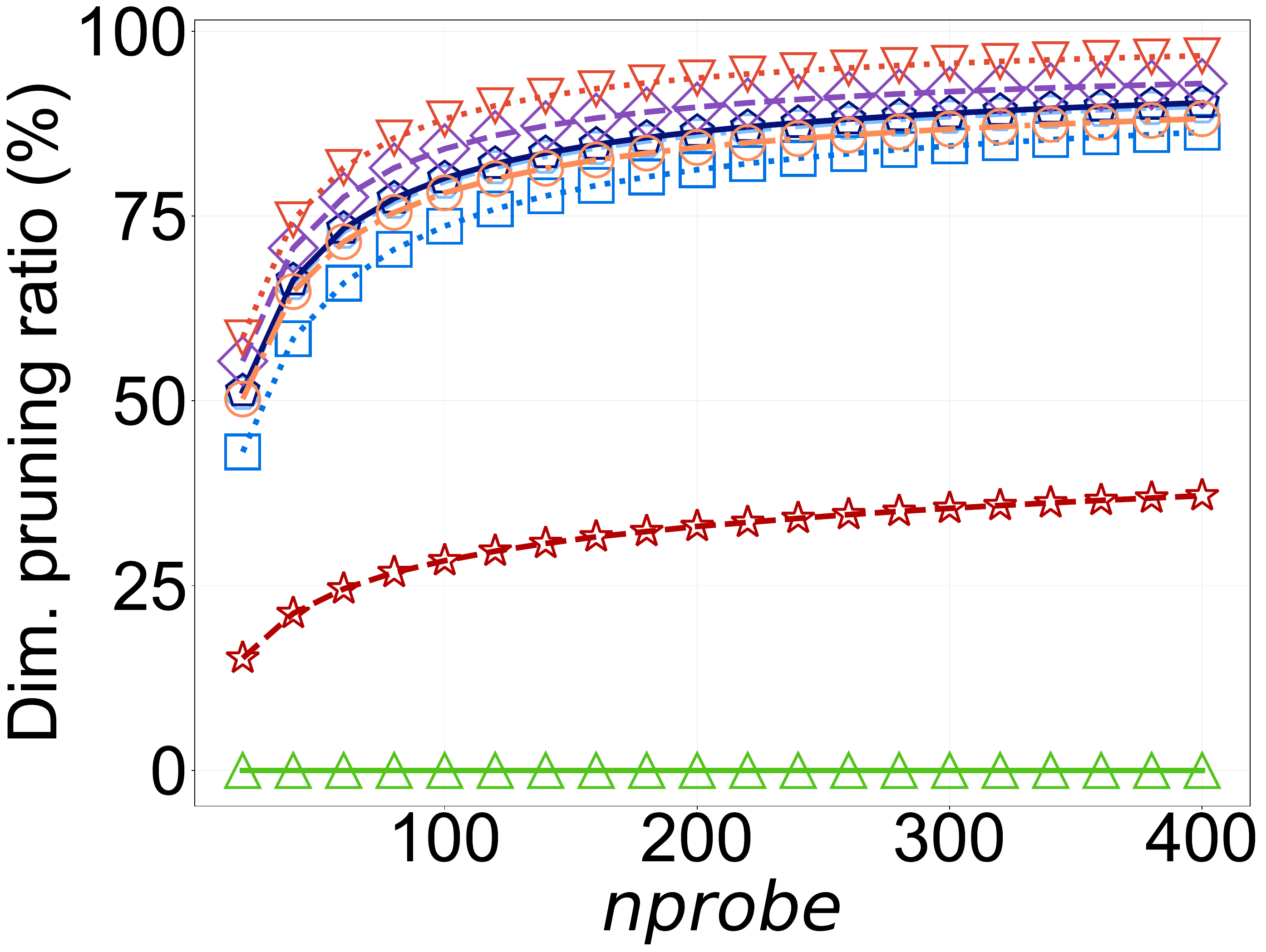}
        \caption{\Gist}\label{fig:pruning-ivf-gist}
    \end{subfigure}
    \begin{subfigure}{0.24\textwidth}
        \centering
        \includegraphics[width=\textwidth]{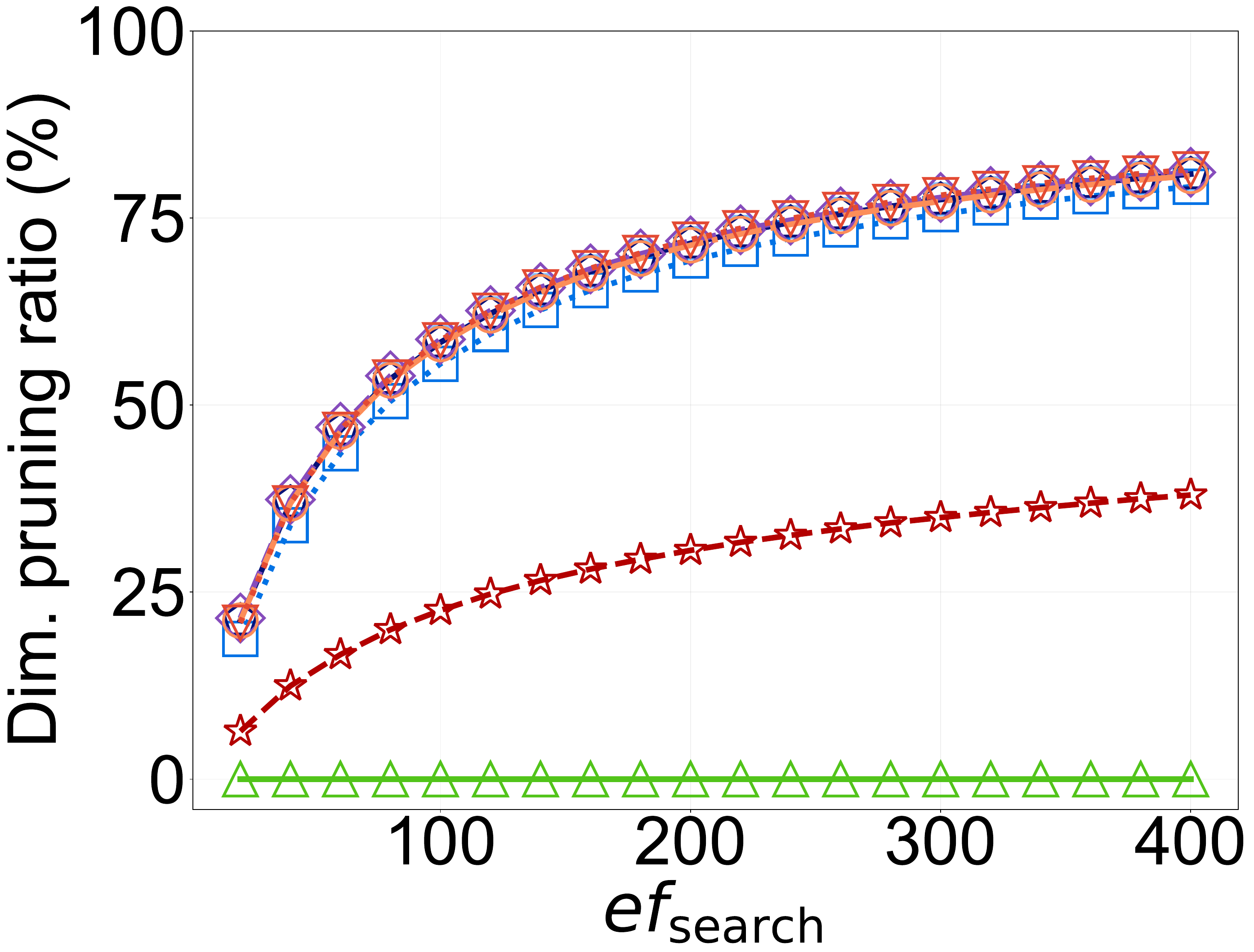}
        \caption{\Trevi}\label{fig:pruning-ivf-trevi}
    \end{subfigure}
    
    \begin{subfigure}{0.24\textwidth}
        \centering
        \includegraphics[width=\textwidth]{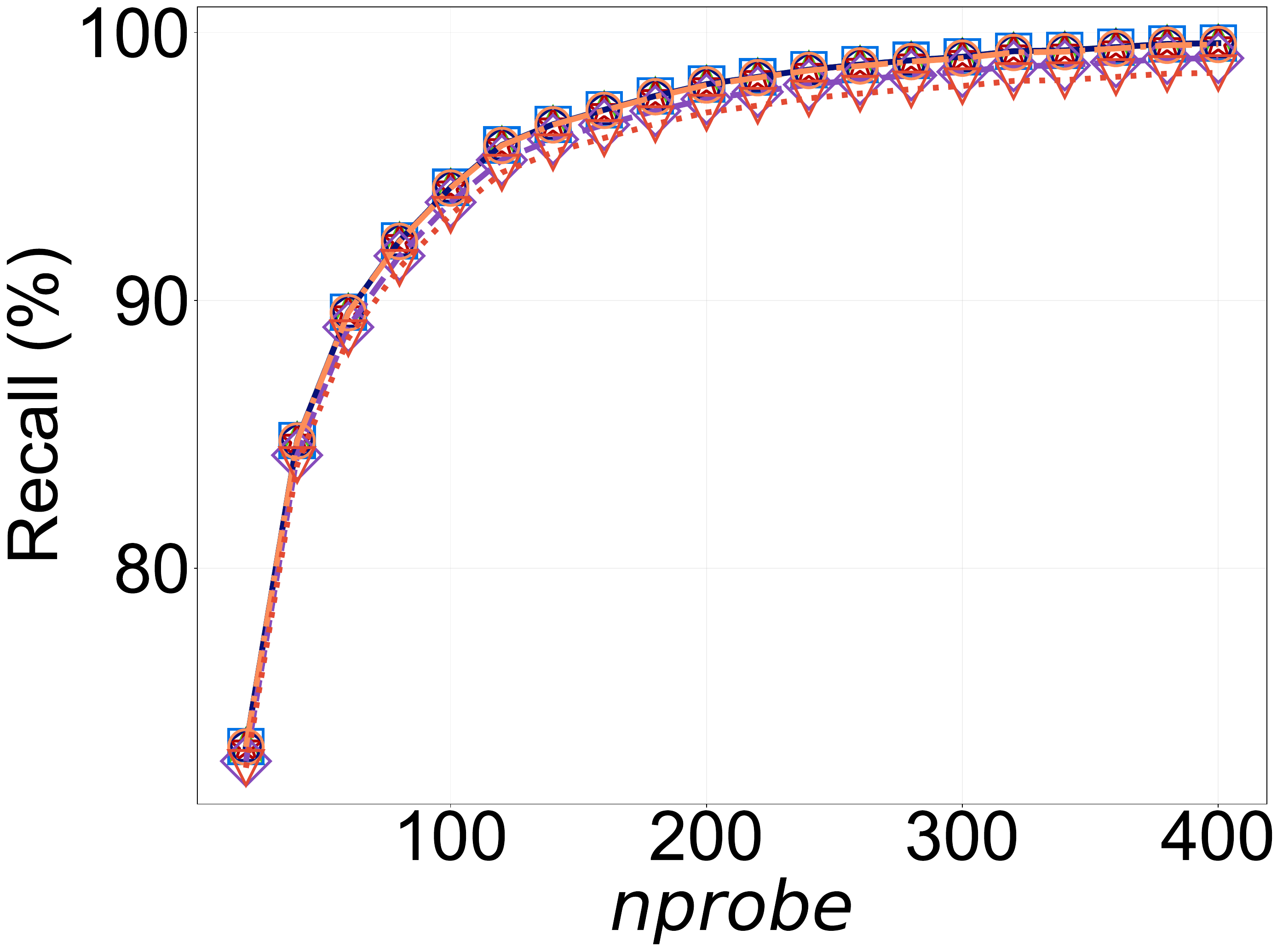}
        \caption{\Glove}\label{fig:pruning-ivf-glove-recall}
    \end{subfigure}
    \begin{subfigure}{0.24\textwidth}
        \centering
        \includegraphics[width=\textwidth]{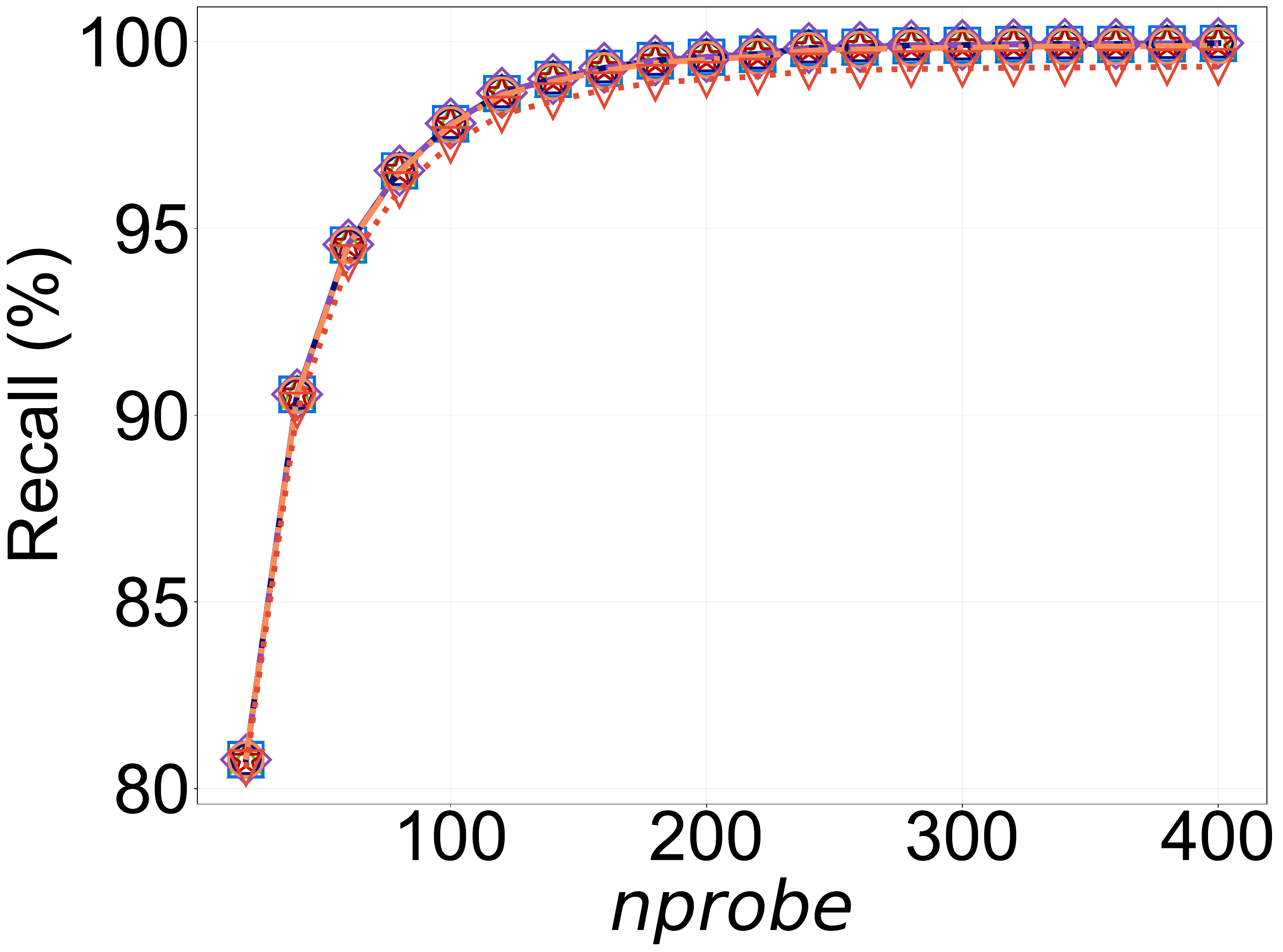}
        \caption{\Sift}\label{fig:pruning-ivf-sift-recall}
    \end{subfigure}
    \begin{subfigure}{0.24\textwidth}
        \centering
        \includegraphics[width=\textwidth]{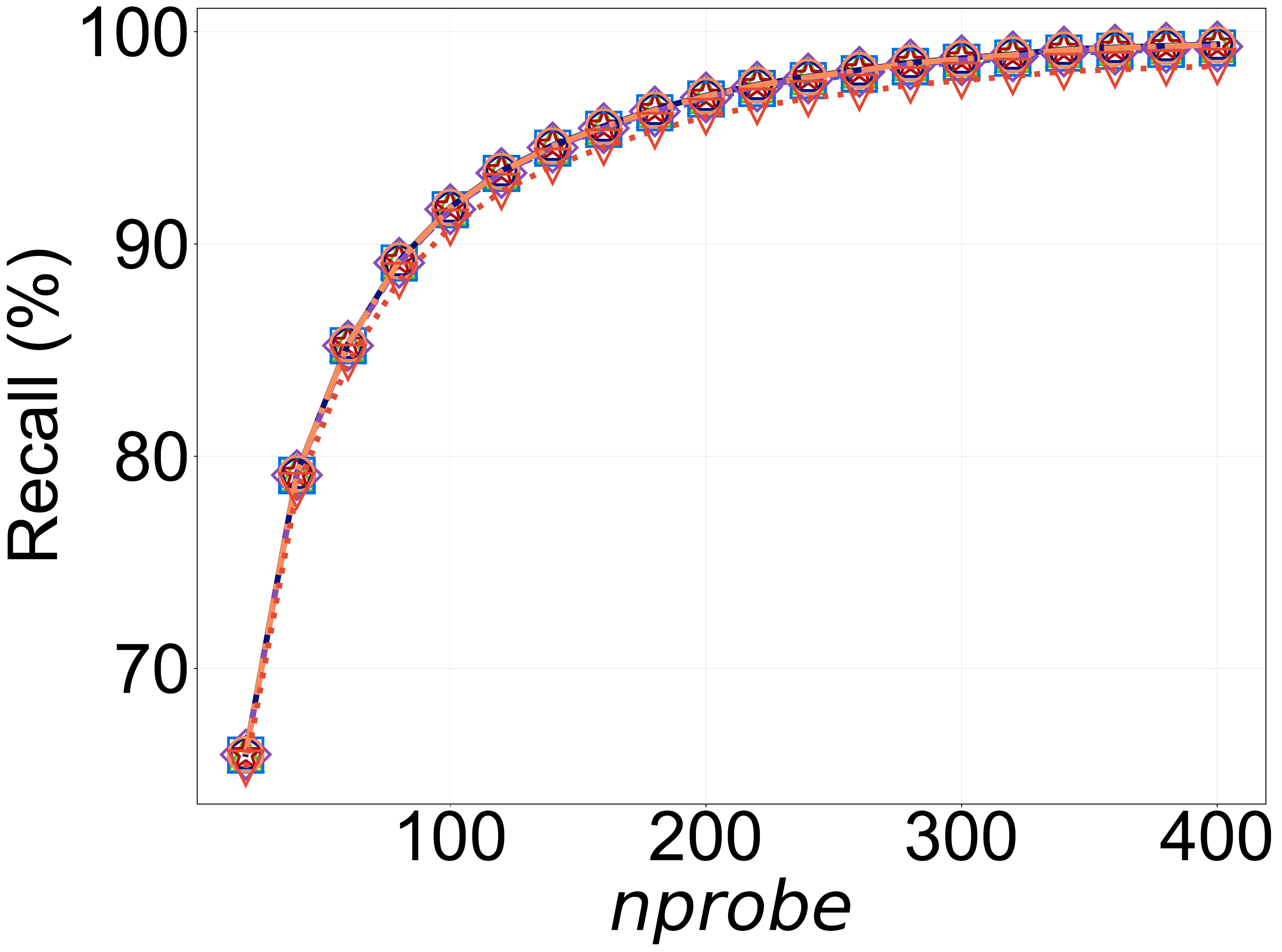}
        \caption{\Gist}\label{fig:pruning-ivf-gist-recall}
    \end{subfigure}
    \begin{subfigure}{0.24\textwidth}
        \centering
        \includegraphics[width=\textwidth]{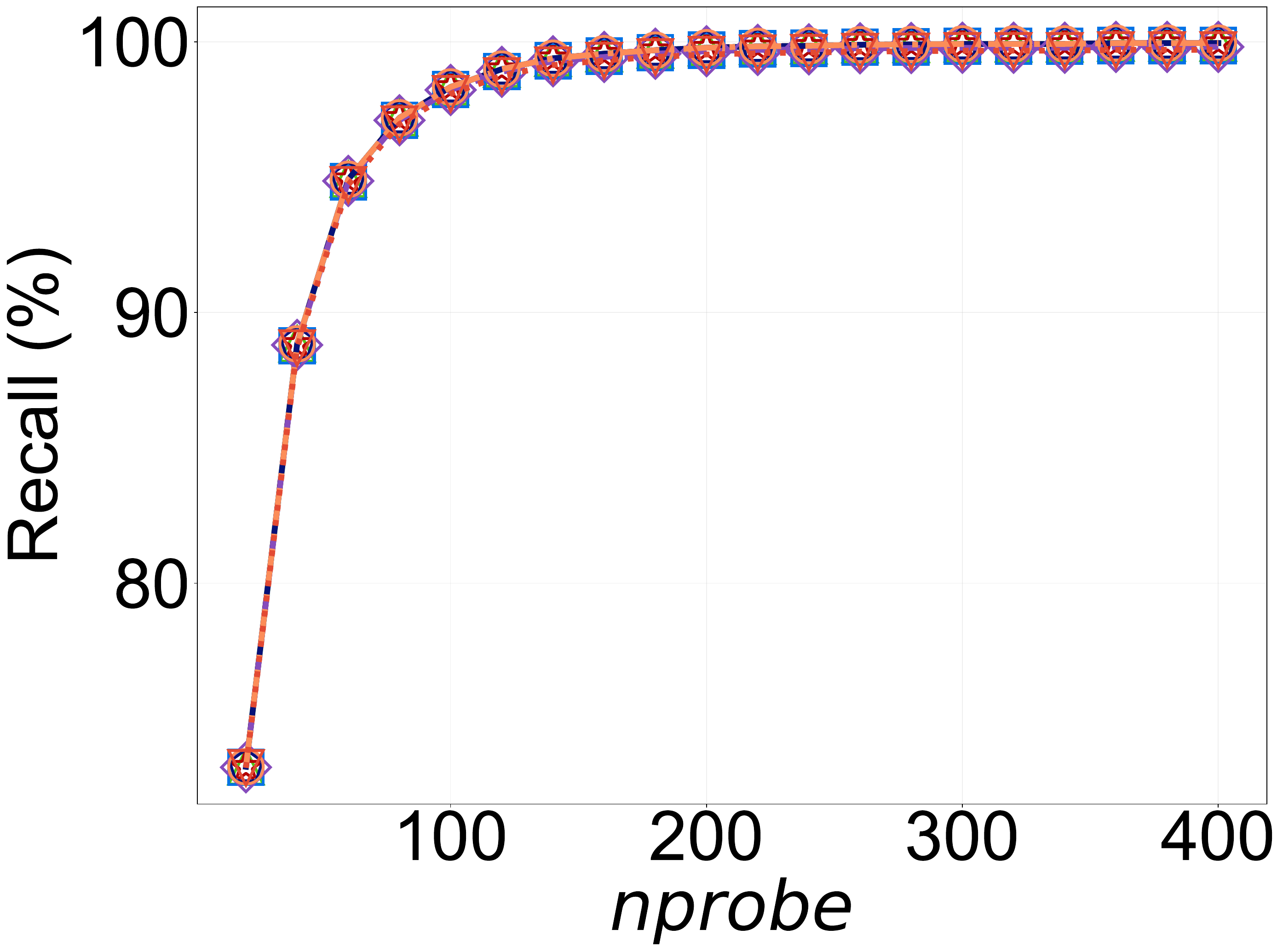}
        \caption{\Trevi}\label{fig:pruning-ivf-trevi-recall}
    \end{subfigure}
    \caption{Dimension pruning via DCO and its impact on recall on IVF}\label{fig:dco-pruning-ivf}
\end{figure*}

\section{Pruning Capability of DCO under the IVF Index}

We evaluate and analyze the capability of DCO on IVF. \figref{fig:dco-pruning-ivf} illustrates the dimension pruning ratios of DCOs and their recall across three datasets of varying dimensionality. 

\fakeparagraph{Pruning Capability of DCO on IVF} The results also demonstrate that the pruning effectiveness of DCOs is dimension-dependent: higher pruning ratios are consistently observed on high-dimensional and ultra-high-dimensional datasets. On these datasets, \DDCopq typically achieves the highest pruning ratio. However, it is worth noting that, in addition to dimension scanning, \DDCopq incurs additional overhead from computing quantized distances. Consequently, despite its strong pruning capability, \DDCopq exhibits poorer query performance on IVF when SIMD is disabled. When SIMD is enabled, thereby significantly accelerating quantized distance computation on IVF, the performance of \DDCopq aligns much more closely with its pruning capability. For other DCO methods, a higher pruning capability generally translates to better acceleration when SIMD is disabled. Moreover, all methods maintain comparable recall, demonstrating that the minor approximation errors introduced by DCOs do not compromise query accuracy on IVF.

\fakeparagraph{Pruning Capability Comparisons: HNSW vs IVF} 
%Additional results on dimension pruning ratios for HNSW are presented in \figref{fig:dco-pruning-hnsw}. 
Combined with the findings in \figref{fig:dco-pruning}, we observe that the pruning capability of DCOs remains relatively stable on HNSW, whereas it increases with $nprobe$ in IVF. 
This behavior can be attributed to the characteristics of candidate selection in each index structure.  
In HNSW, the retrieved candidates are typically close to the query vector.  
In contrast, IVF retrieves many candidates that are far from the query vector; for these, DCO methods can quickly determine irrelevance using only a few scanned dimensions. As a result, the pruning ratio in IVF rises with increasing $nprobe$.

% \begin{figure*}[h]
%     \centering
%     \includegraphics[width=0.9\textwidth]{figure/query_performance/legend.pdf}
%     \begin{subfigure}{0.24\textwidth}
%         \centering
%         \includegraphics[width=\textwidth]{figure/appendix/prunning/dimensionality-HNSW-SIFT.pdf}
%         \caption{\Sift}\label{fig:pruning-hnsw-sift}
%     \end{subfigure}
%     \begin{subfigure}{0.24\textwidth}
%         \centering
%         \includegraphics[width=\textwidth]{figure/appendix/prunning/dimensionality-HNSW-SIFT-recall.pdf}
%         \caption{\Sift}\label{fig:pruning-hnsw-sift-recall}
%     \end{subfigure}
%     \begin{subfigure}{0.24\textwidth}
%         \centering
%         \includegraphics[width=\textwidth]{figure/appendix/prunning/dimensionality-HNSW-OpenAI.pdf}
%         \caption{\Openai}\label{fig:pruning-hnsw-opeanai}
%     \end{subfigure}
%     \begin{subfigure}{0.24\textwidth}
%         \centering
%         \includegraphics[width=\textwidth]{figure/appendix/prunning/dimensionality-HNSW-OpenAI-recall.pdf}
%         \caption{\Openai}\label{fig:pruning-hnsw-opeanai-recall}
%     \end{subfigure}

%     \caption{Dimension pruning via DCO and its impact on recal on HNSW}\label{fig:dco-pruning-hnsw}
% \end{figure*}

\section{Parameter Study}
This experiment investigates two primary parameters of DCOs: 
the \textit{initial step size} $\Delta_0$ (\ie the initial number of dimensions to scan),
and the \textit{incremental step size} $\Delta_d$ used for each subsequent scan.
We systematically evaluate their effects on both HNSW and IVF indexes using the \Gist dataset.
Our study primarily reports QPS, as recall remains largely unaffected.
Notice that, prior work \cite{jayaram2019diskann, DBLP:journals/pvldb/WangXY021, DBLP:conf/nips/ZhangWCCZMHDMWP23} usually sets these parameters as $\Delta_d = \Delta_0 = 32$.

\fakeparagraph{Impact of Initial Step Size $\Delta_0$}
\figref{fig:initial-dimension} presents the QPS of each method on both HNSW and IVF indexes.
In \figref{fig:initial-dimension-hnsw}, the QPS of \Loop methods remains relatively stable, whereas \Hypothesis and \Class methods achieve a notable QPS increase by tuning $\Delta_0$ appropriately.
A similar trend is observed on the IVF index.
\ADSampling, \DADE, \DDCres, and \DDCpca improve QPS by up to 1.3, 1.5, 1.2, and 1.8$\times$ respectively, in \figref{fig:initial-dimension-ivf}.

The efficiency gain of tuning $\Delta_0$ is more pronounced on IVF than on HNSW.
This is because IVF stores each data partition contiguously in the main memory, a data layout that is cache-friendly.
In contrast, the graph-based index HNSW lacks such a compact data layout.
Consequently, even \RPDScanning achieves a considerable QPS improvement by adjusting $\Delta_0$ in \figref{fig:initial-dimension-ivf}.

\fakeparagraph{Impact of Incremental Step Size $\Delta_d$}
\figref{fig:sampling-step} reports the QPS when varying $\Delta_d$.
Most algorithms are highly sensitive to this parameter, with \FDScanning being the only exception. 
The optimal $\Delta_d$ is 64, 96 or 160 for HNSW, and 64 for IVF.
Compared to the default $\Delta_d = 32$ in prior work \cite{jayaram2019diskann, DBLP:journals/pvldb/WangXY021, DBLP:conf/nips/ZhangWCCZMHDMWP23}, \ADSampling, \DADE, and \DDCres achieve speedups of up to 1.1, 1.2, and 1.1$\times$.

Overall, jointly using the optimal parameter combinations, \ADSampling, \DADE, and \DDCres can boost QPS by up to 1.5, 1.8, and 1.8$\times$ compared to the settings in prior work.
\stepcounter{takeaway}
\vspace{-0.5ex}
\begin{tipbox}
\textbf{Takeaway: While tuning $\Delta_0$ and $\Delta_d$ can substantially boost QPS, their optimal values are determined by datasets, indexes, and specific DCO methods.}  
\end{tipbox}

\end{document}